\pdfoutput=1
\documentclass[usenatbib]{mn2e}
\usepackage{apjfonts}
\usepackage{amssymb}
\usepackage{amsmath}
\usepackage{ctable}
\usepackage{url}
\usepackage{fixltx2e} 
\usepackage[implicit=false,breaklinks,colorlinks,citecolor=blue]{hyperref}

\newcommand{\Hn}{\rm H0}
\newcommand{\Hp}{\rm HI}
\newcommand{\Hen}{\rm He0}
\newcommand{\Hep}{\rm HeI}
\newcommand{\Hepp}{\rm HeII}
\newcommand{\be}{\begin{equation}}
\newcommand{\ee}{\end{equation}}

\newcommand{\etal}{et al.}

\newcommand{\msun}{M_{\sun}}

\newcommand{\paperone}{Paper {\small II}}
\newcommand{\papertwo}{Paper {\small III}}

\newcommand{\rDMconv}{r_{\rm DM}^{\rm conv}}

\newcommand{\demofigrestart}{Fig.~\ref{fig:sf.z0.mass.resolution}}
\newcommand{\demofigcosmo}{Fig.~\ref{fig:demo}}

\newcommand{\ICsurl}{\href{http://www.tapir.caltech.edu/~phopkins/publicICs}{\url{http://www.tapir.caltech.edu/~phopkins/publicICs}}}
\newcommand{\FIREurl}{\href{http://fire.northwestern.edu}{\url{http://fire.northwestern.edu}}}
\newcommand{\gizmourl}{\href{http://www.tapir.caltech.edu/~phopkins/Site/GIZMO.html}{\url{http://www.tapir.caltech.edu/~phopkins/Site/GIZMO.html}}}
\newcommand{\movieurl}{\href{http://www.tapir.caltech.edu/~phopkins/Site/animations}{\url{http://www.tapir.caltech.edu/~phopkins/Site/animations}}}
\newcommand{\uvburl}{\href{http://galaxies.northwestern.edu/uvb}{\url{http://galaxies.northwestern.edu/uvb}}}
\newcommand{\coolingurl}{\href{http://www.strw.leidenuniv.nl/WSS08}{\url{http://www.strw.leidenuniv.nl/WSS08}}}

\newcommand{\apjl}{ApJL}
\newcommand{\apjs}{ApJS}

\newcommand{\aap}{A\&A}

\newcommand\plotonesize[2]
 {\centering \leavevmode \includegraphics[width={#2\columnwidth}]{#1}}
\newcommand{\plotsidesize}[2]
 {\centering \leavevmode \includegraphics[width={#2\textwidth}]{#1}}
\newcommand{\acknowledgments}{\begin{small}\section*{Acknowledgments}\end{small}}
\newcommand\altaffilmark[1]{$^{#1}$}
\newcommand\altaffiltext[1]{$^{#1}$}
\title[FIRE-2: Numerics vs.\ Physics]{FIRE-2 Simulations: Physics versus Numerics in Galaxy Formation
\vspace{-0.5cm}}

\vspace{-0.2cm}
\author[Hopkins \etal]{
\parbox[t]{\textwidth}{ 
Philip F.~Hopkins\thanks{E-mail:phopkins@caltech.edu}\altaffilmark{1},
Andrew Wetzel\altaffilmark{1,2,3}\thanks{Caltech-Carnegie Fellow},
Du\v{s}an Kere\v{s}\altaffilmark{4}, 
Claude-Andr{\'e} Faucher-Gigu{\`e}re\altaffilmark{5}, 
Eliot Quataert\altaffilmark{6}, 
Michael Boylan-Kolchin\altaffilmark{7}, 
Norman Murray\altaffilmark{8}, 
Christopher C.\ Hayward\altaffilmark{9},
Shea Garrison-Kimmel\altaffilmark{1},
Cameron Hummels\altaffilmark{1}, 
Robert Feldmann\altaffilmark{6,10},
Paul Torrey\altaffilmark{11},
Xiangcheng Ma\altaffilmark{1},
Daniel Angl\'es-Alc\'azar\altaffilmark{5},
Kung-Yi Su\altaffilmark{1},
Matthew Orr\altaffilmark{1},
Denise Schmitz\altaffilmark{1},
Ivanna Escala\altaffilmark{1},
Robyn Sanderson\altaffilmark{1},
Michael Y. Grudi{\'c}\altaffilmark{1},
Zachary Hafen\altaffilmark{5},
Ji-Hoon Kim\altaffilmark{12}, 
Alex Fitts\altaffilmark{7},
James S.~Bullock\altaffilmark{13}, 
Coral Wheeler\altaffilmark{1}, 
T.~K.\ Chan\altaffilmark{4}, 
Oliver D.~Elbert\altaffilmark{13}, 
Desika Narayanan\altaffilmark{14}
} 
\vspace*{6pt} \\
\altaffiltext{1}{TAPIR, Mailcode 350-17, California Institute of Technology, Pasadena, CA 91125, USA} \\
\altaffiltext{2}{Carnegie Observatories, Pasadena, CA 91101, USA} \\
\altaffiltext{3}{Department of Physics, University of California, Davis, CA 95616, USA} \\
\altaffiltext{4}{Department of Physics, Center for Astrophysics and Space Science, University of California at San Diego, 9500 Gilman Drive, La Jolla, CA 92093, USA} \\ 
\altaffiltext{5}{Department of Physics and Astronomy and CIERA, Northwestern University, 2145 Sheridan Road, Evanston, IL 60208, USA} \\ 
\altaffiltext{6}{Department of Astronomy and Theoretical Astrophysics Center, University of California Berkeley, Berkeley, CA 94720} \\
\altaffiltext{7}{Department of Astronomy, The University of Texas at Austin, 2515 Speedway, Stop C1400, Austin, TX 78712, USA} \\
\altaffiltext{8}{Canadian Institute for Theoretical Astrophysics, 60 St. George Street, University of Toronto, ON M5S 3H8, Canada}\\
\altaffiltext{9}{Center for Computational Astrophysics, Flatiron Institute, 162 Fifth Avenue, New York, NY 10010, USA}\\
\altaffiltext{10}{Institute for Computational Science, University of Zurich, Zurich CH-8057, Switzerland}\\
\altaffiltext{11}{Department of Physics, MIT, 77 Massachusetts Avenue, Cambridge, MA 02139, USA}\\ 
\altaffiltext{12}{Kavli Institute for Particle Astrophysics and Cosmology, Department of Physics, Stanford University, Stanford, CA, USA} \\
\altaffiltext{13}{Department of Physics and Astronomy, University of California, Irvine, CA 92697, USA}\\
\altaffiltext{14}{Department of Astronomy, University of Florida, Gainesville, FL 32611, USA} 
\vspace{-0.5cm}
}

\date{Submitted to MNRAS, February 2017\vspace{-0.6cm}}
\begin{document}
\maketitle
\label{firstpage}

\vspace{-0.2cm}
\begin{abstract}
\vspace{-0.2cm}

The Feedback In Realistic Environments (FIRE) project explores feedback in cosmological galaxy formation simulations. Previous FIRE simulations used an identical source code (``FIRE-1'') for consistency. Motivated by the development of more accurate numerics -- including hydrodynamic solvers, gravitational softening, and supernova coupling algorithms -- and exploration of new physics (e.g.\ magnetic fields), we introduce ``FIRE-2'', an updated numerical implementation of FIRE physics for the {\small GIZMO} code. We run a suite of simulations and compare against FIRE-1: overall, FIRE-2 improvements do not qualitatively change galaxy-scale properties. We pursue an extensive study of numerics versus physics. Details of the star-formation algorithm, cooling physics, and chemistry have weak effects, provided that we include metal-line cooling and star formation occurs at higher-than-mean densities. We present new resolution criteria for high-resolution galaxy simulations. Most galaxy-scale properties are robust to numerics we test, provided: (1) Toomre masses are resolved; (2) feedback coupling ensures conservation, and (3) individual supernovae are time-resolved. Stellar masses and profiles are most robust to resolution, followed by metal abundances and morphologies, followed by properties of winds and circum-galactic media (CGM). Central ($\sim$kpc) mass concentrations in massive ($>L_{\ast}$) galaxies are sensitive to numerics (via trapping/recycling of winds in hot halos). Multiple feedback mechanisms play key roles: supernovae regulate stellar masses/winds; stellar mass-loss fuels late star formation; radiative feedback suppresses accretion onto dwarfs and instantaneous star formation in disks. We provide all initial conditions and numerical algorithms used.

\end{abstract}

\begin{keywords}
galaxies: formation --- galaxies: evolution --- galaxies: active --- 
stars: formation --- cosmology: theory --- methods: numerical\vspace{-0.5cm}
\end{keywords}

\vspace{-1.1cm}
\section{Introduction}
\label{sec:intro}

\begin{figure*}
\begin{tabular}{cc}
\hspace{-0.27cm}
\includegraphics[width=0.5\textwidth]{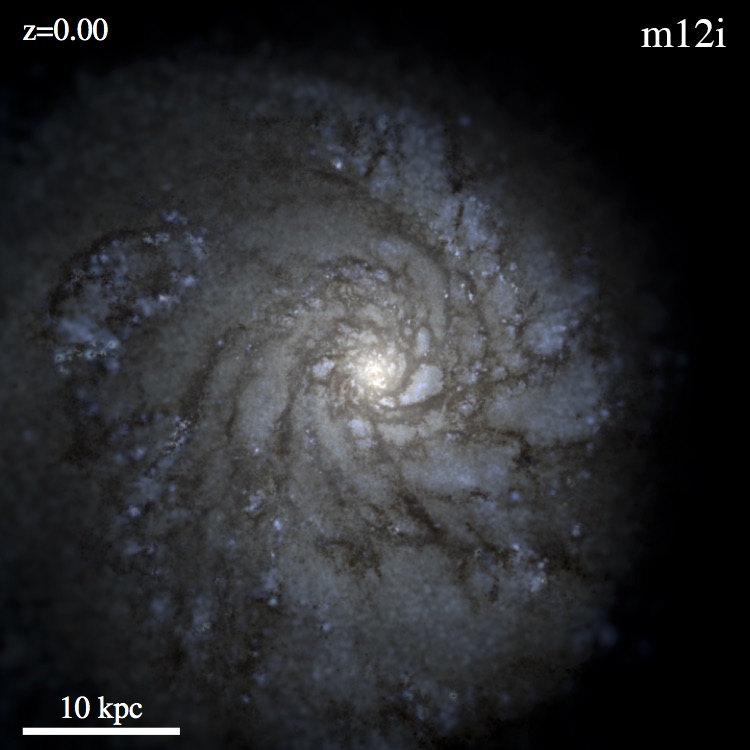} &
\hspace{-0.45cm}
\includegraphics[width=0.5\textwidth]{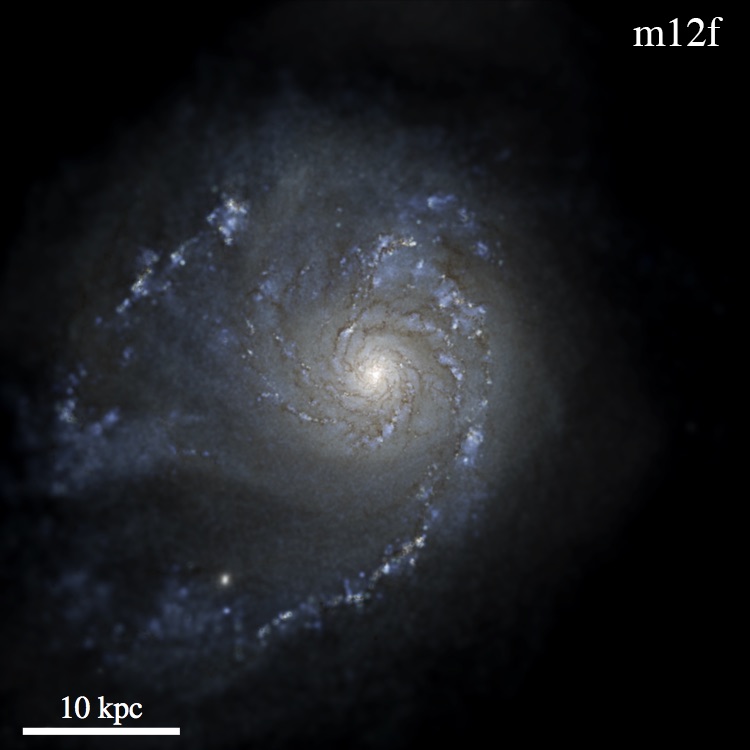} \\
\hspace{-0.27cm}
\includegraphics[width=0.5\textwidth]{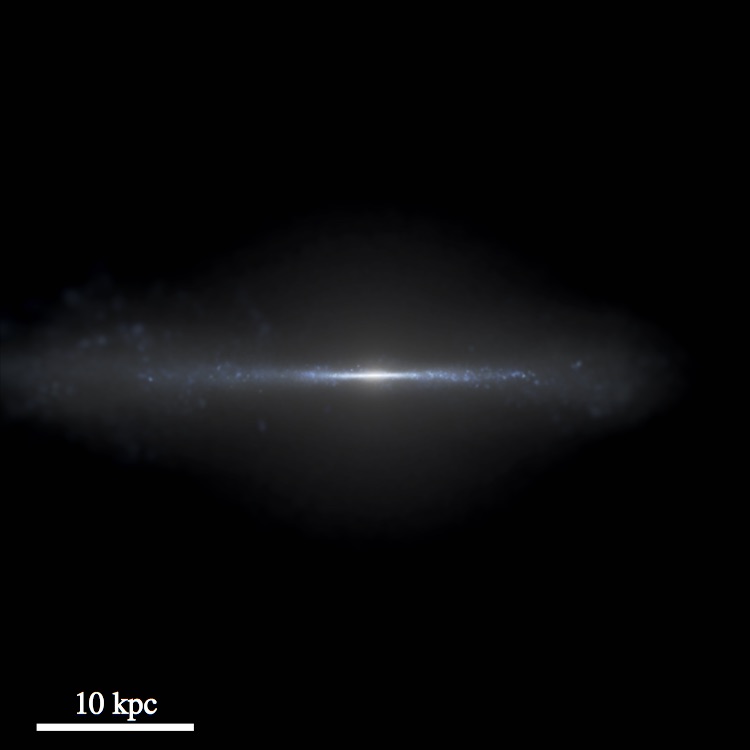} &
\hspace{-0.45cm}
\includegraphics[width=0.5\textwidth]{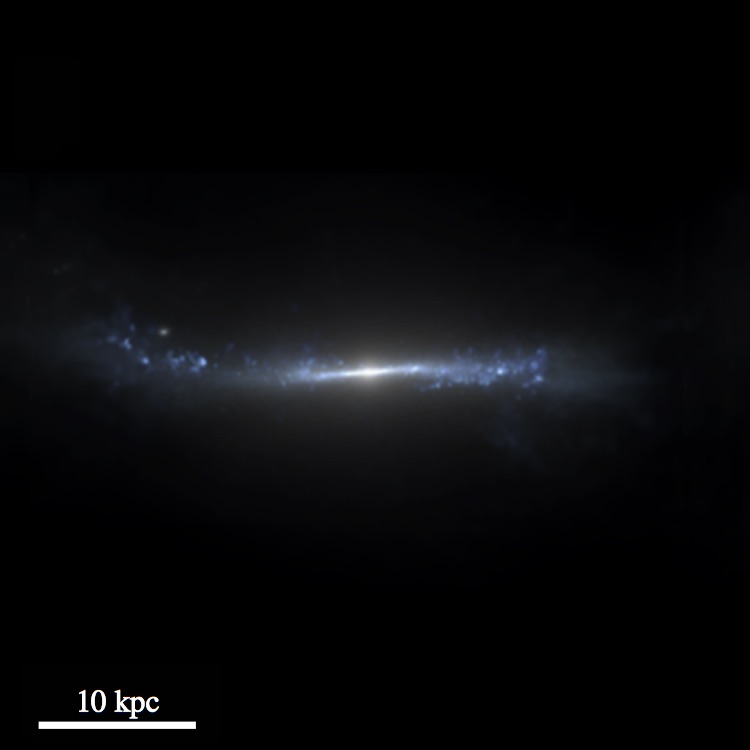} 
\end{tabular}
    \vspace{-0.25cm}
    \caption{Mock images of two Milky Way (MW)-mass galaxies at $z=0$ simulated using FIRE-2: ({\bf m12i} and {\bf m12f}). Each image is a mock Hubble Space Telescope $u/g/r$ composite with a logarithmic stretch, using {\small STARBURST99} to determine the SED of each star based on its age and metallicity and ray-tracing following \citet{hopkins:lifetimes.letter} with attenuation using a MW-like reddening curve with a dust-to-metals ratio of $0.4$. We show face-on ({\em top}) and edge-on ({\em bottom}) images. Both form thin disks, with clear spiral structure, clear dust lanes, and visibly resolved star-forming regions. Properties of each galaxy are in Table~\ref{tbl:sims}. 
    \label{fig:images.m12}}
\end{figure*}

\begin{figure*}
\begin{tabular}{c}
\includegraphics[width={1.0\textwidth}]{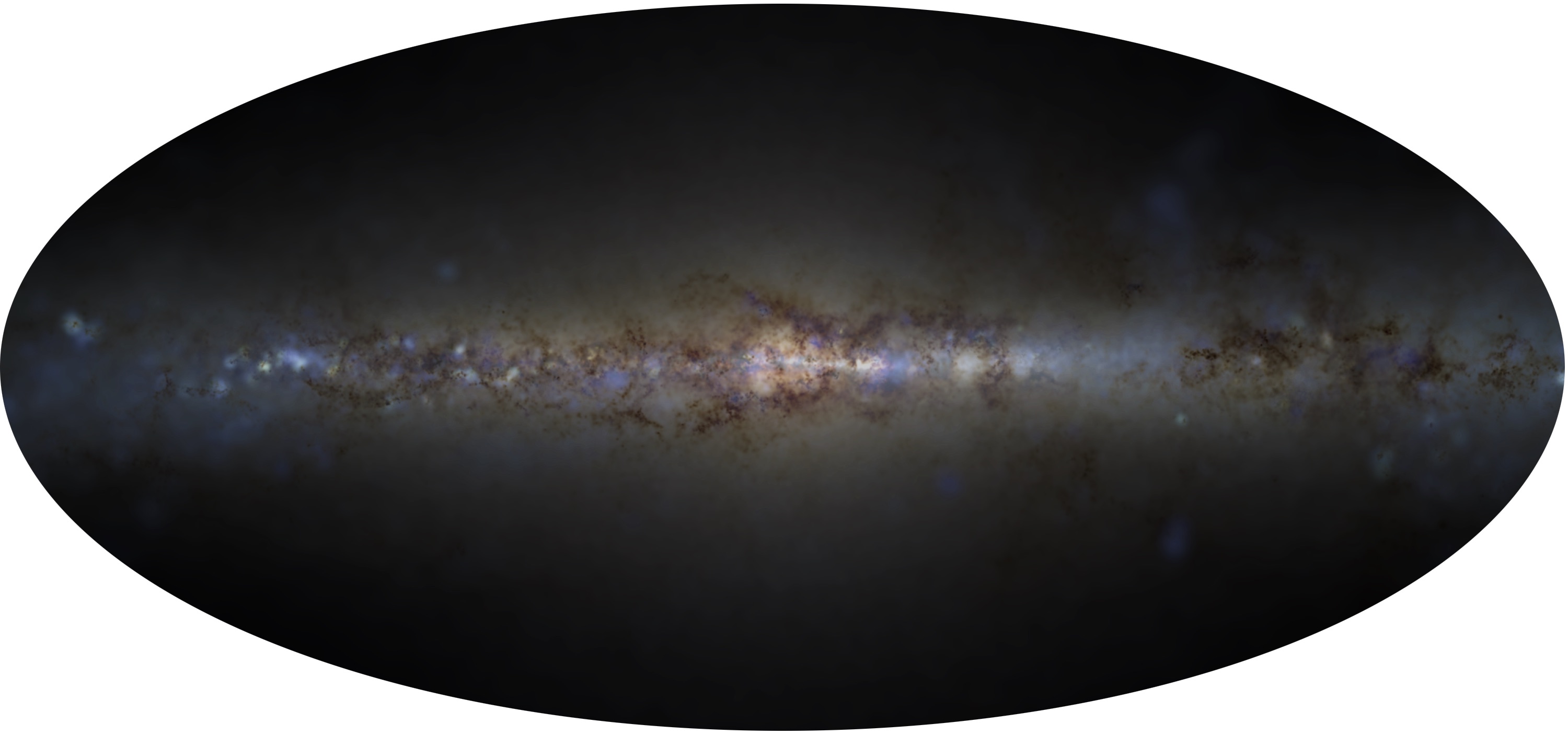}\\
\includegraphics[width={1.0\textwidth}]{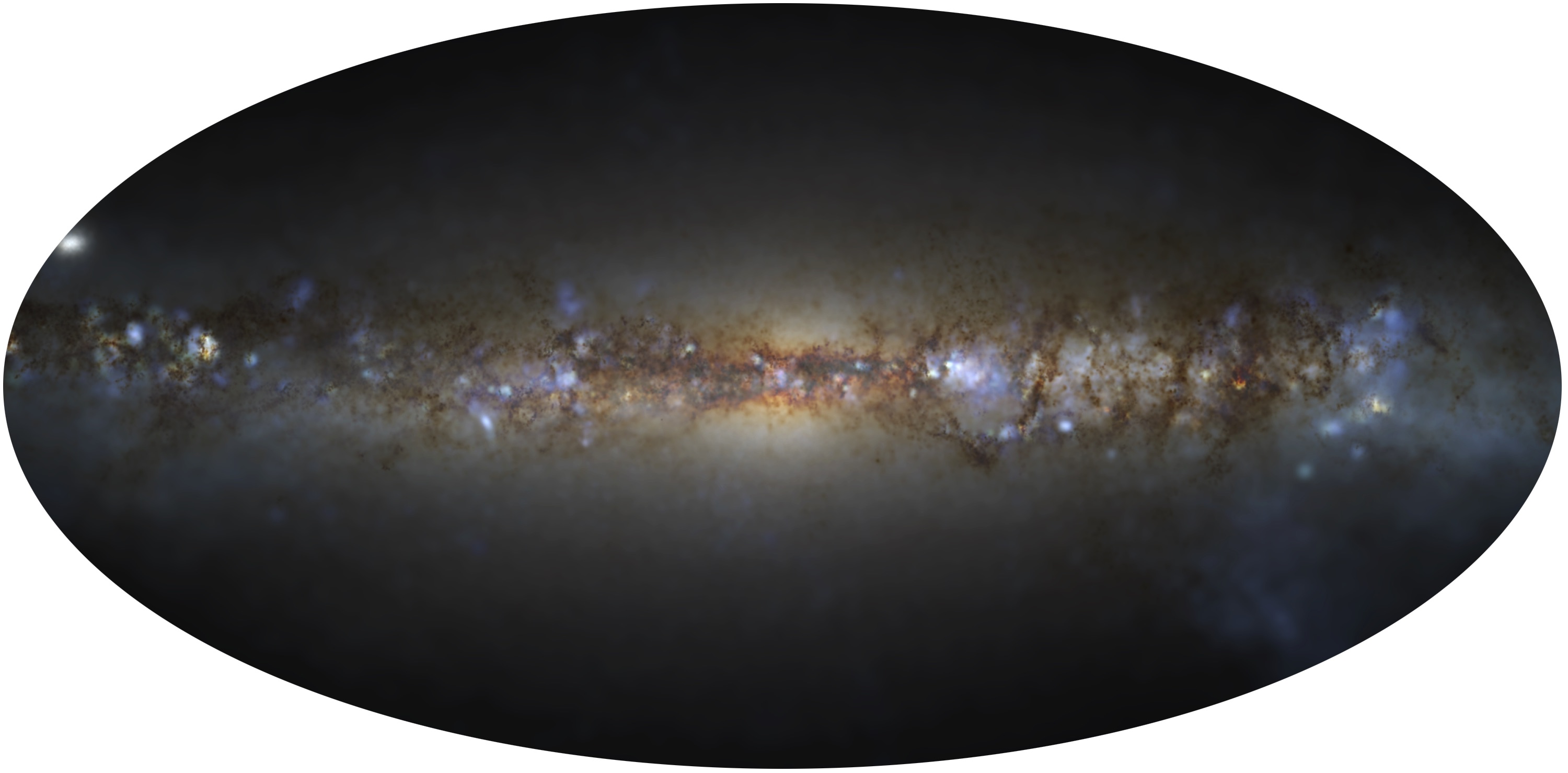}
\end{tabular}
    \vspace{-0.25cm}
    \caption{Mock galactic map, similar to Fig.~\ref{fig:images.m12}, but as seen from within the galaxy, for {\bf m12i} ({\em top}) and {\bf m12f} ({\em bottom}). We ray-trace a Galactic (Aitoff) projection, as seen from a random star $\sim 10\,$kpc from the galactic center. Individual, filamentary giant molecular cloud (GMC) complexes and young star clusters are visible, and both galaxies have a clear thin disk plus bulge morphology.
    \label{fig:images.fisheye.m12f}}
\end{figure*}

\begin{figure}
\begin{tabular}{cc}
\hspace{-0.25cm}
\includegraphics[width=0.5\columnwidth]{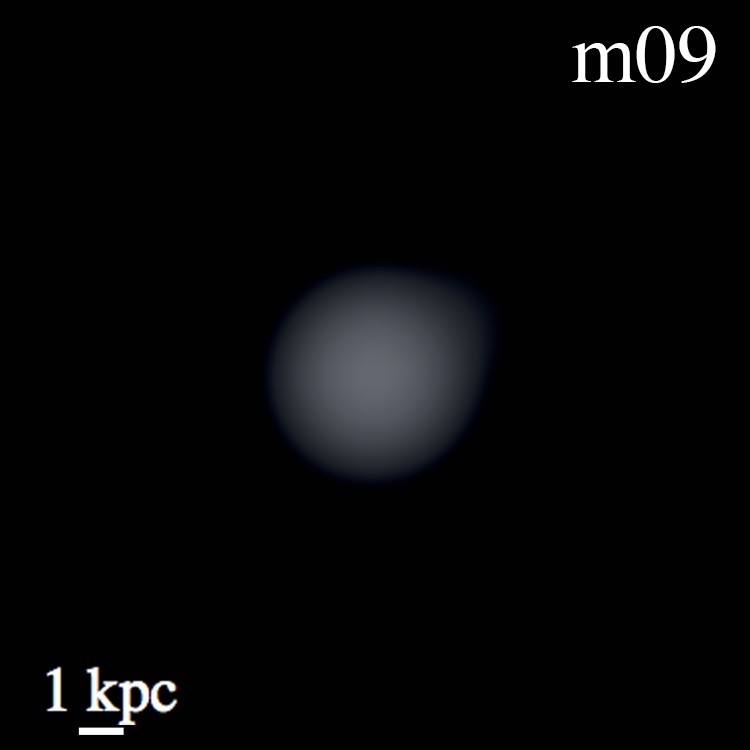} &
\hspace{-0.4cm}
\includegraphics[width=0.5\columnwidth]{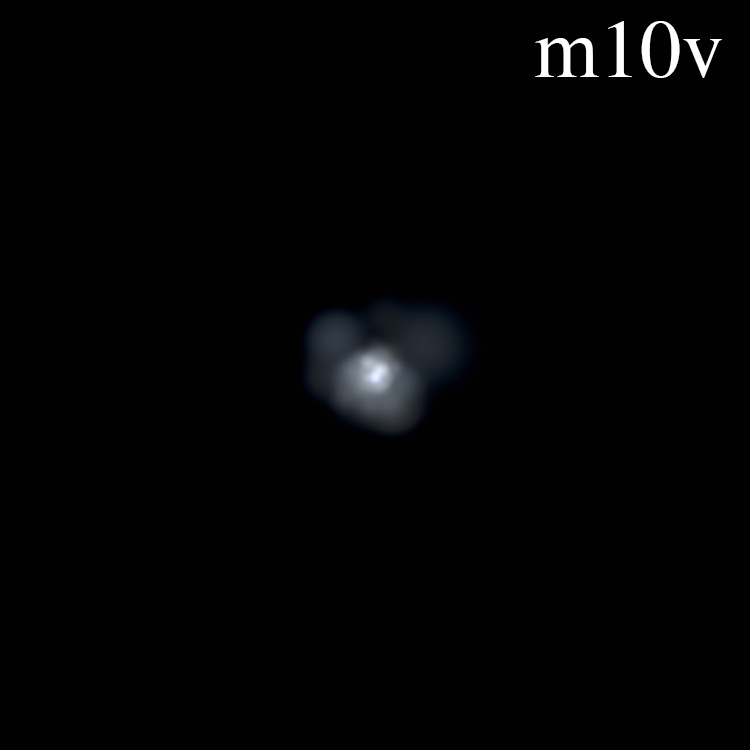} \\
\hspace{-0.25cm}
\includegraphics[width=0.5\columnwidth]{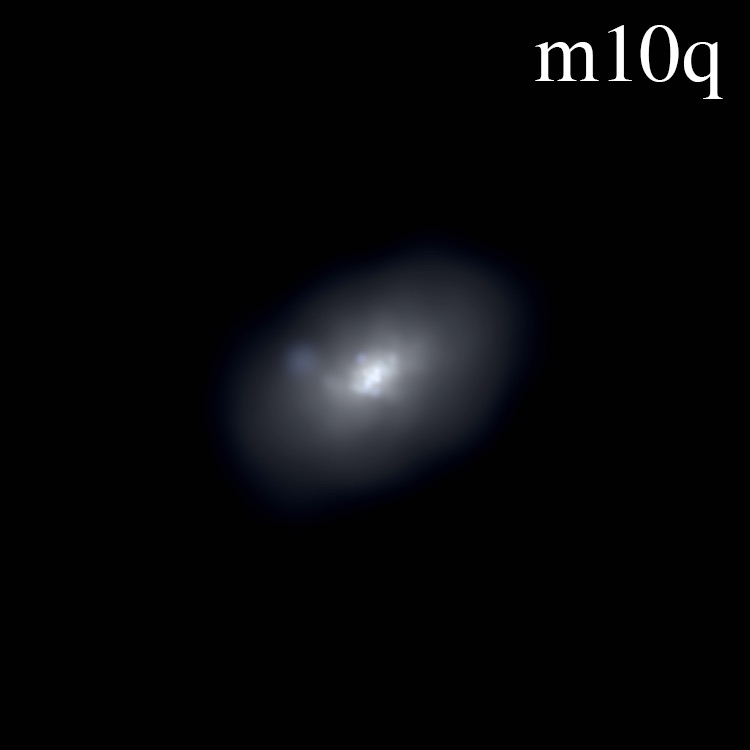} &
\hspace{-0.4cm}
\includegraphics[width=0.5\columnwidth]{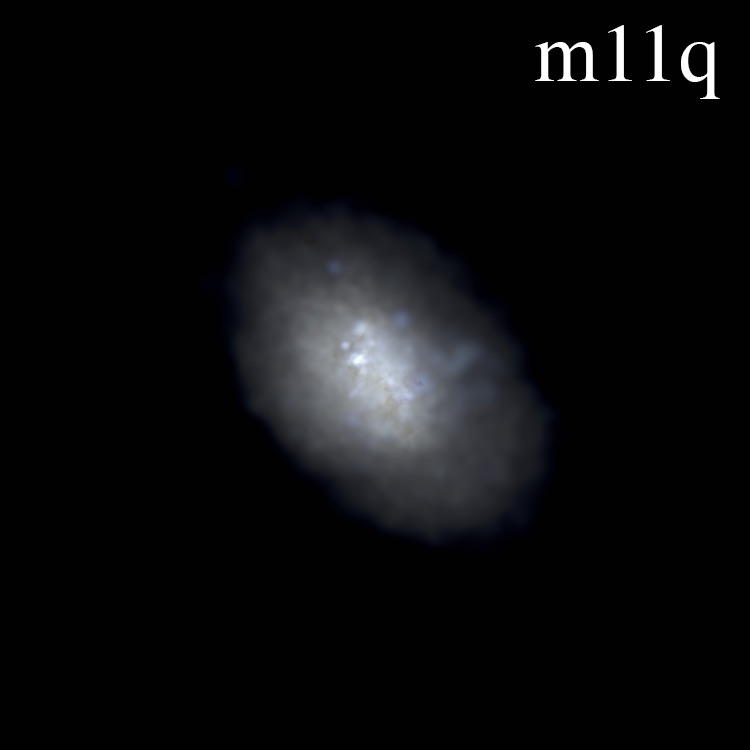} \\
\hspace{-0.25cm}
\includegraphics[width=0.5\columnwidth]{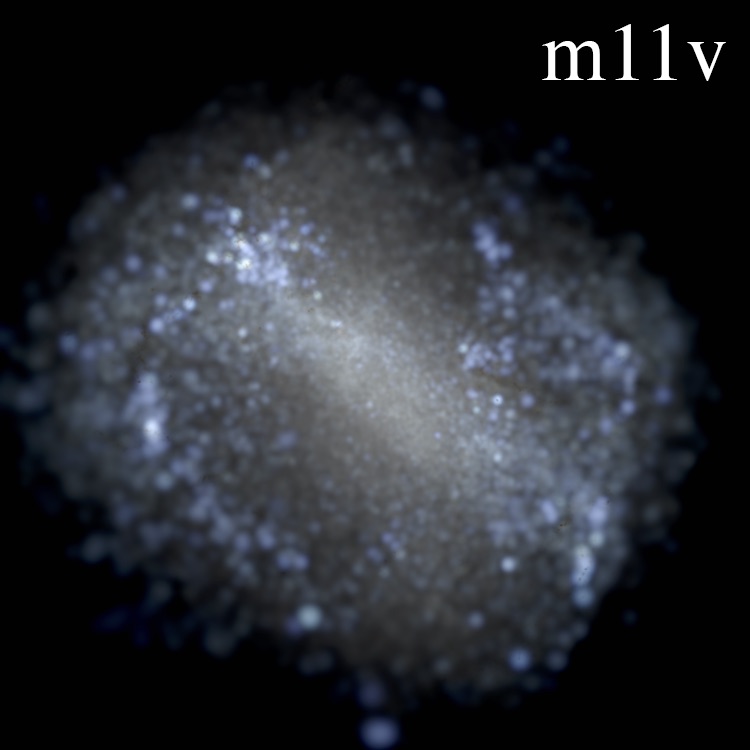} &
\hspace{-0.4cm}
\includegraphics[width=0.5\columnwidth]{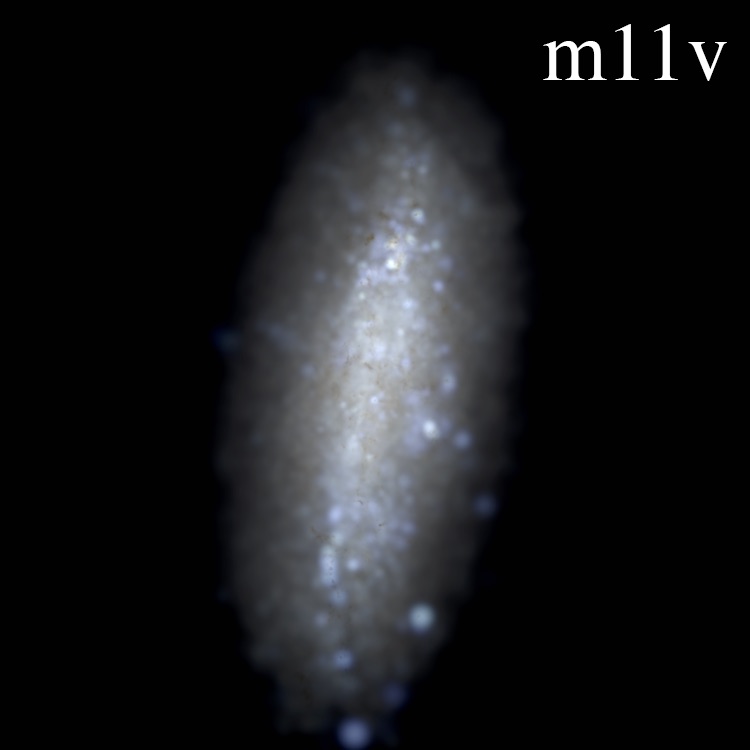} \\
\end{tabular}
    \vspace{-0.25cm}
    \caption{Mock images, as in Fig.~\ref{fig:images.m12}, but for a subset of dwarf galaxies in our sample: {\bf m09}, an ultra-faint with $M_{\ast}\sim 10^{4}\,\msun$ (similar to Coma Berenices, Leo IV, or Canes Venatici II); {\bf m10v}, a faint dwarf with $M_{\ast}\sim 10^{5}\,\msun$ (similar to Hercules or Leo T); {\bf m10q}, an intermediate-mass dwarf with $M_{\ast}\sim 10^{6}\,\msun$ (similar to Sextans, Carina, or Leo II); {\bf m11q}, an SMC-mass dwarf with $M_{\ast}\sim 10^{9}\,\msun$; and {\bf m11v}, an LMC-mass galaxy with $M_{\ast} \sim 2\times10^{9}\,\msun$. Most have spheroidal morphologies, as is observed and as was seen in FIRE-1 \citep{wheeler.2015:dwarfs.isolated.not.rotating}. We show, {\bf m11v}, the LMC-mass galaxy, both face-on and edge-on to illustrate the dramatic bar and elongated/flattened structure, similar to the actual LMC. Note that the surface-brightness scale is not the same in each image (an ultra-faint has $\sim 1000\times$ lower mean surface brightness than the LMC, so it would be invisible on the same scale).
    \label{fig:images.dwarfs}}
\end{figure}


\begin{footnotesize}
\ctable[
mincapwidth=\textwidth,
  caption={{\normalsize Initial suite of FIRE-2 simulations run to redshift $z=0$}\label{tbl:sims}},center,star,
  ]{lcccccccr}{
\tnote[ ]{Parameters describing the initial suite of FIRE-2 simulations in this paper. Each simulation contains several (in some, several dozen) galaxies in the high-resolution zoom-in region; halo and stellar properties listed refer only to the main ``target'' halo around which the high-resolution zoom-in region is centered. All properties refer to our highest-resolution simulation using each initial condition. All units are physical. \\
{\bf (1)} Name of simulation. \\
{\bf (2)} $M_{\rm halo}^{\rm vir}$: Virial mass \citep[following][]{bryan.norman:1998.mvir.definition} of the ``target'' halo at $z=0$ in simulation with baryons.\\ 
{\bf (3)} $R_{\rm vir}$: Virial radius (spherical) of the main halo at $z=0$. \\
{\bf (4)} $M_{\ast}$: Stellar mass (within $\le 3\,R_{1/2}$) of the central galaxy in the target halo at $z=0$. \\
{\bf (5)} $R_{1/2}$: Half-mass radius of stars in the central galaxy at $z=0$ (see \S~\ref{sec:results:overview}). \\
{\bf (6)} $m_{i,\,1000}$: the mass of baryonic (gas or star) particles, in units of $1000\,\msun$. Dark matter (DM) particle masses are $\approx 5\times$ larger, according to the universal baryon fraction. \\
{\bf (7)} $\epsilon_{\rm gas}^{\rm MIN}$: Minimum gravitational force softening reached by gas in the simulation. For gas, gravitational force resolution is {\em identical} to hydrodynamic spatial resolution (the same spatial gas distribution appears in gravity and hydrodynamic equations at all times). The gravitational force softening, $\epsilon_{i}$, therefore corresponds to the gas inter-particle separation, $h_{i}$: $\epsilon_{i}=h_{i}$. See \S~\ref{sec:resolution} for other definitions of ``spatial resolution''. Forces become exactly Keplerian (point-mass like) at $> 1.95\,\epsilon_{\rm gas}$ from a particle; the ``Plummer-equivalent'' softening is $\approx 0.7\,\epsilon_{\rm gas}$. \\
{\bf (8)} $\rDMconv$: Radius of convergence in DM properties (based on the \citealt{power:2003.nfw.models.convergence} criterion, with our calculation of where DM-only simulations converge from \S~\ref{sec:resolution}). This is approximately the radius enclosing $>200$ DM particles. We show below that convergence in DM profiles can in fact extend to much smaller radii in runs with baryons. \S~\ref{sec:resolution} shows that the DM force softening $\epsilon_{\rm DM}$ is much less important, as long as it is $\lesssim \rDMconv$. In our default runs, $\epsilon_{\rm DM}$ is fixed with values $=40\,{\rm pc}$ for our dwarfs and $=30\,{\rm pc}$ in our MW-mass (m12) systems. \\
{\bf (9)} Notes: Additional information on each simulation.
}
}{
\hline\hline
Simulation & $M_{\rm halo}^{\rm vir}$ & $R_{\rm vir}$ & $M_{\ast}$ & $R_{1/2}$ & $m_{i,\,1000}$ & $\epsilon_{\rm gas}^{\rm MIN}$ & $\rDMconv$ & Notes \\
Name \, & $[\msun]$ & $[{\rm kpc}]$ & $[\msun]$  & $[{\rm kpc}]$ & $[1000\,\msun]$ & $[{\rm pc}]$ & $[{\rm pc}]$ & \, \\ 
\hline 
\multicolumn{9}{c}{Ultra-Faints ($M_{\rm halo}\lesssim 10^{10}\,\msun$):} \\
\hline 
{\bf m09} & 2.4e9 & 35.6 & 9.4e3 & 0.29 & 0.25 & 1.1 & 65 & early-forming, ultra-faint field dwarf \\
%
%
{\bf m10a} & 7.0e9 & 51.5 & -- & -- & 0.50 & -- & 88 & very late-forming, dark halo (no stars) \\
%
\hline 
\multicolumn{9}{c}{Low-Mass Dwarf ($M_{\rm halo}\sim 10^{10}\,\msun$) Survey:} \\
\hline 
{\bf m10q} & 8.0e9 & 52.4 & 1.8e6 & 0.63 & 0.25 & 0.52 & 73 & isolated dwarf, early-forming halo \\
%
%
{\bf m10g} & 8.0e9 & 52.4 & 5.3e6 & 0.89 & 0.50 & 2.0 & 110 & early-forming \\
%
{\bf m10v} & 8.3e9 & 53.1 & 1.0e5 & 0.31 & 0.25 & 0.73 & 65 & isolated dwarf, late-forming halo \\
%
{\bf m10d} & 8.5e9 & 53.6 & 1.3e6 & 0.50 & 0.50 & 2.0 & 77 & intermediate-forming \\
%
%
{\bf m10c} & 9.0e9 & 54.6 & 4.7e5 & 0.30 & 0.50 & 2.0 & 92 & late-forming \\
%
{\bf m10b} & 9.4e9 & 55.4 & 4.5e5 & 0.33 & 0.50 & 2.0 & 77 & late-forming \\
%
{\bf m10e} & 1.0e10 & 57.1 & 1.8e6 & 0.58 & 0.50 & 2.0 & 120 & late-forming \\
%
{\bf m10i} & 1.1e10 & 57.8 & 7.1e6 & 0.52 & 0.50 & 2.0 & 75 & early-forming \\
%
{\bf m10l} & 1.1e10 & 57.8 & 1.2e7 & 0.72 & 0.50 & 2.0 & 110 & early-forming \\
%
{\bf m10j} & 1.1e10 & 58.5 & 8.4e6 & 0.70 & 0.50 & 2.0 & 86 & late-forming, dense environment \\
%
%
{\bf m10k} & 1.2e10 & 59.3 & 9.5e6 & 1.0 & 0.50 & 2.0 & 140 & early-forming \\
%
{\bf m10m} & 1.2e10 & 59.4 & 1.3e7 & 0.89 & 0.50 & 2.0 & 120 & early-forming \\
%
{\bf m10h} & 1.3e10 & 61.6 & 6.8e6 & 0.74 & 0.50 & 2.0 & 96 & intermediate-forming \\
%
{\bf m10f} & 1.3e10 & 62.3 & 1.1e7 & 1.1 & 0.50 & 2.0 & 150 & early-forming \\ 
%
{\bf m10y} & 1.4e10 & 63.9 & 1.0e7 & 0.96 & 0.26 & 0.21 & 74 & early-forming, large core \\
\hline 
\multicolumn{9}{c}{Intermediate-Mass Dwarfs ($10^{10}\,\msun \lesssim M_{\rm halo} \lesssim 10^{12}\,\msun$):} \\
\hline 
%
{\bf m10z} & 3.5e10 & 85.6 & 3.5e7 & 2.1 & 0.26 & 0.21 & 130 & ultra-diffuse galaxy \\
%
{\bf m11a} & 4.1e10 & 90.5 & 1.2e8 & 2.7 & 2.1 & 4.3 & 310 & diffuse, large core \\
%
{\bf m11b} & 4.3e10 & 92.2 & 1.1e8 & 2.4 & 2.1 & 2.9 & 250 & intermediate-forming \\
%
{\bf m11q} & 1.4e11 & 136 & 4.1e8 & 2.7 & 0.88 & 0.71 & 120 & early-forming, large core \\
{\bf m11c} & 1.4e11 & 138 & 8.1e8 & 2.7 & 2.1 & 0.40 & 250 & intermediate-forming \\
%
%
%
{\bf m11v} & 3.2e11 & 177 & 2.4e9 & 2.5 & 7.0 & 1.3 & 310 & multi-merger ongoing ($z=0$) \\
{\bf m11f} & 5.0e11 & 208 & 2.4e10 & 2.6 & 12 & 0.9 & 280 & quiescent late history \\
%
\hline
\multicolumn{9}{c}{Milky Way-Mass ``Latte'' ($M_{\rm halo} \sim 10^{12}\,\msun$) Halos:} \\
\hline
{\bf m12i} & 1.2e12 & 275 & 6.5e10 & 2.9 & 7.0 & 0.38 & 150 & ``Latte'' primary halo \\  
%
{\bf m12f} & 1.6e12 & 306 & 8.0e10 & 4.0 & 7.0 & 0.51 & 130 & MW-like halo \\ 
%
%
{\bf m12m} & 1.5e12 & 301 & 1.2e11 & 5.6 & 7.0 & 0.27 & 180 & earlier-forming halo, boxy bulge  \\ 
%
\hline
\multicolumn{9}{c}{``Low''-Resolution Milky Way-Mass Halo Survey:} \\
\hline
{\bf m12i\_LowRes} & 1.2e12 & 278 & 1.0e11 & 2.3 & 56 & 1.4 & 290 & Low-resolution ``Latte'' halo \\
%
{\bf m12f\_LowRes} & 1.6e12 & 310 & 1.3e11 & 3.1 & 56 & 1.4 & 310 & Low-resolution MW-like halo\\
%
{\bf m12b\_LowRes} & 1.4e12 & 291 & 9.8e10 & 1.5 & 56 & 1.4 & 300 & early-forming halo\\
%
{\bf m12c\_LowRes} & 1.3e12 & 285 & 9.2e10 & 1.6 & 56 & 1.4 & 310 & late-forming halo\\
%
{\bf m12m\_LowRes} & 1.5e12 & 302 & 1.4e11 & 5.0 & 56 & 1.4 & 360 & early-forming halo \\
%
%
{\bf m12q\_LowRes} & 1.6e12 & 308 & 1.2e11 & 1.9 & 56 & 1.4 & 240 & early-forming halo \\
%
{\bf m12z\_LowRes} & 8.7e11 & 251 & 4.3e10 & 6.0 & 33 & 8.0 & 520 & little/no bulge, merger at $z\approx0$ \\
%
{\bf m12\_ELVIS\_Robin} & 1.6e12 & 306 & 6.7e10 & 3.4 & 56 & 1.5 & 400 & late-forming, gas-rich in pair \\
{\bf m12\_ELVIS\_Batman} & 2.0e12 & 325 & 1.2e11 & 1.0 & 56 & 1.5 & 210 & compact, early-forming in pair \\
%
{\bf m12\_ELVIS\_Thelma} & 1.1e12 & 272 & 7.0e10 & 3.6 & 32 & 2.0 & 260 & MW-like in Local Group pair \\
{\bf m12\_ELVIS\_Louise} & 1.5e12 & 297 & 1.3e11 & 4.2 & 32 & 2.0 & 300 & M31-like in Local Group pair \\
%
{\bf m12\_ELVIS\_Romeo} & 1.3e12 & 285 & 8.1e10 & 6.5 & 28 & 1.0 & 280 & M31-like in Local Group pair \\
{\bf m12\_ELVIS\_Juliet} & 1.1e12 & 267 & 6.0e10 & 5.0 & 28 & 1.0 & 260 & MW-like in Local Group pair \\
\hline\hline
}
\end{footnotesize}

\begin{footnotesize}
\ctable[
  caption={{\normalsize Physics \&\ Numerics Explored in This Paper (and Papers {\small II} \&\ {\small III})}\label{tbl:summary}},center,star]{lclc}{
\tnote[ ]{A cursory outline of the physics and numerics explored in this paper. All ``standard'' FIRE-2 simulations, including all in Table~\ref{tbl:sims}, are run with the {\em identical} simulation code and physics. However, to understand how physical and numerical changes influence our results, we systematically ``turn off'' different physics and vary the numerical method and/or resolution in the sections listed here.
\\
{\bf (1)} Physics/Numerics: what we consider. \\
{\bf (2)} \S: Section where we pursue a detailed study of the effects of each numerics/physics on galaxy formation. \\
{\bf (3)} Effects in FIRE-2 Simulations: Overall summary of the effects of variation in the relevant physics or numerics, insofar as it is relevant (or not) for the predictions of our simulations. This applies only {\em for quantities discussed in this paper}, that is, global galaxy properties such as SFRs, stellar masses, sizes, and morphologies. For example, although we show that arbitrarily removing molecular chemistry from our cooling networks has no effect on predicted galaxy properties or star formation (because other cooling channels are available and molecular gas is primarily a tracer, not a causal driver of star formation), molecular chemistry is obviously fundamentally important if one wishes to predict molecular lines. Furthermore, we do not examine detailed properties of the CGM or IGM, where different physics may dominate. \\
{\bf (4)} Guidelines: Approximate ``rules of thumb'' for the relevant physics or numerics in the context of our ``zoom-in'' galaxy simulations. In the text, we provide more detailed guidelines. For example, for numerical resolution and other numerical parameters, we provide equations that approximately determine whether or not key physics should be resolved.
}
}{
\hline\hline
\multicolumn{1}{c}{Physics/Numerics} &
\multicolumn{1}{c}{\S} &
\multicolumn{1}{c}{Effects in FIRE-2 Simulations} &
\multicolumn{1}{c}{Guidelines / Default Choice} \\
\hline 
\hline
\multicolumn{4}{c}{Resolution:} \\
\hline 
\hline 
Mass Resolution & \ref{sec:resolution:mass} & Most results robust after resolving the Toomre scale, some (e.g.\ & 
Resolution criteria in \\
\, & \, & massive galaxy morphology) depend on resolved winds/hot gas  & \S~\ref{sec:resolution:mass:firephysics} (Eq.~\ref{eqn:mtoomre}-\ref{eqn:mbursty}) \\
\hline 
Collisionless (DM/Stellar) & \ref{sec:resolution:spatial} & Irrelevant unless {extremely} small or very large values used, & 
Optimal range of values  \\
\ \ Force Softening & \, & adaptive collisionless softenings require additional timestep limiters & in \S~\ref{sec:resolution:spatial:optimal} \\
\hline 
Gas Force Softening & \ref{sec:resolution:spatial} & Forcing fixed softening generally has no effect, unless too large, & 
Fully-adaptive softenings \\
\, & \, & then fragmentation \&\ SF are artificially suppressed  & (matching gas) should be used \\
\hline 
Timestep Criteria & \ref{sec:resolution:time} & Provided that standard stability criteria are met, this has no effect. & 
Standard limiters + Stellar (Eq.~\ref{eqn:stellar.dt.limiter}) \\
\, & \, & Additional limiters needed for stellar evolution \&\ adaptive softening  & + Adaptive softening (Eq.~\ref{eqn:dt.ags}) \\
\hline 
\hline 
\multicolumn{4}{c}{(Magneto)-Hydrodynamics:} \\
\hline 
\hline 
Hydro Method & \ref{sec:hydro} & Irrelevant for dwarfs. Important for massive galaxies with hot halos. & Newer methods recommended \\ 
\ \ (MFM vs.\ SPH) & \, & SPH may suppress cooling \&\ artificially allows clumpy winds to vent & \, \\ 
\hline 
Artificial Pressure & \ref{sec:hydro:artificial.pressure} & Unimportant unless set too large, then prevents real fragmentation.
& Do not use with \\
\ \ ``Floors'' & \, & Double-counts ``sub-grid'' treatment of fragmentation with SF model
& self-gravity based SF models \\
\hline
Magnetic Fields,  & \ref{sec:additional.physics.methods} & Weak effects on sub-galactic scales (dense gas, morphology, turbulent ISM) & See \citet{su:2016.weak.mhd.cond.visc.turbdiff.fx} \\ 
\ \ Conduction, Viscosity & \, & (Not studied here, but in \citealt{su:2016.weak.mhd.cond.visc.turbdiff.fx}; effects in CGM could be larger) & \, \\ 
\hline
Metal Diffusion  & \ref{sec:turbulent.diffusion.tests} \&\ \ref{sec:additional.physics.methods} & Small effects on galaxy properties \&\ dynamics, & Best practice depends \\
\ \ (sub-resolution mixing) & \, & but potentially important for abundance distributions of stars & \, on numerical hydro method \\
\hline 
\hline 
\multicolumn{4}{c}{Cooling:} \\
\hline 
\hline 
Molecular Chemistry/Cooling & \ref{sec:cooling} \&\ \ref{sec:cooling.approximations} & No effect on galaxy properties or star formation (just a tracer). & May be relevant at {\small [Z/H]}\,$\ll -3$, can be\\ 
\, & \, & Not important star formation criterion if fragmentation is resolved & important for observational tracers \\
\hline
Low-Temperature Cooling & \ref{sec:cooling} \&\ \ref{sec:cooling.approximations} & Details have no dynamical effects because $t_{\rm cool} \ll t_{\rm dyn}$ in cold gas & {\em Some} needed to form cold clouds, \\ 
\ \ ($T \ll 10^{4}\,$K) & \, &  to opacity limit ($\sim 0.01\,\msun$). Relevant for observables in cold phase & details dynamically irrelevant \\
\hline
Metal-Line Cooling & \ref{sec:cooling} \&\ \ref{sec:cooling.approximations} & Dominates cooling in metal-rich centers of ``hot halos'' around massive 
 & Needed: important in \\ 
\ \ ($T \gtrsim 10^{4}\,$K) & \, & galaxies, and of individual SNe blastwaves & super-bubbles \&\ ``hot halos''  \\  
\hline
Photo-Heating (Background) & \ref{sec:cooling} \&\ \ref{sec:cooling.approximations} & Significantly suppresses star formation in small 
($M_{\rm halo}\lesssim 10^{10}\,\msun$) dwarfs & Needed: dwarfs \&\ CGM/IGM \\ 
\hline
\hline 
\multicolumn{4}{c}{Star Formation:} \\
\hline 
\hline 
Self-Gravity (Virial) Criterion & \ref{sec:star.formation} \&\ \ref{sec:sf.algorithm} & Negligible effect on galaxy properties (SF is feedback-regulated). More & Recommended; \\
\, & \, & accurately identifies collapsing regions in high-dynamic range situations & see Appendix~\ref{sec:sf.algorithm} for implementation \\
\hline 
Density Threshold & \ref{sec:star.formation} \&\ \ref{sec:sf.algorithm} & Negligible effect on galaxy properties (SF is feedback-regulated) & Should exceed galactic mean density;  \\
\, & \, & Can be arbitrarily high with adaptive gas softenings & ideally, highest resolved densities \\
\hline 
Jeans-Instability Criterion & \ref{sec:star.formation} \&\ \ref{sec:sf.algorithm}  & Negligible effect on galaxy properties (SF is feedback-regulated). & Not necessary \\
\, & \, & Automatically satisfied in high-density, self-gravitating gas & \, \\
\hline 
Self-shielding/Molecular & \ref{sec:star.formation} \&\ \ref{sec:sf.algorithm}  & Negligible effect on galaxy properties (SF is feedback-regulated). & Not necessary \\
\ \ Criterion & \, & Automatically satisfied in high-density, self-gravitating gas & \, \\
\hline 
``Efficiency'' (Rate) & \ref{sec:star.formation} \&\ \ref{sec:sf.algorithm}  & Negligible effect on galaxy properties (SF is feedback-regulated). & $\sim 100\%$ per free-fall \\
\ \ at Resolution Limit & \, & If artificially lowered, more dense gas ``piles up'' until same SFR achieved & in {\em locally-self-gravitating} gas \\
\hline 
\hline 
\multicolumn{4}{c}{Stellar Feedback:} \\
\hline 
\hline 
Continuous Mass-Loss & \ref{sec:feedback} \&\ \ref{sec:stellar.evolution.approximations}  & Primarily important as a late-time fuel source for SF & Couple as Appendix~\ref{sec:mechanical.fb.implementation}. \\
\ \ (OB \&\ AGB) & \, & Relatively weak ``primary'' feedback effects on galactic scales & Rates given in \S~\ref{sec:stellar.evolution.approximations} \\
\hline
Supernovae (Ia \&\ II) & \ref{sec:stellar.evolution.approximations} \&\ \ref{sec:mechanical.fb.implementation} & 
Type-II: Dominant FB mechanism on cosmological scales. Need to account  & Couple as Appendix~\ref{sec:mechanical.fb.implementation}. \\ 
\ \ (``How to Couple'') & \paperone & 
for $PdV$ work if Sedov phase un-resolved. Subgrid models should reproduce & Validation \&\ convergence tests \\ 
\, & \, & exact solutions, conserve mass, energy, \&\ momentum, and converge & \&\ criteria in \paperone \\ 
\hline
Radiative Feedback & \ref{sec:stellar.evolution.approximations} \&\ \ref{sec:radiative.fb.implementation} & ``Smooths'' SF in dwarfs (less bursty) \&\ suppresses SF in dense gas.  & Need photo-heating \&\ single-  \\ 
\ \ (Photo-Heating \&\  & \papertwo & UV background dominates in dwarfs. Photo-electric heating unimportant. & scattering rad.\ pressure (\papertwo). \\ 
\ \ Radiation Pressure) & \, & IR multiple-scattering effects weak, except in massive galaxy nuclei. & Rad.-hydro algorithm sub-dominant \\
\hline\hline\\
}
\end{footnotesize}

Feedback from stars is an essential and still poorly-understood component of galaxy formation. In the absence of stellar feedback, most gas accreted into galaxies should cool rapidly on a timescale much shorter than the dynamical time, collapse, fragment, and turn into stars \citep{bournaud:2010.grav.turbulence.lmc,tasker:2011.photoion.heating.gmc.evol,hopkins:rad.pressure.sf.fb,dobbs:2011.why.gmcs.unbound,krumholz:2011.rhd.starcluster.sim,harper-clark:2011.gmc.sims}. Cosmologically, efficient cooling inevitably results in most of the baryons turning into stars, producing galaxies much more massive than observed \citep{katz:treesph,somerville99:sam,cole:durham.sam.initial,springel:lcdm.sfh,keres:fb.constraints.from.cosmo.sims}, regardless of the details of star formation in the simulation \citep{white:1991.galform,keres:fb.constraints.from.cosmo.sims}. 

However, the observed Kennicutt-Schmidt (KS) relation implies that gas consumption timescales are long ($\sim 50$ dynamical times; \citealt{kennicutt98}), and giant molecular clouds (GMCs) appear to turn just a few percent of their mass into stars before they are disrupted \citep{zuckerman:1974.gmc.constraints,williams:1997.gmc.prop,evans:1999.sf.gmc.review,evans:2009.sf.efficiencies.lifetimes}. Observed galaxy mass functions and the halo mass-galaxy mass relation require that galaxies incorporate or retain only a small fraction of the universal baryon fraction in stars and the ISM \citep{conroy:monotonic.hod,behroozi:mgal.mhalo.uncertainties,moster:stellar.vs.halo.mass.to.z1}. Observations of the intergalactic medium (IGM) and circum-galactic medium (CGM) require that many of those baryons must have been accreted into galaxies, enriched, and then expelled in galactic super-winds with mass loading $\dot{M}_{\rm wind}$ many times larger than the galaxy SFR \citep{aguirre:2001.igm.metal.evol.sims,pettini:2003.igm.metal.evol,songaila:2005.igm.metal.evol,martin:2010.metal.enriched.regions,oppenheimer:outflow.enrichment}, and indeed such winds are ubiquitously observed \citep{martin99:outflow.vs.m,martin06:outflow.extend.origin,heckman:superwind.abs.kinematics,newman:z2.clump.winds,sato:2009.ulirg.outflows,chen:2010.local.outflow.properties,steidel:2010.outflow.kinematics,coil:2011.postsb.winds}. 

Until recently, numerical simulations treated stellar feedback in highly-simplified fashion and have had difficulty reproducing these observations. This is especially true of models which invoke only energetic feedback (thermal injection) via supernovae (SNe), which typically found the energy was efficiently radiated away \citep{katz:1992.sne.fb.initial.implementation,guo:2010.hod.constraints,powell:2010.sne.fb.weak.winds,brook:2010.low.ang.mom.outflows,nagamine:2010.dwarf.gal.cosmo.review,bournaud10}. By ``turning off cooling'' for some adjusted duration, as in \citet{stinson:2006.sne.fb.recipe,governato:2010.dwarf.gal.form,maccio:2012.cuspcore.outflows,teyssier:2013.cuspcore.outflow,stinson:2013.new.early.stellar.fb.models,crain:eagle.sims}, or directly putting in winds ``by hand'' as in \citet{springel:multiphase,dave.2006.winds.at.reionization.enrichment,angles.alcazar:zoom.sims.with.winds.2014,vogelsberger:illustris.nature}, it is possible to reproduce some of the observed galaxy properties. But this obviously does not prove that known stellar feedback mechanisms actually act in this way, nor can it predict many ISM and CGM-scale properties that depend explicitly on e.g.\ the phase-structure of feedback-driven outflows \citep[see][]{hummels:2013.cgm.vs.obs}. 

Accurate treatment of star formation and galactic winds ultimately requires realistic treatment of the stellar feedback processes that maintain the multi-phase ISM. Observationally, many stellar feedback processes -- SNe, protostellar jets, photo-heating, stellar mass loss (O-star and AGB winds), and radiation pressure -- act efficiently on the ISM \citep[see][and references above]{evans:2009.sf.efficiencies.lifetimes,lopez:2010.stellar.fb.30.dor}. Simulations of either single molecular clouds/star clusters or the ``first stars,'' which resolve individual stars and can treat these microphysics in detail, have {\em universally} found that the non-linear interaction of these feedback mechanisms successfully suppresses star formation, pre-process giant molecular clouds before SNe explosions (so that SNe occur in rarified environments), and generate galactic chimneys and super-bubbles that generate fountains and super-winds \citep[e.g.][]{krumholz:2007.rhd.protostar.modes,krumholz:2011.rhd.starcluster.sim,offner:2009.rhd.lowmass.stars,offner:2011.rad.protostellar.outflows,harper-clark:2011.gmc.sims,bate:2012.rmhd.sims,wise:2012.rad.pressure.effects,pawlik:2013.rad.feedback.first.stars,muratov:2013.popIII.star.feedback}. A new generation of high-resolution galaxy-scale simulations has since emerged, which reach resolution sufficient to begin directly incorporating these physics, and to begin to resolve the multi-phase structure of the ISM \citep{tasker:2011.photoion.heating.gmc.evol,hopkins:rad.pressure.sf.fb,hopkins:fb.ism.prop,kannan:2013.early.fb.gives.good.highz.mgal.mhalo,agertz:2013.new.stellar.fb.model}. For example, in these works and a series of related papers focused on isolated galaxy simulations \citep{narayanan:2012.mw.x.factor,hopkins:dense.gas.tracers,hopkins:clumpy.disk.evol,hopkins:2013.accretion.doesnt.drive.turbulence,hopkins:virial.sf,hopkins:stellar.fb.mergers,hopkins:2013.merger.sb.fb.winds}, the authors showed in isolated galaxy simulations that the combination of multiple feedback processes together produce a quasi-steady state ISM, in which GMCs form and disperse rapidly, with turbulence, phase structure, GMC properties, a KS law, and galactic winds in reasonable agreement with observations.

Motivated by the success and predictive power of these simulations, in \citet{hopkins:2013.fire} we introduced the Feedback In Realistic Environments (FIRE) project.\footnote{\label{foot:movie}See the {\small FIRE} project website:\\
\FIREurl \\
For additional movies and images of FIRE simulations, see:\\
\movieurl}
The FIRE code synthesized the physics and numerical methods developed in the previous work (with relevant improvements) into a single code suitable for high-resolution cosmological simulations of galaxy formation. These simulations explicitly treat the multi-phase ISM with heating and cooling physics from gas at a range of temperatures $T\sim10-10^{10}\,$K, star formation restricted only to self-gravitating, self-shielding, molecular, high density ($n_{\rm H}\gtrsim 5-50\,{\rm cm^{-3}}$) gas, resolution reaching $\sim 250\,M_{\sun}$ or $\sim 0.5\,$pc, and (most importantly) explicit treatment of stellar feedback including the energy, momentum, mass, and metal fluxes from SNe Types Ia \&\ II, stellar mass-loss (O-star and AGB), radiation pressure (UV and IR), and photo-ionization and photo-electric heating. All stellar evolution and feedback inputs are taken directly from stellar evolution models, without subsequent ``parameter tuning.'' 

In a series of papers, we have subsequently shown that cosmological zoom-in simulations incorporating these physics can reproduce a diverse range of galaxy properties at a wide range of redshifts, including stellar masses, star formation histories (SFHs) and the galactic ``main sequence'' \citep{hopkins:2013.fire,sparre.2015:bursty.star.formation.main.sequence.fire, feldmann.2016:quiescent.massive.highz.galaxies.fire}; metallicities and metal abundance ratios in both ``standard'' and r-process elements \citep{ma:2015.fire.mass.metallicity,van-de-voort:2015.rprocess}; detailed morphological and kinematic structure of thin and thick disks \citep{ma:2016.disk.structure}; rotation curves and morphologies of Milky Way-mass galaxies \citep{wetzel.2016:latte}; observed satellite mass functions, rotation curves/kinematics, and cusp/core structure of dwarfs \citep{onorbe:2015.fire.cores,chan:fire.dwarf.cusps,wheeler:dwarf.satellites,wheeler.2015:dwarfs.isolated.not.rotating,wetzel.2016:latte}; abundance gradients \citep{elbadry.2015:core.transformation.stellar.kinematics.gradients.in.dwarfs,ma:radial.gradients}; neutral hydrogen absorption in the CGM \citep{faucher-giguere:2014.fire.neutral.hydrogen.absorption,faucher.2016:high.mass.qso.halo.covering.fraction.neutral.gas.fire,hafen:2016.lyman.limit.absorbers}; galactic outflows \citep{muratov:2015.fire.winds,muratov:2016.fire.metal.outflow.loading,angles.alcazar:particle.tracking.fire.baryon.cycle.intergalactic.transfer}; star-formation properties of galactic nuclei \citep{torrey.2016:fire.galactic.nuclei.star.formation.instability}; escape fractions of ionizing photons needed for reionization \citep{ma:2015.fire.escape.fractions,ma.2016:binary.star.escape.fraction.effects}; and at least some of the diversity of star-forming and quiescent massive galaxies at high redshifts \citep{narayanan.2015:smg.from.gas.accretion,feldmann.2016:quiescent.massive.highz.galaxies.fire}. There are of course a number of areas where the simulations fail to reproduce the observations: most notably, the bimodality of galaxy colors at both $z=0$ \citep{hopkins:2013.fire} and $z=2$ \citep{feldmann:colors.highz.quiescent.massivefire.gals} -- these are likely clues to important physics missing from the FIRE-1 simulations.

For the sake of consistency and clarity, all FIRE simulations have used an {\em identical} source code  -- what we will now refer to as ``FIRE-1.'' This ensured $100\%$ identical physics and numerical choices (up to the simulation resolution and choice of the specific halo simulated) in all runs, necessary for simulation comparisons. Unfortunately, this ignores development of new, more accurate hydrodynamic solvers and gravitational force softening algorithms \citep[see e.g.][]{hopkins:gizmo}, improvements to the numerical accuracy of feedback coupling algorithms (i.e.\ ways to ensure machine-accurate momentum conservation in SNe coupling to gas), code optimizations that would allow higher-resolution simulations, and physics neglected in FIRE-1 such as magnetic fields, cosmic rays, conduction, viscosity, optically thick radiative cooling, and more. These effects could, in principle, have large consequences for galaxy formation. For example, FIRE-1 used an improved version of the smoothed-particle hydrodynamics (SPH) method to solve the hydrodynamic equations; but it is well-known that SPH has certain low-order errors that do not converge accurately, add noise, and artificially suppress phenomena such as fluid mixing and sub-sonic turbulence \citep{agertz:2007.sph.grid.mixing,bauer:2011.sph.vs.arepo.shocks,sijacki:2011.gadget.arepo.hydro.tests}, potentially leading directly to large differences in cooling in massive galaxies \citep{keres:2011.arepo.gadget.disk.angmom,torrey:2011.arepo.disks}. There has been considerable effort to ``fix'' SPH, and FIRE-1 used the improved P-SPH methods from \citet{saitoh:2012.dens.indep.sph} and \citet{hopkins:lagrangian.pressure.sph} to reduce these errors, but some (e.g.\ the zeroth-order errors) cannot be entirely eliminated in SPH without de-stabilizing the method \citep[see][]{price:2012.sph.review}. As a result, especially for fluid mixing problems, newer moving-mesh or mesh-free Godunov methods provide still greater accuracy and more rapid convergence \citep[see][]{springel:arepo,hopkins:gizmo}.

We therefore introduce the ``FIRE-2'' simulations: an update of the FIRE physics modules in the code {\small GIZMO}. This includes a new, more accurate hydrodynamics solver that resolves the main known issues of SPH, as well as more accurate treatments of cooling and recombination rates, gravitational force softening, and numerical feedback coupling. In this paper we present a large suite of cosmological zoom-in simulations, and compare these to our FIRE-1 results and to some basic observed galaxy properties. We find that the qualitative results from the FIRE-1 simulations are reproduced in FIRE-2. We then use these simulations to extensively explore numerical and algorithmic choices in the simulation setup, and whether these have any effect on the predictions. Some first science results from these FIRE-2 simulations have been presented in \citet{wetzel.2016:latte,su:2016.weak.mhd.cond.visc.turbdiff.fx,fitts:fire.dwarf.concentration.mass}.

The goals of this paper are twofold. First, this is a methods and numerical/physical tests paper for the FIRE-2 simulations; we present the simulations, extensive tests of the methods, and explicitly detail all aspects of the numerical methods and algorithms. Second, we survey numerical and physical effects, e.g.: resolution (mass, spatial, and temporal), hydrodynamic solvers (SPH vs.\ modern Godunov methods), criteria for star formation, details of the cooling physics, and stellar feedback from radiation, winds, and SNe.  For each effect we present an extensive study in simulations to understand which effects are physical, and which numerical, and where our simulations should and should not be trusted. Because we will show that feedback is the most important property determining the galaxy's formation history, a pair of companion papers will separately explore the details of the numerical implementation and physics of mechanical/SNe feedback (\citealt{hopkins:sne.methods}, henceforth \paperone; this paper is Paper {\small I}) and radiative feedback (Hopkins et al., in prep., henceforth \papertwo).

Table~\ref{tbl:sims} presents the initial suite of FIRE-2 simulations studied here; Table~\ref{tbl:summary} provides an ``executive summary'' of our study and key conclusions; Table~\ref{tbl:res} provides a high-level overview of what ``numerical resolution'' actually means in our simulations. \S~\ref{sec:methods} summarizes our methods: we direct the reader to the appropriate appendices where the complete algorithmic details are presented in detail. \S~\ref{sec:results:overview} presents a basic overview of the resulting simulations and specifically examines any differences between FIRE-1 and FIRE-2 predictions. \S~\ref{sec:resolution} extensively studies the effects of resolution, in mass (\S~\ref{sec:resolution:mass}), space (\S~\ref{sec:resolution:spatial}), and time (\S~\ref{sec:resolution:time}). \S~\ref{sec:hydro} examines the effects of the hydrodynamic methods, including SPH vs.\ finite-volume methods and \S~\ref{sec:hydro:artificial.pressure} considers the effects of so-called ``artificial pressure'' terms used in some (non-FIRE) simulations. \S~\ref{sec:cooling} studies the details of cooling, chemical yields, and numerical metal-mixing terms; and \S~\ref{sec:star.formation} considers the star formation algorithm. \S~\ref{sec:feedback} considers the effects of different stellar feedback physics, turned on and off in turn, to provide an indication of which feedback processes dominate, and to provide a way of quantifying the relative importance of numerical and physical (feedback) uncertainties for our results. The Appendices present additional cooling/feedback tables and algorithmic information necessary for implementing the FIRE-2 simulations.

\vspace{-0.5cm}
\section{Methods}
\label{sec:methods}

Here we describe our numerical methods in the FIRE-2 simulations. For further details, at the end of each sub-section below, we direct interested readers to the appendix, paper, or public code where an exact algorithmic breakdown is provided. We will study each aspect in more detail below. 

Before FIRE-1, a series of papers developed the numerical methods, and tested each physical addition individually using higher-resolution ISM-scale simulations and analytic solutions where possible \citep[we refer the interested reader to][]{hopkins:rad.pressure.sf.fb,hopkins:fb.ism.prop,hopkins:stellar.fb.winds,hopkins:clumpy.disk.evol,hopkins:stellar.fb.mergers,hopkins:2013.accretion.doesnt.drive.turbulence,hopkins:2013.merger.sb.fb.winds,hopkins:dense.gas.tracers,hopkins:virial.sf}. These developments and improvements were then synthesized into the physics implemented in FIRE-1, described in \citet{hopkins:2013.fire}. 

In our FIRE-2 runs, the ``core'' or ``baseline'' physics is the same as in FIRE-1: we simply seek to improve the numerical accuracy with which we solve the relevant equations. However, the switch to a new hydrodynamics method in FIRE-2 also makes it possible {\em in principle} to include new physics such as magnetic fields: these are described here as ``additional'' physics, and will be studied in separate work.

As with FIRE-1, all runs denoted as ``FIRE-2'' here or in any other papers \citep[e.g.][]{wetzel.2016:latte,su:2016.weak.mhd.cond.visc.turbdiff.fx,fitts:fire.dwarf.concentration.mass} use the {\em identical} source code and physical parameters, unless explicitly labeled otherwise for comparison purposes. Of course certain numerical parameters (e.g.\ force softening) scale explicitly with resolution; these are provided for each simulation.

\vspace{-0.5cm}
\subsection{Hydrodynamics}
\label{sec:methods:hydro}

A major motivation for our introduction of ``FIRE-2'' is to take advantage of a new generation of accurate, mesh-free Godunov hydrodynamics methods that have been recently developed \citep[see][]{gaburov:2011.meshless.dg.particle.method,hopkins:gizmo}. Because we enforced the strict requirement that all FIRE-1 simulations use the identical source code, all FIRE-1 runs used the older ``P-SPH'' method \citep{hopkins:lagrangian.pressure.sph}, an improved ``pressure-energy'' variant of smoothed-particle hydrodynamics (SPH). We will explore the effects of the hydrodynamic solver in our simulations in \S~\ref{sec:hydro}. More importantly, however, the new Godunov methods allow us to accurately include more complicated plasma physics such as magnetic fields and anisotropic diffusion, which were not possible to solve accurately in P-SPH \citep[see][]{hopkins:mhd.gizmo,hopkins:gizmo.diffusion}. 

We therefore employ the meshless finite-mass (MFM) magneto-hydrodynamics solver in {\small GIZMO}.\footnote{A public version of this code is available at \gizmourl } This is a mesh-free, Lagrangian finite-volume Godunov method designed to capture advantages of both grid-based and particle-based methods, built on the gravity solver and domain decomposition algorithms of {\small GADGET-3} \citep{springel:gadget}. In a series of methods papers \citep{hopkins:gizmo,hopkins:mhd.gizmo,hopkins:cg.mhd.gizmo,hopkins:gizmo.diffusion}, {\small GIZMO} has been tested extensively (involving $\sim100$ distinct test problems) compared to state-of-the-art fixed grid Godunov codes (e.g.\ {\small ATHENA} and {\small RAMSES}; \citealt{teyssier:2002.RAMSES,stone:2008.athena}), moving-mesh codes (e.g.\ {\small AREPO}; \citealt{springel:arepo}), and ``modern'' SPH methods (e.g.\ {\small P-SPH}; \citealt{hopkins:lagrangian.pressure.sph,rosswog:2014.sph.accuracy,hu:2014.psph.galaxy.tests}). 

We emphasize that in essentially every test problem we find MFM gives more accurate results (at fixed particle number or CPU time) and faster convergence compared to state-of-the-art SPH methods, and demonstrates accuracy and convergence in good agreement with well-studied fixed-grid and moving-mesh codes. Most importantly, this includes areas where SPH has had historical difficulty, including sharp shock-capturing, fluid-mixing instabilities, magneto-rotational instabilities, and anisotropic diffusion \citep{ritchie.thomas:2001.egy.wtd.sph,agertz:2007.sph.grid.mixing,price:2008.sph.contact.discontinuities,wadsley:2008.sph.mixing.cosmology,read:2012.sph.w.dissipation.switches,saitoh:2012.dens.indep.sph}. For some problems relevant in cosmological simulations, e.g.\ those with moving contact discontinuities, orbiting thin disks, supersonically shearing fluid-mixing instabilities, poorly resolved explosions, hydrostatic gravitational equilibrium or gravitational collapse, the Lagrangian nature of the method here also allows us to converge at much lower resolution compared to fixed-grid methods \citep{muller:1995.grid.code.gravity.problems,zingale:2002.grid.hydro.eqm.issues,oshea:sph.tests,heitmann:2008.cosmic.code.comparison,hopkins:gizmo} and provides excellent angular momentum conservation \citep[avoiding ``grid alignment'' and spurious torques common in grid-based codes;][]{hahn:2010.disk.gal.orientations.ramses,byerly:2014.hybrid.cartesian.scheme.for.ang.mom,hopkins:gizmo}. 

As discussed in \citet{hopkins:gizmo}, this increased accuracy and convergence rate effectively makes our simulations effectively higher-resolution (at least in terms of the spatial resolution of the hydrodynamics and its convergence), compared to FIRE-1 simulations at the same particle number. 

For reasons discussed in \S~\ref{sec:hydro:artificial.pressure}, we do not adopt an artificial ``pressure floor'' of any kind for hydrodynamics; unresolved fragmentation is instead explicitly treated via our star formation model.

Readers interested in further details of the hydrodynamic solver should consult \citet{hopkins:gizmo} and the public {\small GIZMO} source code. Tests and comparisons of different hydrodynamic methods are in \S~\ref{sec:hydro}.

\begin{figure*}
\plotsidesize{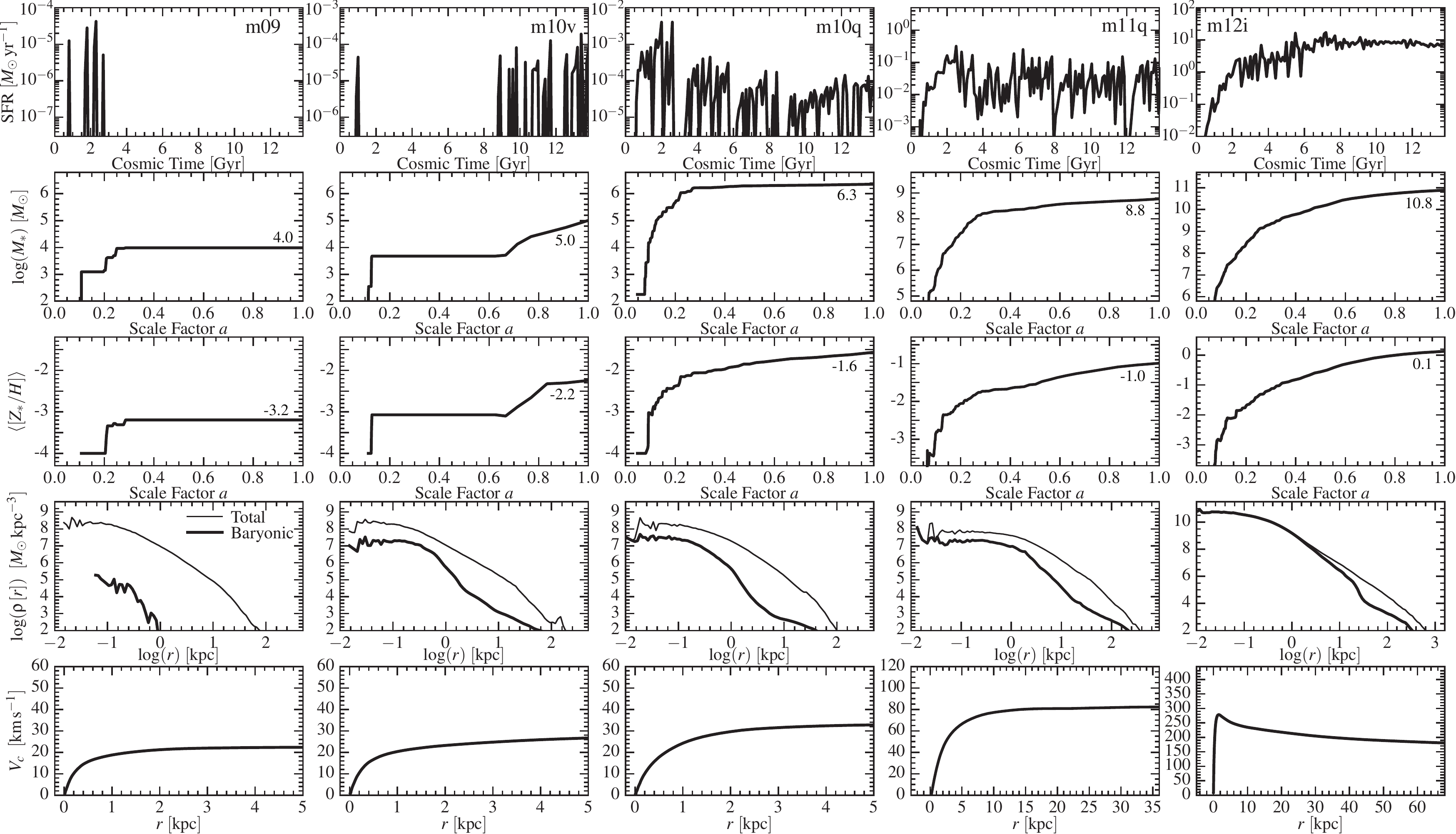}{1.0}
    \vspace{-0.6cm}
    \caption{Basic galaxy-scale properties in a subset of FIRE-2 simulations. Each column shows a different galaxy from Fig.~\ref{fig:images.m12} and Table~\ref{tbl:sims}. 
     {\em Top:} Star formation rate (averaged in $100\,$Myr intervals) of the primary (most massive) galaxy in each simulation versus cosmic time. Ultra-faint galaxies ($V_{c}\lesssim20\,{\rm km\,s^{-1}}$) are quenched after reionization. In more massive dwarf galaxies, SF is highly ``bursty'', but it becomes less so at even higher masses and at later times \citep[as seen in our FIRE-1 runs;][]{sparre.2015:bursty.star.formation.main.sequence.fire}.
     {\em Second from Top:} Total stellar mass versus scale factor, $a=1/(1+z)$, in the zoom-in region (we show against scale factor so that the rate of growth at early times is more clear). The value at $z=0$ appears on the plot.
     We show all stellar mass within $R_{\rm vir}$, but this is dominated by the main galaxy, so it evolves similarly. Growth occurs rapidly at high redshifts then settles into a more steady state at late times, allowing galaxy structure to relax \citep[][]{elbadry.2015:core.transformation.stellar.kinematics.gradients.in.dwarfs}. 
     {\em Middle:} Stellar Mass-Weighted Average Metallicity versus scale factor (value at $z=0$ also shown). This rise is similar to the stellar mass, because these galaxies evolve on a redshift-dependent stellar mass-metallicity relation; the metallicities at each mass and redshift are nearly identical to FIRE-1 galaxies \citep[see][]{ma:2015.fire.mass.metallicity}. 
     {\em Second from Bottom:} Baryonic ({\em thick}) and total ({\em thin}) mass density profiles as a function of radius around the most massive galaxy at $z=0$. Profiles are averaged in spherical shells. The dwarfs with stellar masses $M_{\ast} \sim 10^{6}-10^{9.5}\,\msun$ exhibit central ``cores'' in their mass profiles, in both stars and dark matter, most prominent at $M_{\ast} \sim 10^{9}\,\msun$, where the cores extend to $\sim$\,kpc scales, consistent with FIRE-1 results \citep{onorbe:2015.fire.cores,wheeler:dwarf.satellites,chan:fire.dwarf.cusps}. 
     {\em Bottom:} Rotation curves of circular velocity, $V_{c}$, versus radius around the most massive galaxy at $z=0$. Dwarfs exhibit slowly-rising rotation curves, while MW-mass systems have flat rotation curves with small-to-modest bulges, as in FIRE-1 \citep{chan:fire.dwarf.cusps}.
     \label{fig:demo}}
\end{figure*}

\vspace{-0.5cm}
\subsection{Gravity}
\label{sec:methods:gravity}

The $N$-body gravity solver is extensively detailed in \citet{hopkins:gizmo}; this is an improved version of the {\small GADGET-3} Tree-PM solver and additional details can be found in \citet{springel:gadget}. This solver is well-tested under a huge range of applications; we will focus here only on how this relates to the spatial or force resolution, and what this means for ``resolved scales'' in FIRE. Detailed discussion and tests of resolution and force softening are presented in \S~\ref{sec:resolution:mass}-\ref{sec:resolution:time}. 

By default, as described in \citet{hopkins:gizmo}, the resolution of gravity and hydrodynamics are equal and the two use the {\em same}, consistent assumptions about the gas mass distribution in the simulations. Specifically, we follow \citet{price:2007.lagrangian.adaptive.softening} and compute gravitational forces from gas particles by assuming the gas is in an extended mass distribution with the same functional form as the interaction kernel centered at the particle.\footnote{\citet{hopkins:gizmo} show this is the leading-order accurate expression for the potential if we integrate Poisson's equation using the exact differential mass distribution assumed in the hydrodynamic equations. Because the separations $h_{i}^{\rm gas}$ change, we must be careful to maintain energy and momentum conservation correctly when elements interact inside the softening; \citet{price:2007.lagrangian.adaptive.softening} show how the relevant expressions can be rigorously derived from the softened $N$-body particle Lagrangian and the expressions used in {\small GIZMO} are presented in \citet{hopkins:gizmo}, Appendix~H2.} This means that, for gas, the gravitational force softening resolution $\epsilon_{i}^{\rm gas}$ identically follows the inter-particle/cell separation $h_{i}^{\rm gas} = \Delta x$, where $\Delta x$ is the equivalent cell-length in a fixed, Cartesian mesh (i.e.\ the average distance between particle centers, around element $i$, is $h_{i}^{\rm gas}$), so $\epsilon_{i}^{\rm gas}\equiv h_{i}^{\rm gas}=16\,{\rm pc}\,m_{i,\,1000}^{1/3}\,(n_{\rm H}/10\,{\rm cm^{-3}})^{-1/3}$ (where $m_{i,\,1000}$ is the mass resolution in units of $1000\,\msun$). Note this definition is {\em independent} of the exact softening kernel shape or ``neighbor number.'' 

As discussed in \S~\ref{sec:resolution:spatial}, a number of studies have shown this produces ``optimal'' softening in terms of (i) physical consistency with the hydrodynamics, (ii) maximizing accuracy and convergence rates, and (iii) reducing $N$-body integration and force errors \citep[e.g.][]{merritt:1996.optimal.softening,bate:1997.sph.res.reqs,romeo.1998:optimal.softening,athanassoula:2000.optimal.force.softening.collisionless.sims,dehnen:2001.optimal.softening,rodionov:2005.optimal.force.softening,price:2007.lagrangian.adaptive.softening,barnes:2012.softening.is.smoothing,hubber:2013.res.criteria.in.star.clusters.strong.scattering.effects}. 
Explicit tests validating the accuracy and convergence of our adaptive self-gravity implementation in {\small GIZMO} are presented in \citet{hopkins:gizmo}, including the \citet{evrard:1988.gas.collapse.problem} polytropic collapse test, the gas (and gas+DM) \citet{zeldovich:1970.pancakes} pancake collapse, the ``Santa Barbara Cluster'' adiabatic zoom-in simulation \citep{frenk:1999.sb.cluster}, and rotating steady-state stable disk tests. 

For dark matter (DM) and stars, the collisionless nature of the fluid makes the ``correct'' softening ambiguous. In \S~\ref{sec:resolution:spatial} we therefore explore and compare a range of choices (both adaptive and constant). In our default simulations we set $\epsilon_{i}^{\rm DM}$ to a constant chosen to give optimal convergence and integration accuracy based on the tests therein (essentially the largest possible value before noticeable ``over-softening'' effects appear), and set $\epsilon_{i}^{\ast}$ to a constant matched to the gas softening at the mean gas density of star formation. However we will show that these choices have little effect on any predictions here, consistent with previous studies \citep{bagla:2009.adaptive.treepm,iannuzzi:2011.collisionless.adaptive.softening.gadget,iannuzzi:2013.no.need.adaptive.softening.for.dm}.

Readers interested in further details of the gravity solver and adaptive softening scheme should consult \citet{hopkins:gizmo} and the public {\small GIZMO} source code. Extensive tests of force softening algorithms and choices are presented in \S~\ref{sec:resolution}.

\vspace{-0.5cm}
\subsection{Cooling}
\label{sec:methods:cooling}

Gas cooling is solved using a standard implicit algorithm, described in \citet{hopkins:2013.fire}, and all details are given in Appendix~\ref{sec:cooling.approximations}. To summarize: heating/cooling rates are computed including free-free, photo-ionization/recombination, Compton, photo-electric, metal-line, molecular, fine-structure, dust collisional, and cosmic ray processes, from $T=10-10^{10}\,$K. We follow $11$ separately-tracked species (H, He, C, N, O, Ne, Mg, Si, S, Ca, and Fe, each with its own yield tables associated directly with the different mass return mechanisms below). The relevant ionization states are tabulated from {\small CLOUDY} simulations including the effects of uniform but redshift-dependent background (from \citealt{faucher-giguere:2009.ion.background}) together with local radiation sources (every star particle is treated as a source; see the feedback description below). We account for self-shielding with a local Sobolev/Jeans-length approximation (Appendix~\ref{sec:cooling.approximations}),  calibrated in full radiative transfer experiments in \citet{faucher-giguere:2010.lya.cooling.selfshield} and \citet{rahmati:2013.selfshield.rt}. 

With this, high-temperature ($>10^{4}\,K$) metal-line excitation, ionization, and recombination rates then follow \citet{wiersma:2009.coolingtables}. Free-free, bound-free, and bound-bound collisional and radiative rates for {\small H} and {\small He} follow \citet{katz:treesph} with the updated fitting functions in \citet{verner.ferland:recombination.rates}. Photo-electric rates follow \citet{wolfire.2003:neutral.atomic.cooling}, accounting for PAHs and local variations in the dust abundance. Compton heating/cooling (off the combination of the CMB and local sources) follows \citet{cafg:2012.egy.cons.bal.winds}. Fine-structure and molecular cooling at low temperatures ($5-10^{4}\,$K) follows a pre-computed tabulation of {\small CLOUDY} runs as a function of density, temperature, metallicity, and local radiation background \citep[see][]{robertson:2008.molecular.sflaw}. Collisional dust heating/cooling rates follow \citet{goldsmith:molecular.dust.cooling.gmcs} with updated coefficients from \citet{meijerink.spaans:xray.cooling.models} assuming a minimum grain size of $10\,$\AA, and a dust temperature of $30\,$K. Cosmic ray heating follows \citet{guo.oh:cosmic.rays} accounting for both hadronic and Compton interactions, with a uniform cosmic ray background of $\sim 5\,{\rm eV\,cm^{-3}}$. At very high densities ($\sim 10^{10}\,{\rm cm^{-3}}$), gas can become optically thick to its own cooling radiation; this is treated self-consistently following \citet{rafikov:2007.convect.cooling.grav.instab.planets}, but this is irrelevant for the simulations here (because they do not reach sufficiently high densities). Hydrodynamic heating/cooling rates (including shocks, adiabatic work, reconnection, resistivity, etc.) are computed in standard fashion by the hydro solver, then included directly into our fully-implicit solution. 
A $10\,$K temperature floor is enforced, but has no detectable effect on our results.

In Appendix~\ref{sec:cooling.approximations}, we provide fitting functions to the complete set of cooling physics above.

\vspace{-0.5cm}
\subsection{Star Formation}
\label{sec:methods:star.formation}

Gas which is locally self-gravitating, self-shielding, Jeans unstable, and above some minimum density is turned into stars using a sink-particle approach. All details are given in Appendix~\ref{sec:sf.algorithm}. Briefly, gas is eligible to turn into stars if and only if it meets the following criteria: 

\begin{enumerate}

\item{\bf Self-Gravitating:} We require the potential energy be larger than the thermal plus kinetic energy within the resolution scale: specifically we use the sink-particle criterion developed in \citet{hopkins:virial.sf}, $\alpha\equiv (\delta v^{2}+c_{s}^{2})\,\delta r/G\,m_{\rm gas}(<\delta r) = [\| \nabla \otimes {\bf v} \|_{i}^{2} + (c_{s,\,i}/h_{i})^{2}] / (8\pi\,G\,\rho_{i}) < 1$, where $\delta v = \| \nabla \otimes {\bf v} \|_{i}\,h_{i}$ and $c_{s}$ give the kinetic and thermal energy, respectively, within the resolution scale $\delta r \rightarrow h_{i}$ around the particle ($\otimes$ is the outer product). \citet{hopkins:virial.sf} and many other studies \citep{li:2005.turb.reg.sfr,federrath:2010.sink.particles,padoan:2011.new.turb.collapse.sims,padoan:2012.sfr.local.virparam} have shown this is more useful than a density criterion (as well as more accurate in converging to the results of higher-resolution simulations), since it actually identifies gas which is collapsing under self-gravity at the resolution scale (i.e.\ the gas which should, physically form stars), {\em independent} of the exact spatial, mass, or density scale. This also does not allow unbound material to form stars (e.g.\ tidally unbound gas, or gas in strong shocks and winds which is dense, but not self-gravitating owing to large internal motions). The exact order-unity coefficient is calibrated from higher-resolution simulations of collapsing clouds/cores \citep[see][]{padoan:2012.sfr.local.virparam,federrath:2012.sfr.vs.model.turb.boxes}, but our results are insensitive to variations. 

\item{\bf Self-Shielding (Molecular):} We estimate the self-shielded (``molecular'') fraction of each gas element following \citet{krumholz:2011.molecular.prescription}, using the local Sobolev approximation and metallicity to estimate the integrated column to dissociating radiation, and allow star formation only from the molecular component ($\rho_{\rm mol} = f_{\rm mol}\,\rho$). This is specifically a requirement that the gas be self-shielding, and therefore able to cool to low temperatures. Given the high $n_{\rm crit}$ (see below), this criterion typically has negligible impact, since the high-density gas is typically all shielded and molecular anyways.

\item{\bf Jeans Unstable:} We require a thermal Jeans mass below the maximum of the particle mass or $10^{3}\,M_{\sun}$ in the element. This is done to ensure that any resolved, massive self-gravitating objects which should collapse coherently (as opposed to fragmenting into stellar mass-scale objects) are followed self-consistently and not simply assigned to stars (the choice of $10^{3}\,\msun$ is designed to separate massive ``clumps'' from normal very massive stars, but is not important). In practice this criterion is always easily met when other criteria are met.

\item{\bf Dense:} To prevent spurious application of the criteria above, we also check that $n_{\rm H}>n_{\rm crit}=1000\,{\rm cm^{-3}}$ (much larger than the mean galaxy density $\langle n \rangle$). This restricts star formation to dense molecular clouds fragmenting out of the background disk.

\end{enumerate}

If gas meets all the criteria above, we assume it turns into stars at a rate $\dot{\rho}_{\ast}=\rho_{\rm mol}/t_{\rm ff}$ where $t_{\rm ff}$ is its free-fall time. This also comes directly from higher-resolution simulations of turbulent clouds \citep{padoan:2011.new.turb.collapse.sims,padoan:2012.sfr.local.virparam}, as well as analytic models for star formation via turbulent fragmentation \citep{hopkins:excursion.imf,hopkins:excursion.ism,guszejnov:cmf.imf,guszejnov:gmc.to.protostar.semi.analytic,guszejnov.2015:feedback.imf.invariance}. We stress that this is an assumption about the rate at which small, {\em locally self-gravitating}  clumps (which may only be a small fraction of the dense gas mass) fragment; it does {\em not} imply the global efficiency of star formation (either on galaxy or GMC scales) is necessarily high -- we find that it is self-regulated by feedback at $\sim1-10\%$ per free-fall time (see \citealt{hopkins:rad.pressure.sf.fb,hopkins:dense.gas.tracers,hopkins:2013.fire}) even in gas with densities $\gtrsim 100\,{\rm cm^{-3}}$, in agreement with observations \citep{lee:mw.cloud.dynamical.sfe}. Recently, similar implementations to ours have also found consistent results on large scales \citep{semenov:non.universal.sfe.on.freefall,agertz:high.sfe.needed.to.explain.galaxies}, consistent with analytic expectations \citep{ostriker.shetty:2011.turb.disk.selfreg.ks,cafg:sf.fb.reg.kslaw}. If particles do not meet all of the criteria above, their SFR is zero. Gas particles which turn into star particles begin life as zero-age main sequence populations, with abundances and total mass inherited from their progenitor gas particle.

We provide the complete set of formulae and detailed algorithmic implementation of star formation in Appendix~\ref{sec:sf.algorithm}.

\vspace{-0.5cm}
\subsection{Stellar Feedback}
\label{sec:methods:stellar.feedback}

Once a star particle forms, it is treated as a single stellar population, with known age $t_{\ast} = t - t_{\rm form}$, metallicity (inherited from its progenitor gas particle), and mass (equal to its progenitor gas particle). All feedback quantities are tabulated directly -- {\em without subsequent adjustment or fine-tuning} -- from standard stellar population models \citep[{\small STARBURST99};][]{starburst99} assuming a \citet{kroupa:2001.imf.var} IMF (the same as FIRE-1).\footnote{Of course, alternative stellar evolution/IMF models may predict different feedback properties, but we will not investigate this here. In general the predicted variation is small, but for some quantities, e.g.\ the escape fraction of ionizing photons at high redshift, we have shown it can be important \citep{ma.2016:binary.star.escape.fraction.effects}.}

Here, we briefly summarize the feedback mechanisms. Because these are the most important and novel aspect of the FIRE simulations, we discuss the exact physics and algorithmic implementation in much greater detail in the companion papers, \paperone\ \&\ \papertwo. These papers present extensive tests of the algorithms, with idealized simulations of e.g.\ individual SNe explosions reaching resolution $<0.01\,\msun$ and experiments using detailed radiation-hydrodynamics simulations, used to validate the exact implementations here. But for the sake of completeness, we summarize them here and provide the complete algorithms in the Appendices.

The physics of stellar feedback in FIRE-2 are the same as in FIRE-1 \citep{hopkins:2013.fire}, and the algorithms are identical up to  improvements in accuracy which we explicitly detail below. 

\begin{enumerate}

\item{\bf Supernovae (Ia \&\ II):} (For details, see Appendix~\ref{sec:mechanical.fb.implementation}.) Every timestep $\Delta t$, for each star particle, the tabulated SN rate as a function of star particle mass, age, and metallicity is used to determine the probability $p$ of an event (Type-Ia and/or Type-II) occurring within the particle within $\Delta t$; our mass and time resolution is such that $p\ll1$, i.e.\ we explicitly treat individual SNe explosions, rather than model their collective effects indirectly. We determine probabilistically if an event occurs within $\Delta t$; if so, the appropriate ejecta mass, metal yields, energy, and momentum (also determined from the stellar evolution tables) are deposited directly in the surrounding gas around the star particle. The algorithm for deposition is constructed to ensure {\em manifest, machine-accurate} conservation of mass, metal mass, energy, and momentum, while also ensuring that the ejecta are distributed isotropically in the rest frame of the star. In \paperone\ we show that this is non-trivial in Lagrangian codes such as ours, where highly anisotropic gas distributions around a star particle can easily bias the momentum distribution and even violate linear momentum conservation, if the algorithm is not carefully designed to prevent this. We properly account for the relative star-gas motion (so e.g.\ the exact shock solution includes the initial stellar motion through the background gas). We determine the coupled momentum by computing the exact Sedov-Taylor solution for an energy-conserving spherical shock, at the coupling location (resolved separation between gas and star); if the resulting momentum exceeds the terminal momentum at which point the shock should have become radiative (equivalently, if the resolved coupling radius is larger than the cooling radius), we deposit only the momentum which would have been present when it reached that cooling radius. \paperone\ shows that this ensures our simulations exactly reproduce the fully-converged solutions (with resolution $<0.01\,\msun$) for individual SNe explosions in high-resolution ISM simulations (once they reach the same radius as our coupling radius), independent of our resolution, for the same ambient density. We stress that we do not turn off cooling or otherwise impose any assumption about ``galactic wind driving.'' 

\item{\bf Continuous Stellar Mass-Loss (OB/AGB-star Winds):} (Details in Appendix~\ref{sec:mechanical.fb.implementation}.) Similarly, stellar mass-loss is injected continuously in the gas surrounding each star particle as a function of stellar age and metallicity, with the appropriate mechanical luminosity, momentum, mass, and metal content, including both fast (O/B-star) and slow (AGB) winds, calculated directly from the stellar evolution models. The algorithm for deposition is exactly the same as for SNe, except there is an ``event'' every timestep with associated ejecta mass $= \Delta t\,\dot{M}_{\rm wind}$. 

\item{\bf Photo-Ionization and Photo-Electric Heating:} (Details in Appendix~\ref{sec:radiative.fb.implementation}.) For computing radiative feedback properties, each star particle is treated as a source with an appropriate age and metallicity-dependent, IMF-averaged spectrum. We approximate the complete spectrum with a five-band treatment that includes ionizing, far-UV (relevant for photo-electric heating), near-UV, optical/near-IR, and mid/far-IR (re-radiated dust emission) photons. The background radiation owing to these sources is then locally extincted by the gas immediately surrounding the star (using a Sobolev approximation to estimate the column integrated to infinity, and extincting each band accordingly), with frequency and metallicity-dependent opacities from dust and neutral gas. The luminosity absorbed by dust (non-ionizing bands) is assumed to re-radiate in the mid/far-IR band. The resulting, post-extinction luminosities are then propagated to long-range distances through an optically thin transport network (using a tree structure), to calculate an incident flux in each band at all positions. We therefore refer to this transport algorithm as the ``Locally Extincted Background Radiation in Optically-thin Networks,'' or LEBRON, approximation. Since we simulate only a single small region of the Universe surrounding one galaxy in our ``zoom-in'' simulations, we add to the diffuse ionizing-band flux a uniform but redshift-dependent meta-galactic background tabulated from \citet{faucher-giguere:2009.ion.background}. The fluxes are then corrected for self-shielding using the same local-extinction Sobolev approximation, at the location of the gas. The resulting incident ionizing and FUV fluxes are then used to self-consistently compute the gas ionization states and radiative heating/cooling rates in our standard cooling algorithms described in Appendix~\ref{sec:cooling.approximations}. 

\item{\bf Radiation Pressure:} (Details in Appendix~\ref{sec:radiative.fb.implementation}.) As photons are tracked according to the algorithm above, each {\em explicitly resolved} absorption transfers the appropriate photon momentum $=L_{\rm abs}\,\hat{n}/c$ (where $\hat{n}$ is the direction of ray propagation) to the gas. This automatically accounts for both direct UV/optical single-scattering, and indirect re-radiated IR photons (which can, in principle, be multiply-scattered, although this rarely occurs on the relatively coarse scales we resolve in the FIRE simulations). We stress that we {\em do not} assume any ``sub-grid'' photon coupling, multiple-scattering, or radiation pressure -- there is no ``boost factor'' anywhere in the model: in both FIRE-2 and FIRE-1, the only radiation pressure in the simulations is from {explicitly resolved} photon absorption. In \papertwo, we show that only $\sim 1/2$ of the total bolometric luminosity of stellar populations is absorbed at all, and, given our numerical resolution (which prevents us from resolving e.g.\ proto-stellar cores), the multiple-scattering IR term accounts for $<10\%$ of the galaxy-averaged radiation pressure (it may be important, however, in dense galactic nuclei corresponding to observed systems like Arp 220). 

\end{enumerate}

We emphasize that while quantities like SNe rates and stellar spectra are IMF-averaged, individual SNe are always discrete events (not continuous energy injection). In future work we will consider the effects of explicitly sampling the spectrum of stellar masses from the IMF \citep{su:discrete.imf.fx.fire}; however our preliminary results indicate the effects on large scales are (unsurprisingly) small compared to our IMF-averaged approach.

For readers interested in reproducing our results, we provide simple fitting functions to all of our stellar evolution tabulations (and yield tables) needed for the feedback mechanisms above, in Appendix~\ref{sec:stellar.evolution.approximations}. All details of the algorithmic implementation of mechanical feedback (SNe and stellar mass-loss) are given in Appendix~\ref{sec:mechanical.fb.implementation}, and all details of the algorithmic implementation of radiative feedback (radiation pressure, photo-ionization, and photo-electric heating) are given in Appendix~\ref{sec:radiative.fb.implementation}.


\vspace{-0.5cm}
\subsection{``Additional'' Physics: Magnetic Fields, Conduction, Viscosity, Diffusion, Cosmic Rays, Black Holes, and more}
\label{sec:methods:additional.physics}

As noted above, a major motivation for our migration to FIRE-2, using the new MFM hydrodynamic solver, is to compare simulations including more complicated plasma physics, e.g.\ magnetic fields.
However, for the sake of clarity and direct comparison with FIRE-1, in this paper we will focus on simulations that include our ``core'' set of FIRE physics (gravity, hydrodynamics, cooling, star formation, and stellar feedback, as described above). This means that our ``default'' or ``core physics only'' FIRE-2 simulations use {\em the same physics} as FIRE-1, just more accurate numerical integration of those physics. 

The effects of additional physics will of course be the subject of their own studies. Some examples include (i) magnetic fields and (ii) anisotropic (Braginskii) conduction \&\ viscosity \citep[both studied in][]{su:2016.weak.mhd.cond.visc.turbdiff.fx}; (iii) passive-scalar turbulent eddy diffusion (e.g.\ metal diffusion), discussed briefly here in \S~\ref{sec:turbulent.diffusion.tests} and in more detail in \citet{escala:turbulent.metal.diffusion.fire}; (iv) cosmic rays (Chan et al., in prep.); (v) alternative radiation-hydrodynamics (using e.g.\ alternative RHD solvers such as the M1 method as implemented in \citealt{hopkins:rhd.momentum.optically.thick.issues} or direct integration following \citealt{jiang:2014.rhd.solver.local}), discussed in detail in \papertwo; (vi) super-massive black hole formation, accretion, and feedback \citep[see e.g.][for a preliminary exploration]{daa:BHs.on.FIRE}.

\vspace{-0.5cm}
\subsection{Timesteps \&\ Integration}
\label{sec:methods:timestepping}

Our time integration scheme is discussed in detail in \citet{hopkins:gizmo}.  Following \citet{springel:gadget} we use an adaptive power-of-two hierarchy for assigning individual timesteps for particles. As shown in \citet{saitoh.makino:2009.timestep.limiter} and \citet{durier:2012.timestep.limiter}, in problems with high Mach number flows, adaptive timesteps can lead to errors if particles with long timesteps interact suddenly mid-timestep with those on much shorter timesteps; this is remedied by requiring that, at all times, any active particle informs its neighbors of its timesteps and none are allowed to have a timestep $>4$ times that of a neighbor. Whenever a timestep is shortened (or energy is injected in feedback of any sort) particles are forced to return to the timestep calculation. This has been tested extensively in \citet{hopkins:lagrangian.pressure.sph} and \citet{hopkins:gizmo}.

The timestep is set by the minimum of various criteria. All particles obey limits $\Delta t < 0.2\,(h_{i} / |{\bf a}_{i}|)^{1/2}$ (see \citealt{power:2003.nfw.models.convergence}; here $h_{i}$ is the minimum of the inter-particle separation or Plummer-equivalent force softening)\footnote{For reference, with our definitions, $\Delta t < 0.2\,(h_{i} / |{\bf a}_{i}|)^{1/2}$ is equivalent to setting the parameter ``ErrTolIntAccuracy'' $\approx0.01$ in {\small GADGET-2}.} and $\Delta t < 0.25/|\nabla \cdot {\bf v}_{i}|$, where  ${\bf v}_{i}$ and ${\bf a}_{i}$ are the total velocity and acceleration of particle $i$ (including all sources of acceleration: e.g.\ feedback and hydrodynamic forces, for gas). For further safety, we always enforce a maximum timestep of $\Delta a < 10^{-4}\,a$ (where $a$ is the scale-factor), but this is almost never important. Gas elements must also obey the Courant (CFL) condition: $\Delta t < 0.4\,h_{i} / v_{{\rm sig},\,i}^{\rm max}$, where $v_{{\rm sig},\,i}^{\rm max}$ is the usual maximum signal velocity between all particles interacting with $i$ (see \citealt{hopkins:gizmo} for tests and details). Additional timestep criteria apply if additional fluid physics (magnetic fields, diffusion, cosmic rays, radiation) are included \citep[see][]{hopkins:gizmo.diffusion}.\footnote{Some physics, such as cooling, photo-ionization, and recombination, are handled in a fully-implicit numerical scheme, which (in the limit where, say, the cooling time is much shorter than the timestep) iteratively solves for the equilibrium temperature balancing all heating and cooling physics over each timestep. This means they do not impose an additional explicit timestep criterion.} In the above equations, note $h_{i}$ is defined by the inter-particle separation as defined in \S~\ref{sec:resolution:spatial}, which is the appropriate value for the prefactors here (see \citealt{hopkins:gizmo}; the pre-factors would need to decrease by a factor $\sim 2$ if we replaced $h_{i}$ with the maximum allowed neighbor distance, for example). If adaptive gravitational softening is used for collisionless (star and dark matter) particles, they must obey additional Courant-like timestep criteria given in \S~\ref{sec:resolution:time:ags}.

For star particles, we additionally impose a restriction $\Delta t < {\rm MAX}(10^{4}\,{\rm yr},\ t_{\ast} / 300)$, where $t_{\ast}$ is the age of the star; this prevents the code from ``skipping'' any significant portion of stellar evolution if, somehow, a star formed in a region where the other timestep criteria allowed long timesteps (although this is very rare), and also ensures that the expectation value of the number of SNe per particle per timestep is always $<1$ at our production resolution.

Readers interested in more details of the time integration scheme should consult the public {\small GIZMO} source code.

\vspace{-0.5cm}
\subsection{Initial Conditions}
\label{sec:methods:ICs}

All simulations in this paper are fully cosmological ``zoom-in'' simulations: a large box is simulated at low resolution to $z=0$, and then the mass within and around the halo(s) of interest is identified, traced back to the starting redshift, and the Lagrangian region containing this mass is re-initialized at much higher resolution for the ultimate simulation \citep{porter:1985.cosmo.sim.zoom.outline,katz:1993.zoomin.technique}. The initial conditions are generated with the {\small MUSIC} code \citep{hahn:2011.music.code.paper}, using second-order Lagrangian perturbation theory to evolve the initial conditions to redshift $z\sim 100$, at which point the {\small GIZMO} simulation begins. In the simulations here, the Lagrangian high-resolution regions are defined by a convex hull including all particles within $\sim 5\,R_{\rm vir}$ of the final ($z=0$) ``primary'' galaxy (most massive galaxy within the high-resolution region); we have used a series of re-simulations with progressively higher resolution, including baryons, to refine the Lagrangian regions more accurately, with a target of zero low-resolution DM particles contaminating the region within $\sim 2\,R_{\rm vir}$ \citep[following][]{onorbe:2013.zoom.methods}. Typically, these regions include a number of smaller galaxies; however, in this paper, we exclude {\em any} galaxy with $>1\%$ contamination (from low-resolution particles) by mass within $R_{\rm vir}$. 

Table~\ref{tbl:sims} describes the initial conditions for the initial set of halos we have simulated to $z=0$. We consider a series of halos with different masses; many of these are chosen to match the halos from our FIRE-1 studies (specifically simulations first presented in \citealt{hopkins:2013.fire} and \citealt{chan:fire.dwarf.cusps}). In all cases, the ICs are re-generated if needed to meet our strict contamination standard above. A couple of FIRE-1 ICs are not re-simulated here, because they were not generated from the {\small MUSIC} code (they were taken from older work); for consistency and clarity we will only include ICs generated in a consistent manner here. We have added new simulations here to increase our statistical sampling of halo growth histories and mass. The specific halos we re-simulate are chosen to represent a broad mass range and be ``typical'' in most properties (e.g.\ sizes, formation times, and merger histories) relative to other halos of the same $z=0$ mass. Simulations labeled ``q'' (e.g.\ {\bf m10q}) have more ``quiescent'' halo growth histories at late times (i.e.\ tend to form earlier) while those labeled ``v'' have more ``violent'' late-time histories (tend to form later); however we stress that these all lie well within the typical scatter in such histories at each mass (for example, each ``q'' history has several major mergers at high redshifts). Other labels (``i'', ``f'') are purely for bookkeeping. Several ICs ({\bf m10q}, {\bf m10v}, {\bf m11q}, {\bf m11v}, {\bf m12q}, {\bf m12i}) are taken from the AGORA project \citep{kim:2013.AGORA,kim:agora.isolated.disk.test}, to enable easy comparisons with a wide range of different codes. We adopt a standard, flat $\Lambda$CDM cosmology with $h\approx0.70$, $\Omega_{\rm M}=1-\Omega_{\Lambda}\approx0.27$, and $\Omega_{b}\approx0.045$ \citep[consistent with current constraints; see][]{planck:2013.cosmological.params}.\footnote{For the sake of comparison with other work, some ICs are matched to simulations which adopted very slightly different cosmological parameters. These differences are at the $\sim1\%$ level and their effects are much smaller than standard halo-to-halo variation.}

We scale the resolution with simulation mass, to achieve the optimal possible mass and force resolution for each halo; we study both mass and force resolution extensively in \S~\ref{sec:resolution}.

For readers interested in more details, or reproducing our results, all initial conditions used here are publicly available.\footnote{For the MUSIC files necessary to generate all ICs here, see:\\
\ICsurl}

\begin{figure*}
\plotsidesize{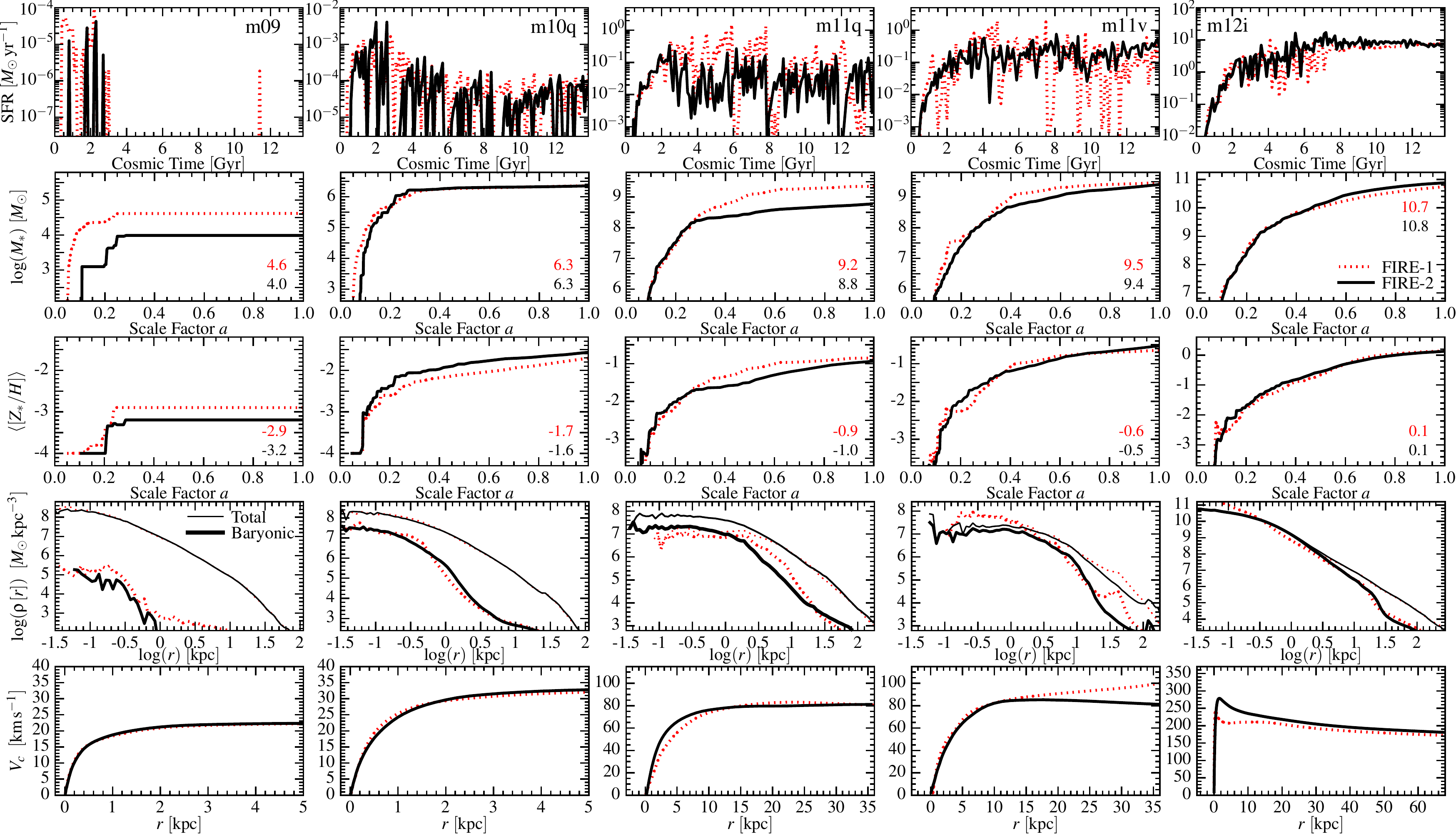}{0.99}
    \vspace{-0.25cm}
    \caption{Comparison of galaxy properties and formation histories in FIRE-1 versus FIRE-2, as in \demofigcosmo. We show galaxies for which the {\em identical} halo is a member of the ``core set'' of both FIRE-1 and FIRE-2 simulations. FIRE-2 combines a more accurate hydrodynamic method, higher resolution, a more accurate numerical algorithm for depositing supernova ejecta into gas around explosions, and updated cooling tables (for a complete list of changes, see \S~\ref{sec:methods:fire12.differences}). Nevertheless the results are qualitatively similar in every property that we examine here. We do see some quantitative differences. For dwarf galaxies, we find slightly lower stellar masses, because of the updated photo-heating tables. Massive galaxies show somewhat higher masses and central rotation velocities, because of enhanced mixing, which occurs because our more accurate hydrodynamic method changes the cooling and efficiency of wind escape in ``hot halos'' at late times. The enhanced ``burstiness'' in FIRE-1 {\bf m11v} occurs because it was run with $\sim10\times$ lower resolution as compared to FIRE-2. We examine each of the numerical aspects of the method in detail below.
    \label{fig:fire1vs2}}
\end{figure*}

\vspace{-0.5cm}
\subsection{Parallelization \&\ Runtime Requirements}
\label{sec:methods:mpi}

The simulations here use a hybrid OpenMP-MPI parallelization scheme with a number of optimizations specific for ``zoom-in'' simulations. These are listed in Appendix~\ref{sec:scaling.details}, together with explicit strong and weak scaling tests. Our improvements allow us to extend good weak scaling on production zoom-in simulations to at least $>16,000$ CPU cores (and $>10^{6}$ cores for large-volume simulations).

With these optimizations, each high-resolution, production-quality simulations of a MW-mass galaxy with particle masses $\sim 7000\,M_{\sun}$ (a few $\times10^{8}$ total particles) typically requires $\sim 10^{6}\,$cpu-hrs; for our smallest dwarfs (particle masses $\sim 250\,M_{\sun}$), their much lower star formation efficiencies and baryonic densities reduce this to $\sim 10^{4}\,$cpu-hrs. All simulations here were run on the XSEDE Stampede, Comet or NASA Pleiades machines.

Details of our code optimizations and scaling tests are presented in Appendix~\ref{sec:scaling.details}.

\begin{figure}
\begin{tabular}{cc}
\hspace{-0.20cm}
\vspace{-0.05cm}
\includegraphics[width=0.48\columnwidth]{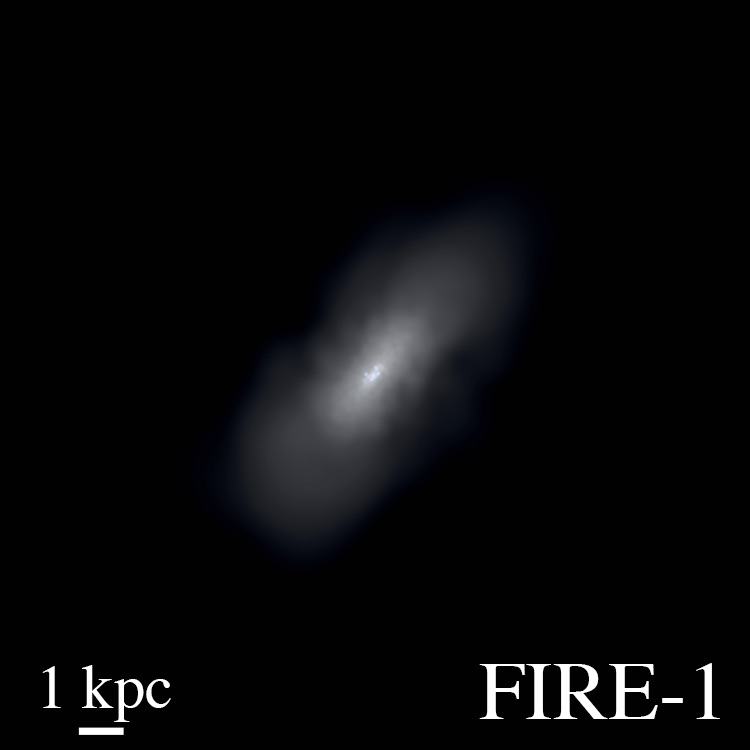} &
\hspace{-0.4cm}
\includegraphics[width=0.48\columnwidth]{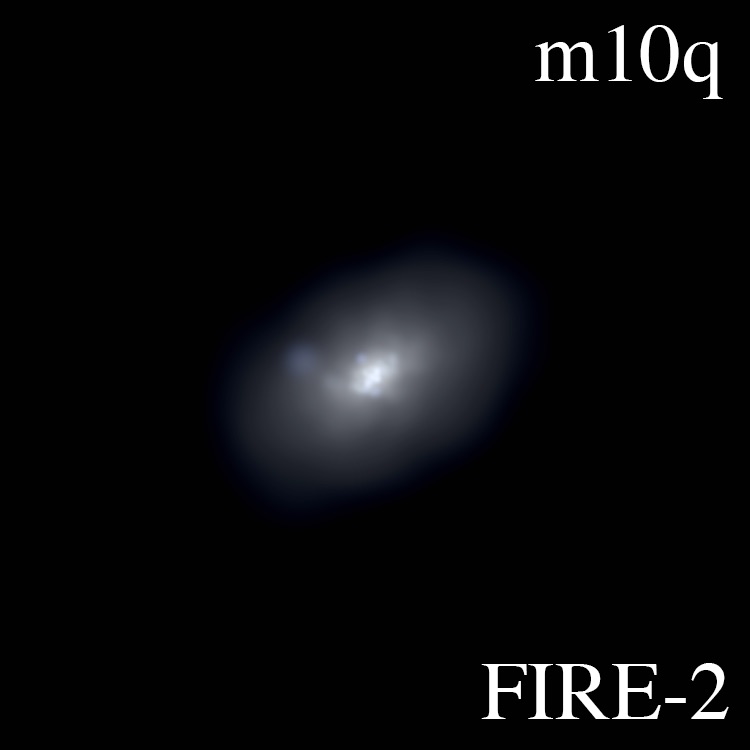} \\
\hspace{-0.20cm}
\vspace{-0.05cm}
\includegraphics[width=0.48\columnwidth]{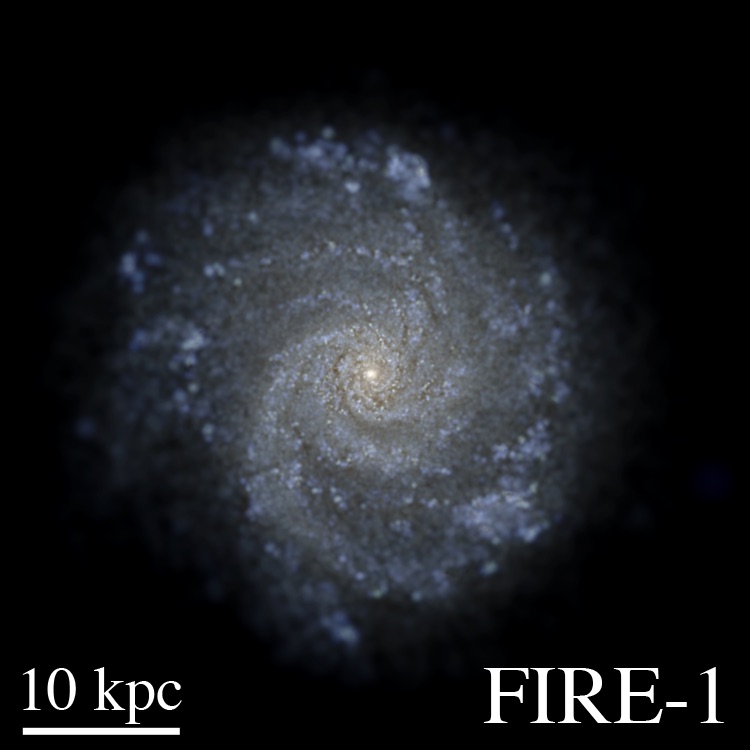} &
\hspace{-0.4cm}
\includegraphics[width=0.48\columnwidth]{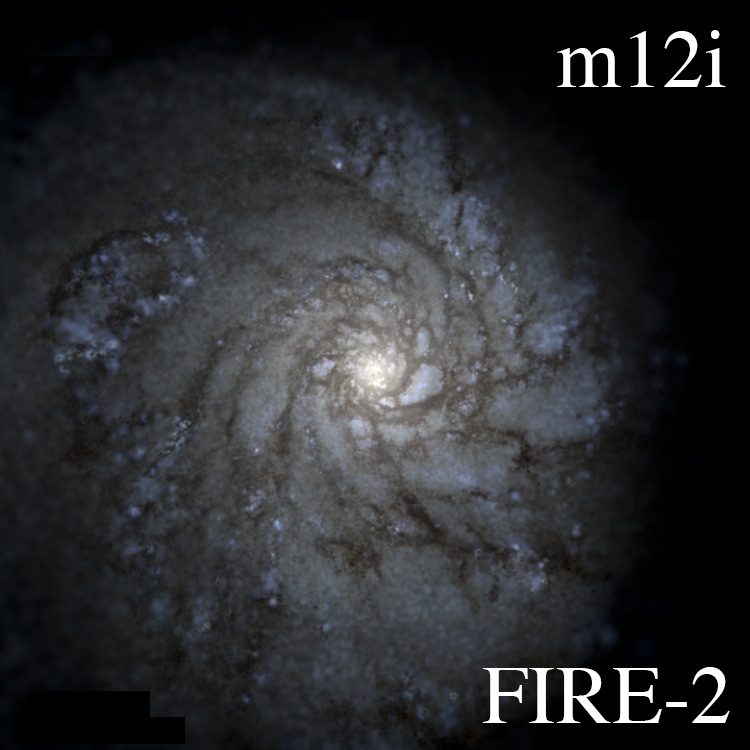} \\
\hspace{-0.20cm}
\includegraphics[width=0.48\columnwidth]{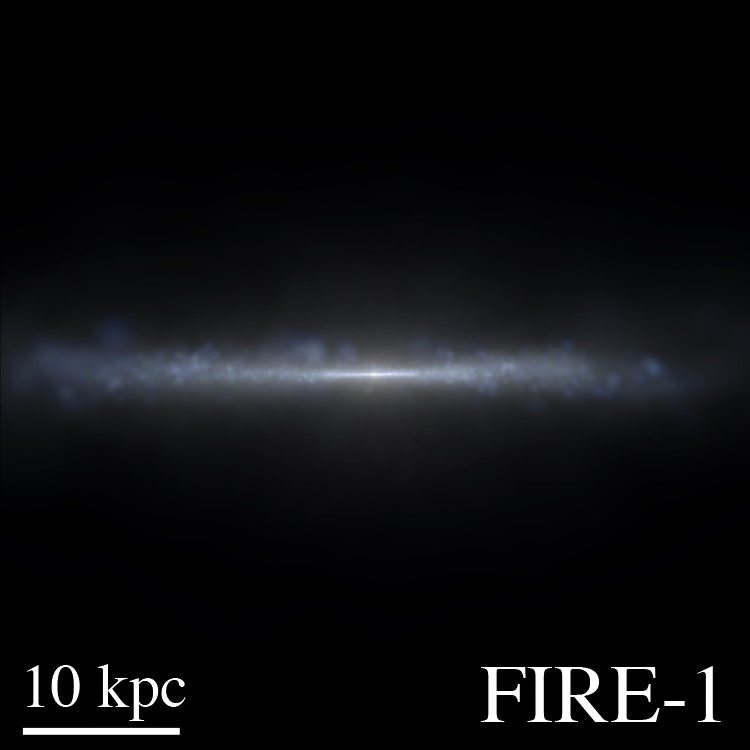} &
\hspace{-0.4cm}
\includegraphics[width=0.48\columnwidth]{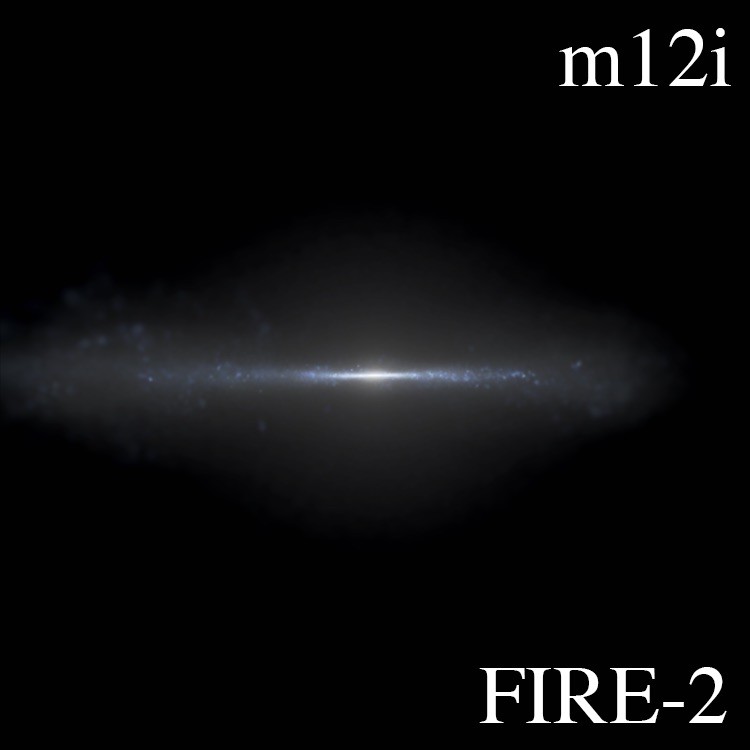} \\
\end{tabular}
    \vspace{-0.25cm}
    \caption{Mock images, as in Fig.~\ref{fig:images.m12}, comparing FIRE-1 ({\em left}) and FIRE-2 ({\em right}) versions of the same galaxy on the same scale.
    {\em Top:} Dwarf galaxy ({\bf m10q}). Because their morphologies are irregular or spheroidal, they are similar independent of numerical details.
    {\em Middle \&\ Bottom:} MW-mass galaxy ({\bf m12i}) seen face-on and edge-on. Qualitatively, the morphologies are similar. The FIRE-2 run is higher-resolution, which translates to a slightly thinner thin disk and a more extended, low surface-brightness outer disk.
    \label{fig:images.fire1vs2}}
\end{figure}

\vspace{-0.5cm}
\subsection{A Complete List of Differences Between FIRE-1 \&\ FIRE-2}
\label{sec:methods:fire12.differences}

Although they are discussed in great detail throughout this paper, for the sake of clarity we here summarize the differences between the FIRE-1 and FIRE-2 simulations, in order from most to least important.

\begin{enumerate}
\item{\bf More Accurate Hydrodynamic Solver:} As described in \S~\ref{sec:methods:hydro}, FIRE-2 uses the newer, more accurate mesh-free finite-volume Godunov-type MFM method to solve the hydrodynamic equations. FIRE-1 used the older ``pressure-energy'' SPH (``P-SPH'') method. In \S~\ref{sec:hydro} we show how this affects our results; while the differences are generally second-order, this appears to be the single change with the largest effects on our predictions. 

\item{\bf Manifestly-Conservative Supernovae Ejecta Distribution:} In Appendix~\ref{sec:mechanical.fb.implementation} and \paperone, we describe in explicit detail how, algorithmically, we distribute the products (mass, metals, momentum, and energy) of mechanical feedback (SNe and continuous stellar mass loss) from star particles into the surrounding gas particles. As discussed there, the FIRE-1 runs used a simpler algorithm, which can, in situations where the gas elements surrounding a star are highly disordered, produce a distorted (anisotropic) deposition (e.g.\ biasing the momentum deposition so it is not deposited symmetrically in the rest frame of the star, violating linear momentum conservation). We stress that the FIRE-1 algorithm still ensured the mass and energy of ejecta were exactly conserved; the issue comes with the spatial/vector distribution of the ejecta (momentum conservation). We have developed a novel scheme in FIRE-2 which eliminates this numerical bias and ensures manifest conservation. This difference generally has small effects, but does appear to influence the central stellar masses/densities of massive galaxies, and because the error term in the older implementation was resolution-independent, it actually can influence galaxy morphologies more dramatically at the highest resolutions.

\item{\bf More Accurate Photo-Ionization Heating:} In Appendix~\ref{sec:radiative.fb.implementation} and \papertwo, we describe our treatment of radiation transport in explicit detail, including UV/optical/IR radiation pressure, photo-electric heating, and photo-ionization heating. The method and source terms are almost entirely identical between FIRE-1 and FIRE-2. In our default treatment of photo-ionization heating specifically, we conduct a gas neighbor search around each star particle, consuming ionizing photons (as we move outward) using a local Stromgren approximation until the photon budget is exhausted. However in FIRE-1, the search was simply terminated at the boundary of the local computational domain -- any remaining photons were lost. In FIRE-2, any remaining photons are propagated via the long-range tree-based radiative transfer method in Appendix~\ref{sec:radiative.fb.implementation}. The fraction of photons affected is small since the vast majority are absorbed locally, and so this produces weak or negligible differences on galactic scales (nearly undetectable except in small dwarfs), but it eliminates the explicit domain-dependence of local HII regions. 

\item{\bf Removal of ``Artificial Pressure'' Terms:} In FIRE-1, we included an artificial numerical ``pressure floor'' for cold gas in the ISM, following the approach in e.g.\ \citet{robertson:2008.molecular.sflaw} (designed to suppress collapse of any gas resolved with $<4$ thermal Jeans lengths). As discussed in \S~\ref{sec:hydro:artificial.pressure}, this term is now (a) redundant with our star formation prescription, and (b) potentially unphysical, as it fails to conserve energy and can introduce noise or suppress real fragmentation.
We therefore include no such terms in FIRE-2, but instead follow standard practice in the star formation community and rely on the sink-particle (star formation) criterion to treat unresolved fragmentation \citep[see][for discussion]{federrath:2010.sink.particles}. In \S~\ref{sec:hydro:artificial.pressure} we show the removal of these terms has no effect except to eliminate some obviously unphysical resolution-scale artifacts in the cold gas, as expected. 

\item{\bf Updated Cooling Tables \&\ SNeII Yields:} The physical mechanisms of stellar feedback, and assumptions about stellar evolution, are the same between FIRE-1 and FIRE-2. This means SNe rates (Ia and II), wind mass loss rates and kinetic luminosities, bolometric luminosities and luminosities in different bands, yields, etc., are the same. We have made one minor update: in FIRE-1, we used the SNe II yields of \citet{woosley.weaver.1995:yields}; however, it is widely known that these older models significantly under-predict the observed yields in {\small Mg} and {\small Ne}, and we confirmed this in \citet{ma:2015.fire.mass.metallicity}. We have therefore updated this to the more recent \citet{nomoto2006:sne.yields} yields, which remedies this issue. We stress, though, that for all other species (especially {\small C} and {\small O}, which constitute most of the metal mass and are the dominant coolants), the IMF-averaged yield is within $\sim 10\%$ of \citet{woosley.weaver.1995:yields}. Since {\small Mg} and {\small Ne} are negligible coolants, this has no detectable effect on our main results. Similarly, the cooling physics is the same in FIRE-1 and FIRE-2. However we have updated some of the actual fitting functions used to compute the cooling functions (specifically for the recombination rates, photo-electric heating including PAHs, optically-thick cooling, and dust cooling), to match more accurate cooling tables made public since FIRE-1 was developed. For the sake of transparency and clarity, a {\em complete} set of fits to the FIRE-2 stellar evolution, yield, and cooling tabulations are presented in Appendices~\ref{sec:stellar.evolution.approximations}-\ref{sec:cooling.approximations}. 

\item{\bf Code Optimization, Higher Resolution:} For FIRE-2, we have made a number of purely numerical optimizations to the {\small GIZMO} code, to improve speed and parallelization efficiency (for details, see Appendix~\ref{sec:scaling.details}). We have also re-compiled some lookup tables and re-fit cooling functions for greater precision. This has no effect on our results, of course, but it has allowed us to run new simulations at even higher resolution compared to FIRE-1. 

\end{enumerate}

\begin{figure}
    \hspace{-0.3cm}
    \includegraphics[width=1.05\columnwidth]{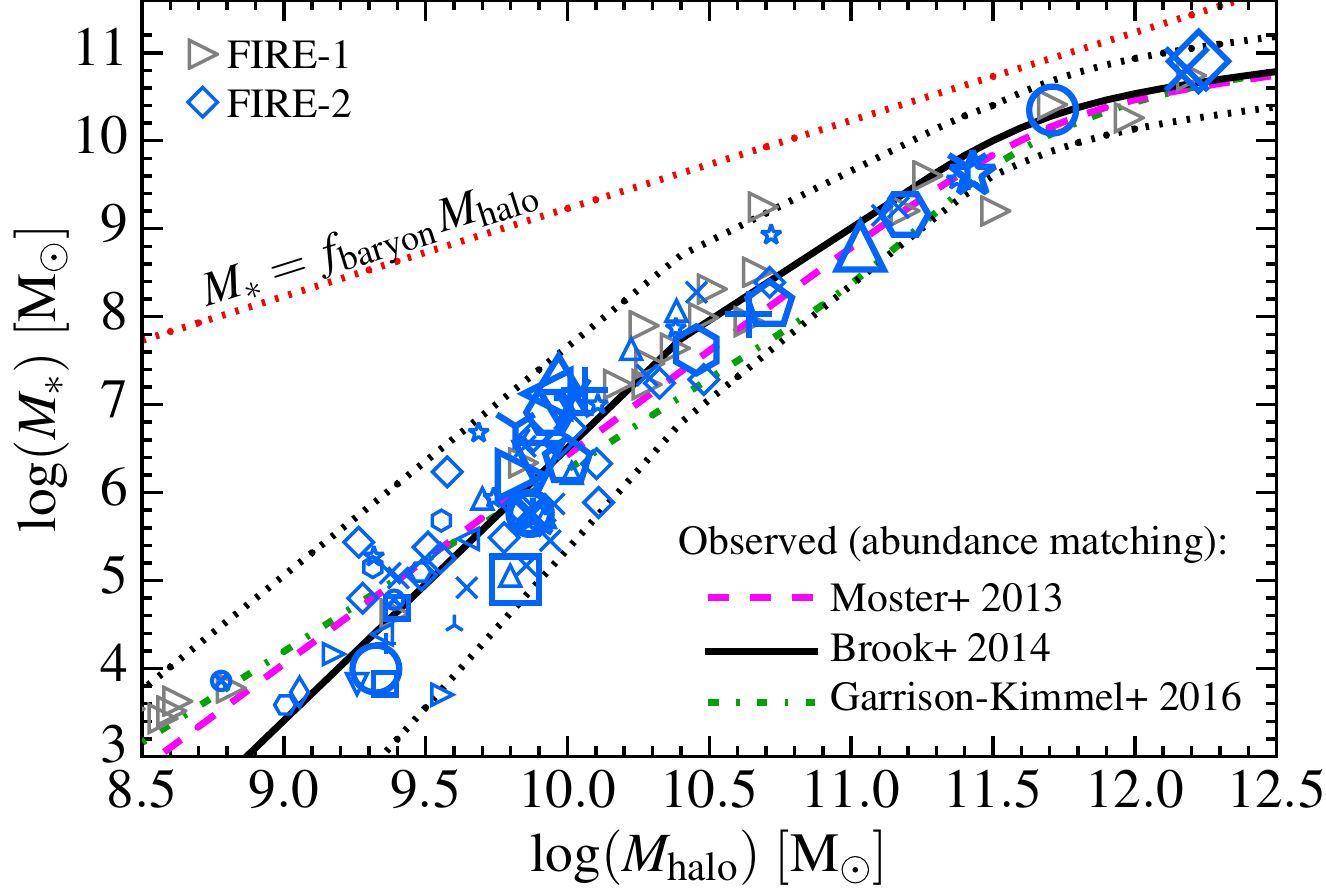}
    \vspace{-0.6cm}
    \caption{Stellar mass-halo mass relation for FIRE-2 simulations (colored points) at $z=0$. Stellar masses and halo virial masses are defined as in Table~\ref{tbl:sims}, for all resolved, uncontaminated halos (116 galaxies total; see text, \S~\ref{sec:results:overview}). Large points show the ``primary'' (most massive) galaxy within the zoom-in region, in each simulation (different point styles). Grey triangles show FIRE-1 simulations. While individual galaxies may differ in mass, the effects are primarily stochastic: the two agree well on average. We compare observational estimates as labeled; black dotted lines show the observationally estimated $\sim 95\%$ intrinsic scatter (see text). Within the scatter and systematic variations between fits, the simulations agree well with the observations at all masses.
    \label{fig:mgal.mhalo}}
\end{figure}

\vspace{-0.5cm}
\section{Basic Results \&\ Comparison Between FIRE-1 \&\ FIRE-2}
\label{sec:results:overview}

Table~\ref{tbl:sims} summarizes all the production-quality FIRE-2 simulations run as of writing this paper. For each, we give the ($z=0$) halo virial mass, virial radius, stellar mass of the ``target'' galaxy (the galaxy used to identify the initial zoom-in region), half-mass radius of the target galaxy, mass resolution of the simulation, and some values describing the ``spatial resolution'' (because our simulations are Lagrangian, mass resolution is well-defined, but ``spatial resolution'' is inherently variable: we discuss this in detail in \S~\ref{sec:resolution:spatial}). We have considered simulations spanning a $z=0$ halo mass range from $M_{\rm halo} \sim 10^{9}-10^{12}$, similar to our FIRE-1 simulations. All the simulations here have been run to redshift $z=0$.

Fig.~\ref{fig:images.m12} shows both face-on and edge-on images of two of our FIRE-2 MW-mass systems ({\bf m12i} and {\bf m12f}), at the highest resolution we have studied ($m_{i,\,1000}=7$). These use {\small STARBURST99} (the same assumptions used {\em in-code}) to compute the stellar spectra as a function of age and metallicity for each star particle, and then ray-trace through the ISM assuming a constant dust-to-metals ratio and physical dust opacities to volume-render the observed images in each band, which we use to construct a mock HST $u/g/r$ composite image as seen by a distant observer. Fig.~\ref{fig:images.fisheye.m12f} shows images from within the galaxy: we select a random star $\sim 10\,$kpc from the galactic center and construct a Galactic Aitoff projection of the ray-traced image from all stars in the galaxy to the mock observer. Fig.~\ref{fig:images.dwarfs} shows images of several dwarf galaxies from the ultra-faint through LMC mass scales. 

Fig.~\ref{fig:demo} shows several properties of a representative subset of our simulations: the star formation rate and stellar mass versus time (archeological formation history of stars within the $z=0$ galaxy); the stellar mass-weighted mean metallicity of those stars versus time; the $z=0$ baryonic and total mass profiles; and the $z=0$ circular velocity curve. Each property is measured for the ``target'' galaxy in the simulation. Essentially all of our high-resolution simulations show qualitative behavior broadly similar to one of the galaxies plotted.

It is not our intention in this paper to explore a quantitative comparison of the simulations and observations: this will be the subject of future work. For example, detailed comparison of the scaling relations of galaxy angular momentum, rotation curves, sizes and the Tully-Fisher relation can be found in \citet{elbadry:fire.morph.momentum,elbadry:HI.obs.gal.kinematics}, while \citet{fitts:fire.dwarf.concentration.mass} and \citet{chan:fire.udgs} compare the size, structural properties, and surface brightness distribution functions of dwarfs, and \citet{garrisonkimmel:fire.morphologies.vs.dm} the sizes of Milky Way-mass systems at $z=0$ (and \citealt{ma:fire.reionization.epoch.galaxies.clumpiness.luminosities} at $z\gtrsim 5$), as a function of halo properties. We will show mass profiles, however, so that one can infer where they are sensitive to the numerical choices explored in this paper.

Fig.~\ref{fig:fire1vs2} compares the galaxies for which we have both production-quality FIRE-1 and FIRE-2 simulations. Fig.~\ref{fig:images.fire1vs2} compares the visual morphology of the same galaxies. Here we can directly compare formation histories and morphologies of the same galaxy, with our improved numerical methods. 

In Fig.~\ref{fig:mgal.mhalo}, we plot the stellar mass-halo mass relation for our FIRE-1 and FIRE-2 simulations, compared to observations. We identify all resolved, un-contaminated halos in the high-resolution region and plot their virial masses and the stellar mass of the {\em central} galaxy in each halo,\footnote{We use the {\small HOP} halo finder \citep{eisenstein:1998.hop.halo.finder} to identify halos in Fig.~\ref{fig:mgal.mhalo}, for the sake of consistency with the FIRE-1 results published in \citet{hopkins:2013.fire}. This combines an iterative overdensity identification with a saddle density threshold criterion to merge subhalos and overlapping halos. We define halo mass $M_{\rm vir}$ and radius $R_{\rm vir}$ as the \citet{bryan.norman:1998.mvir.definition} virial mass/radius. We discard any halo outside the fully high-resolution region ($>1\%$ contamination by mass, from low-resolution particles), as well as unresolved halos (with $<5\times10^{4}$ DM, $<100$ baryonic, or $<10$ star particles), and subhalos (any halo within $<2\,R_{\rm vir}$ of a more massive halo center). The exact value of these cuts makes no difference to our conclusions. 
We define central stellar mass as in Table~\ref{tbl:sims} iteratively by first measuring the half-mass radius of all stars within a large cut (inside $15\%$ of $R_{\rm vir}$), then taking all stars within $3\times$ this radius (and then re-defining the half-stellar mass radius on these stars). This eliminates all satellites we identify by visual inspection and gives results reasonably close to fitting mass profiles of the central system (a detailed mock observational study is presented in \citealt{price:FIRE.size.mass.recovery.z2}; but for an exponential disk this recovers $97\%$ of the mass). Using a simpler cut of all stars at $<0.1\,R_{\rm vir}$ gives similar results, except one case with a $z\approx0$ merger; but for massive galaxies the $0.1\,R_{\rm vir}$ cut includes stars that are clearly part of the halo.
} which is (as expected) always smaller than the total mass in $R_{\rm vir}$ plotted in Fig.~\ref{fig:demo}. The FIRE-1 results here are taken directly from \citet{hopkins:2013.fire}. We compare these predictions to recent observational constraints, from a combination of abundance matching and mass modeling. The observational fit from \citet{moster:2013.abundance.matching.sfhs} only includes galaxies with $M_{\ast} \gtrsim 10^{9}\,\msun$ (so it is extrapolated here), but this extrapolation has been shown to reproduce well the observed local group dwarf luminosity functions to $\sim 10^{4}\,\msun$ \citep{sgk:2016.mgal.mhalo.lowmass.scatter}, so we consider it as well. \citet{brook:2014.mgal.mhalo.local.group} combine the constraints from \citet{behroozi:2012.abundance.matching.sfhs} at high masses ($M_{\ast}>10^{9}\,\msun$) with local group and field dwarf constraints. \citet{sgk:2016.mgal.mhalo.lowmass.scatter} perform a similar exercise, allowing the scatter below $M_{\rm halo}<10^{11}\,\msun$ to vary; we plot their best-fit median relations for a constant scatter below this mass $\approx1\,$dex (but note that for any scatter in the range $0.5-2\,$dex, the results are similar). We expect the scatter to vary continuously with mass, so we show the $95\%$ inclusion contour if we take the model from \citet{sgk:2016.mgal.mhalo.lowmass.scatter} where the scatter is constant at $0.2\,$dex above $M_{\rm halo}>10^{11.5}\,\msun$ (the value favored by \citealt{behroozi:2012.abundance.matching.sfhs} and \citealt{moster:2013.abundance.matching.sfhs}), and varies linearly with halo mass as $\sigma_{\rm dex} = 0.2 -0.2\,\log_{10}(M_{\rm halo}/10^{11.5}\,\msun)$ at lower masses (rising to $\approx 0.5\,$dex at $M_{\rm halo}\approx 10^{10}\,\msun$).\footnote{At ultra-faint stellar masses $\lesssim 1000\,M_{\sun}$, it is likely that details of first star (Pop III) formation and primordial molecular (metal-free) cooling, not treated explicitly in FIRE, become important. This will be the subject of future study, but we consider these stages un-resolved given our mass resolution here, and simply initialize a metallicity ``floor'' of $10^{-4}\,Z_{\sun}$.}

We emphasize that although matching the full observed (2D) distribution of galaxies in stellar mass-halo mass space is equivalent to matching the observed stellar mass function (SMF), with the limited sample here, we can only test whether our simulations are consistent with being drawn from this distribution (we do not have a sufficiently large ensemble of halos to forward-model the entire SMF). However, in \citet{wetzel.2016:latte} and with a much larger sample in Garrison-Kimmel et al.\ (in prep), we have sufficient statistics to forward-model the observed dwarf galaxy SMF and compare to the local (Milky Way and M31) observed satellite and local field SMFs, at stellar masses $\sim 10^{4}-10^{9}\,M_{\sun}$. And in \citet{ma:fire2.reion.gal.lfs} we use an expanded sample of simulations run to high-redshift to compare the SMF (and luminosity functions) at $\sim 10^{6}-10^{11}\,M_{\sun}$ to observations at $z\gtrsim 5$. However more quantitative comparison to the SMF of massive galaxies at lower redshifts ($z\sim0-2$) will require larger statistical volumes.

In any case, in Figs.~\ref{fig:images.m12}-\ref{fig:mgal.mhalo}, we confirm the conclusions of our previous FIRE-1 studies. Although it is impossible to be exhaustive at this point, we have yet to identify any area in which the FIRE-2 predictions differ at the order-of-magnitude level from FIRE-1 predictions. In future work, we will examine detailed properties of the CGM (e.g.\ column density distributions of different absorbers) where the hydrodynamic solver could, in principle, have a larger effect.

As in FIRE-1, in FIRE-2 the dwarfs tend to have spheroidal morphologies, with relatively little rotation \citep[see][]{wheeler.2015:dwarfs.isolated.not.rotating}. Especially around $M_{\rm halo}\sim 10^{11}\,\msun$, repeated bursts of star formation driving cycles of outflow, subsequent infall, and repeated star formation leads to ``puffing up'' of the dark matter and stellar orbits, generating large cores in the dark matter profiles \citep{onorbe:2015.fire.cores,chan:fire.dwarf.cusps}. This also leads to expansion of the galaxy size and low surface brightness in their stellar distribution \citep{elbadry.2015:core.transformation.stellar.kinematics.gradients.in.dwarfs}. ``Bursty'' star formation predominates in low-mass dwarfs and high-redshift progenitors of massive galaxies \citep{sparre.2015:bursty.star.formation.main.sequence.fire, cafg:bursty.sf.toymodel}, and in galactic nuclei \citep{torrey.2016:fire.galactic.nuclei.star.formation.instability}. In FIRE-1 and FIRE-2, MW-mass galaxies develop a clear disk+bulge morphology, with large {\em thin} stellar disks (despite the presence of strong feedback; \citealt{ma:2016.disk.structure}), with clear spiral structure, pronounced radial metallicity and age gradients \citep{ma:radial.gradients}. The stellar (and gas phase) metallicities agree well with the observed mass-metallicity relation both as a function of stellar mass at $z=0$ and as a function of redshift \citep{ma:2015.fire.mass.metallicity}; to the extent that galaxies differ in metallicity between FIRE-1 and FIRE-2, they primarily move {\em along} the stellar mass-metallicity relationship. In both sets of simulations, galaxies drive strong winds, with higher mass-loading in low mass galaxies, sufficient to place galaxies on the observed relationship between stellar mass and halo mass \citep{hopkins:2013.fire,muratov:2015.fire.winds,hayward.2015:stellar.feedback.analytic.model.winds}. Initial examination of our FIRE-2 runs shows these winds produce galactic outflow rates and covering factors of neutral hydrogen similar to our FIRE-1 simulations, similar to observations at a wide range of galaxy masses \citep{faucher-giguere:2014.fire.neutral.hydrogen.absorption,faucher.2016:high.mass.qso.halo.covering.fraction.neutral.gas.fire}, although this will be studied in more detail in the future. 

There are some modest quantitative differences between FIRE-2 and FIRE-1; most of this manuscript will explore the origin of these differences, but they appear to primarily owe to the change in the hydrodynamic solver. On average, dwarfs are slightly lower-mass in FIRE-2; this owes to both hydrodynamics and the fact that (unlike in FIRE-1) we do not artificially ``ignore'' ionizing photons when they pass outside numerical domains (hence they can still heat gas). The difference in Fig.~\ref{fig:fire1vs2} can be as large as a factor of $\sim 3$ for a single galaxy, but this is largely stochastic (since the star formation histories are dominated by a few bursts, a small perturbation to the formation history or feedback strength can lead to factor $\sim2$ changes in mass) -- Fig.~\ref{fig:mgal.mhalo} makes this clear, as the {\em systematic} offset between FIRE-1 and FIRE-2 appears to be a factor $<2$, well within the systematic uncertainties in the $M_{\ast}-M_{\rm halo}$ relation at $M_{\rm halo} \le 10^{11}\,\msun$. We have also examined this relation at $z=(0.5,\,1,\,2,\,4,\,6)$ and find similar agreement, so we refer to \citet{hopkins:2013.fire} for an extensive analysis. Metallicities at the same mass are slightly higher, by $<30\%$; this owes primarily to updated yield tables; this is far smaller than the factor of $\sim 2-5$ systematic uncertainty in the calibration of observed galaxy metallicities. For MW-mass galaxies, the stellar masses are slightly higher in FIRE-2, and the bulges slightly more concentrated (the rotation curves have stronger peaks, by a modest amount). We will show that this is a direct consequence of the hydrodynamic treatment, but is also sensitive to simulation resolution and the SNe coupling algorithm. 


\begin{footnotesize}
\ctable[
  caption={{\normalsize Illustrative Examples of ``Resolution'' Scales in a MW-Mass Halo ({\bf m12i})}\label{tbl:res}},center,star]{ccccccc}{
\tnote[ ]{Several ``spatial resolution'' and ``time resolution'' properties of the simulations discussed in \S~\ref{sec:resolution}. We focus on our resolution study of the {\bf m12i} system, but results for other MW-mass galaxies are nearly identical, and results for dwarf galaxies are qualitatively similar at similar mass resolution. Because our simulations are Lagrangian, only mass resolution is truly ``fixed.'' Spatial (both hydrodynamic and gravitational, which are the same always) and time resolution are both adaptive, and in principle can reach arbitrarily small values, but in practice reach minimum values based on the densest mass-resolved structures in the simulation.}
}{
\hline\hline
\multicolumn{1}{c}{\,} &
\multicolumn{1}{c}{\,} &
\multicolumn{4}{c}{Resolution Level Run to $z=0$} \\ 
\multicolumn{1}{c}{Property} &
\multicolumn{1}{c}{Notation} &
\multicolumn{1}{c}{$m_{i,\,1000} = 450$} &
\multicolumn{1}{c}{$m_{i,\,1000} = 56$} &
\multicolumn{1}{c}{$m_{i,\,1000} = 7$} &
\multicolumn{1}{c}{$m_{i,\,1000} = 0.88$ (DM-only)} \\ 
\hline
& & & & & \\
Particle Number & $N_{\rm tot}$ & $2.5\times10^{6}$ & $2.0\times10^{7}$ & $1.4\times10^{8}$ & $5.6\times 10^{8}$ \\ 
Baryonic Particle Mass ($\msun$) & $m_{i}$ &  $4.4\times10^{5}$ &  $5.6\times10^{4}$ & $7070$ & -- \\ 
Minimum Timestep (yr) & $\Delta t_{\rm min}$ & $600$ & $260$ & $120$ & $1000$ (no gas) \\ 
& & & & & \\
\hline 
\multicolumn{6}{c}{Star-Forming Densities:} \\
\hline 
& & & & & \\
Minimum Density of Star Formation (${\rm cm^{-3}}$) &  $n_{\rm SF,\,min}$ & $100$ & $1000$ & $1000$ & -- \\ 
Mean Density of Star Formation (${\rm cm^{-3}}$) &  $\langle n_{\rm SF} \rangle$ & $700$ & $1900$ & $3400$ & --\\
& & & & & \\
\hline 
\multicolumn{6}{c}{Gas Resolution (Inter-Particle Separation $=$ Force Softening) at Star-Forming Densities:} \\
\hline 
  & & & & & \\
Minimum Spatial Resolution (pc) & $h_{i}^{\rm min}$ & $5.0$ & $1.4$ & $0.38$ & -- \\
Spatial resolution at $n_{\rm SF,\,min}$ (pc) & $h_{i}^{\rm threshold}$ & $57$ & $13$ & $7$ & -- \\ 
Spatial resolution at $\langle n_{\rm SF} \rangle$ (pc) & $h_{i}^{\langle SF \rangle}$ & $30$ & $10$ & $4.6$ & --\\
& & & & & \\
\hline 
\multicolumn{6}{c}{Jeans Scales in Warm ($10^{4}$\,K) ISM, Corresponding to Marginally-Resolved (10-element) Structures:}\\
\hline 
& & & & & \\
Minimum Jeans Radius $=L_{\rm Jeans}/2$ (pc) & $\lambda^{J,\,WIM}$ & $130$ & $17$ & $2.1$ & -- \\ 
Maximum Density with Resolved $M_{\rm Jeans}$ (${\rm cm^{-3}}$) & $n_{\rm max}^{J,\,WIM}$ & $20$ & $1200$ & $7.4\times10^{4}$ & -- \\ 
& & & & & \\
\hline 
\multicolumn{6}{c}{Turbulent Jeans Scales in Cold Clouds ($T=10\,$K, $\Sigma_{\rm GMC} = 300\,M_{\sun}\,{\rm pc^{-2}}$), Corresponding to Marginally-Resolved (10-element) Structures:}\\
\hline 
& & & & & \\
Minimum Turbulent Jeans Radius (pc) & $\lambda^{{\rm turb},\,CNM}$ & $42$ & $15$ & $5.3$ & --\\
Maximum Density with Resolved $M_{\rm Jeans}^{\rm turb}$ (${\rm cm^{-3}}$) & $n_{\rm max}^{{\rm turb},\,CNM}$ & $570$ & $1600$ & $4600$ & --\\
& & & & & \\
\hline 
\multicolumn{6}{c}{Dark Matter Resolution:} \\
\hline 
& & & & & \\
Particle Mass ($\msun$) & $m_{\rm DM}$ &  $2.7\times10^{6}$ & $3.4\times10^{5}$ & $4.3\times10^{4}$ & $5400$ \\ 
Minimum Inter-Particle Separation (pc) & $h_{i}^{\rm DM,\,min}$ &  $66$ & $32$ & $16$ & $8.2$ \\ 
RMS Separation within galaxies at $z=0$ (pc) & $h_{i}^{\rm DM,\,core}$ &  $330$ & $150$ & $70$ & $38$ \\ 
DM Convergence Radius, $t_{\rm relax}=0.06\,t_{\rm circ}(R_{200})$ (pc) & $r_{0.06}$ & $670$ & $300$ & $150$ & $75$ \\ 
& & & & & \\
\hline 
\multicolumn{6}{c}{Typical ``N-Body Heating'' Rates:} \\
\hline 
& & & & & \\
Gas-Gas Scattering (${\rm erg\,cm^{3}\,s^{-1}}$) & $\langle Q_{\rm heat}^{\rm gas-gas} \rangle$ &  $8\times10^{-30}$ & $1\times10^{-30}$ & $1\times10^{-31}$ & --\\
Gas-DM Scattering (${\rm erg\,cm^{3}\,s^{-1}}$) & $\langle Q_{\rm heat}^{\rm gas-DM} \rangle$ &  $2\times10^{-27}$ & $2\times10^{-28}$ & $5\times10^{-30}$ & --\\
& & & & & \\
\hline\hline\\
}
\end{footnotesize}

\vspace{-0.5cm}
\section{Resolution in the FIRE-2 Simulations}
\label{sec:resolution}

We now discuss our mass, spatial, and time resolution. For each, we will present a series of tests, and summarize these with a set of resolution criteria.

Throughout this paper, when we refer to ``convergence'' of some property, we more precisely mean to test whether the property is strongly, systematically sensitive to resolution (at our highest resolution). Because even pure $N$-body integration (let alone galaxy formation, with explicitly stochastic effects such as SNe included) is a fundamentally chaotic problem, and has no known exact solution, we of course cannot define convergence in the more formal sense (of e.g.\ some error norm relative to said solution). Some properties in nature (e.g.\ halo mass functions, or turbulence in the ISM) extend down to scales vastly smaller than any simulation could resolve -- for these, convergence must be defined ``down to'' some minimum resolvable scale (e.g.\ the mass function for objects larger than some minimum number of particles). This also means that ``convergence'' in one quantity should not be taken to necessarily imply convergence in another.

\vspace{-0.5cm}
\subsection{Mass Resolution}
\label{sec:resolution:mass}

\begin{figure*}
\plotsidesize{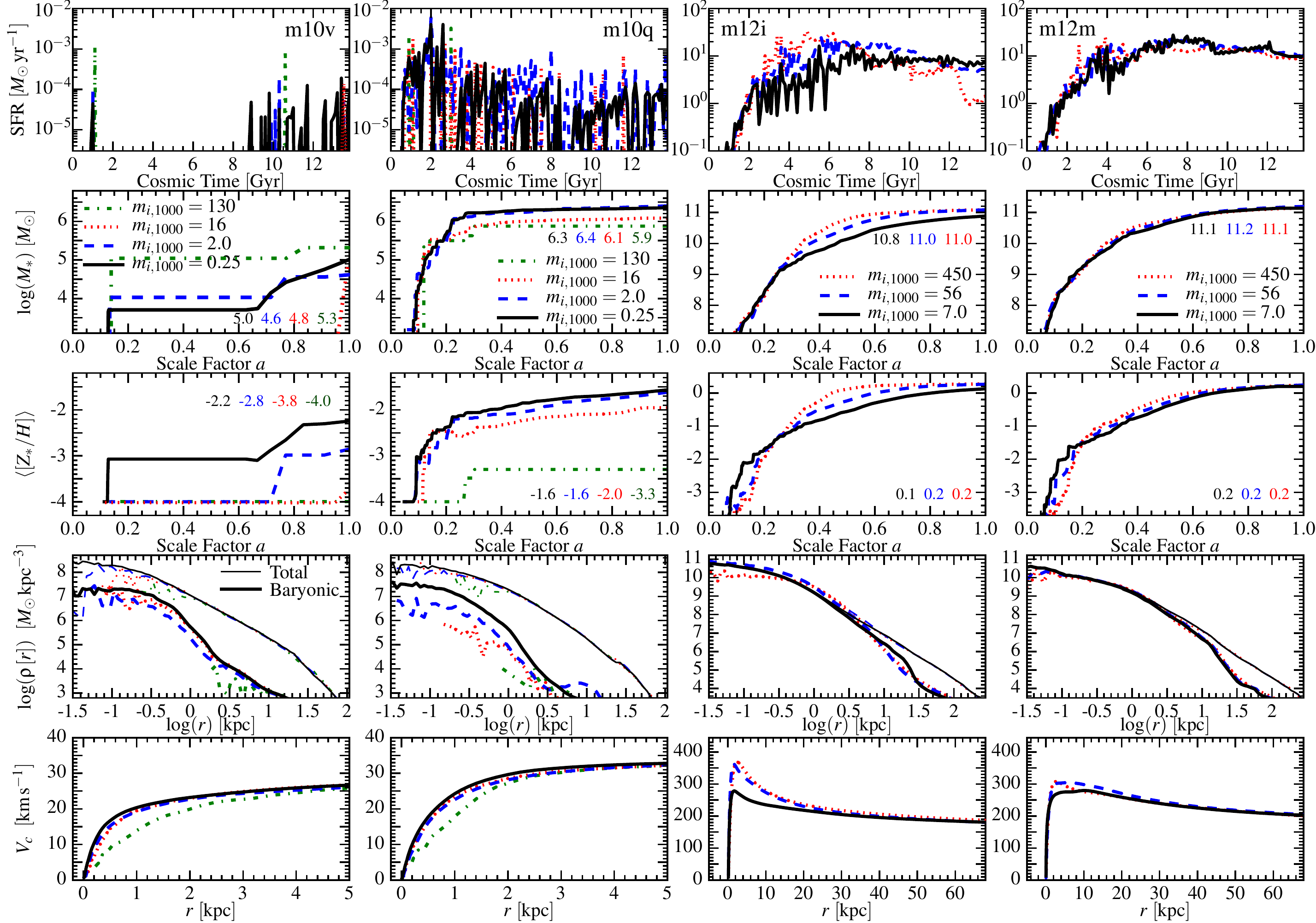}{1.0}
    \vspace{-0.5cm}
    \caption{Resolution study, as in Fig.~\ref{fig:demo}. Each column is one galaxy, and each line shows a different mass resolution ($m_{i,\,1000}$) up to our production resolution.
    {\em Top:} Star formation rate versus cosmic time. In dwarfs ({\bf m10v,\,q}), low resolution leads to artificially large ``burstiness'': a few bursts dominate the history and eject the remaining baryons. At higher resolution, star formation is still ``bursty'', but it becomes robust to changes in resolution. In MW-mass galaxies, higher resolution shifts star formation to slighly later times by more efficiently regulating the low-mass projenigtor at high redshifts.
    {\em Second from Top:} Stellar mass versus scale factor. Note that the $m_{i,\,1000}=16$ ($130$) simulations of {\bf m10v} have only $10$ ($2$) star particles in the main halo, and yet they are within a factor $\sim 3$ of the highest-resolution stellar mass; by the time $\sim 100$ particles are in the main galaxy, the mass is stable to within $\sim 20\%$. 
         {\em Middle:} Average stellar metallicity versus scale factor. This converges more slowly than stellar mass; with $<100$ star particles, dwarf galaxies show artificially suppressed metallicity, because low resolution under-samples the enrichment history and leads to artificially bursty SFH that blows out metals completely. Massive galaxies show smaller differences that match differences in their SFHs.
     {\em Second from Bottom:} Mass density profile at $z = 0$: results are robust down to radii enclosing $\sim100-200$ particles of the ``type'' of interest (details below).
     {\em Bottom:} Circular velocity profile at $z = 0$. For dwarfs, this is dark-matter dominated and therefore under-resolved only at our lowest resolution (where the DM ``convergence radius'' discussed below is $>$\,kpc). For the MW-mass galaxy, the modest difference in the SFH at $z \sim 0.5-2$ in {\bf m12i} translates to a factor $\sim 2$ difference in the central bulge mass, which leads to a more strongly peaked central $V_{c}$ at low resolution. {\bf m12f} (not shown) shows nearly identical behavior to {\bf m12i}; {\bf m12m} shows better agreement at all resolution levels.
    \label{fig:res.summary}}
\end{figure*}

\begin{figure*}
\begin{tabular}{ccccc}
\hspace{-0.25cm}
\includegraphics[width=0.2\textwidth]{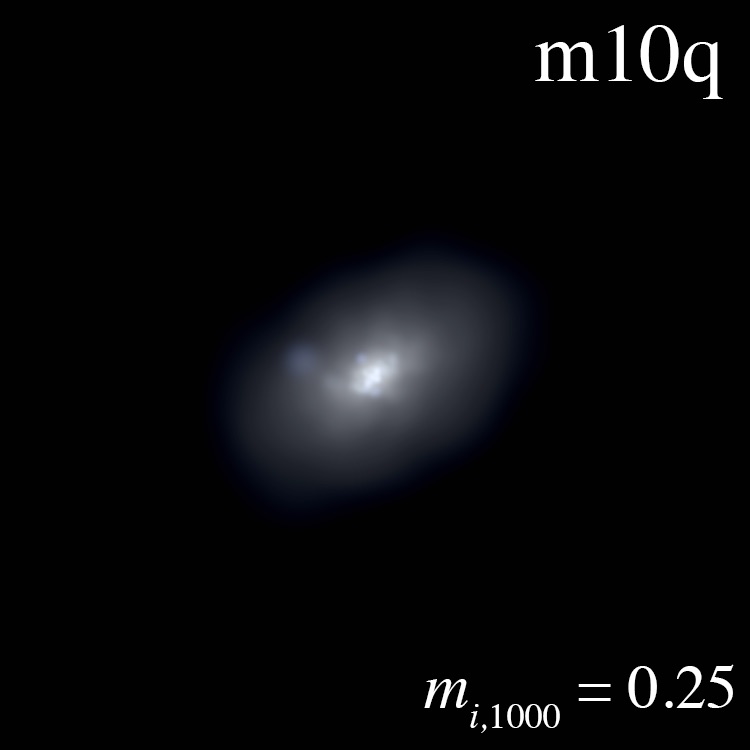} &
\hspace{-0.40cm}
\includegraphics[width=0.2\textwidth]{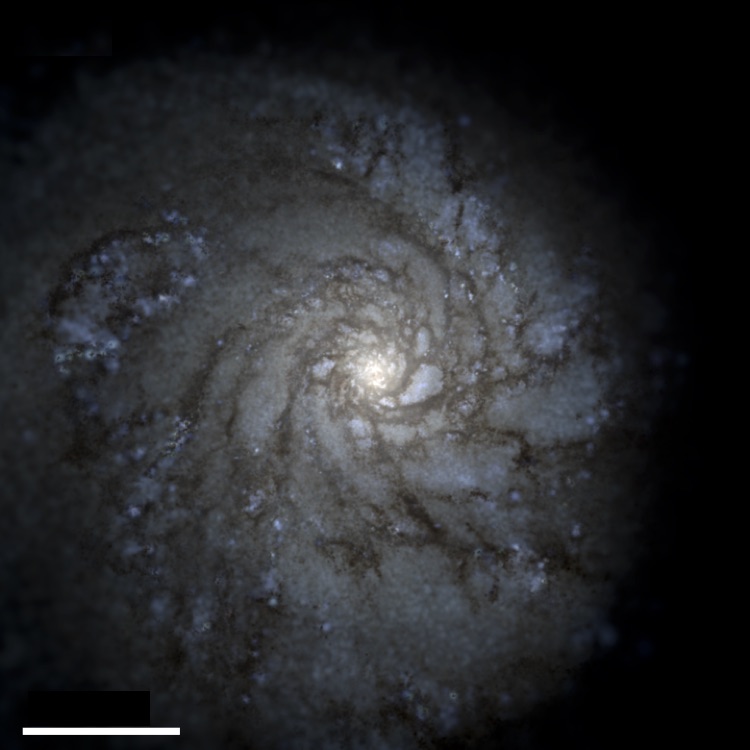} &
\hspace{-0.60cm}
\includegraphics[width=0.2\textwidth]{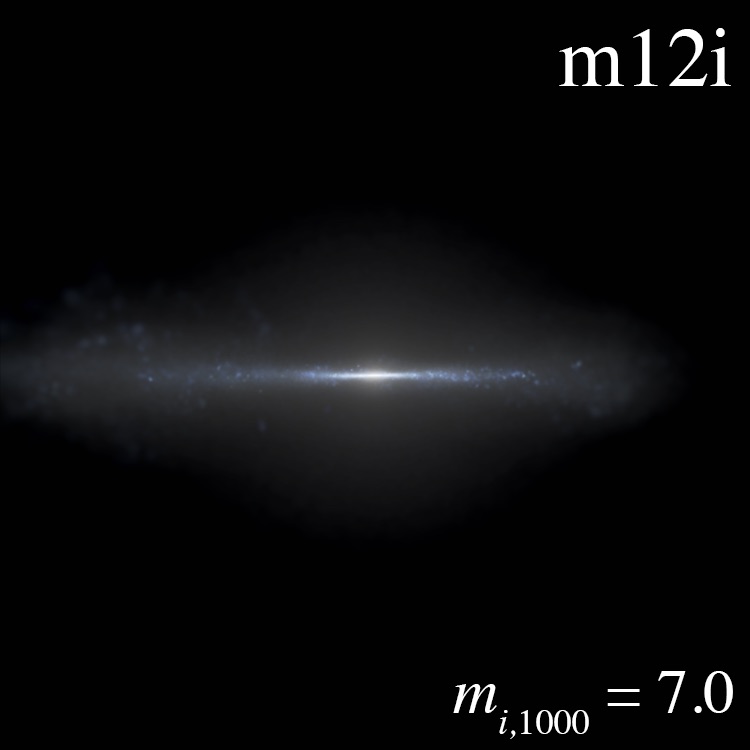} &
\hspace{-0.40cm}
\includegraphics[width=0.2\textwidth]{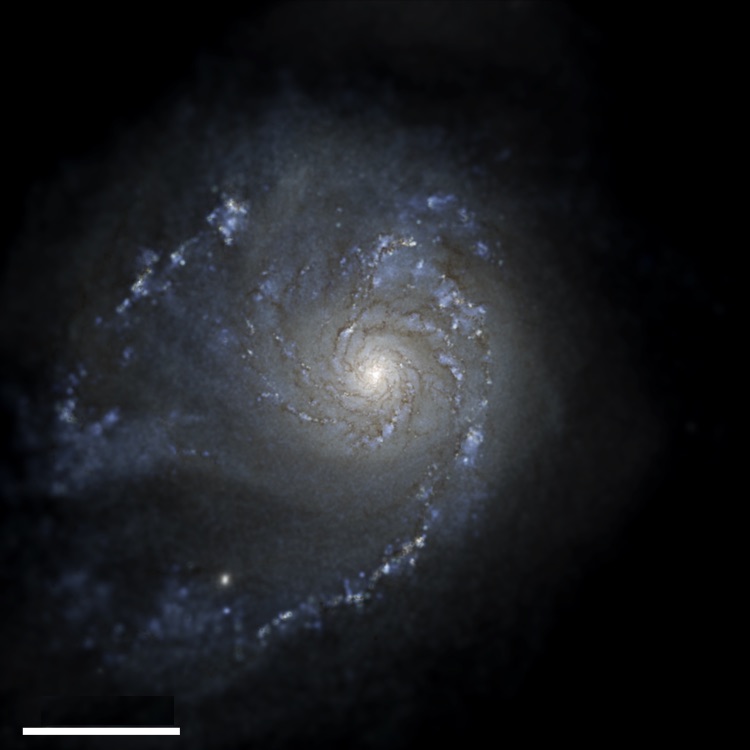} &
\hspace{-0.60cm}
\includegraphics[width=0.2\textwidth]{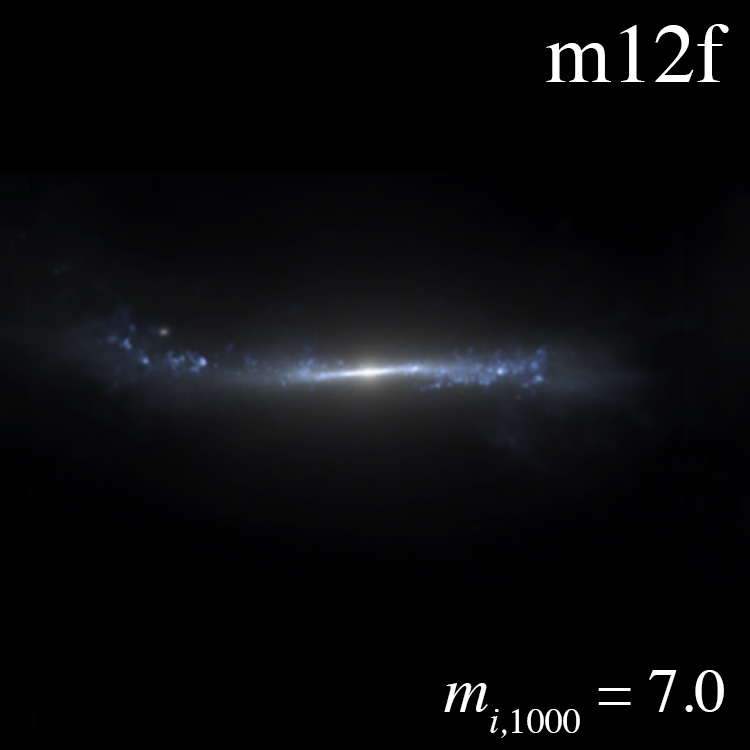} \\
\hspace{-0.25cm}
\includegraphics[width=0.2\textwidth]{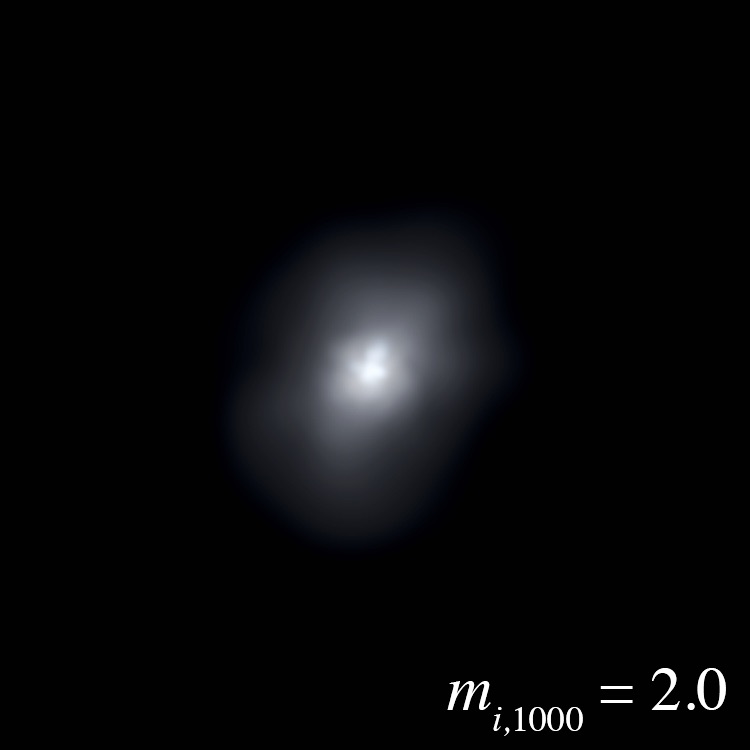} &
\hspace{-0.40cm}
\includegraphics[width=0.2\textwidth]{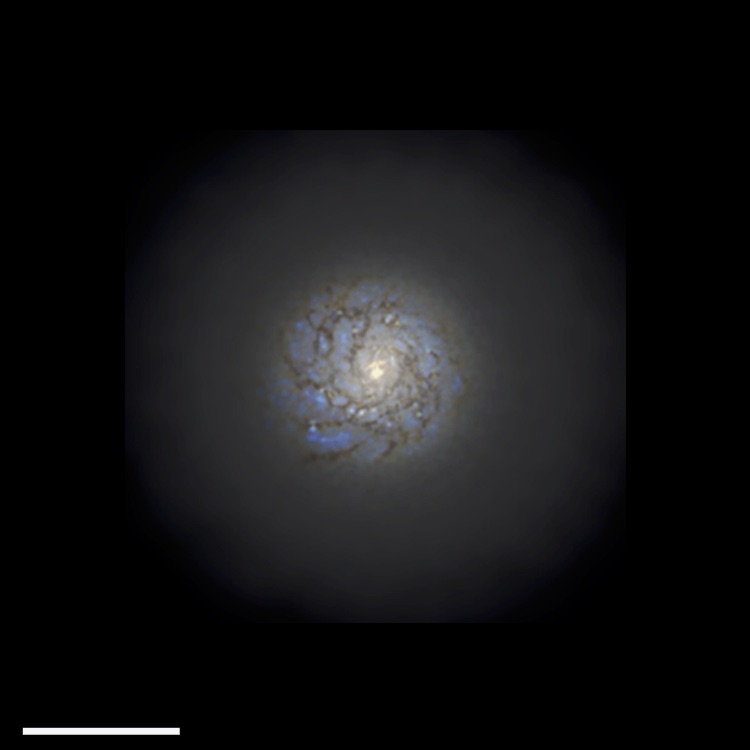} &
\hspace{-0.60cm}
\includegraphics[width=0.2\textwidth]{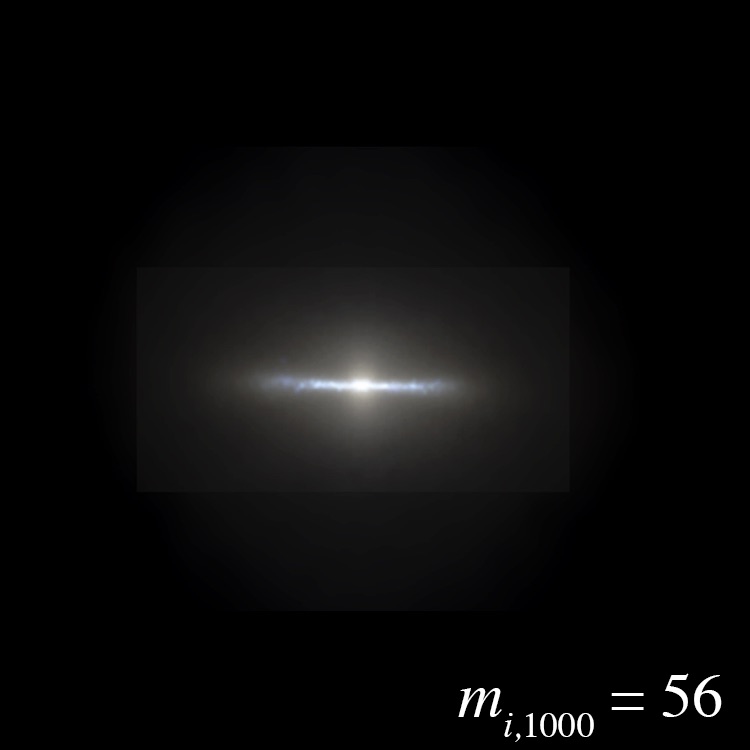} &
\hspace{-0.40cm}
\includegraphics[width=0.2\textwidth]{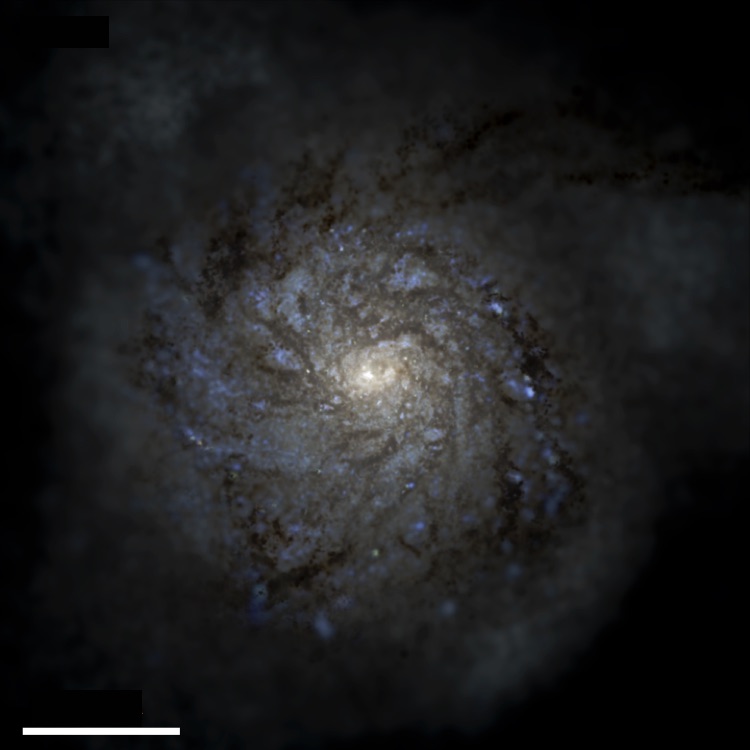} &
\hspace{-0.60cm}
\includegraphics[width=0.2\textwidth]{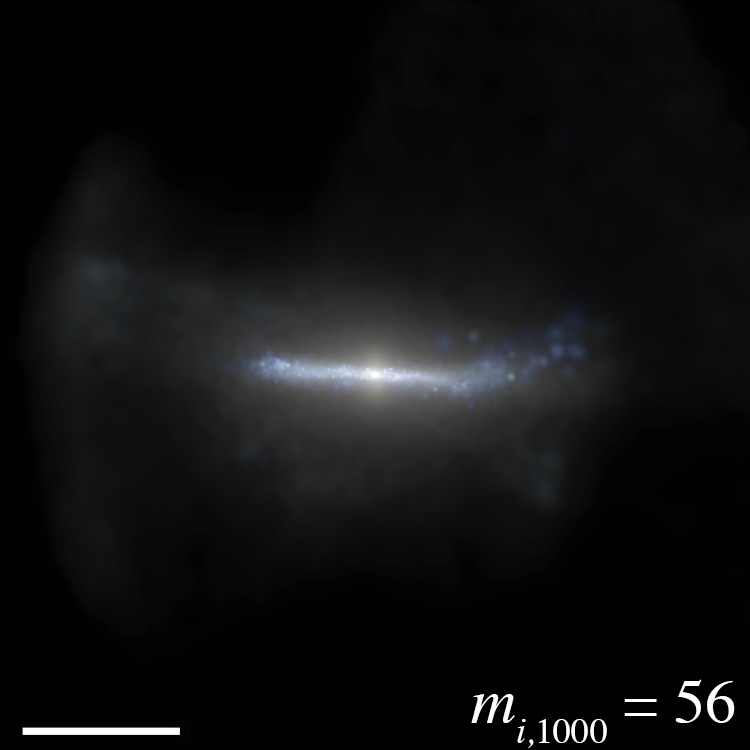} \\
\hspace{-0.25cm}
\includegraphics[width=0.2\textwidth]{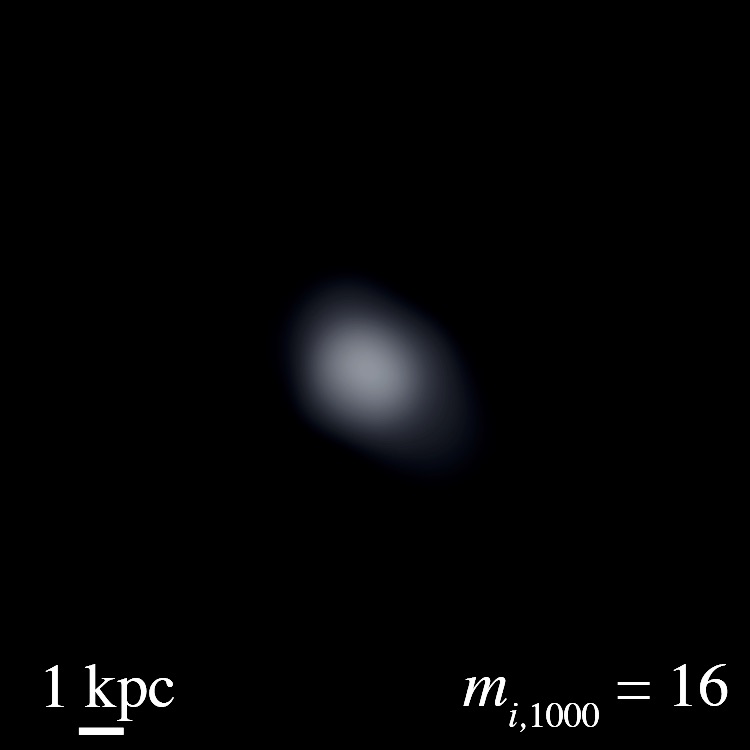} &
\hspace{-0.40cm}
\includegraphics[width=0.2\textwidth]{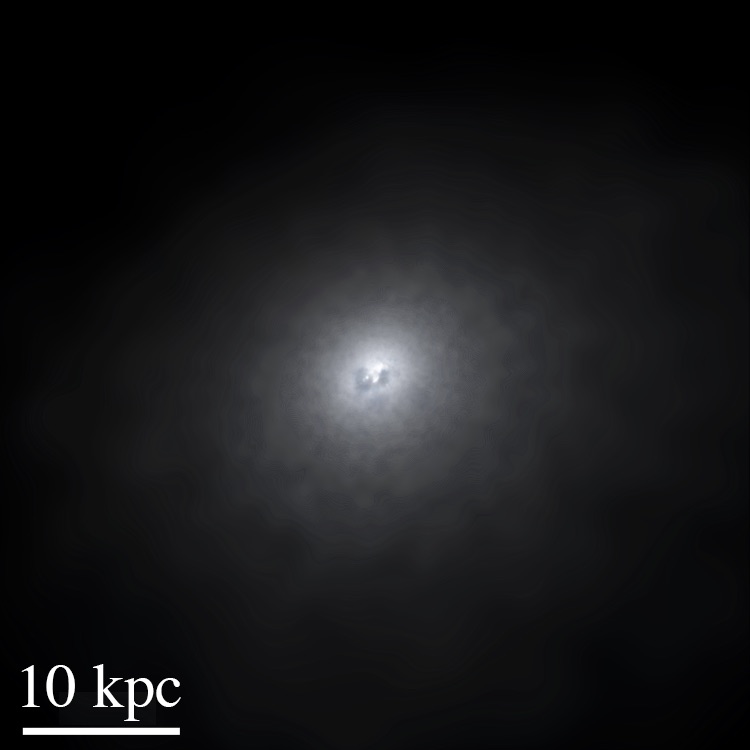} &
\hspace{-0.60cm}
\includegraphics[width=0.2\textwidth]{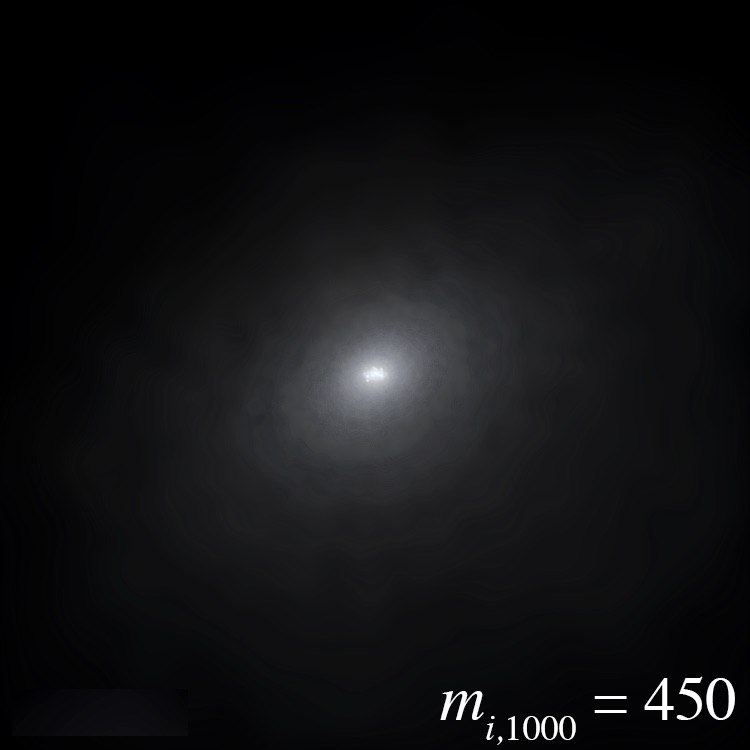} &
\hspace{-0.40cm}
\includegraphics[width=0.2\textwidth]{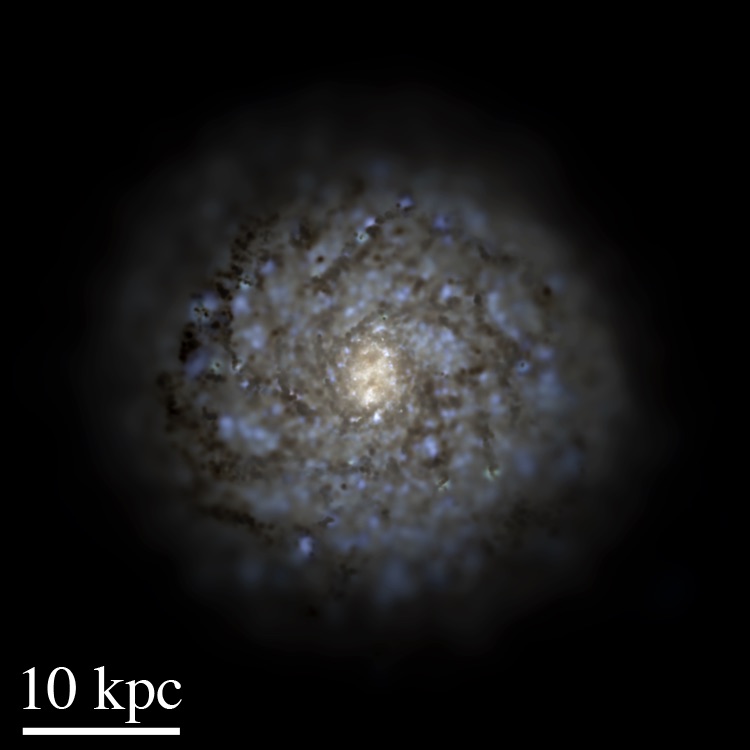} &
\hspace{-0.60cm}
\includegraphics[width=0.2\textwidth]{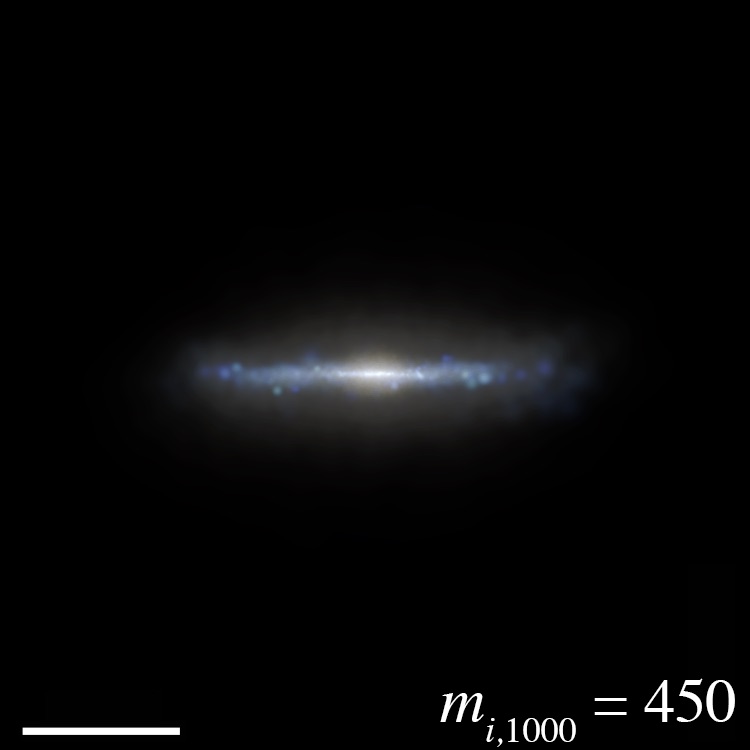} \\
\end{tabular}
    \vspace{-0.25cm}
    \caption{Mock images, as in Fig.~\ref{fig:images.dwarfs}, showing the effects of resolution on the morphology of the simulated galaxies at $z=0$, for a subset of the resolution series from Fig.~\ref{fig:res.summary}.
    {\em Left:} Dwarf galaxy ({\bf m10q}): because dwarf morphologies are disordered, their qualitative morphology does not depend sensitively on resolution (this is true for all our dwarfs).
    {\em Center:} MW-mass galaxy {\bf m12i}: In {\bf m12i} we see a trend towards a more extended thin disk component dominating as we increase resolution. The effective radius also increases, but much less dramatically (just by $\sim 40\%$ from $m_{i,\,1000}=56$ to $m_{i,\,1000}=7.0$); the bulge and central $\sim 2-3$\,kpc remain similar in each case, but the extended, gas-rich disk is much more prominent. At the lowest resolution ($m_{i,\,1000}=450$), we see little disk-like structure at all (although there is a rotating gas+stellar disk $\sim 1\,$kpc in size). At intermediate resolution, a clear disk with $\sim 5\,$kpc radius appears -- there is also an extended, smooth-light component out to $\sim 10\,$kpc. At our highest resolution, this extended component exhibits spiral structure and the gas+young stellar disk extends to $>10\,$kpc from the galaxy center (albeit at low surface brightness: all figures here use an $8$-magnitude stretch).
    {\em Right:} MW-mass galaxy {\bf m12f}: In this slightly more-massive galaxy, an extended disk is present at all resolution levels (the trends seen in {\bf m12i} are still present, but much weaker). {\bf m12m} (not shown) similarly shows a disk at every resolution level.
    \label{fig:images.resolution}}
\end{figure*}

\begin{figure}
\plotonesize{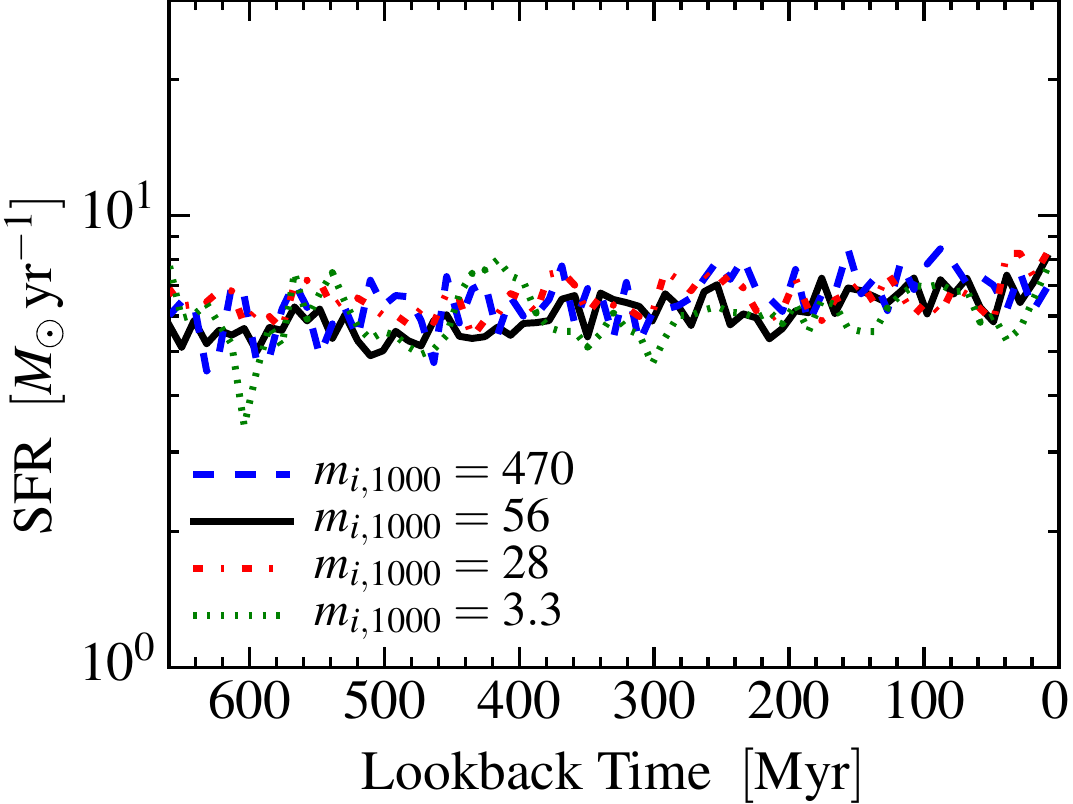}{1.0}
    \vspace{-0.5cm}
    \caption{Star formation rate versus lookback time, in our MW-mass {\bf m12i} simulations, re-started at late times. In each case, we re-start using the snapshot at $z=0.06$ from our run with $m_{i,\,1000}=56$ (in Fig.~\ref{fig:res.summary}) as our initial condition, so the late-time ICs of the different realizations are identical. We then run from $z \approx 0.06-0$ ($\sim 700\,$Myr physical time), to study how the SFR varies in a massive galaxy given the same initial galaxy properties (that is, factoring out how variations affect the earlier phases when the progenitor was low-mass). Here, we use particle splitting/merging before running to vary the mass resolution. We see that for the same ICs, the SFR is almost completely insensitive to resolution; variations in the SFR with resolution in Fig.~\ref{fig:res.summary} are dominated by (1) early stages when the progenitor galaxy was much lower mass, hence much less well resolved, and (2) less prominently, weak resolution dependence of wind mixing versus escape from the outer halo, which changes the inflow rate back into the galaxy at later times.
    \label{fig:sf.z0.mass.resolution}}
\end{figure}

\begin{figure*}
\begin{tabular}{cc}
\includegraphics[width=0.45\textwidth]{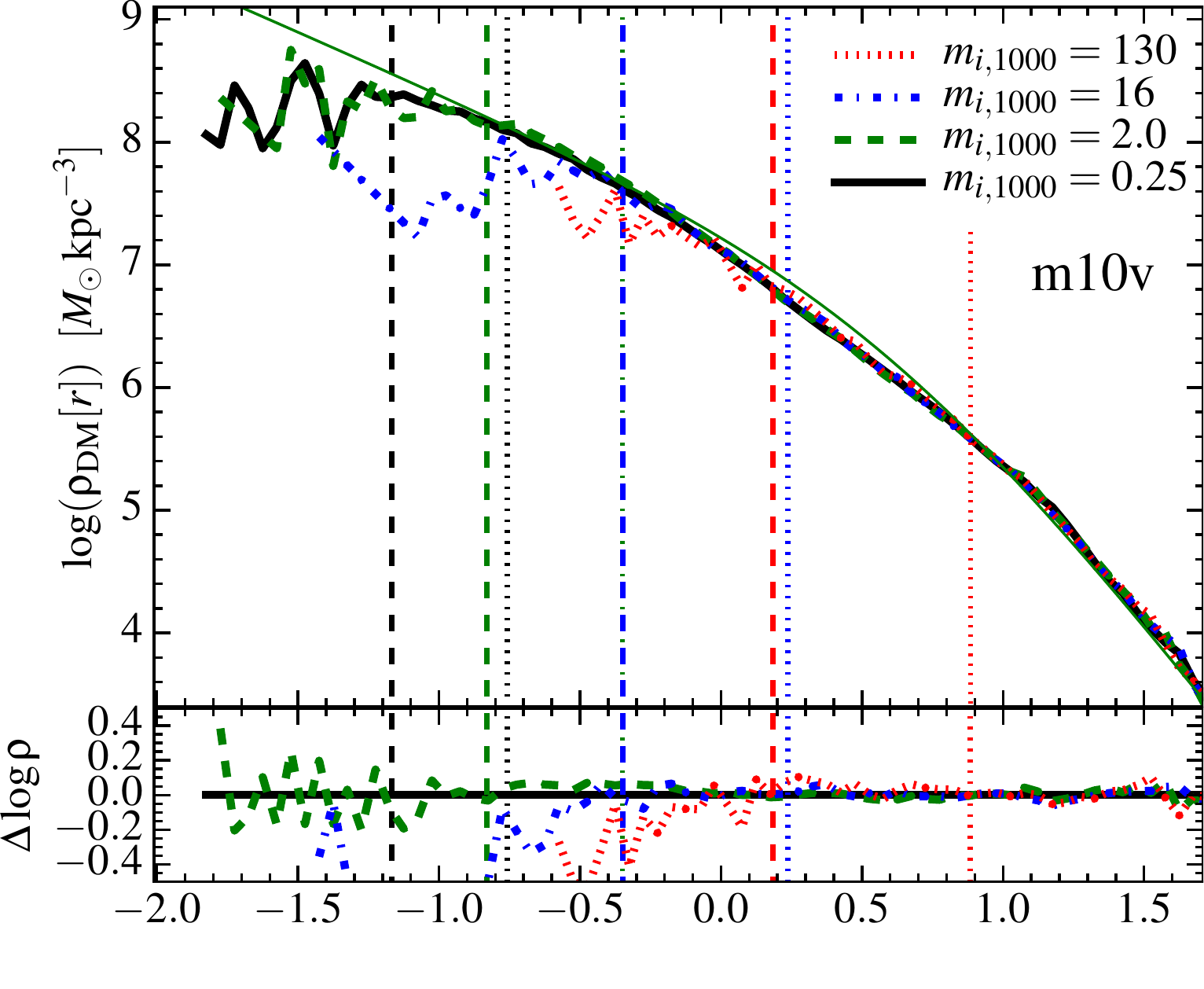} & 
\hspace{-0.7cm}
\vspace{-0.5cm}
\includegraphics[width=0.45\textwidth]{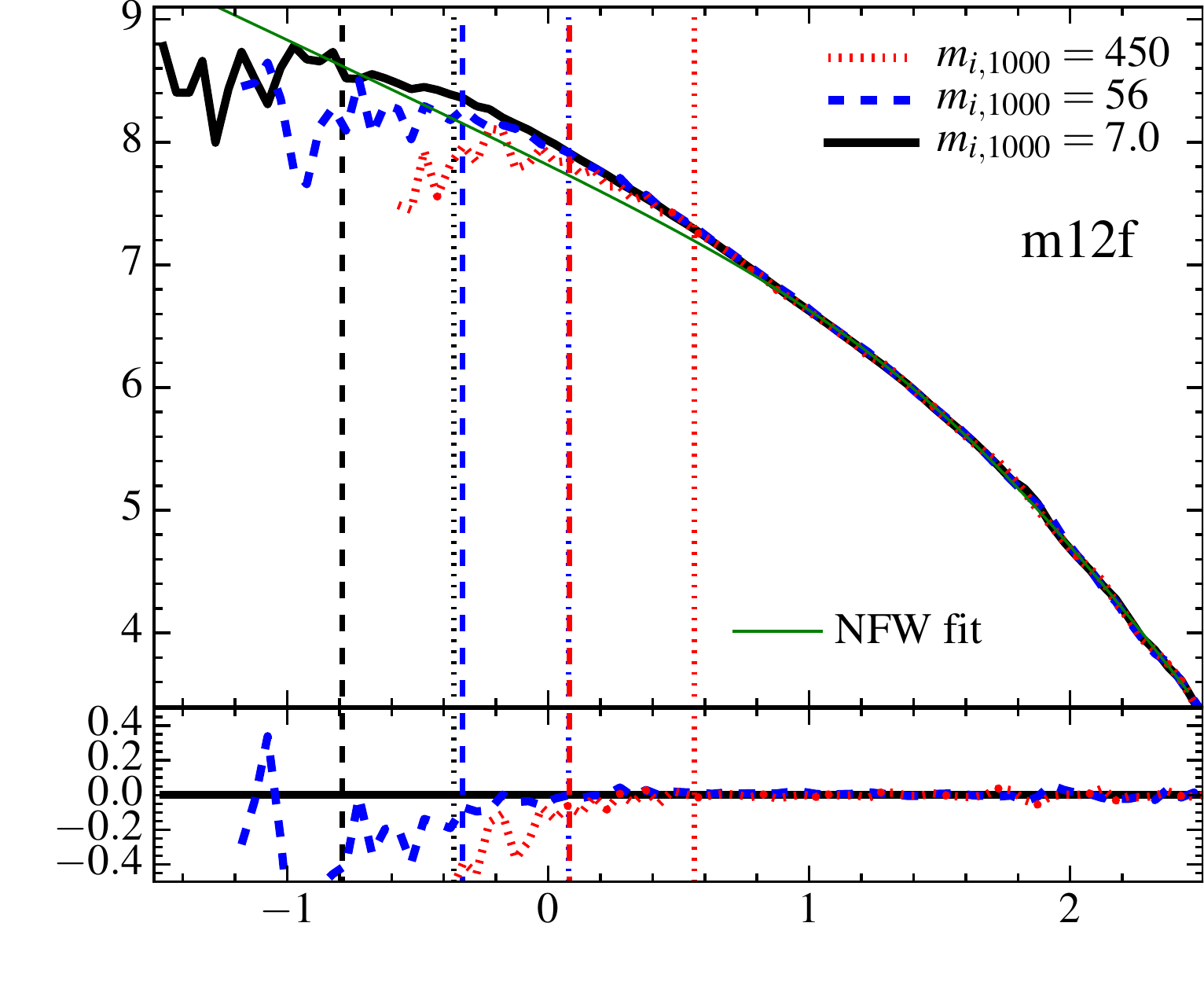} \\
\includegraphics[width=0.45\textwidth]{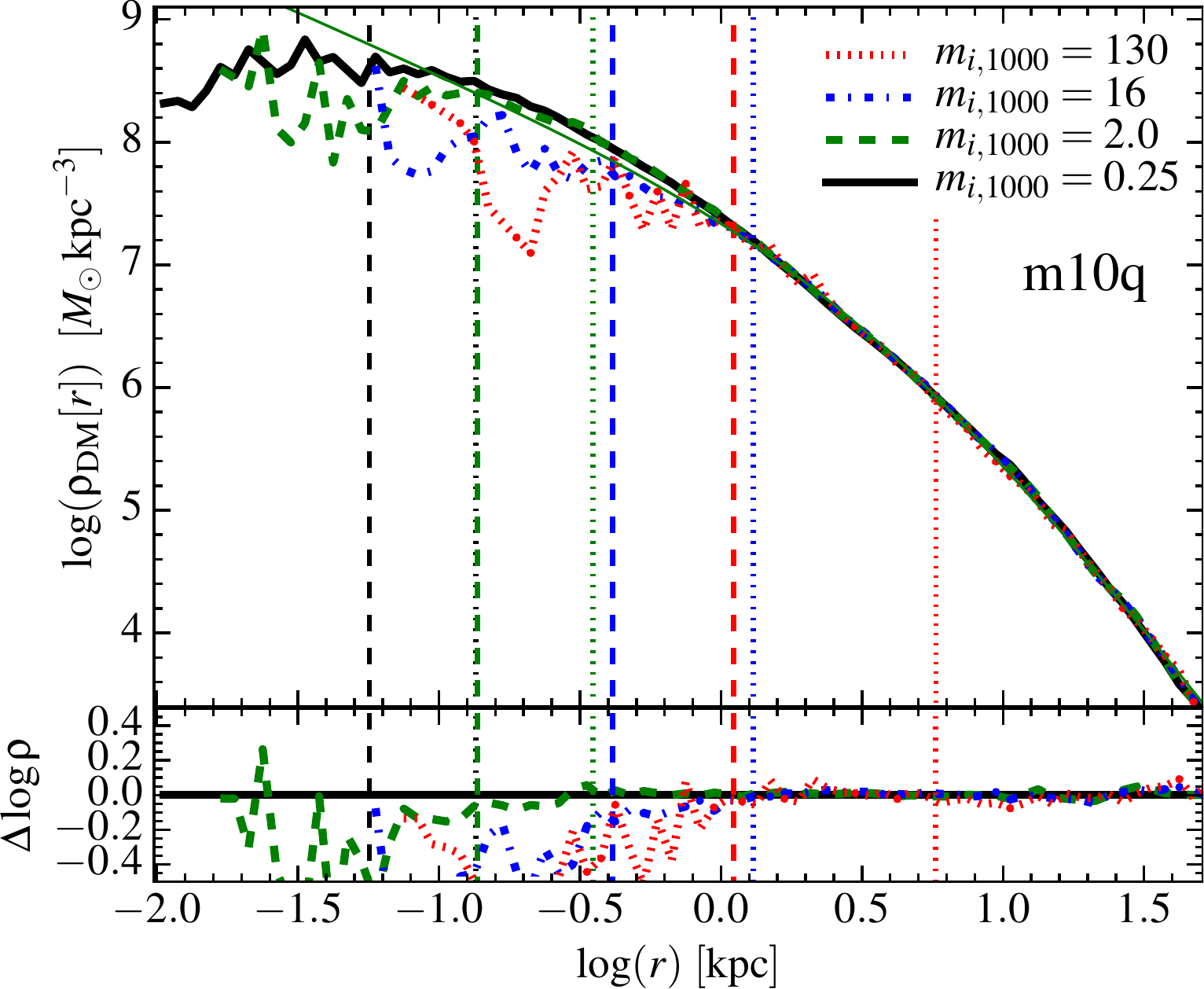} &
\hspace{-0.7cm}
\includegraphics[width=0.45\textwidth]{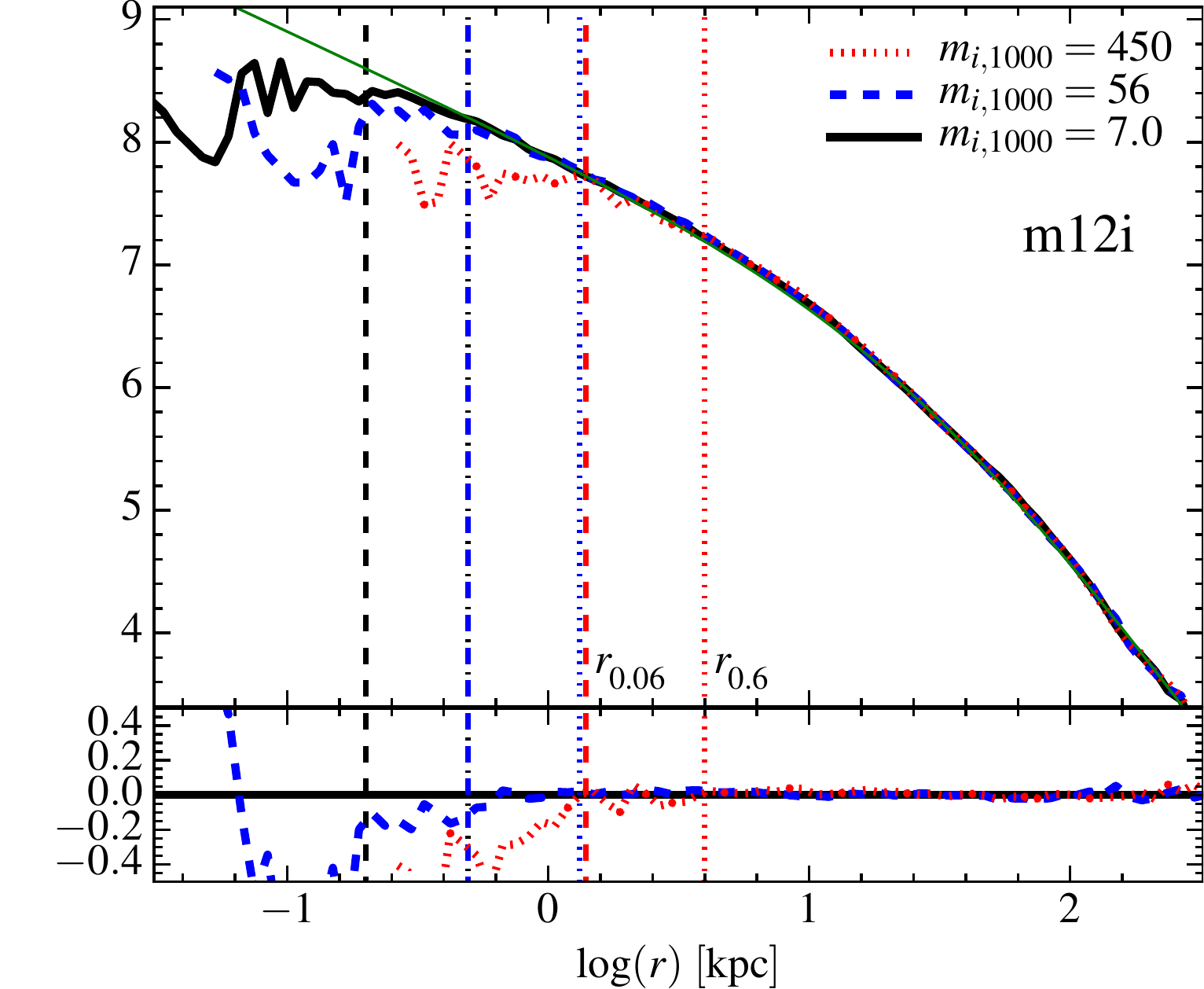} 
\end{tabular}
    \vspace{-0.25cm}
    \caption{Resolution studies \textit{in DM-only simulations} of the dwarf and MW-mass halos from Fig.~\ref{fig:res.summary}. We show the radially averaged DM mass profile at $z=0$ around the main halo, varying the mass resolution. Inset shows the difference plot ($\Delta \log{\rho}$ in dex) relative to the ``reference'' highest-resolution run available. For each factor of $8$ in mass resolution, we change the DM force softening by a factor of $2$. With increasing mass resolution we converge more and more closely to an NFW-like profile at small radii (thin green line). At radii larger than the \citet{power:2003.nfw.models.convergence} radius, $r_{0.6}$. where the relaxation time $t_{\rm relax}\approx 0.6\,t_{\rm circ}(R_{200})$, which encloses approximately $\approx2200$ particles (dotted vertical lines), the agreement is near-perfect. However, even at radius, $r_{0.06}$, where the $t_{\rm relax} \approx 0.06\,t_{\rm circ}(R_{200})$, which encloses just $\approx 200$ particles (dashed vertical lines), the agreement is still quite good: densities are under-estimated by at most $\sim 0.05-0.15$\,dex. We define the latter as our DM ``convergence radius'' throughout this paper.
    \label{fig:dm.mass.resolution}} 
\end{figure*}

\begin{figure*}
\begin{tabular}{cc}
\includegraphics[width=0.45\textwidth]{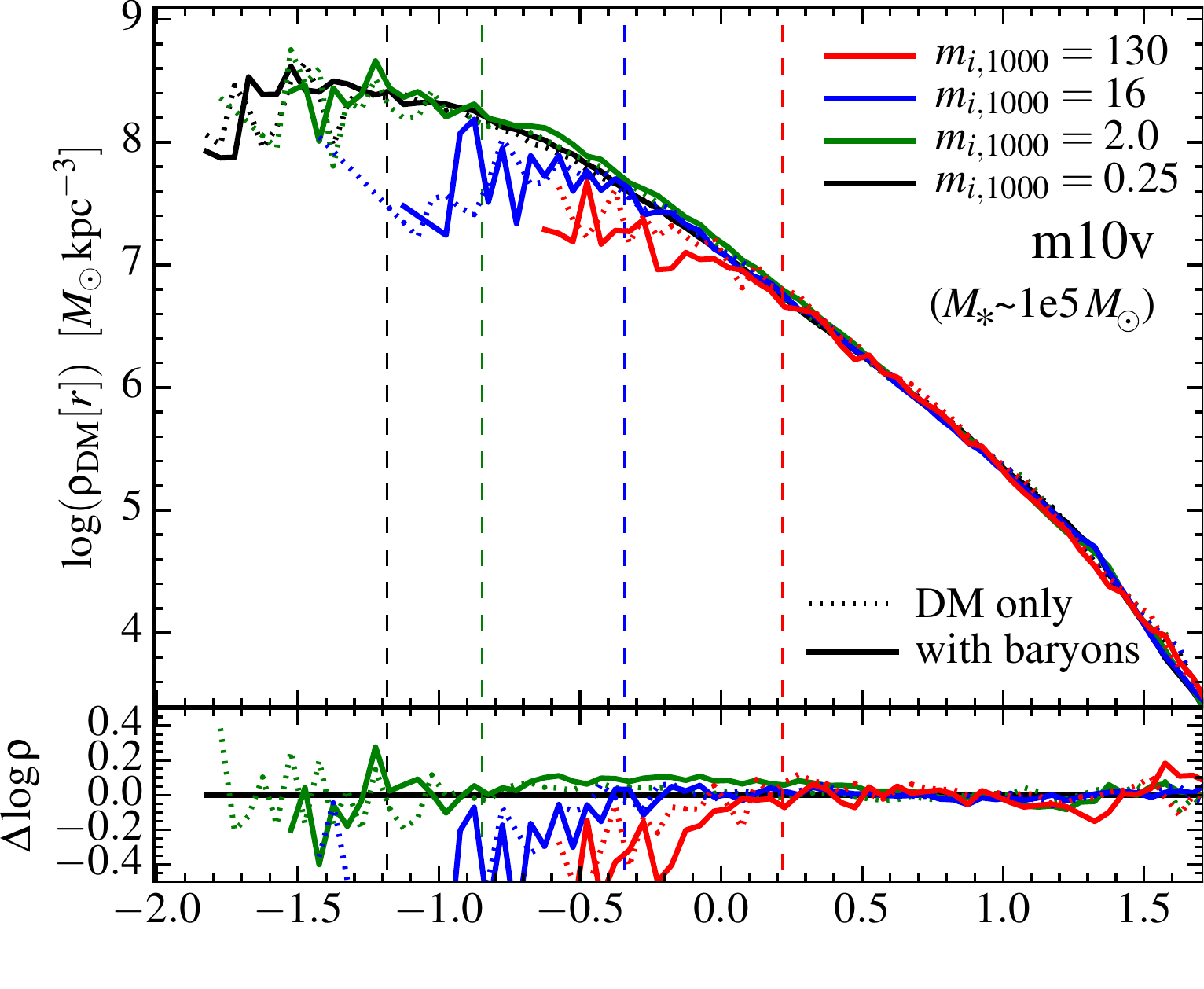} & 
\hspace{-0.7cm}
\vspace{-0.5cm}
\includegraphics[width=0.45\textwidth]{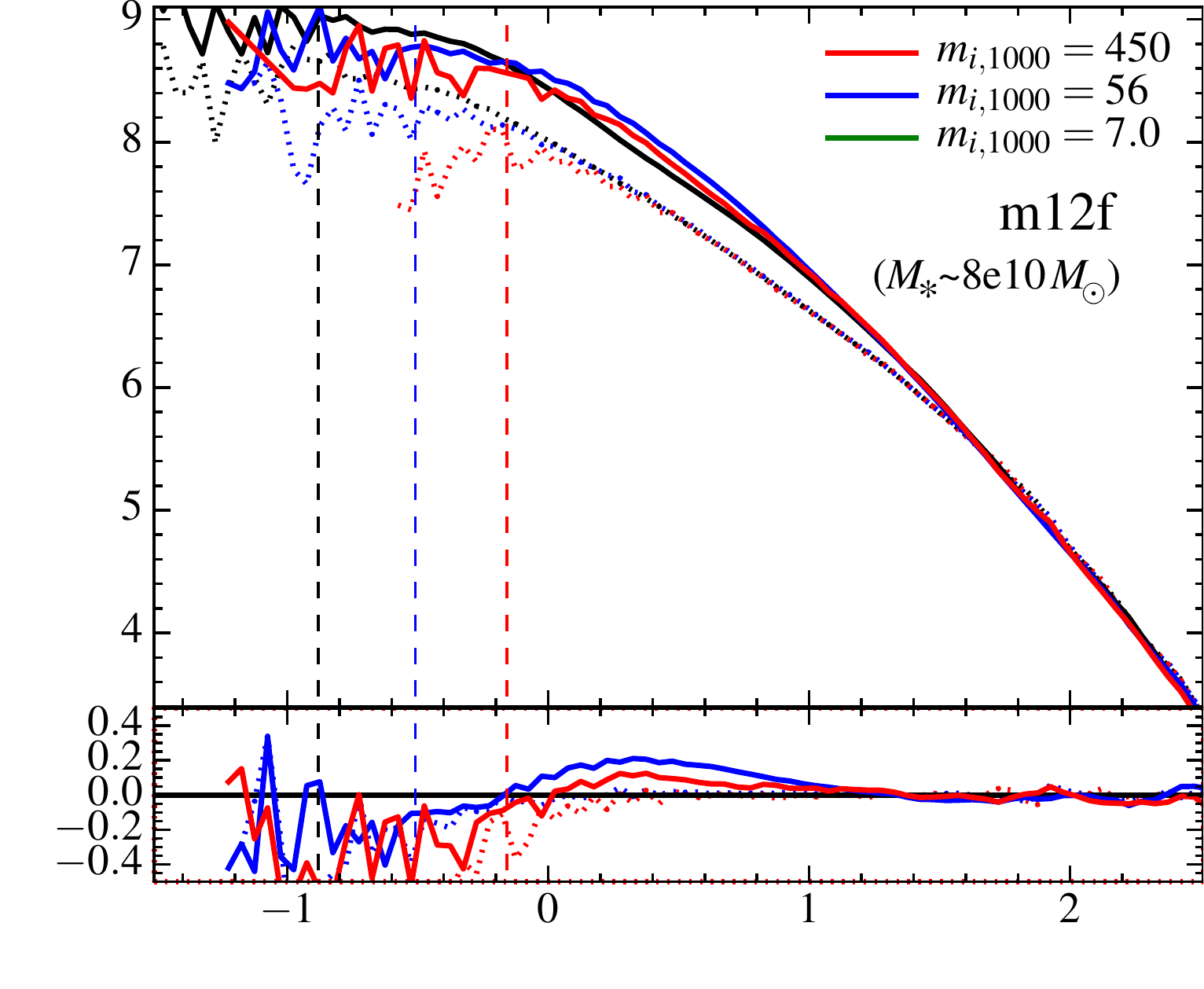} \\
\includegraphics[width=0.45\textwidth]{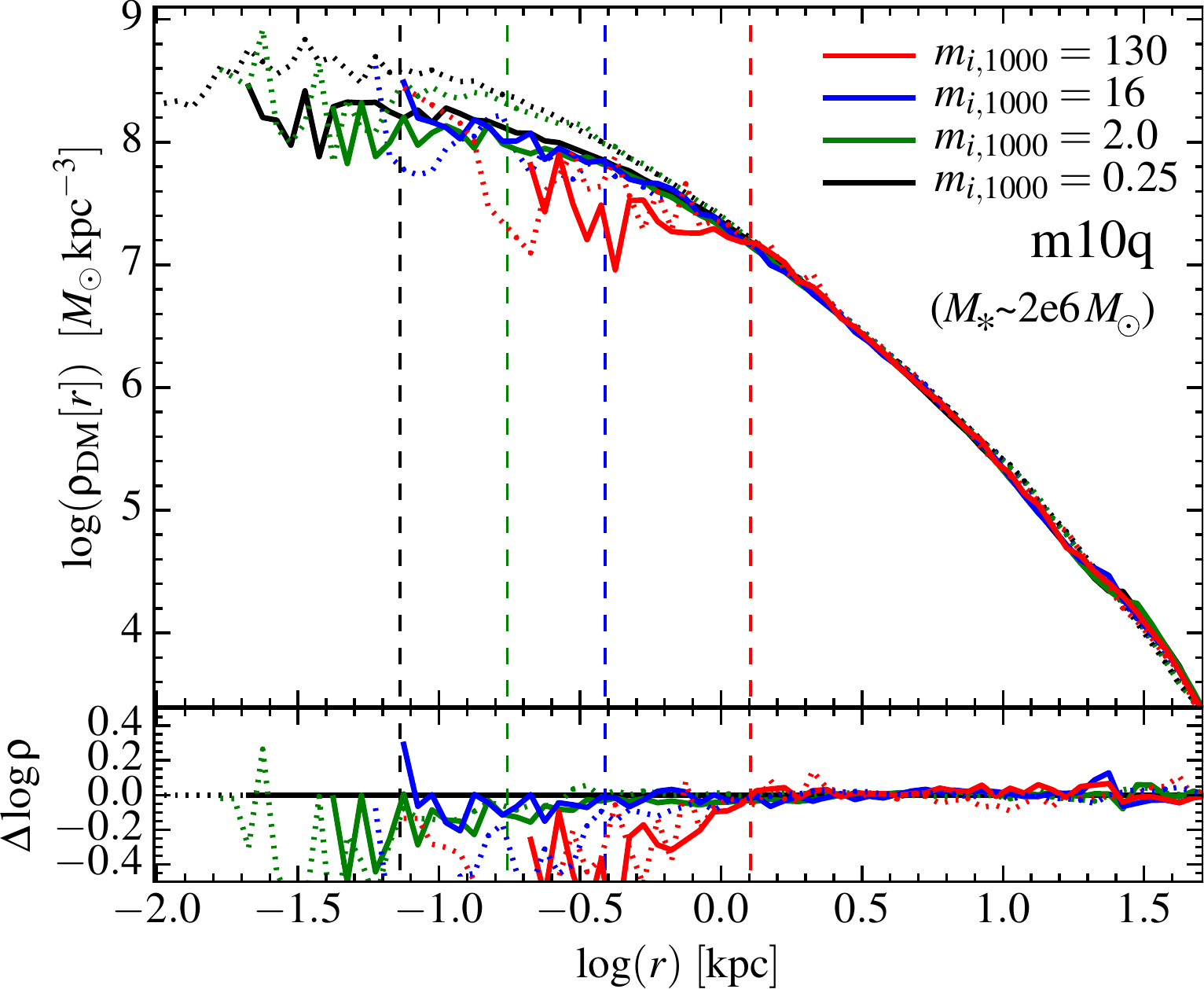} &
\hspace{-0.7cm}
\includegraphics[width=0.45\textwidth]{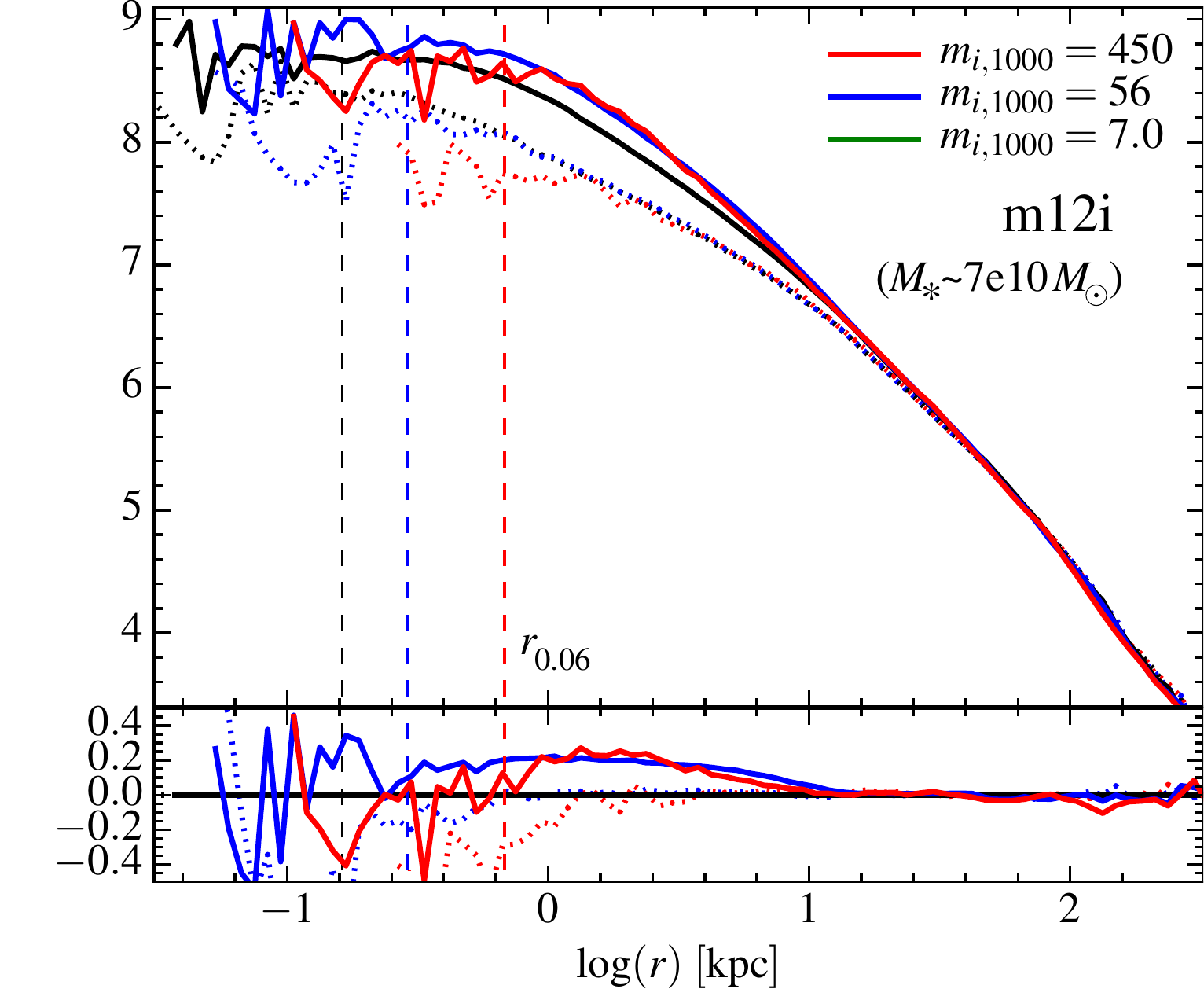} 
\end{tabular}
    \vspace{-0.25cm}
    \caption{Resolution studies of DM mass profiles in the dwarf and MW-mass halos from Figs.~\ref{fig:res.summary}-\ref{fig:dm.mass.resolution}, in simulations with baryons (solid) compared to DM-only simulations (dotted). We re-normalize the simulations with baryons by $\Omega_{m}/(\Omega_{m}-\Omega_{b})$ so that if the baryons traced the DM perfectly, the DM-only and baryonic simulations would agree exactly. 
    For clarity, we label the ``convergence radius'' ($r_{0.06}$, enclosing $\approx 200$ DM particles; vertical dashed lines) for just the simulation with baryons. In {\bf m10v}, the galaxy is sufficiently low mass ($M_{\ast}\sim 10^{5}\,\msun$) that it is DM dominated and has little or no core, so DM and baryonic simulations agree well. We also see good agreement down to radii $\sim 30\,$pc enclosing just $\sim 10$ particles. In {\bf m10q}, the more massive dwarf is still DM-dominated but forms a core in the baryonic runs (suppression of $\rho_{\rm DM}$ to $\sim 600\,$pc); the core is actually more robust to resolution than the ``cusps'' in the DM-only runs, down to similar radii $\sim 30\,$pc. In MW-mass runs, the baryons dominate the central mass and cause some contraction of the DM (although less than would be expected from pure adiabatic contraction given the galaxy mass). Differences in the central DM profile are, in these galaxies, dominated by differences in the baryonic mass (high-resolution runs can be {\em less} dense, owing to slightly smaller stellar masses), not by traditional DM resolution considerations such as $N$-body relaxation.
    \label{fig:dm.plus.baryons.mass.resolution}} 
\end{figure*}

\begin{figure}
\begin{tabular}{c}
\hspace{-0.2cm}\includegraphics[width=0.99\columnwidth]{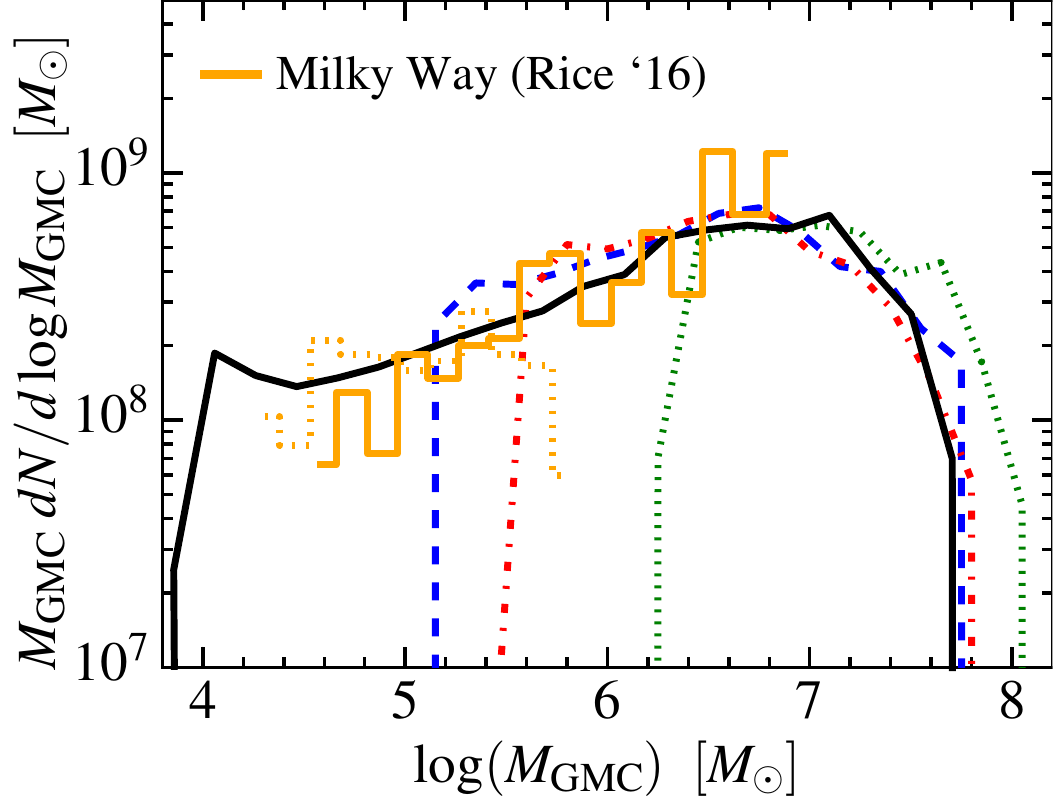} \\
\hspace{-0.2cm}\includegraphics[width=0.99\columnwidth]{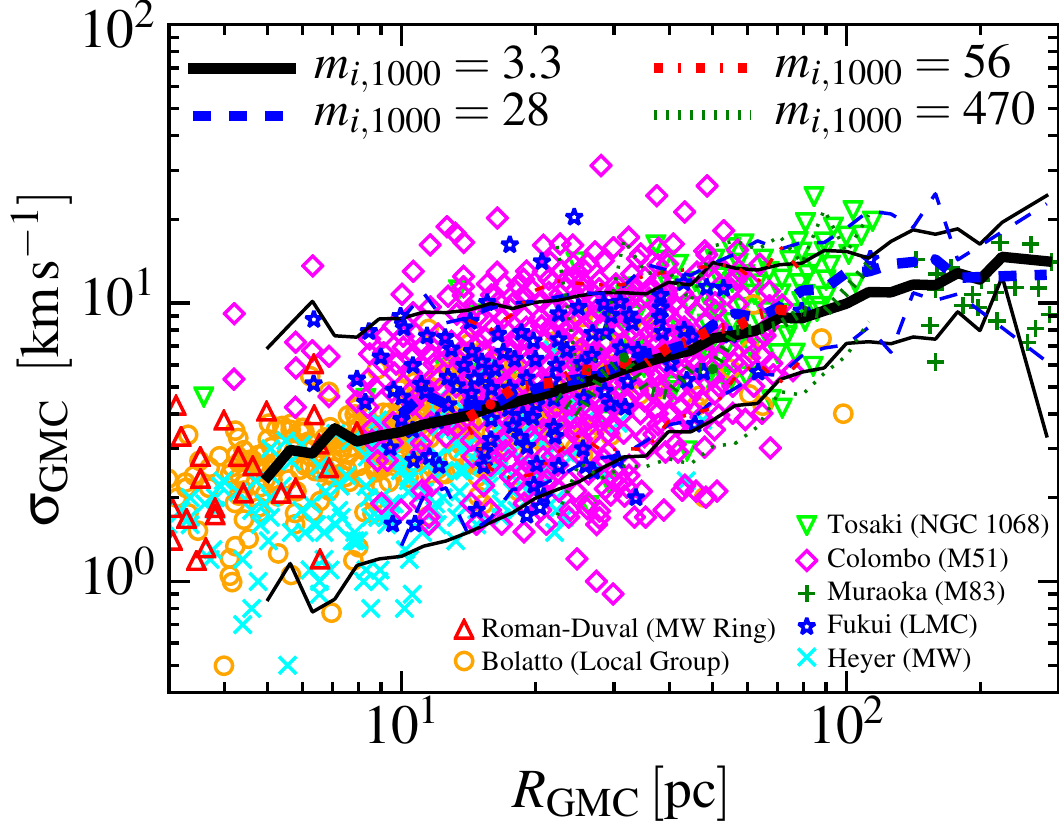} 
\end{tabular}
    \vspace{-0.3cm}
    \caption{Resolution study of ISM structure, in the runs from \demofigrestart\ (re-running the MW-mass {\bf m12i} from $z=0.07-0$ with different mass resolution). 
    {\em Top:} Time-averaged mass function (MF) of ``GMCs'': dense, cold gas clouds (identified with a friends-of-friends algorithm; see \S~\ref{sec:resolution:mass:tests}). We plot the total mass in clouds, per logarithmic interval in cloud mass ($M_{\rm GMC}\,dN_{\rm GMC}/d\log M_{\rm GMC}$), as a function of mass ($M_{\rm GMC}$). All identified structures as small as $3$ gas particles are shown. We compare the observed GMC MF in the MW (\citealt{rice:2016.gmc.mw.catalogue}; inside/outside the solar circle as solid/dotted), normalized to the same total mass. 
    The MF has power-law slope $dN/dM\propto M^{-(1.6-1.8)}$ and exponential cutoff around the Toomre mass ($\sim 10^{7}\,\msun$), similar to both observations and analytic predictions from turbulent fragmentation theories. 
    {\em Bottom:} Linewidth (internal velocity dispersion) vs.\ size relation (for the same clouds; median and $5-95\%$ interval in thick/thin lines), compared to observations (labeled). 
    Down to $\sim3-5$ particles per cloud, the predictions (both MF and linewidth-size) agree well; higher resolution simply samples smaller clouds. Most of the GMC mass is around the Toomre mass, so the total cloud mass is within $\sim 30\%$ in all runs shown. Only the lowest-resolution run cannot resolve the peak, biasing the mean $M_{\rm GMC}$ higher by a factor $\sim 2$. 
    \vspace{-0.5cm}
    \label{fig:clumps.res}}
\end{figure}

\begin{figure}
\begin{tabular}{c}
\includegraphics[width=0.9\columnwidth]{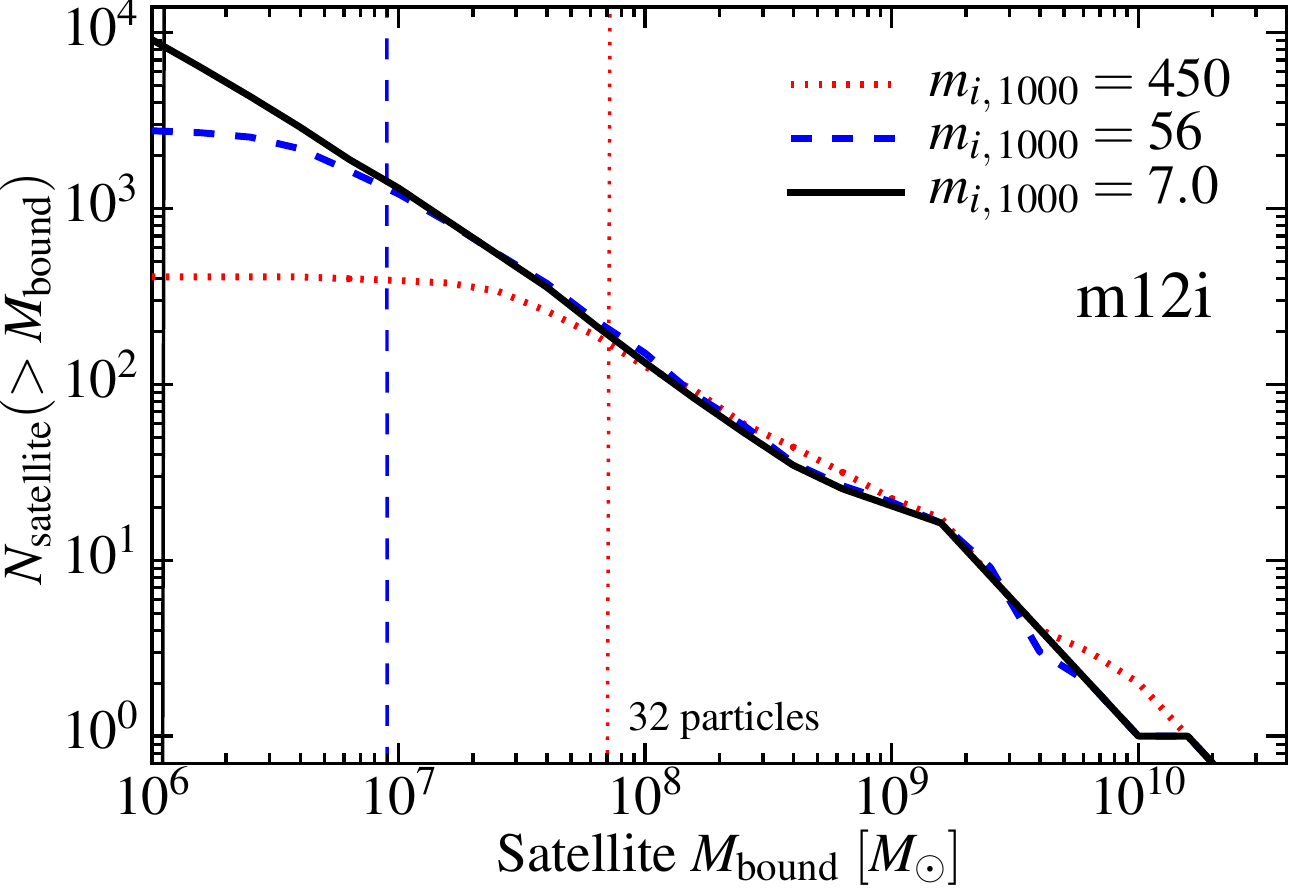} \\
\includegraphics[width=0.9\columnwidth]{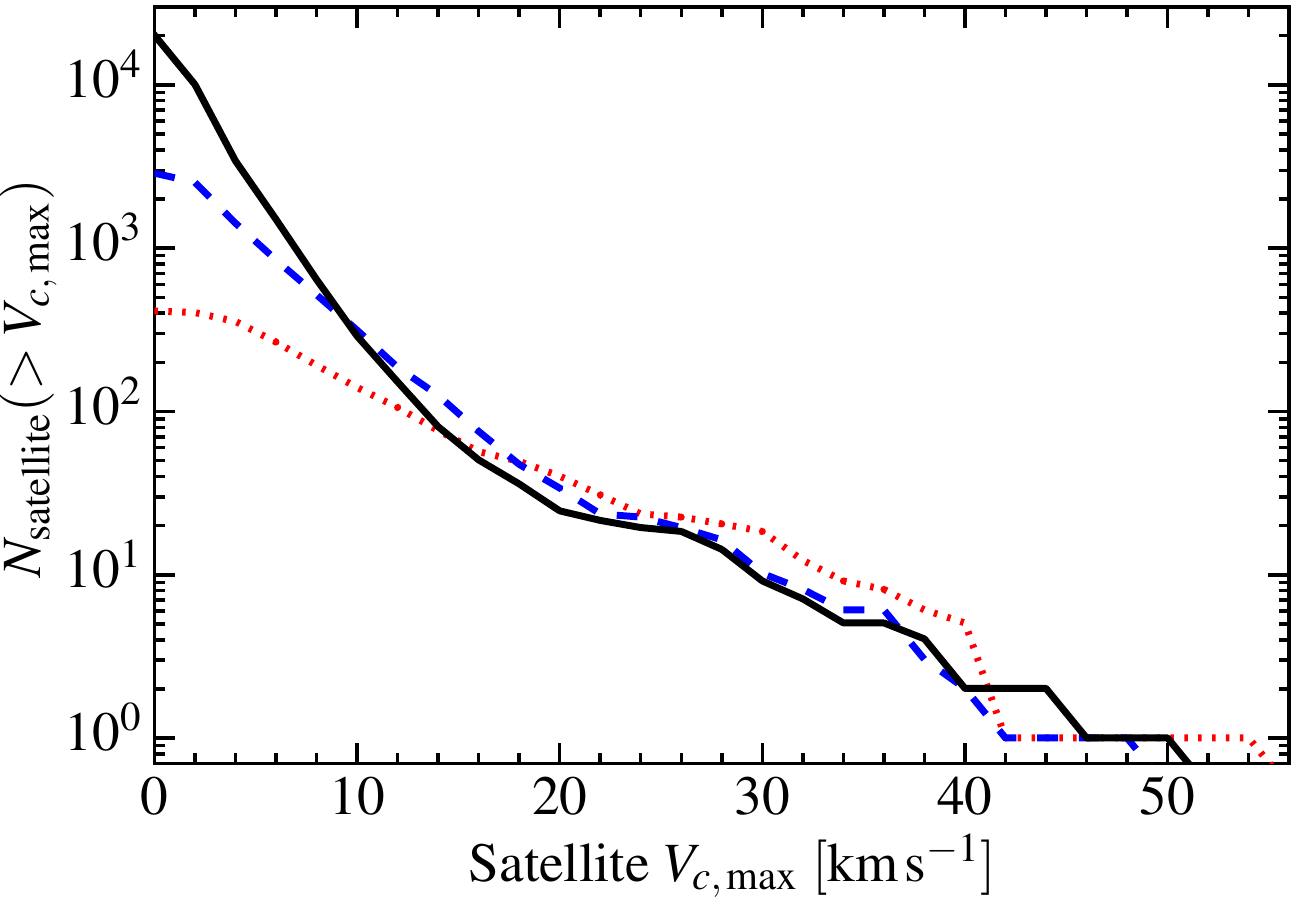} \\
\includegraphics[width=0.94\columnwidth]{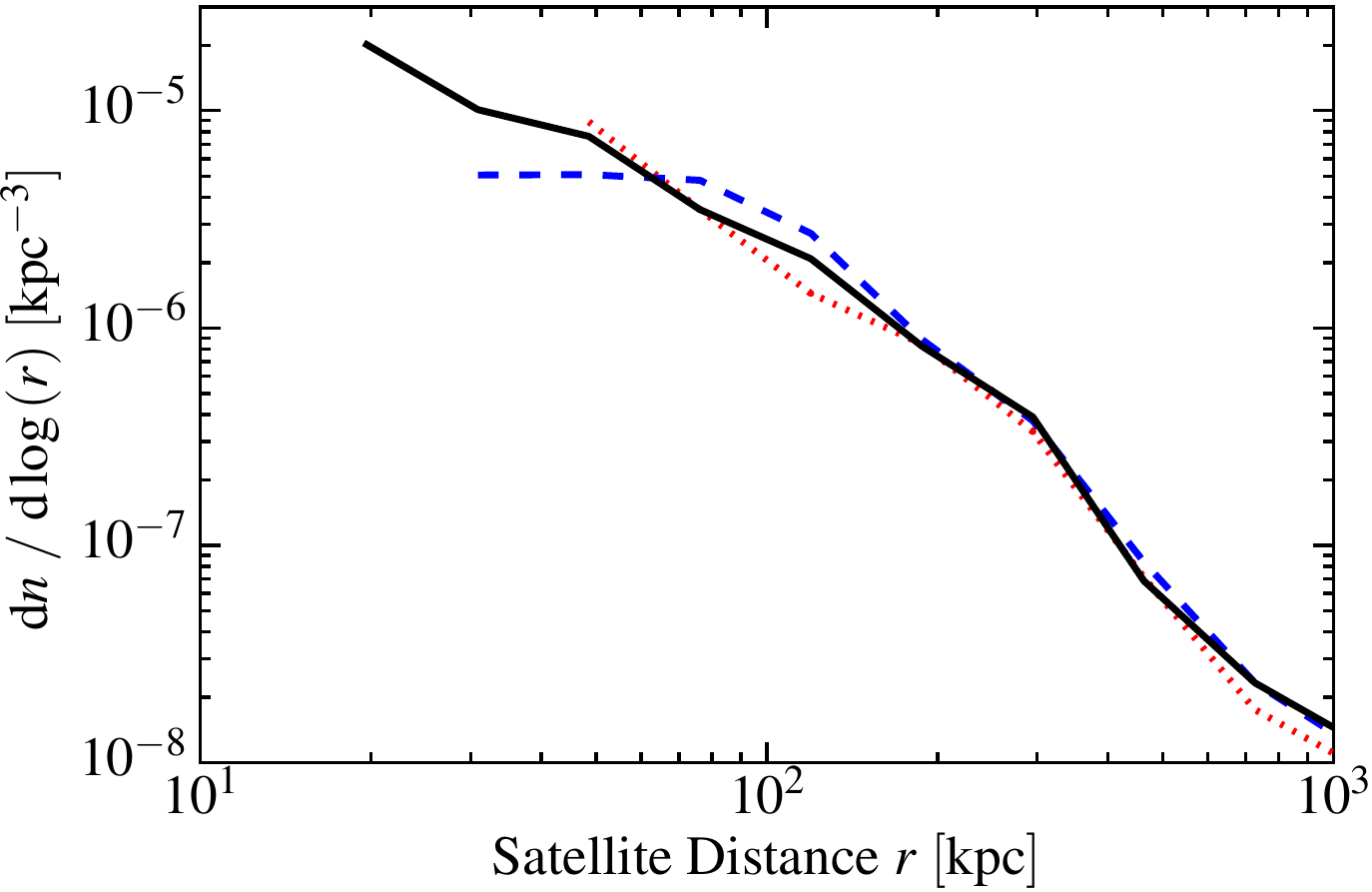} 
\end{tabular}
    \vspace{-0.25cm}
    \caption{Mass function ({\em top}), circular velocity distribution ({\em middle}), and spatial distribution ({\em bottom}) of satellites (DM subhalos) within the primary halo in DM-only runs of our {\bf m12i} simulation (at $z=0$), at varied mass resolution (as labeled). 
    {\em Top:} Number of subhalos versus bound subhalo mass. Vertical lines show $32$ DM particles; agreement is good down to structures containing $\sim10-30$ particles.
    {\em Middle:} Number of subhalos versus maximum circular velocity. Resolution dependence is similar to the mass function (deviations occur only below $\sim10-15\,{\rm km\,s^{-1}}$, corresponding to the un-resolved subhalos above). 
    {\em Bottom:} Distribution of subhalos in radial distance from the center of the primary halo. Up to shot noise in the exact position of the individual subhalos in their orbits, the distributions agree well.
    As the DM is approximately self-similar, we find qualitatively identical results comparing our other MW-mass ({\bf m12f}) or dwarf mass ({\bf m10q}, {\bf m10v}) simulations.
    \vspace{-0.7cm}
    \label{fig:dm.mass.resolution.substructure}} 
\end{figure}

In Lagrangian or $N$-body methods such as ours, there is a well defined mass-resolution given by the particle mass $m_{i}$ in the high-resolution Lagrangian region.\footnote{The low-resolution regions of the box, outside several virial radii of the main galaxies, are populated by lower-resolution collisionless particles, stepping up in a powers-of-eight hierarchy. This is sufficient for resolution of long-range tidal forces from these regions but we do not consider any halo contaminated ($>1\%$ by mass) by these particles to be ``resolved.''}

To maintain a well-defined mass resolution scale and minimize $N$-body integration errors, single gas particles are converted into single star particles with the same masses (rather than, for example, spawning star particles with much smaller/larger masses). However, mass loss from stars to gas in O-star/AGB winds and SNe means that particles will not have perfectly equal masses, so to prevent pathological behavior in very rare circumstances (if e.g.\ a single gas particle sees many SNe that increase its mass) we use the standard particle splitting and merging routine from \citet{hopkins:gizmo} to ensure no particle ever deviates by more than a factor of $3$ from the median particle mass. This affects only a tiny number of particles (one in $\sim10^{6}$). Averaged over our runs at $z=0$, $\sim 90\%$ ($99\%$, $99.99\%$) of all gas particles are within $<0.005$ ($0.1$, $0.2$) dex of the median particle mass. We will therefore refer only to the median baryonic particle mass $m_{i}$ in this paper.

Dark matter particles are always more massive by the universal baryon fraction, $m_{\rm dm} \approx 5\,m_{i}$; this ensures halos and baryonic  galaxies are comparably resolved at initial collapse. Of course, since many galaxies retain only a small fraction of their baryons, and dark matter does not cluster on small scales, dark matter structures tend to be vastly larger than baryonic structures and so are far better mass-resolved. 

We define, for convenience, the baryonic particle mass in units of $1000\,\msun$, and note that this specifies both baryonic and dark matter particle masses:
\begin{align}
\label{eqn:m1000.def} m_{i,\,1000} &\equiv \frac{m_{i}^{\rm baryonic}}{1000\,\msun}\\
\label{eqn:mdm.def} m_{\rm DM} &= \frac{\Omega_{m}-\Omega_{b}}{\Omega_{b}}\,m_{i}^{\rm baryonic} \approx 5000\,\msun\,m_{i,\,1000}
\end{align}

\vspace{-0.5cm}
\subsubsection{Requirements for ``Resolved'' Self-Gravitating Structures}
\label{sec:resolution:mass:gravity}

The mass resolution ``required'' to accurately model different phenomena depends, of course, on the question being asked and desired level of accuracy. A wide range of studies have shown that structures with masses of $\sim 5-100$ times the element mass are ``believable'' in the sense that they exist and can be identified as self-gravitating in higher-resolution re-simulations \citep{klypin:subhalos.vs.resolution,springel:cluster.subhalos,kravtsov:subhalo.mfs,nurmi:subhalo.mf,boylankolchin:millenium2.res.tests,wetzel:subhalo.disruption}. This corresponds to halos, stellar galaxies, or gas clouds of mass: 
\begin{align}
\label{eqn:resolved.halo.min} M_{\rm halo}^{\rm min} &\sim 0.6\times10^{5}\,\msun\,N_{10}\,m_{i,\,1000} \\ 
\label{eqn:resolved.stellar.min} M_{\ast}^{\rm min} = M_{\rm cloud}^{\rm min} &\sim 1\times10^{4}\,\msun\,N_{10}\,m_{i,\,1000} 
\end{align}
where $m_{i,\,1000} \equiv m_{i}/1000\,M_{\sun}$ is the {\em baryonic} particle mass, to which we reference all quantities, and $N_{10} = N_{\rm desired} / 10$ reflects the ``desired'' number of particles ($N_{\rm desired}$). 

We will show below for DM halos and subhalos, stellar galaxies, gas clumps and GMCs within galaxies, only a few particles are sufficient for robust prediction of masses and mass functions. 

Of course, the presence of even a few baryonic particles might require a quite massive halo. The observed relationship between galaxy stellar mass and halo mass implies that galaxies with $M_{\ast}\sim 10^{4}\,\msun$ typically live in halos of mass $M_{\rm halo} \sim 10^{9}\,\msun$; using the $M_{\ast}-M_{\rm halo}$ relation at low masses from \citet{moster:2013.abundance.matching.sfhs}, we estimate 
the minimum halo mass with a ``believable'' baryonic relic is $\sim 10^{9}\,\msun$ at $m_{i,\,1000}\sim 1$. This is actually quite well-resolved in dark matter, with $\gtrsim 10^{5}$ particles. This also means that $\gtrsim 10^{5}$ {baryonic} particles have {\em participated} in the formation history of the halo and have ``cycled through it'' (assuming something like the Universal baryon fraction is associated with the halo). It is just that the star formation efficiency is so low that the {\em residual} mass in stars is small. There is no question, then, that such objects are ``real'' in the simulations and, if they have such small stellar masses, that feedback had a real effect (it had to prevent $> 10^{5}$ particles from cooling and forming stars!). 

Moreover, because dark matter dominates the gravitational forces in these small galaxies, the orbital dynamics of the surviving stars (determined by the dark matter potential, not the negligible self-gravity) can be believed so long as the dark-matter structure is resolved (i.e.\ the stars are just tracer particles). This is easily satisfied in halos with $M_{\rm halo}\gtrsim 10^{9}\,\msun$, {\em independent of the number of stars} (given a realistic $M_{\ast}-M_{\rm halo}$ relation). 

We will show below that even the {\em internal} dynamics, evolution over a Hubble time, and mass profile of DM structures, are robust to resolution down to radii enclosing just $\sim 200$ elements. This is much more demanding than ``existence'' of halos or gaseous structures (which requires just a few elements), as expected.

\vspace{-0.5cm}
\subsubsection{Requirements for Well-Behaved Internal Hydrodynamics}
\label{sec:resolution:mass:hydro}

In \citet{hopkins:gizmo}, we show that, for the MFM method here, a few hundred resolution elements are sufficient to capture the orbital dynamics of thin Keplerian disks for $\sim 10-100$ orbital times (for a galactic disk, roughly equivalent to a Hubble time), the shock structure of strong blastwaves and/or implosions (the Sedov and Noh tests), and all the correct qualitative behaviors of self-gravitating polytropic sphere collapse (the Evrard test) and cosmological structure formation (the Zeldovich test), all to within a factor much better than $\sim 2$ of the exact solution. This means that a single star particle generating SNe in a halo and blowing out $\sim 100-1000\,$ surviving gas particles (total baryonic mass $\gtrsim 10^{6}\,\msun\,m_{i,\,1000}$ or, at the Universal baryon fraction, halo mass $\gtrsim 10^{7}\,\msun\,m_{i,\,1000}$) is at least hydrodynamically well-behaved, if not converged.\footnote{For reference, {\bf m10v}, the lowest-baryonic mass system in Fig.~\ref{fig:res.summary}, contains $\sim 10^{4}$ ($10^{7}$) gas elements inside $<1$\,kpc ($R_{\rm vir}$) at $z=0$, at our fiducial resolution.} This is consistent, roughly, with the resolution dependence of properties such as the enrichment history presented below, which require similar particle numbers in the relic baryonic galaxy to resolve the entrainment/recycling of SNe ejecta and its re-incorporation into subsequent generations of star formation.

\vspace{-0.5cm}
\subsubsection{Requirements for Capturing ISM Properties, Star Formation, and Stellar Feedback}
\label{sec:resolution:mass:firephysics}

The most demanding mass resolution requirements come from the physics of the ISM and CGM. 

\begin{enumerate}

\item{\bf The Toomre Mass \&\ ISM Structure:} Many previous studies have shown that reliably capturing star formation (both rates and spatial distribution) requires the ability to resolve at least the {\em existence and self-gravity} of the largest self-gravitating gas structures (i.e.\ fragmentation) in a galactic disk \citep[see][]{saitoh:2008.highres.disks.high.sf.thold,hopkins:rad.pressure.sf.fb,hopkins:fb.ism.prop,hopkins:stellar.fb.winds,hopkins:2013.fire,kim:tigress.ism.model.sims}. These studies showed that once this criterion is met, the SFR predicted in the simulations becomes independent of the numerical star formation model (see \S~\ref{sec:star.formation}), because it is regulated by feedback. This means we do not strictly need to resolve the internal dynamics of such clouds to capture galaxy-scale dynamics. And although there will always be clouds and sub-structure below our resolution limits, both observations and simulations have repeatedly shown that almost all star formation occurs in the largest GMC complexes in galaxies \citep[see][]{evans:1999.sf.gmc.review,rosolowsky:2003.gmc.rotation,blitz:2004.gmc.mf.universal,bolatto:2008.gmc.properties,
elmegreen:rapid.gmc.collapse.disruption,tasker:2009.gmc.form.evol.gravalone,feldmann:2011.gmc.sfe.eff.vs.time,murray:2010.sfe.mw.gmc,harper-clark:2011.gmc.sims,hopkins:excursion.ism}. This mass $M_{\rm GMC}^{\rm max}$ is set by the Toomre mass, $M_{\rm Toomre} \sim f_{\rm gas}^{2}\,M_{\rm gas}$ where $f_{\rm gas}$ is the gas fraction inside the relevant effective radius of the baryonic galaxy. So, if we require $ M_{\rm cloud}^{\rm min} \lesssim M_{\rm Toomre}$, we obtain the desired resolution criterion $m_{i,\,1000} \lesssim m_{i,\,1000}^{\rm Toomre}$ where 
\begin{align}
\label{eqn:mtoomre} m_{i,\,1000}^{\rm Toomre} \lesssim \frac{100}{N_{10}}\,\frac{M_{\rm Toomre}}{10^{6}\,\msun} \sim 
100\,\left(\frac{f_{\rm gas}^{2}\,M_{\rm gas}}{10^{6}\,M_{\sun}\,N_{10}}\right)
\end{align}
This defined the target resolution of the original FIRE-1 massive-galaxy simulations ($10^{12}\,\msun$ halos). 

\item{\bf SNe Cooling Radii:} A still more demanding criterion is set by SNe physics. As discussed in detail in \paperone\ (as well as \citealt{cioffi:1988.sne.remnant.evolution,thornton98,martizzi:sne.momentum.sims,walch.naab:sne.momentum}), a single SN remnant radiates its thermal energy rapidly upon reaching a nearly-invariant ``swept up'' mass $M_{\rm cool} \sim 3000\,\msun$; if a number $N_{\rm SNe}$ occur before the cooling time expires, the ``cooling mass'' simply scales $M_{\rm cool}\propto N_{\rm SNe}$. If SN energy is injected in a kernel-weighted fashion over $N_{\rm NGB}$ elements as we do here, then almost all of the energy goes into the nearest $N_{\rm NGB}^{1/3}$ neighbors; requiring they have a total mass $<M_{\rm cool}$ in turn requires: 
\begin{align}
\label{eqn:mcool.sne} m_{i,\,1000}^{\rm SNe} \lesssim 0.9\,(N_{\rm NGB}/32)^{-1/3}\,N_{\rm SNe}
\end{align}
This set the target mass resolution for the original FIRE-1 dwarf galaxies ($<10^{10}\,\msun$ halos). As we show in \paperone, unless $m_{i,\,1000} \ll m_{i,\,1000}^{\rm SNe}$, it is necessary to properly account for the conversion of energy into momentum in the unresolved Sedov-Taylor (S-T) phases. Failure to do so will significantly under-estimate the effects of feedback.\footnote{Note that Eq.~\ref{eqn:mcool.sne} does not mean we cannot resolve hot gas and/or overlapping SNe bubbles at lower resolution. If any mass of stars $m_{0}$ forms and the corresponding $N_{\rm SNe} \sim m_{0}/100\,\msun$ go off in an overlapping resolution element within a cooling time, then they can heat a mass $\sim 60\,m_{0}$ to $>10^{6}\,$K. So because star formation is clustered, we can still resolve super-bubbles and galactic chimneys at relatively modest resolution. However, resolving the full hot gas content, venting, momentum contribution from confined blastwaves, and early evolution of SNe explosions requires a criterion like Eq.~\ref{eqn:mcool.sne}.}

\item{\bf Dwarf Galaxy ``Burstiness'':} In previous papers we have shown that for small dwarf galaxies, star formation is robustly ``bursty'';\footnote{For the sake of quantitative comparison, we define a specific measure of ``burstiness'' as the standard deviation in the quantity $\log{( \dot{M}_{\ast}(\Delta t_{1})  /  \dot{M}_{\ast}(\Delta t_{0}) ) }$ where $ \dot{M}_{\ast}(\Delta t) $ is the SFR averaged over a time interval $\Delta t$, and we compare a short interval $\Delta t_{1} = 10\,$Myr and longer interval $\Delta t_{0} =$\,Gyr.} however, at low resolution, the ``burstiness'' is artificially enhanced by numerical effects \citep[see e.g.][]{sparre.2015:bursty.star.formation.main.sequence.fire}. This owes to the fact that stars form in units of star particles. At sufficiently low resolution, a single star particle implies a massive, co-eval star cluster, therefore a large number of approximately co-eval SNe. Although we know SNe are clustered in reality, this artificial numerical clustering could easily ``overshoot'' reality, producing too many synchronized SNe in the same location, which in turn leads to an over-large super-bubble that could heat the entire galaxy gas supply to super-virial temperatures. If we consider the ``unit'' of star formation (minimum resolved cloud) to be, say $\sim 10$ particles (since a single lone gas particle cannot be at much higher density than its neighbors), then taking the IMF-average $N_{\rm SNe}\sim m_{i}/100\,\msun$ per particle, the entire baryonic mass in the galaxy can be heated to $T\gg 10^{6}\,$K (``super-bubble'' temperatures) if that baryonic mass is below $\sim 600\,m_{i}$. So resolving ``venting'' of individual super-bubbles (much smaller than the entire galaxy) requires $\gtrsim 1000$ gas particles in the galaxy. Assuming further a gas fraction of $\sim 1$ during the gas-rich phases of star formation and/or typical dwarf galaxies, and using the stellar mass-halo mass relation above, this implies a particle mass target: 
\begin{align}
\label{eqn:mbursty} m_{i,\,1000}^{\rm burstiness} \lesssim 5\,\left( \frac{M_{\rm halo}}{10^{10}\,\msun} \right)
\end{align}
Equivalently, we can say that for a simulation with fixed mass resolution, once halos exceed a mass $M_{\rm halo} \gtrsim 2\times10^{9}\,\msun\,m_{i,\,1000}$ the numerical ``excess clustering/burstiness'' is not important. 

\item{\bf The Jeans Mass:} A common mis-conception is that one needs to resolve the Jeans mass in order to capture basic fragmentation physics. This is incorrect, for two important reasons. First, the warm/cold ISM is super-sonically turbulent, so the ``fragmentation cascade'' (hierarchical structure of fragmentation into GMCs, clumps, cores, etc.) is determined by the {\em turbulent} Jeans mass, not the {\em thermal} Jeans mass, at least down to the sonic scale ($R \lesssim R_{\rm sonic} \sim 0.1\,{\rm pc}$ and mass scale $\sim 1\,\msun$). We discuss this below, but the turbulent Jeans mass of a super-sonically turbulent GMC with virial parameter $\alpha \sim 1$ is of order the GMC mass itself ($\sim 10^{7}\,\msun$), whereas the thermal Jeans mass is $\sim 0.1\,\msun$. Second, in a homogeneous medium, the Jeans mass defines the {\em smallest} scales of the fragmentation cascade -- by definition, all larger scales are also unstable. Therefore, failure to resolve the Jeans scale simply means all {\em resolved} scales fragment as they should -- the fragmentation cascade is just truncated at the resolution scale, instead of the (smaller) Jeans scale. This is analogous to ``resolving turbulence'' in the ISM; the actual Kolmogorov (``termination'') scale of the turbulent cascade ($\sim {\rm au}$) is far smaller than achievable resolution, but that does not mean turbulence cannot be captured (it simply limits the dynamic range of the cascade that can be followed). In both cases (for the same physical reason, in fact; see \citealt{hopkins:excursion.ism}), the power in the cascade is dominated by the largest-scale structures, so integrated quantities converge very quickly with increasing resolution \citep{padoan:2012.sfr.local.virparam,federrath:2012.sfr.vs.model.turb.boxes,hopkins:frag.theory,guszejnov:cmf.imf,guszejnov:gmc.to.protostar.semi.analytic}. We show this explicitly below -- the mass function of dense, cold clouds is well behaved regardless of resolution, it simply extends to smaller and smaller sub-structures as we increase the resolution. Thus, the full cascade is not followed, but this is not problematic, provided that the goal of our simulation is not to resolve individual brown dwarf or star formation (the actual ``end point'' of the fragmentation cascade in the ISM). This is where the sub-grid model for star formation enters -- it is, explicitly, a sub-grid model for the un-resolved fragmentation cascade from a resolved, self-gravitating cold gas clump/cloud all the way down to an aggregate of individual stars with some IMF. The Toomre scale, on the other hand, is important to resolve, because it is the {\em largest} unstable scale in a disk -- in other words, failure to resolve the Toomre scale means that no fragmentation will occur (when it physically should). 

\end{enumerate}

\vspace{-0.5cm}
\subsubsection{Mass Resolution Tests}
\label{sec:resolution:mass:tests}

We now explore in detail how mass resolution can alter our conclusions. Figs.~\ref{fig:res.summary}-\ref{fig:dm.mass.resolution.substructure} present a series of explicit mass resolution tests of our full-physics FIRE-2 simulations. Table~\ref{tbl:res} gives some typical values for mass, spatial, and time resolution for {\bf m12i} runs at different resolution. 

Fig.~\ref{fig:res.summary} shows our full-physics cosmological simulations at varying mass resolution. We compare two low-mass dwarfs and two MW-mass galaxies, to bracket the range of behavior. Fig.~\ref{fig:images.resolution} compares the visual morphologies of the galaxies.

For the dwarfs, the total stellar mass (and circular velocity profile) becomes robust to better than a factor $<2$, with just $\gtrsim 2-16$ star particles in the $z=0$ galaxy. This is because the total stellar mass is set by an integral balance between feedback energetics and gravity binding the baryons. Also, given the low particle masses, even these low-resolution dwarf runs easily satisfy our Toomre-mass criterion (Eq.~\ref{eqn:mtoomre}). However, at such low resolution ($m_{i,\,1000}=16-130$) they do {\em not} satisfy our ``numerical burstiness'' or ``SNe cooling'' criteria (Eq.~\ref{eqn:mbursty} and \ref{eqn:mcool.sne}, respectively). As a consequence, the SFHs are visibly more ``bursty'' (dominated by just a couple large bursts), and the metallicities are systematically under-predicted (compared to our high-resolution runs). The latter occurs because the single bursts blow out nearly all the baryons (and metals) from the galaxy: they fail to capture partial entrainment/mixing/incomplete blowout that would keep the metals in the galaxy. The metallicity and burstiness of the SFH appear to robust to a factor $\sim 2$ ($\sim10\%$) when the number of stars reaches $>100$ ($>500$) at $m_{i,\,1000}=2$, as expected from our criterion in Eq.~\ref{eqn:mbursty}. More subtle properties, such as the internal SF structure of bursts and hot gas properties of the galaxy, and escape fraction of ionizing photons \citep{ma:2015.fire.escape.fractions,ma.2016:binary.star.escape.fraction.effects}, require still higher resolution, hence our highest-resolution runs, with $m_{i,\,1000}=0.25$, satisfying Eq.~\ref{eqn:mcool.sne}. These results are also supported by the analysis of dwarf satellite galaxies of our massive, high-resolution {\bf m12i} halo in \citet{wetzel.2016:latte}; the convergence in the stellar mass function of dwarfs appears good down to $\sim 5-10$ star particles per galaxy; but the metallicities of the satellites with $\lesssim 100$ star particles are suppressed relative to our higher-resolution isolated-dwarf simulations and observations \citep[compare][]{ma:2015.fire.mass.metallicity}. Not surprisingly, because the gross morphology of the dwarfs is irregular, it is essentially resolution-independent. Galaxy sizes (effective radii) are also nearly independent of resolution (because of the more ``bursty'' star formation, the sizes tend to be $\sim 10-20\%$ larger at the lowest resolutions here). In \paperone, we show the star formation histories of dwarf galaxy simulations reaching $\sim 30\,M_{\sun}$ resolution (which will be studied in more detail in Wheeler et al., in prep), and show that they agree well with the $\sim 250\,M_{\sun}$-resolution runs here.

For the massive, MW-mass system, even the lowest-resolution ($m_{i\,1000}=450$) run has $\gg 10^{5}$ particles in the baryonic galaxy and easily satisfies the Toomre and ``burstiness'' criteria (although its progenitor dwarf galaxies at high redshifts may not). As such the SFH and metallicity are more robust to resolution (compared to the dwarf runs). However we do see higher resolution systematically shifts the SFH from early to later times, as higher resolution allows better resolution of two key physics. First, the generation of winds via resolution of the hot gas ``channels'' and their escape (``venting'') from a multi-phase halo \citep[our ``SNe'' criterion; for explicit studies see][]{hopkins:stellar.fb.winds,hopkins:2013.merger.sb.fb.winds,muratov:2015.fire.winds,martizzi:sne.momentum.sims,fielding:sne.vs.galaxy.winds}. At low resolution, hot gas is necessarily ``dragged'' by a large mass of implicitly-coupled gas, because mass is locked into massive particles: at our lowest resolution, for example, a single SN cannot affect less mass than $\sim 10^{6}\,\msun$ -- this means much lower ``launch velocities'' and temperatures, unless a huge number of SNe explode simultaneously (see \paperone). Second, high resolution allows better resolution of the ``burstiness'' and SFHs within the smaller {\em progenitor} galaxies of the more massive $z=0$ MW-mass system (note that the largest differences appear at early times, when the galaxy is a progenitor dwarf), which reduces their stellar and gas masses, leading to more gas expelled to large enough radii where its recycling times are long. The gas is still re-incorporated eventually, evident in the similar late-time masses and SFRs, but appears to re-accrete later.

As a result of these effects, at higher resolution the final $z=0$ MW-mass galaxy is slightly lower-mass, but more importantly, because it shifts star formation and re-accretion of recycled material to later times, that material carries larger angular momentum, and the galaxy is less compact. For {\bf m12i}, the effective radius in our high-resolution $m_{i,\,1000}=7$ run is $\sim 1.4$ times larger than the $m_{i,\,1000}=450$ run, which translates to a factor $\sim 1.8$ lower mass inside $<5\,$kpc; this in turn lowers the peak in the rotation curve from $\sim 370\,{\rm km\,s^{-1}}$ to $\sim 270\,{\rm km\,s^{-1}}$. Thus the galaxy rotation curve is noticeably ``less bulgy'' at high resolution. The differences in {\bf m12i} are more dramatically evident in the low surface-brightness outer disk morphology, which goes from being entirely absent at very low resolution to quite prominent at high resolution. Interestingly, however, in {\bf m12m} there is much weaker dependence on resolution (the effective radius changes by $<15\%$). Also in {\bf m12f}, although the change in effective radius and the rotation curve with resolution is similar to {\bf m12i}, the visual morphology changes much less dramatically -- even at $m_{i,\,1000}=450$ there is still a prominent, extended thin disk. This may owe to the fact that {\bf m12m} and {\bf m12f} have a somewhat larger mass, but likely also owes to the specific fact that {\bf m12i} has a series of mergers around $z\sim1$, which launch strong fountains and change the angular momentum of the gas that will form its disk (whereas {\bf m12f} and {\bf m12m} grow more smoothly). So it appears {\bf m12i} is simply more sensitive to resolution effects. Clearly, it is important to push to even higher resolution (in progress), to test whether or not this is actually converged. 

If the dependence on resolution in our MW-mass runs owes to a combination of (1) better-resolving the progenitor (dwarf) galaxies, and (2) resolving channels by which hot gas outflows can escape, we should see similar SFRs independent of resolution at late times {\em if we start from the same ICs}. In other words, it is useful to test whether indeed the difference with resolution comes from progenitor influence on the late-time galaxy, or whether it is present for fixed conditions in a massive MW-mass galaxy. We therefore consider a series of simulations where we use our {\bf m12i} simulation with $m_{i,\,1000}=56$, re-started at $z\approx 0.06$ from a snapshot of the simulation in Fig.~\ref{fig:res.summary}, but using our particle splitting/merging routine to first split/merge the ICs until they are re-sampled with a desired target resolution. Fig.~\ref{fig:sf.z0.mass.resolution} shows that when we do this, the SFR is nearly identical over $\sim 2.5$\,dex in mass resolution (small deviations at the highest resolution owe mostly to some artifacts of our very aggressive particle splitting/up-sampling routine applied for this specific test). 

This is consistent with our argument above, that we only need to marginally resolve the Toomre scale to achieve a robust prediction for the SFR, {\em given} a specific gas disk initial condition. Quantities such as the Kennicutt-Schmidt relation are thus extremely robust to mass resolution. However, non-linear, long-timescale effects in cosmological simulations (e.g.\ recycling) shift the SFR {\em by changing the supply or loss rate of gas in the CGM}.

To better understand how much of the resolution dependence owes to purely gravitational physics, Figs.~\ref{fig:dm.mass.resolution}-\ref{fig:dm.plus.baryons.mass.resolution} consider how the dark matter mass profiles of the galaxies change with mass resolution. First we consider DM-only simulations (i.e.\ gravity-only simulations), in Fig.~\ref{fig:dm.mass.resolution}. We see near-ideal convergence in the mass profiles of the DM halos (as well as their substructure mass functions, shown below, and halo formation/growth histories). As expected, the $z=0$ DM mass profiles are well-fit by a \citet{nfw:profile} (NFW)-like profile, down to some minimum scale where numerical effects flatten the profile. Because of the Lagrangian nature of our code, improving mass resolution also improves the effective spatial resolution/force resolution; in fact, we will show below (consistent with many previous studies) that the nominal spatial force softening is generally much less important than mass resolution. \citet{power:2003.nfw.models.convergence} argue that the central ``flattening'' in DM profiles is dominated by $N$-body relaxation, and that robust results should be obtained outside a radius where the $N$-body relaxation time $t_{\rm relax} \sim 0.6\,t_{0}$ (where $t_{0}\equiv t_{\rm circ}(R_{200})=2\pi\,R_{200}/V_{200}$) is comparable to the Hubble time. Because $N$-body relaxation depends most strongly on $N$, this is effectively a requirement on the number of particles -- more exactly $(5/\sqrt{8})\,(N[<r]/\ln{N[<r]})\,(\rho_{\rm crit}/\bar{\rho}[<r])^{1/2} < 0.6$, satisfied for the densities in Fig.~\ref{fig:dm.mass.resolution} when $N \gtrsim 2200$. Not surprisingly, we see almost excellent agreement at these radii. In fact, as others have shown, because of more accurate integration criteria, shorter timesteps, and a smoother spline for gravitational softening (which reduces the $N$-body relaxation time below the equation for strict point masses assumed by the argument in \citealt{power:2003.nfw.models.convergence}), if we are willing to tolerate slightly larger errors, within a factor $\sim 1.3$ ($0.1\,$dex) of the converged (NFW) solution we find excellent agreement independent of resolution, down to radii containing just $N \gtrsim 200$ particles.

Fig.~\ref{fig:dm.plus.baryons.mass.resolution} compares the DM-only result to the full baryonic physics runs from Fig.~\ref{fig:res.summary}. Following standard practice, we correct the profiles from baryonic runs by the cosmic mean baryon fraction, so that if the baryons behaved identically to the DM, the curves would lie exactly on top of one another. With or without baryons, in both our dwarf and MW-mass runs, agreement is good to $\lesssim0.15\,$dex outside the radii enclosing $\sim 200$ particles. In {\bf m10v}, the $z=0$ galaxy is sufficiently low-mass that baryons have a negligible effect on the DM profile \citep[see][]{chan:fire.dwarf.cusps}, so the two sets of runs track each other closely. In {\bf m10q}, the galaxy starts to reach masses where stellar feedback generates a ``core'': even though small (factor $<2$) differences in the stellar mass have large effects on core creation at these masses \citep[see][]{onorbe:2015.fire.cores}, the resolution dependence is similar in baryonic and DM runs. In the MW-mass runs, we see that runs with baryons have higher central densities than DM-only runs, because the large galaxy baryonic mass has caused some halo contraction. This necessarily leads to better ``convergence'' in an $N$-body sense, but it means that the DM profile is more sensitive to changes in the galaxy stellar mass (here, the highest-resolution run has a lower $M_{\ast}$ and correspondingly lower central DM density). In all  cases, our highest-resolution baryonic simulations reach an approximate ``convergence radii'' of $\sim 50-100$\,pc {\em in the dark matter}. We of course achieve much higher effective spatial resolution in the baryons dense enough to form stars.

Note that there is no analogous \citet{power:2003.nfw.models.convergence} criterion for baryons. However, Fig.~\ref{fig:clumps.res} and various self-gravitating baryonic collapse tests in \citet{hopkins:gizmo} show that self-gravitating structure can be captured across just a couple inter-particle separations. Larger numbers of particles are needed for convergence in the mass profile in dark matter because the $N$-body orbits must be integrated for a Hubble time. In the gas, however, self-gravitating sub-structures (e.g.\ GMCs) survive for only a few dynamical times. Thus the appropriate \citet{power:2003.nfw.models.convergence}-equivalent criterion is $t_{\rm relax} \gtrsim {\rm couple} \times t_{\rm dyn}$, which is easily satisfied even for point masses (un-softened gravity) whenever an object is comparable in mass (or larger) than a kernel. With adaptive softening for gas (our default choice), this is {\em by definition} true even for ``structures'' of $\sim 2$ particles. 

Fig.~\ref{fig:clumps.res} specifically considers the resolution dependence of structure {\em within} the ISM; we use the same simulations from Fig.~\ref{fig:sf.z0.mass.resolution} of the MW-mass system at $z\sim 0$, but study the mass function of dense, cold gas structures within the ISM.\footnote{For simplicity, we use a friends-of-friends group finder with a linking length $=0.2$ times the mean inter-particle separation within the galaxy, to identify substructures in the gas at temperatures below $\le 8000\,$K and densities $n > 5\,{\rm cm^{-3}}$, within $0.1$ virial radii of the center of the main galaxy, at $\sim 30$ uniformly-spaced snapshots in time between $z=0.06-0$, and plot the time-averaged mass function of gas structures, $M\,dN/d\log_{10}{M}$ for each simulation.} 
A more detailed discussion of the cloud properties in these simulations is presented in \citet{guszejnov:imf.var.mw}; for our purposes here we simply desire a proxy for the GMC mass function to understand the resolution-dependence of the simulations. In each case, the shape of the mass function is, as expected, a power-law with slope $dN/dM \propto M^{-(1.6-1.8)}$ \citep[very similar to observed; see][]{blitz:2004.gmc.mf.universal,rice:2016.gmc.mw.catalogue}, and turnover at a maximum mass about the Toomre mass $M_{\rm Toomre}$ defined above \citep[also as expected from observations and analytic theory;][]{murray:2010.sfe.mw.gmc,hopkins:excursion.ism}. The lower limit is purely a resolution effect: we only keep structures with $ \ge 3$ particles. We also compare the {\em internal} properties of the clouds, specifically the linewidth-size relation (the one-dimensional velocity dispersion of gas within clouds, versus their projected mass-weighted rms radius), to observations of nearby galaxies (\citep{bolatto:2008.gmc.properties,fukui:2008.lmc.gmc.catalogue,heyer:2009.gmc.trends.w.surface.density,Muraoka_2009_M83,RomanDuval_2010_clouds,Colombo_2014_PAWS_survey,Heyer_Dame_2015,Tosaki_2017_linewidth_size_data}; note our definition of $R_{\rm cloud}$ is equivalent to their $\sigma_{r}$). The generic power-law scaling here is also predicted in turbulent fragmentation models (references above) and similar to observations.

Remarkably, the mass function and linewidth-size relation appear independent of resolution down to clouds with just a few gas particles -- we simply sample further and further down the mass function as we increase the mass resolution. Since most of the mass is in Toomre-mass structures, the total mass in clouds is also identical to within $\sim 30\%$ in all the runs plotted (likewise for most of the turbulent power/kinetic energy). The only case which may be biased is the lowest-resolution example ($m_{i\,1000}=470$), where the most massive clouds appear slightly more massive: this is because the clouds with $\sim 10^{6}\,\msun$ (not quite the peak of the MF, but just below it) contribute a significant total mass but cannot be resolved (this is $\sim 2$\,particles), so that mass is ``shifted'' into more massive structures numerically. But even in this marginal case, the gas-mass weighted mean $M_{\rm GMC}$ is only over-estimated by a factor $\sim 2$. In our best-resolved case, $>75\%$ of the GMC population gas mass (and $>75\%$ of the SFR of the entire galaxy) is contained in clouds with $>10^{6}\,\msun$, similar to what is observed in the Milky Way \citep{williams:1997.gmc.prop}.

Note that \citet{hopkins:fb.ism.prop}, using the same feedback physics (but in non-cosmological simulations using a different code), consider a much more detailed study of simulated GMC mass functions as well as size-mass relations, linewidth-size relations, virial parameters, internal column density distribution functions, lifetimes, and star formation efficiencies. Their resolution studies (and agreement with observations) are all consistent with our simple comparison here.

This further demonstrates, as we argued above, that resolving the thermal Jeans mass in cold gas is not important to cloud-or-larger scale dynamics (it only should matter if we are trying to resolve individual proto-stars). We will further demonstrate this below, when we consider adding artificial pressure floors to the ISM, or remove the low-temperature ($\ll 10^{4}\,$K) cooling physics: because of the strong dependence of Jeans mass on pressure, these changes in turn change the thermal Jeans mass in the cold phase gas by several orders of magnitude (factors $\sim 10^{3}-10^{5}$), yet they have no appreciable systematic effect on any results we measure.

Finally, Fig.~\ref{fig:dm.mass.resolution.substructure} compares the mass function, internal structure, and spatial distribution of subhalos within our $z=0$ {\bf m12i} halo, as a function of mass resolution. Down to a subhalo mass (and corresponding maximum rotation velocity) corresponding to $\sim 10-30$ DM particles, the predictions are independent of resolution, consistent with other studies.\footnote{Compare, for example, Fig.~\ref{fig:dm.mass.resolution.substructure} here to Fig.~9 in \citet{springel:2008.aquarius} (our runs with $m_{i,\,1000}=450,\,56,\,7$ correspond approximately to the mass resolution in their runs Aq-A-5, Aq-A-4, Aq-A-3). The results are similar, although the low-resolution mass functions accurately track the high-resolution solution down to somewhat smaller masses here (likely owing to a combination of somewhat different force softening, and a more restrictive but accurate timestep and force-tree node opening criterion).} Given the strong dependence of galaxy baryonic mass on halo mass, and much stricter resolution criteria for galaxy baryonic properties, the DM substructure mass functions and orbital distributions are always extremely well-resolved when we consider a galaxy resolved.

\vspace{-0.5cm}
\subsection{Spatial Resolution}
\label{sec:resolution:spatial}

Because our simulations are Lagrangian, there is no single meaningful definition of the spatial and/or force resolution. Here we discuss different criteria, and study their importance for our conclusions.

\begin{figure*}
\begin{tabular}{cc}
\includegraphics[width=0.50\textwidth]{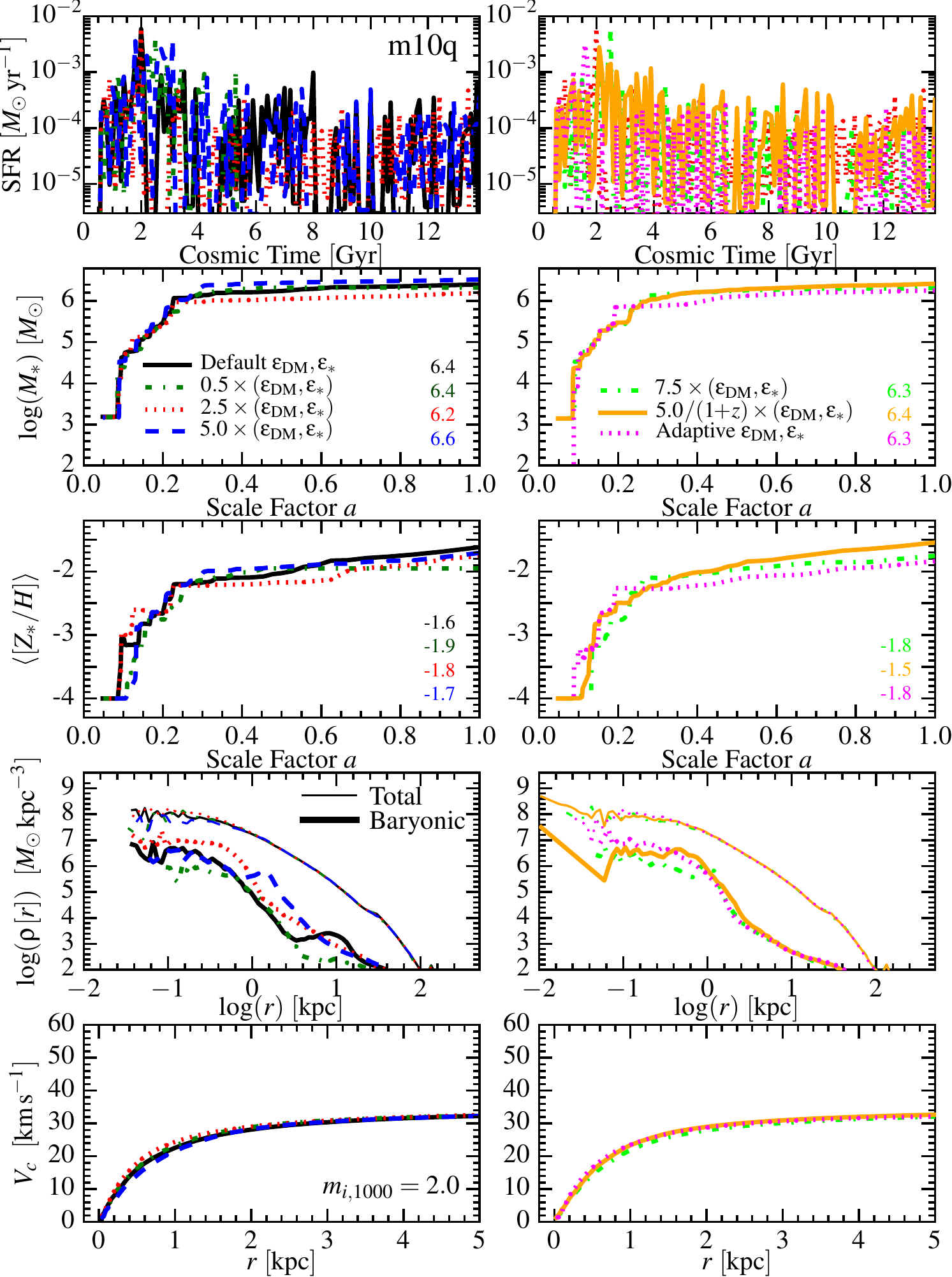} 
\hspace{-0.5cm} 
\includegraphics[width=0.50\textwidth]{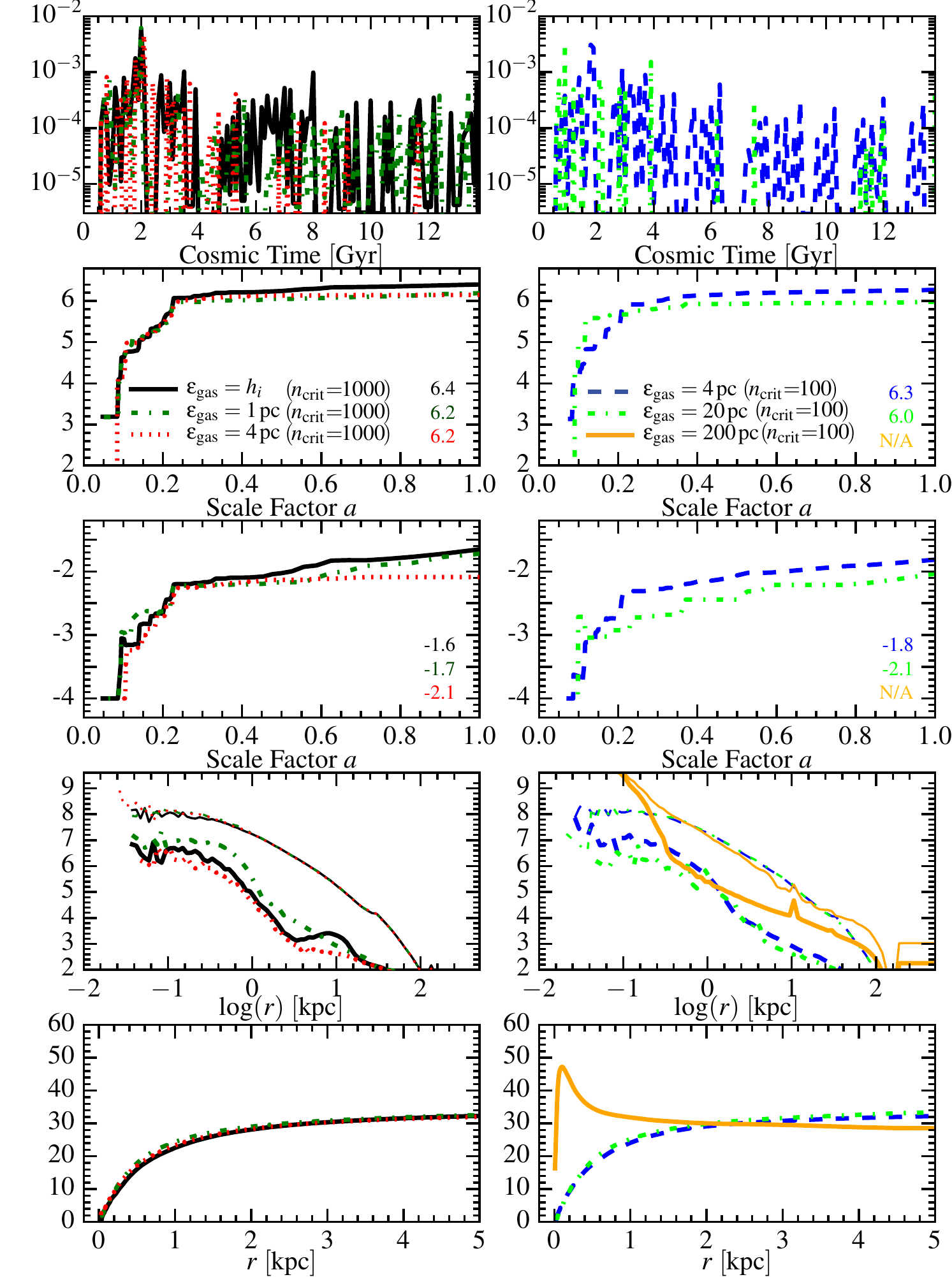} 
\end{tabular}
    \vspace{-0.25cm}
    \caption{Spatial resolution study, as Fig.~\ref{fig:res.summary}. We take the {\bf m10q} simulation with fixed mass resolution $m_{i,\,1000}=2$, and re-run with different gravitational force softening. 
    {\em Left (and middle-left):} Varying force softening for collisionless particles (DM $\epsilon_{\rm DM}$ and stars $\epsilon_{\ast}$). We multiply the default values from Fig.~\ref{fig:res.summary} at this mass resolution (constant $\epsilon_{\rm DM}=40\,{\rm pc},\ \epsilon_{\ast}=10\,{\rm pc}$) by constant values from $\sim 0.5-7.5$; we also compare softening fixed in co-moving units (default values $\times 5/(1+z)$); and we consider fully adaptive $\epsilon_{\rm DM}$ and $\epsilon_{\ast}$. These changes have no systematic effect (small differences in mass are consistent with stochastic variations, given the bursty SFH). We have verified the same in more limited surveys of $\epsilon_{\rm DM}$ and $\epsilon_{\ast}$ in simulations {\bf m10y}, {\bf m11q}, {\bf m11v}, {\bf m12i}.
    {\em Right (and middle-right):} Varying force softening for gas. We compare our default, fully-adaptive softening ($\epsilon_{\rm gas}=h_{i}$, the inter-particle spacing), to simulations with fixed gas softening set to $1-200\,{\rm pc}$. Our default (adaptive) run and those with $\epsilon_{\rm gas}=1,\,4\,{\rm pc}$ enforce a minimum gas density for star formation $n_{\rm crit}=1000\,{\rm cm^{-3}}$ (as labeled). With $1\,{\rm pc}$ softening, this produces identical results to our adaptive-softening run; but with $4\,{\rm pc}$ softening, the maximum gravitationally-resolved density $n^{\rm max} = (m_{i}/\epsilon_{\rm gas}^{\rm min})/m_{p} \approx 1000\,{\rm cm^{-3}}\,(m_{i,\,1000}/2)\,(\epsilon_{\rm gas}^{\rm min}/4\,{\rm pc})^{-3}$ barely reaches $n_{\rm crit}$, so our self-gravity criterion cannot properly resolve if gas is self-gravitating at $n>n_{\rm crit}$ and SF is artificially suppressed (slightly lower $M_{\ast}$ and $[Z_{\ast}/H]$). Taking $\epsilon_{\rm gas}=20$\,pc ($n^{\rm max}\approx 10\,{\rm cm^{-3}}$) with $n_{\rm crit}=100$ shows this problem  more severely. 
    With fixed (non-adaptive) softening, it is necessary to choose $n_{\rm crit} < n^{\rm max}$; lowering $n_{\rm crit}=100$ for $\epsilon_{\rm gas}=4\,$pc shows excellent agreement with our default (adaptive) run. 
    With $\epsilon_{\rm gas}=200\,{\rm pc}$, $n^{\rm max}=0.01\,{\rm cm^{-3}}$ is much less than $n_{\rm crit}$ {\em and} the mean galaxy gas density -- this means we cannot resolve Toomre-scale structures (large GMCs), so star formation is completely suppressed ($M_{\ast}=0$, exactly, here), and a super-dense (but non-fragmenting) gas disk forms.
     \label{fig:spatial.res.history}}
\end{figure*}

\begin{figure}
\plotonesize{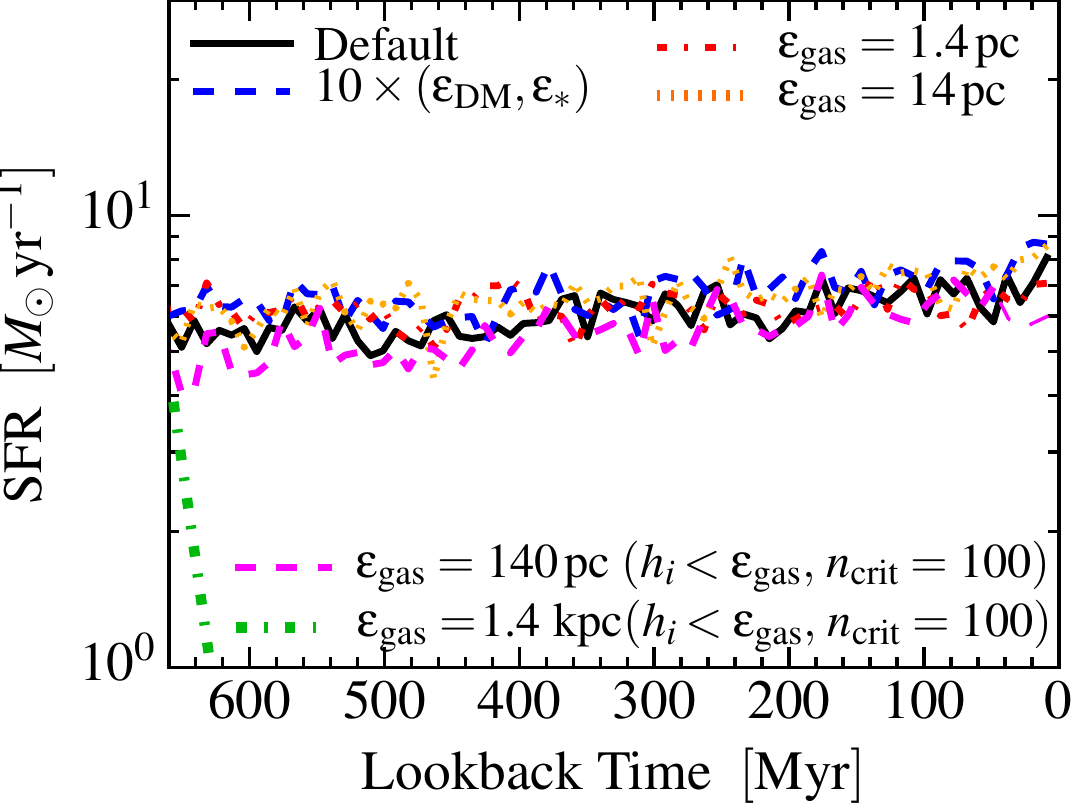}{1.0}
    \vspace{-0.25cm}
    \caption{Spatial resolution study in re-starts of a MW-mass galaxy ({\bf m12i}) at late times, as Fig.~\ref{fig:sf.z0.mass.resolution}, keeping fixed mass resolution $m_{i,\,1000}=56$ but varying the spatial resolution, as in Fig.~\ref{fig:spatial.res.history}. All runs unless otherwise labeled take $n_{\rm crit}=1000\,{\rm cm^{-3}}$. Our default run uses adaptive softening for gas, fixed softening for DM and stars ($\epsilon_{\rm DM},\,\epsilon_{\ast}$); the run labeled $10\times(\epsilon_{\rm DM},\,\epsilon_{\ast})$ increases these by $10\times$, but keeps the adaptive gas softening -- this has no effect on our prediction. Runs labeled $\epsilon_{\rm gas}$ fix the gas softening; for $\epsilon_{\rm gas}=1.4\,$pc and $14\,$pc ($n^{\rm max}=10^{6},\,10^{3}\,{\rm cm^{-3}}$, respectively), this has little or no effect. For $\epsilon_{\rm gas}=140\,{\rm pc}$ ($n^{\rm max}\approx1\,{\rm cm^{-3}}$), we see no star formation unless we lower $n_{\rm crit}$ and allow the hydrodynamic resolution $h_{i}$ to decrease to values much smaller than $\epsilon_{\rm gas}$ ($<0.1\,\epsilon_{\rm gas}$, here) -- then shocks driven by SNe produce gas at high densities (i.e.\ hydrodynamic, not gravitational, effects allow the gas to reach high densities), but the most massive GMCs are still only marginally resolved. For $\epsilon_{\rm gas} \sim \,$kpc, the galaxy never forms locally self-gravitating structures, even lowering $n_{\rm crit}=1$ and allowing $h_{i}=0.01\,\epsilon_{\rm gas}$.
    \label{fig:sf.z0.spatial.res}}
\end{figure}

\begin{figure}
\begin{tabular}{c}
\vspace{-0.4cm}
\includegraphics[width=0.95\columnwidth]{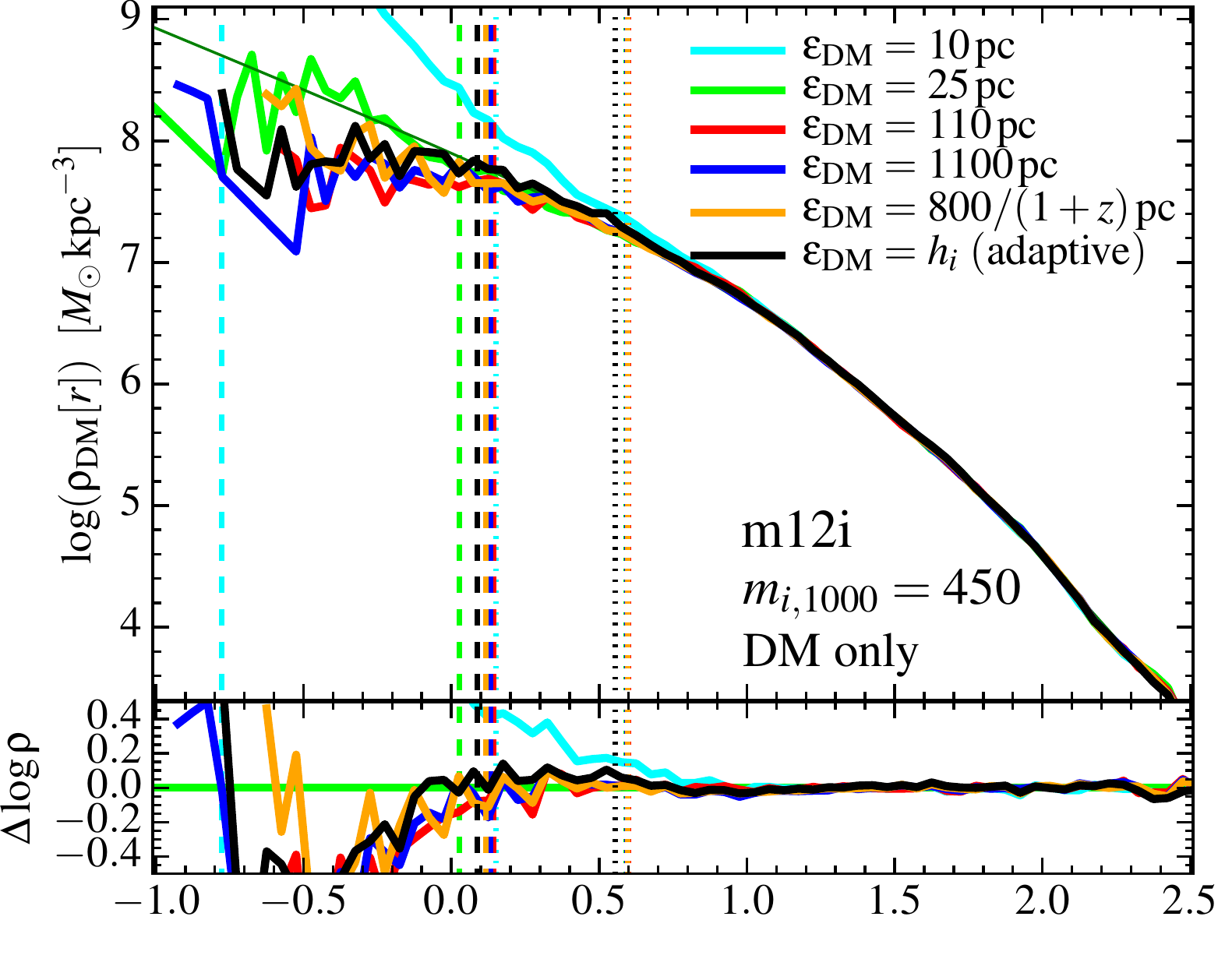} \\
\vspace{-0.4cm}
\includegraphics[width=0.95\columnwidth]{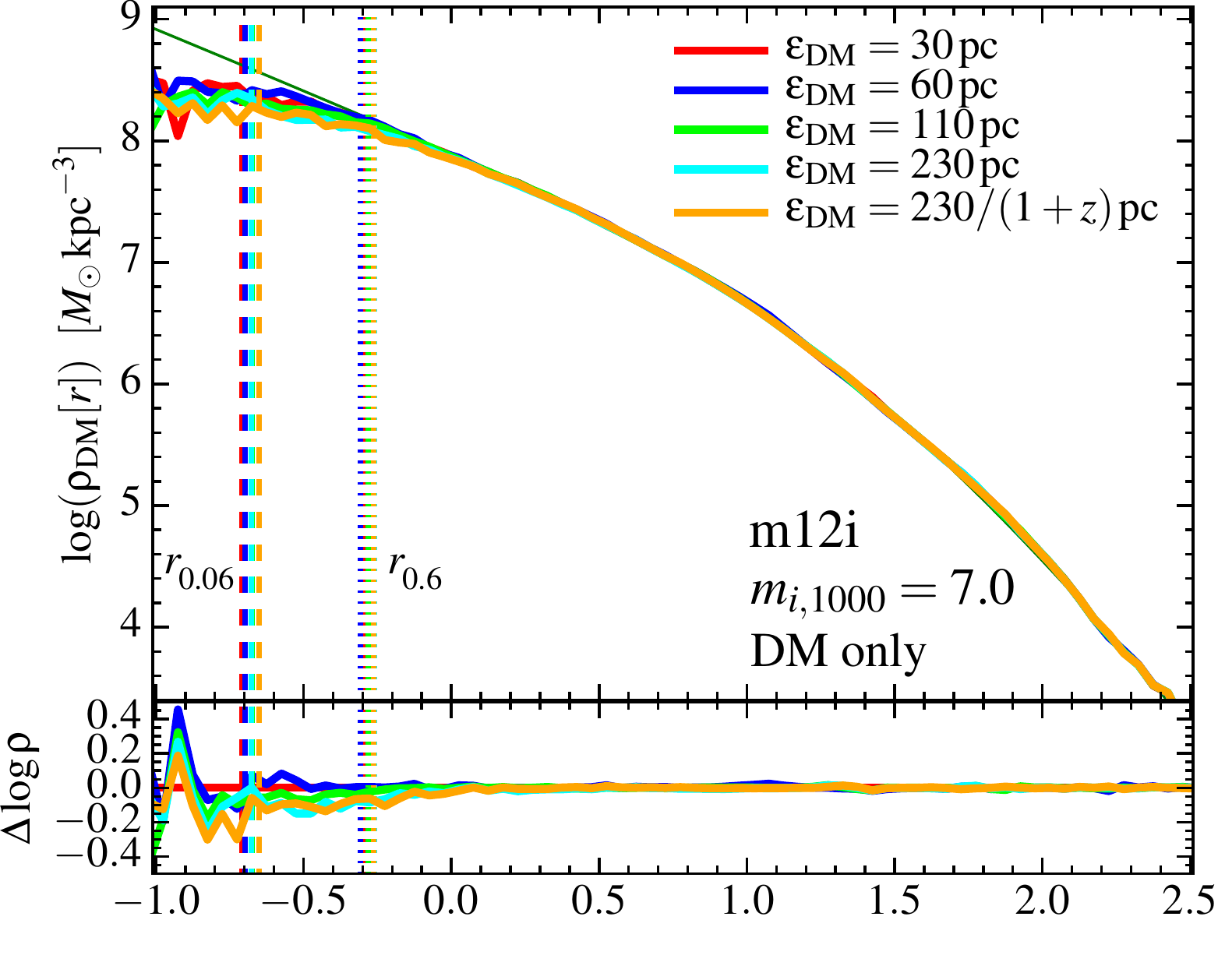} \\  
\hspace{-0.1cm}\includegraphics[width=0.925\columnwidth]{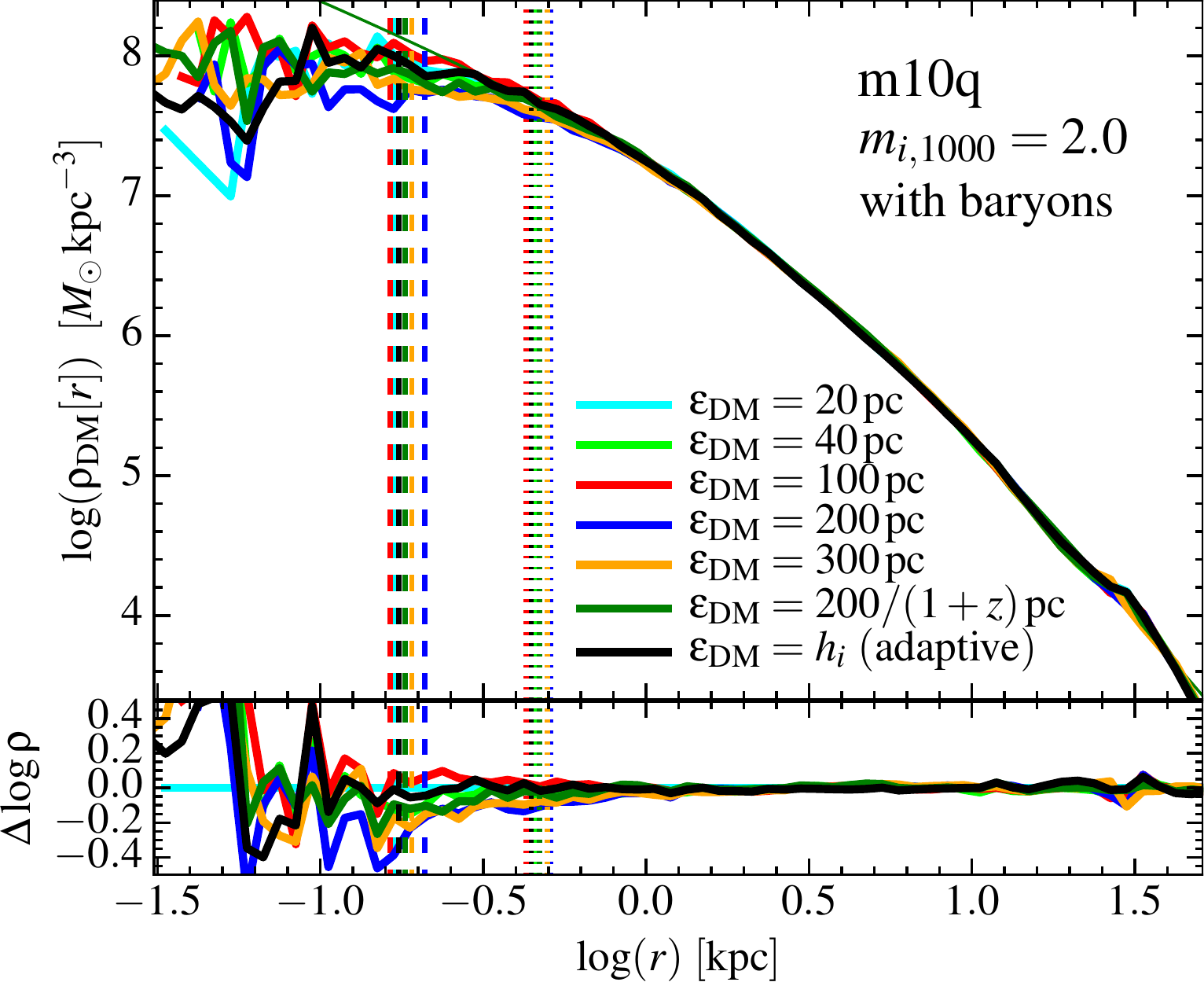} \\  
\end{tabular}
    \vspace{-0.25cm}
    \caption{Effects of force softening on the DM profile at $z=0$, at fixed {\em mass} resolution. {\em Top:} DM-only simulation of {\bf m12i}, with poor mass resolution ($m_{i,\,1000}=450$). 
    {\em Middle:} DM-only {\bf m12i}, with better mass resolution ($m_{i,\,1000}=7$). 
    {\em Bottom:} Full-physics simulation of {\bf m10q}, with mass resolution $m_{i,\,1000}=2$. 
    Vertical dashed (dotted) lines show ``convergence radii'' $r_{0.06}$ ($r_{0.6}$), as Fig.~\ref{fig:dm.mass.resolution}; thin green line is the best-fit NFW profile in the highest-resolution simulation we have run.
    DM softenings $\epsilon_{\rm DM}\sim 20-1000\,$pc ($\sim0.05-5$ times the mean inter-particle spacing inside $\sim1\,$kpc) have almost no effect at $>r_{0.6}$ and only small ($<0.2\,$dex) effects within $r_{0.06}$ (suppressing the densities when $\epsilon_{\rm DM}$ is too large, $\gtrsim 0.5\,r_{0.06}$). Mass resolution is clearly much more important, compared to force softening (compare {\em top} vs.\ {\em middle}). Only the smallest $\epsilon_{\rm DM}=10\,$pc at very poor mass resolution $m_{i,\,1000}=450$ shows hard-scattering effects (the spurious cusp) -- this occurs when $\epsilon_{\rm DM} \lesssim 0.01\,r_{0.06}$.
    \label{fig:dm.spatial.resolution}}
\end{figure}

\begin{figure}
\begin{tabular}{l}
\vspace{-0.5cm}
\includegraphics[width=0.953\columnwidth]{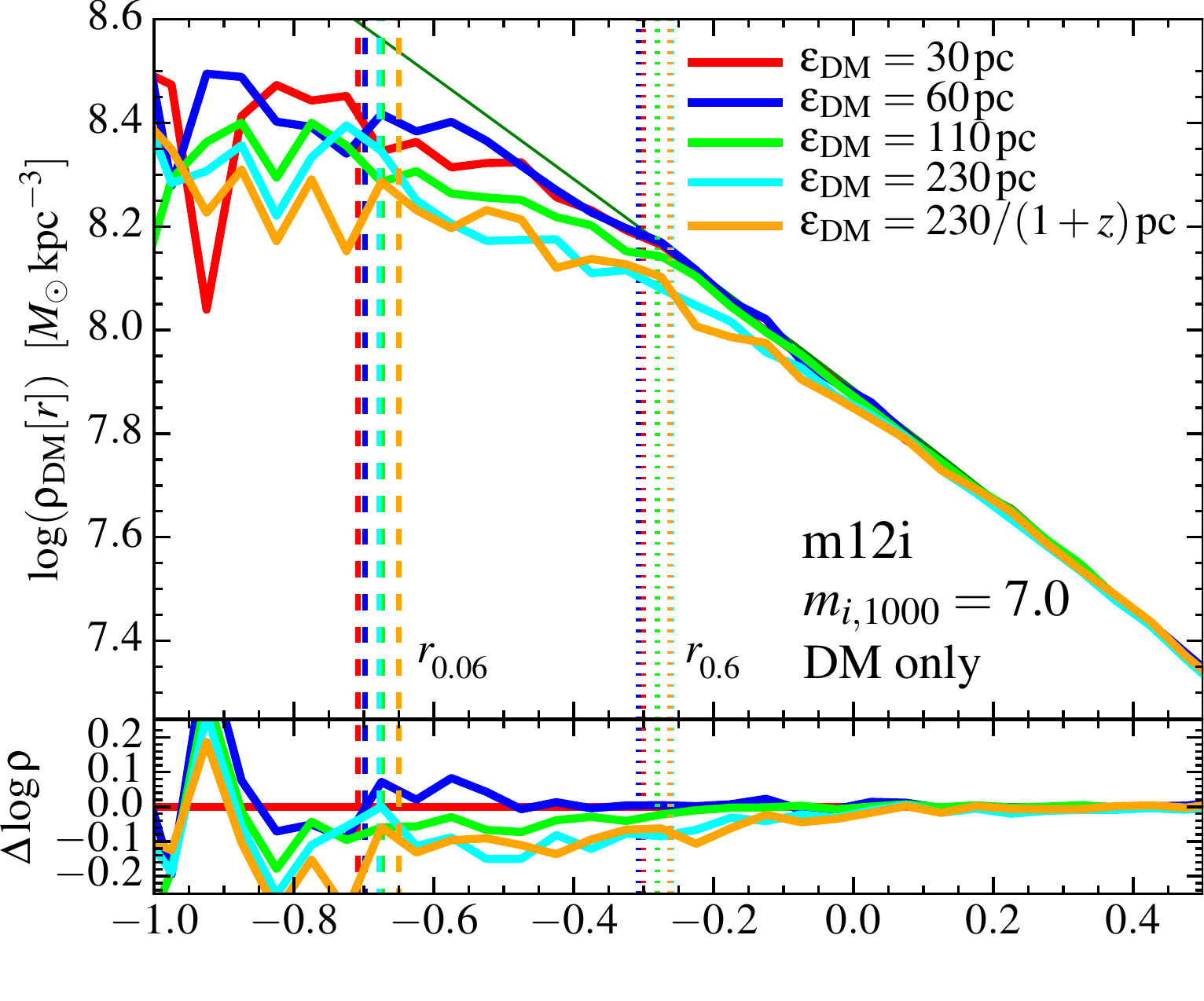} \\  
\hspace{-0.05cm}
\includegraphics[width=0.95\columnwidth]{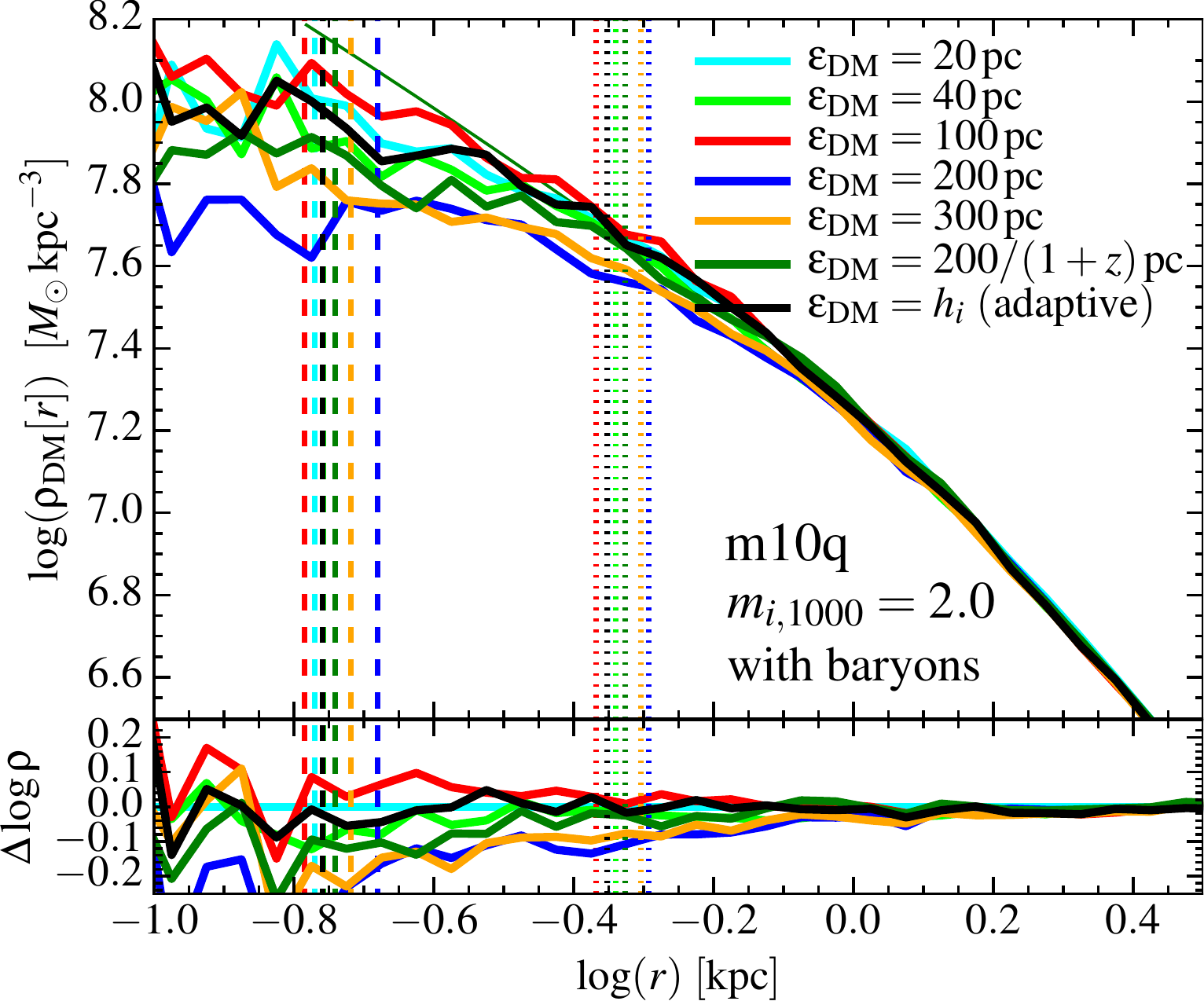} \\  
\end{tabular}
    \vspace{-0.25cm}
    \caption{Close-up of the two higher-resolution simulations from Fig.~\ref{fig:dm.spatial.resolution}, showing the DM density profile in the halo center in more detail. We can now clearly see that while the profiles agree very well at large radii ($\gtrsim$\,kpc), at small (sub-kpc) radii, the runs with excessively large force softening $\epsilon_{\rm DM}$ suppress the density profile relative to the higher-resolution solutions, by a modest factor $\sim1.2$ ($\sim1.5$) at $\sim500\,$pc ($\sim100\,$pc). The suppression can even extend to radii of order $r_{0.6}$ ({\em dotted vertical line}). The excess suppression appears in all cases we have tested when $\epsilon_{\rm DM} \gtrsim 0.5\,r_{0.06}$ (our ``convergence radius'' $r_{0.06}$ is shown as the vertical dashed line) -- approximately $\epsilon_{\rm DM} \gtrsim 100\,$pc in both cases shown here. 
    Once $\epsilon_{\rm DM}$ is less than $\sim 0.25\,r_{0.06}$, we see no evidence for improved accuracy in the simulations (differences between these runs and adaptive-softening runs are consistent with shot noise). 
    \label{fig:dm.spatial.resolution.zoom}}
\end{figure}

\begin{figure}
\begin{tabular}{c}
\includegraphics[width=0.9\columnwidth]{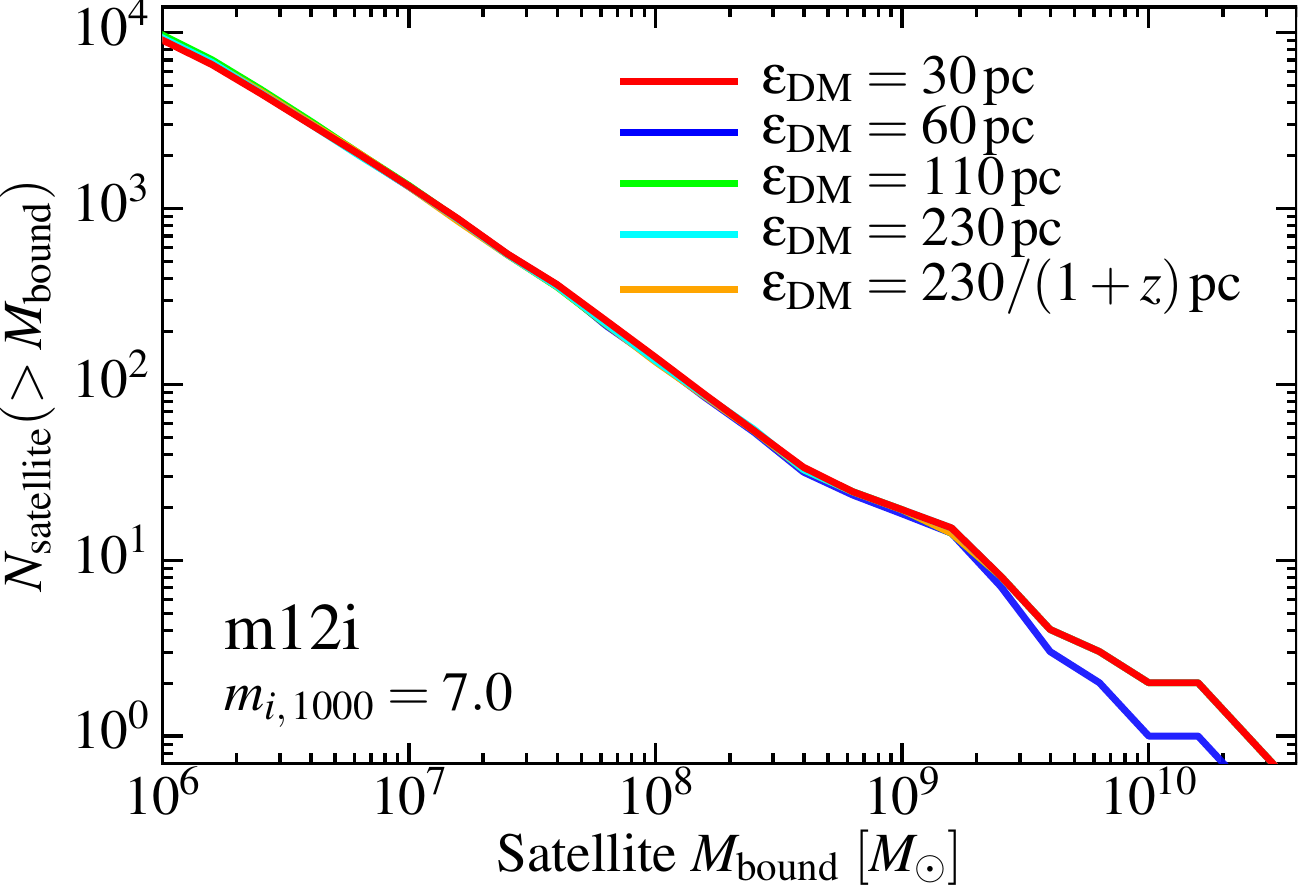} \\
\includegraphics[width=0.9\columnwidth]{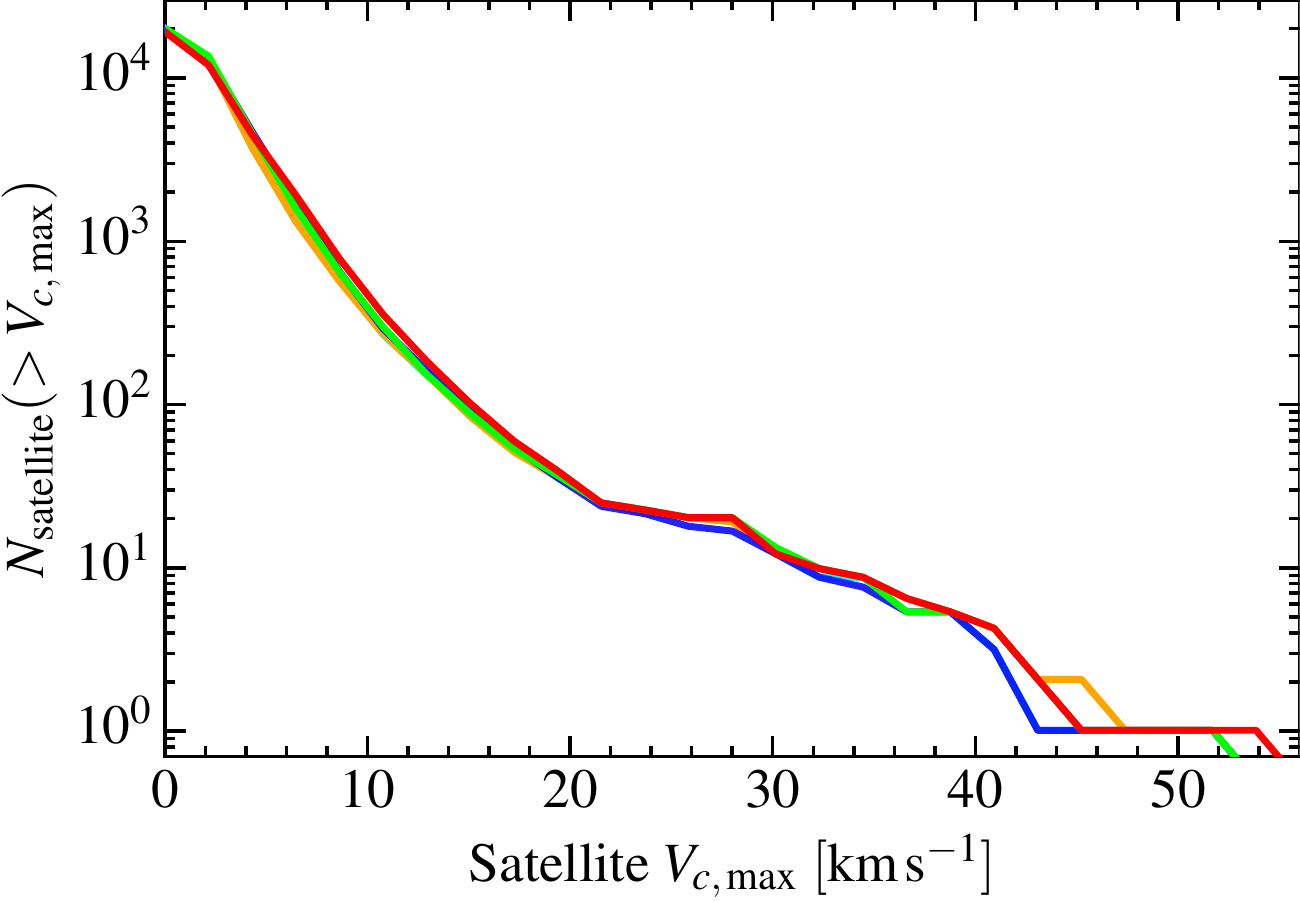} \\
\includegraphics[width=0.94\columnwidth]{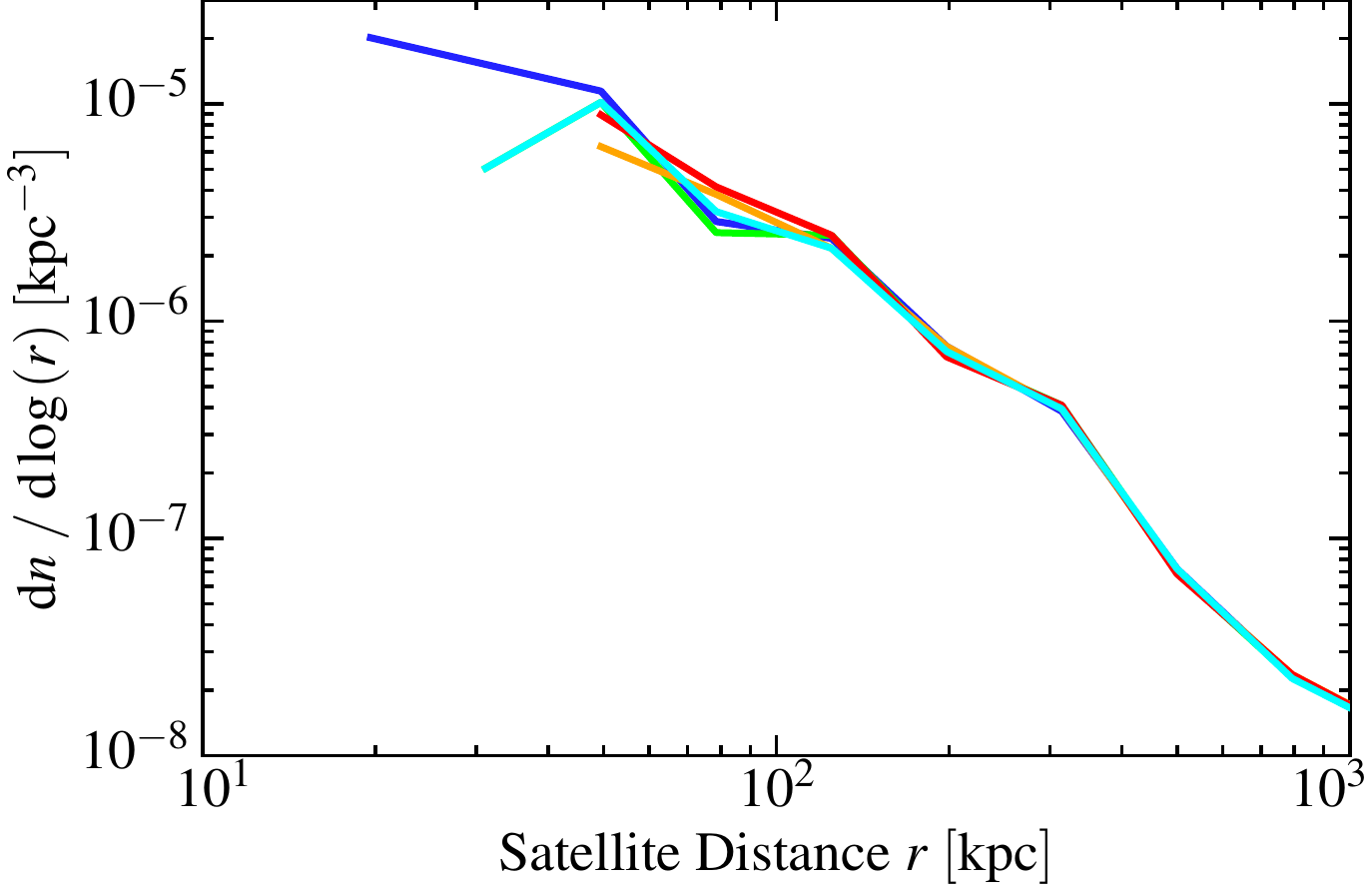} 
\end{tabular}
    \vspace{-0.25cm}
    \caption{Mass function ({\em top}), circular velocity distribution ({\em middle}), and spatial distribution ({\em bottom}) of satellites (DM subhalos) within the primary halo in DM-only runs of our {\bf m12i} simulation (at $z=0$), as Fig.~\ref{fig:dm.mass.resolution.substructure}, at fixed mass resolution ($m_{i,\,1000}=7$) but varied force softening $\epsilon_{\rm DM}$. 
    {\em Top:} Number of subhalos versus bound subhalo mass. Objects are plotted down to $<30$ DM particles.
    {\em Middle:} Number of subhalos versus maximum circular velocity. 
    {\em Bottom:} Distribution of subhalos in radial distance from the center of the primary halo. All the distributions agree well, independent of force resolution (the small deviation in the mass function of $\epsilon_{\rm DM}=60\,{\rm pc}$ owes to a single subhalo which falls just outside, instead of inside, the radius cut used to identify substructures at $z=0$).
    \label{fig:dm.force.resolution.substructure}} 
\end{figure}

\vspace{-0.5cm}
\subsubsection{Hydrodynamic Resolution}
\label{sec:resolution:spatial:hydro}

The meaning of hydrodynamic resolution in our mesh-free Godunov methods is discussed extensively with dozens of numerical examples in the methods papers \citep{hopkins:gizmo,hopkins:cg.mhd.gizmo,hopkins:gizmo.diffusion,hopkins:mhd.gizmo}. We briefly review it here. For hydrodynamics the resolution is fully adaptive, set by the inter-particle spacing $h_{i}$ -- it can, in principle, become arbitrarily small. In our Lagrangian method, the mean inter-particle spacing is directly related to the gas density $n_{\rm H}$, via
\begin{align}
\label{eqn:spatial.res.gas} h_{i}^{\rm gas} &= 16\,{\rm pc}\ m_{i,\,1000}^{1/3}\, \left( \frac{n_{\rm H}}{10\,{\rm cm^{-3}}} \right)^{-1/3}
\end{align}

In \citet{hopkins:gizmo,hopkins:mhd.gizmo} we show that the ``effective'' resolution for sound waves is identical to second-order grid methods such as {\small ATHENA}, meaning one element (``particle'') is equivalent to one cell, and the appropriate ``resolution'' should be taken to be the inter-particle spacing, as opposed to the extended ``kernel search radius'' (maximum distance to any interacting neighbor cell), which is more akin to the gradient stencil in grid-based codes (and depends on the detailed kernel shape). This is different from SPH methods, where properties are (by definition) ``smoothed'' over a kernel. In fact, because of the higher-order reconstruction, sound waves can be accurately re-constructed to scales about $\sim 1/2$ the inter-particle separation (\citealt{hopkins:gizmo} Fig.~2, \citealt{hopkins:mhd.gizmo} Fig.~1). Contact discontinuities can be captured and conserved across $\sim2-3$ elements (two inter-element spacings), much sharper than non-Lagrangian codes if the discontinuity is moving (see \citealt{hopkins:gizmo} Figs.~16-23). Strong shocks are resolved over a similar $\sim 3$ elements, close to {\small ATHENA} (\citealt{hopkins:gizmo} Figs.~3,\,10-15,\,29-30). Intrinsically multi-dimensional problems, such as conserving vorticity of a rotating, pressure-equilibrium disk for a few orbits are more demanding; this requires a few hundred total elements (about $\sim 10$ elements ``per radian'' around the structure; \citealt{hopkins:gizmo}, Fig.~4-5). However that is easily satisfied for any disk with a resolved vertical scale-height $H\gtrsim 1-2\,h_{i}$. In multi-dimensional super-sonic turbulence, numerical dissipation and noise truncates the inertial range at wavelengths approximately $\sim5$ times the inter-particle spacing (comparable to {\small AREPO} and {\small ATHENA}; \citealt{hopkins:gizmo} Fig.~26-28).

\vspace{-0.5cm}
\subsubsection{Force Softening: Definitions \&\ Optimal Choices}
\label{sec:resolution:spatial:optimal}

The meaning of ``gravitational spatial resolution'' is somewhat ambiguous. Commonly, however, this is used to refer to the force softening $\epsilon$. As discussed in \S~\ref{sec:methods:gravity}, we solve gravitational forces for the {\em same} gas-mass distribution as hydrodynamic forces: this means setting the gravitational force softening for gas adaptively, 
\begin{align}
\label{eqn:force.softening.gas} \epsilon_{i}^{\rm gas} \equiv h_{i}^{\rm gas} &= 16\,{\rm pc}\ m_{i,\,1000}^{1/3}\, \left( \frac{n_{\rm H}}{10\,{\rm cm^{-3}}} \right)^{-1/3}
\end{align}
For clarity, we use the same definition of force softening and spatial resolution, corresponding to the inter-particle separation $h_{i}=\Delta x$ (defined such that $\rho_{i} = m_{i}/h_{i}^{3}$). For our default kernel, this means that the commonly quoted ``Plummer equivalent'' softening is $\epsilon_{\rm plummer} \approx (2/3)\,\epsilon = (2/3)\,h_{i}$. For the reasons in \S~\ref{sec:methods:gravity} and \S~\ref{sec:resolution:spatial:optimal} below, we adopt constant force-softening parameters for DM ($\epsilon_{\rm DM}$) and stars ($\epsilon_{\ast}$), with their values set as described below (for specific values, see Table~\ref{tbl:sims}). We define these for clarity the same way as for gas, so that they are softened according to the same kernel shape (in other words, a gas, star or DM particle with the same mass and $\epsilon$ will exert identical gravitational forces at all radii).

We will explore variations below, but first summarize our best estimate of ``optimal'' parameters here:



\begin{itemize}
\item{\bf Gas:} For gas, the optimal force softening is un-ambiguous: it should be set adaptively to match the hydrodynamic resolution. This is optimal for several reasons. (1) It is the physically correct set of equations for a collisional fluid. The hydrodynamic solver describes a mass distribution (not just at particle locations, but everywhere in the domain, according to the reconstruction of the method). 
We should therefore solve the Poisson equation {\em for the same mass distribution} -- anything else is fundamentally ill-posed and can produce unphysical outcomes \citep{bate:1997.sph.res.reqs}.\footnote{Strictly speaking, it is only possible to solve the Poisson equation for a mass distribution accurate to the same order as the reconstruction order of the hydrodynamic method. But up to truncation errors, this is exact in matching the gas distribution.} (2) Many authors have shown that this provides an optimal numerical softening, in the sense that it automatically provides the most accurate solution (converges most rapidly) while minimizing $N$-body integration errors \citep{merritt:1996.optimal.softening,bate:1997.sph.res.reqs,romeo.1998:optimal.softening,athanassoula:2000.optimal.force.softening.collisionless.sims,dehnen:2001.optimal.softening,rodionov:2005.optimal.force.softening,price:2007.lagrangian.adaptive.softening,barnes:2012.softening.is.smoothing,hubber:2013.res.criteria.in.star.clusters.strong.scattering.effects}. (3) It guarantees that numerical hard gas-gas scattering can {\em never} dominate over physical self-gravitating motions. (4) It removes ambiguity about the meaning of force/spatial/hydrodynamic resolution (it makes the mass resolution the un-ambiguous resolution scale). (5) In Lagrangian codes, spatially, self-gravitating objects will be able to collapse (correctly) to arbitrarily high densities, where some star formation criterion should identify them. This makes it essentially impossible to artificially suppress star formation by ``tuning'' some star formation threshold. (6) It automatically behaves correctly in all density regimes, and naturally removes any ambiguity about co-moving or physical softenings in cosmological integrations.\footnote{Adaptive gas softening is the standard in most grid-based codes \citep{kravtsov:1997.ART,teyssier:2002.RAMSES,bryan:2014.enzo} and moving-mesh codes \citep[][]{springel:arepo} and in particle-based codes as well in e.g.\ the fields of star and planet formation \citep[see e.g.][]{bate:1997.sph.res.reqs}. Particle-based codes in galaxy formation have proved a surprising historical exception.} 

If constant gravitational softenings for gas must be used, they must be chosen sufficiently small to resolve the vertical scale-heights of the cold gas disk, and Toomre lengths of the most massive Toomre-mass objects discussed in our mass-resolution criteria ($\epsilon_{\rm gas} \ll 100\,$pc).

\item{\bf Dark Matter:} For dark matter, we will show that the force softening makes little or no difference to any conclusions, so long as it is not extremely small ($\sim100\times$ smaller than our default, which would trigger hard-scattering effects), or extremely large (which would over-soften the central DM cusp). One option is to use fully adaptive softening. However, (1) this requires stronger timestep constraints (\S~\ref{sec:resolution:time:ags}), which considerably increase computational expense for no apparent improvement in accuracy; and (2) for a collisionless fluid, the ``correct'' adaptive scaling is {\em physically} ambiguous. It is not simply the case that the spatial domain of the DM represented by particle ``$a$'' must shrink, if the DM particles $b$ around $a$ move inwards (i.e.\ if $\nabla \cdot {\bf v}_{\rm DM}^{a} < 0$), because those particles can move {\em through} particle $a$ without compressing it (formally, there is no unique relation between the real-space DM $N$-body particle distribution and their phase-space distribution; for discussion see \citealt{abel:phase.space.sheets.for.dm}). 

Therefore, fixed softenings appear preferable. We adopt softenings fixed in physical units at $z<10$ (co-moving above this), since the halo centers do not change significantly in density over this redshift range, so this maintains an approximately fixed ratio of $\epsilon_{\rm DM}$ to the inter-particle spacing $h_{i}^{\rm DM} \sim 17\,{\rm pc}\,m_{i,\,1000}^{1/3}\,(\rho_{\rm DM}/10^{9}\,M_{\sun}\,{\rm kpc^{-3}})^{-1/3}$.

For fixed DM softenings, avoiding over-softening in the central DM profile -- i.e.\ converging as accurately as possible to the solution of higher-resolution simulations -- requires a force softening {\em smaller} than the 200-particle Power-like radius (where $t_{\rm relax}\approx 0.06\,t_{\rm circ}(R_{200})$) by a factor of at least a couple,  i.e.\ $\epsilon_{\rm DM} \lesssim 0.5\, r_{0.06}$. This is approximately ensured if $\epsilon_{\rm DM} < 30\,{\rm pc}\,m_{i,\,1000}^{1/2}\,(M_{\rm vir}/10^{12}\,\msun)^{-0.2}$ (we estimate this by taking $M_{\rm enc}(<r_{0.06}) \approx 220\,m_{i}$, and assuming NFW profiles with concentration $c\approx 10\,(M_{\rm vir}/10^{12}\,\msun)^{-0.15}$). The minimum softening is given by the value where we begin to see hard scattering effects cause excessive $N$-body relaxation in the profile and/or velocity distribution function, which is roughly $\epsilon_{\rm DM} > 0.03\,{\rm pc}\,m_{i,\,1000}\,(M_{\rm vir}/10^{12}\,\msun)^{-2/3}$ (obtained by comparing the hard-scatter $\Delta v$ to $V_{c}$; \S~\ref{sec:resolution:spatial:nbody.heating}). Note the large dynamic range in between (e.g.\ for our default runs, this gives $1\,{\rm pc} \lesssim \epsilon_{\rm DM} \lesssim 75\,{\rm pc}$). We will show we can vary $\epsilon_{\rm DM}$ by multiple orders of magnitude without changing out results.\footnote{Note that it becomes impossible to satisfy both criteria for particle masses $m_{i}$ corresponding to $\lesssim 40$ DM particles in the halo; of course this is because the halos then cannot be internally spatially resolved.} To be conservative we choose softenings about $\sim 2-3$ times smaller than the upper limit above (giving $\epsilon_{\rm DM} \sim h_{i}^{\rm DM}$ in halo centers). Various tests have shown this is optimal to reduce both noise and over-softening errors \citep[see, e.g.][]{merritt:1996.optimal.softening,athanassoula:2000.optimal.force.softening.collisionless.sims,dehnen:2001.optimal.softening,barnes:2012.softening.is.smoothing,hubber:2013.res.criteria.in.star.clusters.strong.scattering.effects}.\footnote{Another common criteria for DM softening is that the maximal 2-body acceleration ($a_{i}\sim G\,m_{i}/\epsilon_{\rm DM}^{2}$) not exceed the bulk acceleration at some radius ($a_{\rm bulk} \sim G\,M_{\rm enc}(<r)/r^{2}$). If we assume an NFW profile, and require $a_{i} < \langle a_{\rm bulk} \rangle$ where  $\langle a_{\rm bulk} \rangle$ is the mass-averaged mean acceleration of DM within the halo, we obtain $\epsilon_{\rm DM} \gtrsim 6\,{\rm pc}\,(c/10)^{-0.45}\,m_{i,\,1000}^{1/2}\,(M_{\rm vir}/10^{12}\,M_{\sun})^{-1/2}$ (we approximate the exact scaling with the halo concentration $c$ by a power-law, good to $\sim 10\%$ over the entire range of interest). Our fixed-$\epsilon_{\rm DM}$ choices agree reasonably well with this scaling. However, we stress that we see no measurable errors or deviations in the mass profile, rate-of-growth of structure, or velocity distribution function, even at/outside $R_{\rm vir}$, using order-of-magnitude smaller $\epsilon_{\rm DM}$. This is because this acceleration criterion is not meaningful when $\epsilon_{\rm DM}$ is below the inter-particle separation, because only very rarely will particles approach within separations $\ll \epsilon_{\rm DM}$, and when they do, the net {\em velocity} imparted by the encounter will be given by the scaling in \S~\ref{sec:resolution:spatial:nbody.heating}: $(\Delta v / \sigma_{v,\,{\rm DM}}) \sim 10^{-4}\,m_{i,1000}\,(20\,{\rm pc}/\epsilon_{\rm DM})\,(\sigma_{v,\,{\rm DM}}/100\,{\rm km\,s^{-1}})^{-2} \ll 1$. Equivalently, the timescale required for $N$-body heating at $\sim R_{\rm vir}$ to perturb a particle orbit by $\sim 10\%$ is much greater than $t_{\rm Hubble}$ for all $\epsilon_{\rm DM} \gtrsim 1\,{\rm pc}$.}

%
%

\item{\bf Stars:} Stars are the most ambiguous softening case. Fortunately, like dark matter, our results appear almost completely independent of how gravity from stars is softened (\S~\ref{sec:resolution:spatial:collisionless.tests}). As with DM, the same physical ambiguities apply to using adaptive softenings for a collisionless fluid. Moreover, if the adaptive softening is based on the inter-star distance, a star born in a pure gas cloud would instantly ``jump'' to a huge force softening; if based on inter-gas distance, a pure-stellar bulge would be over-softened out to the gaseous halo. In runs with adaptive DM+stellar softening, we therefore set $\epsilon_{\ast}$ to scale with the inter-baryon distance (gas+stars), which at least has the advantage that it handles both extremes above correctly. Even in this case, however, a ``real'' (self-gravitating, resolved) star cluster can artificially expand once gas is blown away, because the neighbor search expands, even though self-gravity should fix the size (hence softenings). With constant softenings, the ambiguity is that stars form with very different densities. We have also tested $\epsilon_{\ast}$ fixed in time but variable particle-to-particle, to equal the softening of the gas from which the particle formed; while this does not change our results, it leads to wildly variable $\epsilon_{\ast}$ for stars in the same location at late times (which formed from gas at different densities at different times), which is numerically problematic. A fixed $\epsilon_{\ast}$ chosen to be large ensures smoothness of the potential but this is not actually physical for stars (because star formation is clustered, the stellar potential is ``lumpy'' on small scales), and we see this again causes sudden, massive expansion of the softening after a star particle forms. A reasonable compromise is to set $\epsilon_{\ast}$ similar to the gas softening at the mean density of star formation, $\epsilon_{\ast} \sim 3.4\,{\rm pc}\,m_{i,\,1000}^{1/3}\,(\langle n \rangle_{\rm SF}/10^{3}\,{\rm cm^{-3}})^{-1/3}$, and our default choice follows this criterion.

\end{itemize}

\vspace{-0.5cm}
\subsubsection{Other ``Spatial Resolution'' Definitions}
\label{sec:resolution:spatial:gravity}

As noted above, there is no single ``spatial resolution'' in our simulations. In Table~\ref{tbl:res}, we therefore provide a number of other ``effective spatial resolution'' values for one of our resolution studies (our {\bf m12i} study) from Fig.~\ref{fig:res.summary}, at each resolution level we have considered. Specifically we quote the ``spatial resolution'' (inter-particle spacing $h_{i}$) at: 
\begin{enumerate}
\item{\bf Maximum Density/Minimum Softening:} This is simply the minimum $\epsilon_{i}^{\rm gas} \equiv h_{i}^{\rm gas}$ reached in the simulation, at any time. This (by definition) represents the extreme in the simulation, not ``typical'' values -- in our highest-resolution simulations, this corresponds to densities $n_{\rm max} \sim 5\times10^{6}\,{\rm cm^{-3}}$, which are reached by the simulation only briefly in the galactic nucleus in an intense, high-redshift starburst.

\item{\bf Star-Forming Densities:} We output from our simulations the gas density at which every individual star particle forms, which maps one-to-one to the equivalent $h_{i}$. We show the mean density, $\langle n_{\rm SF} \rangle$, weighted by total star formation, integrated over the entire galaxy history to $z=0$, and corresponding $h_{i}^{\langle SF \rangle}$. We emphasize that because star formation is only allowed in self-gravitating gas which is also self-shielding and molecular, the mean $\langle n_{\rm SF} \rangle$ is always significantly larger than the {\em minimum} density at which star formation is allowed ($n_{\rm SF,\,min}$), even when the latter is set to $n_{\rm SF,\,min}=1000\,{\rm cm^{-3}}$. 

\item{\bf Thermal Jeans Mass (Warm/Hot Gas):} As discussed in \S~\ref{sec:resolution:mass:firephysics}, the smallest resolveable Jeans mass simply tells us where the ``fragmentation cascade'' in self-gravitating objects will be truncated (and our sub-grid star formation model will take over). We can calculate this minimum scale, but the correct ``effective Jeans mass'' (the actual characteristic mass of structures) depends on turbulence, {\em not} on the sound speed, in a super-sonically turbulent medium \citep[see e.g.][]{padoan:2002.density.pdf,hennebelle:2008.imf.presschechter,hopkins:excursion.imf,hopkins:excursion.ism,hopkins:excursion.clustering,hopkins:excursion.imf.variation,hopkins:frag.theory,hopkins:2013.turb.planet.direct.collapse,guszejnov:cmf.imf,guszejnov:gmc.to.protostar.semi.analytic}. To get some insight, however, for the warm ($\gtrsim 10^{4}\,$K) gas, the turbulence is expected to be trans-sonic, so we can reasonably simplify by considering the thermal Jeans properties. We calculate the smallest possible Jeans mass resolvable with at least $\sim10$ elements at these temperatures, and the corresponding resolved Jeans length (with $\sim 3\,h_{i}$) and gas density.\footnote{Define the Jeans length $\lambda^{J}$ and mass $m_{J}=(4\pi/3)\,\rho\,(\lambda^{J}/2)^{3}$. Following \S~\ref{sec:resolution:mass:gravity}, a ``resolved'' structure has some number $N\sim 10\,N_{10}$ elements, so mass $\sim N\,m_{i}$. By definition of the inter-particle spacing, we also have $\rho = m_{i}/h_{i}^{3}$. This gives $\lambda_{J} \approx 3\,h_{i}\,N_{10}^{1/3}$, always -- in other words, our adaptive softening ensures the Jeans length $\lambda_{J}$ is always spatially-resolved {\em provided} the Jeans mass $m_{J}$ is mass-resolved. Now use this and the definition of Jeans length, $\lambda^{J}=c_{s}/\sqrt{G\,\rho}$, to solve for the Jeans radius $\lambda^{J}/2 \approx 9\,G\,m_{i}\,N_{10}/c_{s}^{2}$; for $T=10^{4}\,K$ (and assuming fully-ionized gas), this becomes $\approx 0.3\,{\rm pc}\,m_{i,\,1000}\,N_{10}$. Taking a minimum resolved size $N_{10}=1$, this gives the minimum resolved $\lambda_{J}$; the corresponding $h_{i}$ gives, in turn, the corresponding maximum density $n_{\rm max}^{J}$.} Any {\em lower} density or {\em larger} Jeans lengths have resolved thermal Jeans fragmentation -- important for our purposes, these values need to be sufficient to capture the largest GMCs (sizes $\sim 100-200\,{\rm pc}$ and densities $\sim 10\,{\rm cm^{-3}}$), which form out of the warm gas. Note that if the gas is hotter, it becomes {\em easier} to resolve Jeans-scale structures.

\item{\bf Turbulent Jeans Mass (Cold Gas):} Once the gas cools below temperatures $\sim 10^{4}\,$K (in e.g.\ GMCs or molecular disks), the turbulence is highly super-sonic -- by definition, thermal pressure does not control the dynamics, and the appropriate Jeans scale for fragmentation is the {\em turbulent} Jeans scale. This replaces $\lambda_{J} = c_{s}/\sqrt{G\,\rho}$ by $\lambda_{\rm turb} \approx \langle v_{\rm turb}^{2}(\lambda_{\rm turb}) \rangle^{1/2} / \sqrt{G\,\rho}$, where $v_{\rm turb}(\lambda_{\rm turb})$ is the rms turbulent velocity measured within the same region. A rigorous definition and derivation of the corresponding turbulent Jeans length, mass, and fragmentation cascade is given in \citet{hopkins:frag.theory}. If we assume a linewidth-size relation seen in the ISM and expected for super-sonic turbulence ($v_{\rm turb} \propto \lambda^{1/2}$), and that the ``parent'' clouds have virial parameter of unity and (according to Larson's Laws) a universal surface density $\Sigma_{\rm cloud} = \Sigma_{300}\,300\,\msun\,{\rm pc^{-2}}$, and take a minimum resolved-object mass $\sim N\,m_{i}$, then following \citet{hopkins:frag.theory} gives a minimum-resolved length $\lambda_{\rm turb} \approx 4.0\,{\rm pc}\,(m_{i,\,1000}\,N_{10}/\Sigma_{300})^{1/2}$ and maximum-resolved density $n_{\rm max}^{\rm turb} \approx 1.2\times 10^{4}\,\Sigma_{300}\,(m_{i,\,1000}\,N_{10}/\Sigma_{300})^{-1/2}$. 

\item{\bf Dark Matter Power-Type Radius:} None of these criteria apply to dark matter. We therefore quote the DM inter-particle separation: both its minimum value $h_{i}^{\rm DM,\,min}$, as well as rms value within twice the effective radius of the baryonic galaxy at $z=0$, $h_{i}^{\rm DM,\,core}$. However for the DM profiles, \S~\ref{sec:resolution:mass:tests} above shows that the internal structure of a collisionless object evolved for a Hubble time is well-converged inside a radius enclosing $\sim 200$ DM particles. We therefore quote this as well, for the DM.

\end{enumerate}

Qualitatively, the results from our other MW-mass series ({\bf m12f}) are nearly identical to those in Table~\ref{tbl:res}. Our dwarfs ({\bf m10v} and {\bf m10q}) have superior resolution owing to their much smaller particle masses.

\vspace{-0.5cm}
\subsubsection{Effects of Collisionless Force Softening: Resolution Studies}
\label{sec:resolution:spatial:collisionless.tests}
    
Figs.~\ref{fig:spatial.res.history}-\ref{fig:sf.z0.spatial.res} consider the effects of varied force softening for both collisional (gas) and collisionless (DM, stars) particles in our cosmological simulations, for both a dwarf and MW-mass galaxy. First, consider the effects of the collisionless softening. We vary both DM and stellar softening simultaneously by multiplying both by a constant ($\sim 0.5-10$) relative to their default values, setting them to be fixed in comoving (instead of physical) units at all redshifts, or setting them purely adaptively (with appropriately careful time-stepping, see \S~\ref{sec:resolution:time:ags}).\footnote{For adaptive DM softening, we determine $\epsilon_{\rm DM}$ using the same methods as for gas, but only counting other DM as ``neighbors.'' In other words, we set $\epsilon_{i}^{\rm DM} = h_{i}^{\rm DM}$, where $h_{i}^{\rm DM}$ is determined from the {\em dark matter} particle neighbor distribution in the identical manner to $h_{i}^{\rm gas}$, ensuring $h_{i}^{\rm DM}$ is the mean (kernel-averaged) inter-particle spacing {\em of dark matter particles} within the kernel. This gives $\epsilon_{i}^{\rm DM} \equiv h_{i}^{\rm DM} = 17\,{\rm pc}\ m_{i,\,1000}^{1/3}\, ({\rho_{\rm DM}}/{10^{9}\,\msun\,{\rm kpc^{-3}}} )^{-1/3}$. For adaptive softening for stars, we follow the same exercise but include all baryonic (gas+star) particles as ``neighbors.''}

To first approximation, we see no effect from these changes. There are some variations in the growth history in Fig.~\ref{fig:spatial.res.history}, but given the bursty nature of dwarf star formation histories, this appears to be primarily stochastic. Still, there is a (weak) tendency, on average, for the runs with larger softenings to produce slightly higher stellar masses (and corresponding metallicities). Although not apparent by-eye in Fig.~\ref{fig:spatial.res.history}, these larger-softening runs appear to have slightly less-bursty SFHs (if we quantify this by measuring the logarithmic variance in the SFR measured in $10\,$Myr bins relative to a rolling Gyr-average value) -- and in previous studies, we have shown that more-bursty SF produces more efficient outflows (because the same feedback is more concentrated; \citealt{muratov:2015.fire.winds}). We have also verified this by re-running our {\bf m11q} and {\bf m11v} simulations (the runs where ``burstiness'' has the most dramatic effect on the DM halo structure) with $5\times$ larger softening $(\epsilon_{\rm DM},\,\epsilon_{\ast}) = (200,\,20)\,{\rm pc}$; the ``burstiness'' is still obviously present but is quantitatively suppressed, and the stellar masses of both increase by a factor $\sim 1.4$. It makes sense that such large softenings suppress burstiness to some extent: they smear out the DM mass profile, meaning the potential is weaker (so SF needs to be less concentrated/dramatic before it can drive outflows) and smear out any star clusters or small merging galaxy stellar components that would otherwise provide more concentrated feedback.  But these are unphysical effects; furthermore, the ``level of burstiness'' in the adaptive softening runs agrees well with our default simulations -- together this gives us confidence in our default choices. But in any case, the magnitude of this effect is much smaller than systematic uncertainties in the stellar mass predictions. Not surprisingly, for the MW-mass system, which has a more smooth SFH, the effect is minimal ($< 10\%$). 

Fig.~\ref{fig:dm.spatial.resolution} confirms the result of previous studies \citep[e.g.][]{power:2003.nfw.models.convergence} that at fixed mass resolution, changing the DM force softening (either in physical or comoving units or using fully-adaptive softenings) has very little effect. We see that outside the radius containing $N\sim 200$ particles, these choices (with fixed $\epsilon_{\rm DM}$ varied from $\sim 10-1000$\,pc) have almost no effect on the mass profile. Of course, if $\epsilon_{\rm DM}$ is too large, it will eventually suppress any smaller-scale structure (e.g.\ artificially flattening the DM cusp) -- this occurs when $\epsilon_{\rm DM}$ is larger than the convergence radius $r_{0.06}$. \citet{power:2003.nfw.models.convergence} show that convergence to the correct solution in the central structure of $N$-body dark matter halos also requires $\epsilon_{\rm DM} < c^{-2}\,[\ln{(1+c)}-c/(1+c)]\,R_{200}\lesssim 0.01\,R_{200}$; this is easily satisfied by any of our simulations that satisfy $\epsilon_{\rm DM} \lesssim r_{0.06}$. Note that most previous studies have considered only pure-DM simulations; we show here the same conclusions apply even in our ``full physics'' runs.

Fig.~\ref{fig:dm.spatial.resolution} does raise one important caveat: if $\epsilon_{\rm DM} \ll h_{i}$ (the inter-particle spacing), and the ``hard scattering'' velocity deflection between two particle is comparable to their velocity dispersion, then runaway $N$-body effects can produce a gravito-thermal catastrophe over a Hubble time. We see this in our test with $m_{\rm DM}=2.8\times10^{6}\,\msun$ and $\epsilon_{\rm DM}=10\,$pc; much smaller force softening for dark matter than used in our production FIRE-2 simulations at comparable mass resolution. Considering the hard-scattering velocity from our $N$-body heating rate calculation below, we estimate that avoiding this requires $\epsilon_{\rm DM} \gtrsim 0.02\,{\rm pc}\,m_{i,\,1000}$, easily satisfied in all our production simulations.

Fig.~\ref{fig:dm.spatial.resolution.zoom} shows the central portion of the profiles from Fig.~\ref{fig:dm.spatial.resolution} in closer detail. Here we can see that there is some systematic suppression of the predicted mass profile, relative to the high-resolution converged solution, in runs with large softenings $\epsilon_{\rm DM} \gtrsim r_{0.06}$. Although the effect is small, this is expected: if the profiles are converged down to $\sim r_{0.06}$ with appropriate softening, then over-softening by setting $\epsilon_{\rm DM} \gtrsim r_{0.06}$ will necessarily smear the profile out at these radii.

Fig.~\ref{fig:dm.force.resolution.substructure} shows the subhalo mass function, $V_{\rm max}$ function, and spatial distribution (as Fig.~\ref{fig:dm.mass.resolution.substructure}), at fixed mass resolution but varying again the DM force softening. These are almost entirely insensitive to the DM force softening for reasonable choices. We have also examined halo formation times and internal kinematics, and find the same result.

\vspace{-0.5cm}
\subsubsection{Effects of Gas Force Softening: Resolution Studies}
\label{sec:resolution:spatial:gas.tests}

As noted previously, \citet{hopkins:gizmo} present a large number of tests demonstrating the accuracy and near-ideal convergence rate of the implementation of adaptive gas softening in {\small GIZMO} on test problems with {\em known} solutions, including self-gravitating polytropic collapse \citep{evrard:1988.gas.collapse.problem}, cosmological collapse of Zeldovich pancakes (with baryons and with/without DM), and steady-state orbit integration of stable (Toomre $Q>1$) Keplerian disks, as well as good agreement with other state-of-the-art codes such as {\small AREPO} on popular code-comparison tests such as the adiabatic ``Santa Barbara Cluster'' \citep{frenk:1999.sb.cluster}. Of course, while necessary, idealized tests do not ensure ideal results in complicated multi-physics simulations like those here, so we explore changes to the gas softening here.

In Figs.~\ref{fig:spatial.res.history}-\ref{fig:sf.z0.spatial.res}, we also considered the effects of changing the gas force softening, replacing our default self-consistent adaptive softening ($\epsilon_{\rm gas}=h_{i}$) with a fixed, constant physical softening ($\epsilon_{\rm gas}$). In both our dwarf and MW-mass simulations, we see, reassuringly, that for sufficiently small fixed $\epsilon_{\rm gas}$, the differences are essentially negligible (entirely consistent with stochastic fluctuations). 

However, recall that our star formation model is based on identifying self-gravitating gas above some density threshold. At fixed mass resolution, a fixed $\epsilon_{\rm gas}$ sets a minimum inter-particle separation, hence maximum gas density, at which self-gravity will be correctly calculated: this density is $n^{\rm max} = m_{p}^{-1}\,\,m_{i}\,(\epsilon_{\rm gas}^{\rm min})^{-3} \approx 1000\,{\rm cm^{-3}}\,m_{i,\,1000}\,(\epsilon_{\rm gas}^{\rm min}/3.5\,{\rm pc})^{-3}$. So not surprisingly, when we make $\epsilon_{\rm gas}$ large enough that $n^{\rm max} \lesssim n_{\rm crit}$ (the minimum density for star formation), we can artificially suppress the SFR. Obviously, one should not therefore, with fixed $\epsilon_{\rm gas}$, choose $n_{\rm crit} \gtrsim  n^{\rm max}$. If we lower $n_{\rm crit}$ sufficiently so that $n_{\rm crit} \ll n^{\rm max}$, then we recover nearly-identical behavior to our default simulations with adaptive $\epsilon_{\rm gas}$ and larger $n_{\rm crit}$. 


This is important: some simulations allow $n_{\rm crit} \ll n_{\rm max}$ (e.g.\ \citealt{guedes:2011.cosmo.disk.sim.merger.survival,shen:2014.seven.dwarfs} where $\epsilon_{\rm gas} \sim 100-200\,{\rm pc}$, $n^{\rm max} \sim 0.1\,{\rm cm^{-3}}$, and $n_{\rm crit}\sim 5-100\,{\rm cm^{-3}}$) -- external forces (e.g.\ shocks) can still produce $n > n_{\rm crit}$ (if $h_{i}$ is allowed to be smaller than $\epsilon_{\rm gas}$), but this means star formation has nothing to do with self-gravity locally. We therefore consider two similar experiments (${\bf m10q}$ with $\epsilon_{\rm gas}=20\,$pc, $n^{\rm max} = 10\,{\rm cm^{-3}}$, and {\bf m12i} with $\epsilon_{\rm gas}=140\,$pc, $n^{\rm max}=1\,{\rm cm^{-3}}$; both with $n_{\rm crit}=100\,{\rm cm^{-3}}$). Surprisingly, these still behave reasonably. Both galaxies have large turbulent motions an external perturbations (e.g.\ mergers) that produce a broad distribution of density fluctuations;\footnote{Gas which can cool efficiently to $T\sim 10\,$K and has turbulent $\delta v \sim 10\,{\rm km\,s^{-1}}$ should produce an approximately lognormal density distribution with $\gtrsim 10\%$ of the mass exceeding $\sim 100\times$ the mean density \citep[see e.g.][]{federrath:2008.density.pdf.vs.forcingtype}.} the overdensities cannot ``detach'' from the turbulent flow and collapse to still higher densities (like real GMCs) if $\epsilon_{\rm gas}$ is too large (which means the properties of the cold, dense gas will be incorrect), but they can still reach $n_{\rm crit}$, at which point the galaxy-averaged SFR is self-regulated by a balance of inflow and feedback. In the {\bf m10q} ($\epsilon_{\rm gas}=20\,$pc) run, the too-large $\epsilon_{\rm gas}$ still suppresses the SFR and stellar mass by a factor $\sim 2.5$, and makes the SFR artificially bursty (more like a much lower-resolution simulation), while the effects are less dramatic in our {\bf m12i} experiment. This is expected as dwarfs have much lower turbulent Mach numbers, so the mechanism above  requires some large-scale perturbation (it appears that gas accumulates until the disk goes globally gravitationally unstable, causing excessively large bursts in the galactic center instead of local fragmentation into clouds). Of course, if we make $\epsilon_{\rm gas}$ still larger, we eventually suppress any SF or ISM substructure/GMCs, giving an unphysical warm gas-pressure supported, non-self-gravitating disk.

\vspace{-0.5cm}
\subsubsection{N-Body Heating Rates}
\label{sec:resolution:spatial:nbody.heating}

One natural concern is that $N$-body heating of gas/star particles by the dark matter (or stars, or other gas particles) could thicken the stellar disk, or inject spurious thermal energy/turbulence into the gas. Consider the worst-case hard-scattering scenario: two particles approach one another in a vacuum (no other particles nearby) and scatter near the particle centers, with encounter velocity $v_{\rm encounter}$. Since the duration of the encounter is short, the impulsive velocity change is $\Delta v = |\Delta {\bf v}| \sim G\,m_{i}/(h_{i}\,v_{\rm encounter})$ (a proper integration over the exact kernel shape used in the simulations gives a maximal deflection $\approx 1.3\,G\,m_{i}/(h_{i}\,v_{\rm encounter}$), which occurs for encounters with impact parameter $\approx h_{i}$). For gas, this is $\Delta v \sim 0.01\,{\rm km\,s^{-1}}\,m_{i,\,1000}^{2/3}\,(n/{\rm cm^{-3}})^{1/3}\,(v_{\rm encounter}/10\,{\rm km\,s^{-1}})^{-1}$. For dark matter, the particle masses are larger, but so are the softenings and typical encounter velocities since the dark matter has approximately isotropic dispersion at the virial velocity, $\sim 100\,{\rm km\,s^{-1}}$. So $\Delta v \sim 1.3\,G\,m_{\rm DM}/(h_{\rm DM}\,v_{\rm encounter}) \sim 0.017\,{\rm km\,s^{-1}}\,m_{i,\,1000}\,(h_{\rm DM}/20\,{\rm pc})^{-1}\,(v_{\rm encounter}/100\,{\rm km\,s^{-1}})^{-1}$, only slightly larger even when we adopt the smallest DM softening seen in our simulations.\footnote{In fact, the $\Delta v$ in the text above is actually a significant over-estimate when we use adaptive gravitational softenings, as particle-particle encounters are never ``in a vacuum.'' While a particle $a$ traverses a domain $h$ within the kernel of particle $b$, it of course feels some acceleration. But provided there is actually mass being represented by $b$, this is completely physical. The numerical noise/error depends not on the absolute magnitude of the acceleration from $b$ onto $a$, but on the {\em deviation} of the potential from $b$, owing to finite sampling, from the smooth potential that would be represented if we had infinite resolution. This is straightforward to estimate. Following \citet{dehnen.aly:2012.sph.kernels}, the acceleration from each particle is constructed assuming its mass distribution follows the kernel function; if we assume the correct background is a uniform density field, then discretize this into particles, it is straightforward to compute the fractional deviation from the correct (infinite-resolution) solution for different particle configurations within the kernel. For our standard cubic-spline kernel, this is essentially what is shown in Fig.~3 of \citet{dehnen.aly:2012.sph.kernels}; the typical deviation considering various different particle configurations is about $|\Delta \rho|/|\rho| \sim 0.005$. Thus the actual hard-scattering amplitudes are suppressed by a factor of $\sim 100$ from their already-small values.} 

We can translate this to an ``N-body heating rate,'' assuming each encounter adds to the velocity incoherently, with particle encounter rate $\sim v_{\rm encounter} / h_{i}$. Accounting for the number of baryons per gas particle ($\sim m_{i}/m_{p}$), this gives a heating rate (rate-of-change of kinetic energy) of: 
\begin{align}
\label{eqn:nbody.heat.gas} \frac{Q_{\rm heat}^{gas-gas}}{{\rm erg\,cm^{3}\,s^{-1}}} &\sim 2\times10^{-32}\,
m_{i,\,1000}\,\left( \frac{10\,{\rm km\,s^{-1}}}{v_{\rm encounter}} \right) \\ 
\label{eqn:nbody.heat.dm} \frac{Q_{\rm heat}^{gas-DM}}{{\rm erg\,cm^{3}\,s^{-1}}} &\sim 6\times10^{-32}\,
m_{i,\,1000}\,\left( \frac{100\,{\rm km\,s^{-1}}}{v_{\rm encounter}} \right)\,\left( \frac{\rho_{\rm DM}}{\rho_{\rm gas}} \right)
\end{align}
These should be compared to the physical heating/cooling rates of the gas, $\Lambda \sim 10^{-23} - 10^{-22}\,{\rm erg\,cm^{3}\,s^{-1}}$ for typical ISM conditions; they are far smaller than almost any other source of error in the baryonic physics, even for the lowest-resolution simulations in our tests.\footnote{We have directly verified this in a series of numerical tests: setting up a periodic box with dark matter particles (matched to the particle mass, velocity dispersion, and space density of the cosmological simulations at the radii within the halo where the $N$-body heating in Eq.~\ref{eqn:nbody.heat.dm} is maximized), and an equilibrium, isothermal, non-radiative gas disk. In tests where the gas feels only DM, we measure $Q^{\rm gas-DM}_{\rm heat}$ a factor of a few {smaller} than Eq.~\ref{eqn:nbody.heat.dm}; in tests where the gas feels self-gravity and is ``stirred'' (so some relative gas-gas motion exists), the heating contribution from $Q^{\rm gas-gas}_{\rm heat}$ is unmeasurably small.}

\begin{figure}
\plotonesize{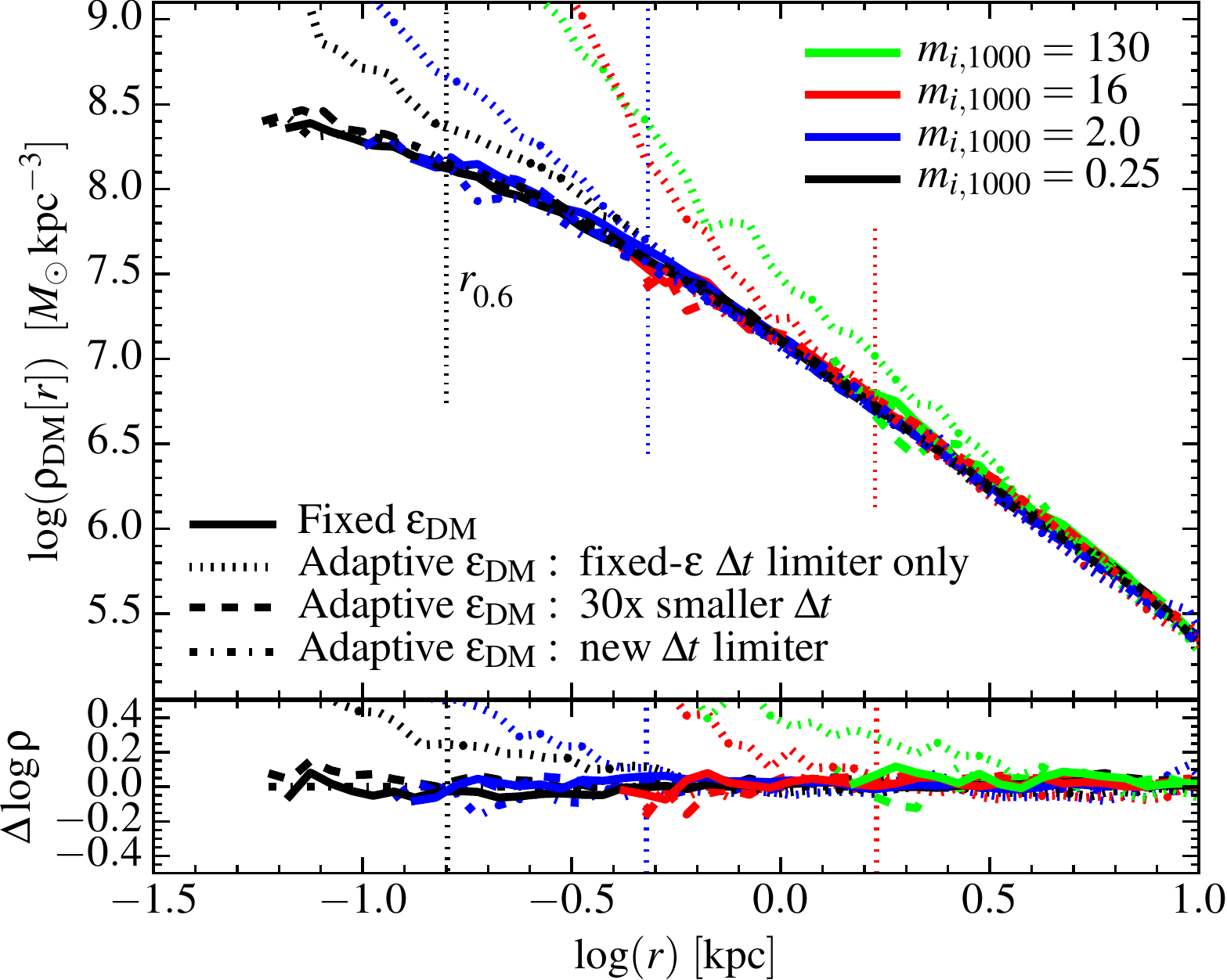}{1}
    \vspace{-0.25cm}
    \caption{Dangers of using adaptive force softening for {\em collisionless} (non-gas) particles, without careful timestepping. We show the $z=0$ mass profile in DM-only simulations (here our {\bf m10v} run, but we repeated a subset of these tests  in {\bf m09}, {\bf m10q}, {\bf m11q}, {\bf m11v}, and {\bf m12i}), at a series of resolution levels (labeled). For our default (fixed-$\epsilon_{\rm DM}$) runs, we show the conservative $\sim 2000$-particle \citet{power:2003.nfw.models.convergence} radius $r_{0.6}$, but plot the full profile down to our ``convergence radius'' $r_{0.06}$ -- the profiles in all previous tests agree well outside this radius, independent of $\epsilon_{\rm DM}$, and we see the same in the fixed-$\epsilon_{\rm DM}$ runs plotted here. We compare our adaptive-softening implementation with three different timestep choices. If we use only the same timestep criteria employed for fixed-softening runs (``fixed-$\epsilon$ $\Delta t$ limiter only''), the simulations clearly exhibit a spurious, non-converged ``cusp'' at the center indicative of integration errors that arise because particles ``move through'' one another in a timestep, preventing the adaptive terms from properly adapting (this cannot occur with gas, owing to the Courant condition). Lowering the timestep by a uniform factor of $30$ resolves the problem, and produces excellent agreement with the fixed-$\epsilon$ runs, but at great computational expense. Our default implementation uses the ``new $\Delta t$ limiter'' described in \S~\ref{sec:resolution:time:ags} (Eq.~\ref{eqn:dt.ags}) to control the timesteps with adaptive softening for collisionless particles: this also produces excellent agreement with fixed-$\epsilon$ runs.
    \label{fig:ags.timesteps}}
\end{figure}

\vspace{-0.5cm}
\subsection{Time Resolution}
\label{sec:resolution:time}

The time resolution in the simulations is set by the timestep $\Delta t$, which is always set according to the minimum of the various criteria in \S~\ref{sec:methods:timestepping}. Together, these criteria ensure that gravitational and fluid dynamics, as well as stellar evolution, are always explicitly time-resolved. Usually, the acceleration-based criterion $\Delta t < 0.2\,(h_{i} / |{\bf a}_{i}|)^{1/2}$ from \citet{power:2003.nfw.models.convergence} is the most demanding, for {\em both} gas and $N$-body particles (because the dense gas is highly super-sonic). Occasionally, however, in gas shock-heated by many SNe, the Courant criterion is most important. 

In either case, the minimum timesteps reached in our simulations are $\approx 100\,$yr (see Table~\ref{tbl:res}). This is reached regularly by some elements from redshifts $z\sim 0-2$, although always by a very small fraction of the total population at any given instant.

\vspace{-0.5cm}
\subsubsection{Standard Criteria \&\ Results of Variations}
\label{sec:resolution:time:standard}

We have tested variations in our standard time-stepping criterion, in a limited sub-set of simulations with both dark matter and baryons: the DM-only runs are run to $z=0$, but baryonic runs are only integrated to $z=4$, because inappropriate timesteps usually lead to catastrophic numerical instability which is evident quickly. Our conclusions are identical to canonical studies in the field. Like \citet{power:2003.nfw.models.convergence}, we find that a coefficient smaller than $\alpha=0.2$ in the timestep limiter $\Delta t < \alpha\,(h_{i} / |{\bf a}_{i}|)^{1/2}$ produces no appreciable gain in accuracy, but much larger values ($\alpha \gtrsim 0.5$) can seriously degrade mass profiles and orbit integration (e.g.\ angular momentum conservation); our gravity solvers are very similar, following \citet{springel:gadget}, so this should not be surprising, especially for DM-only tests. Even though this acceleration-based criterion usually dominates for gas (recall ${\bf a}_{i}$ includes all accelerations, i.e.\ hydrodynamic and gravitational), a Courant-type condition is still necessary, with coefficients $\alpha > 0.8$ in $\Delta t < \alpha\,h_{i} / v_{{\rm sig},\,i}^{\rm max}$ giving rise to numerical instability, while values $\alpha<0.4$ (our preferred value) do not produce any obvious improvement. This is consistent with standard hydrodynamic tests and analytic numerical stability analysis (see \citealt{hopkins:gizmo} for discussion). 

\vspace{-0.5cm}
\subsubsection{Stellar-Evolution Timestep Limits}
\label{sec:resolution:time:stellar}

Recall, we do not allow the timestep for star particles to exceed 
\begin{align}
\label{eqn:stellar.dt.limiter} \Delta t_{\ast} < {\rm MAX}\left( 10^{4}\,{\rm yr},\ \frac{t_{\ast}}{300} \right)
\end{align}
where $t_{\ast}$ is the age of the star, to ensure stellar evolution is time-resolved and the expected number of SNe (per star particle per timestep) is always $<1$ at our production resolution. We have experimented with weakening this limiter, increasing it by a factor of $\approx 3$ (which produced no measureable effect), or most radically using $\Delta t_{\ast} < 20\,{\rm Myr}\,m_{i,\,1000}^{-1}$ -- for $m_{i,\,1000}\gtrsim1$, this is equivalent to allowing as many as $\sim 10\,$SNe per timestep in an extremely young star particle. We did not see any significant difference running tests of {\bf m10q} and {\bf m12i} at low resolution (the difference in timestep is maximized at low resolution) to $z=2$; however, we found that usually the number of SNe exploding at once was much smaller ($\sim 2-3$), because the youngest stars are in dense regions where other timestep limits (e.g.\ the acceleration criterion above) impose much stricter limits than $\sim $Myr. However, if we removed the limiter entirely ($\Delta t_{\ast}\rightarrow \infty$), we saw clear (albeit rare) pathological activity: for example lone star particles formed in poorly-resolved dwarfs in the outskirts of the high-resolution region or ejected via tidal interactions might be assigned extremely long dynamical timesteps and have a huge single-timestep injection of SNe, continuous stellar mass loss, and radiation (giving for example a small number of gas elements with artificially high metallicity). 

\vspace{-0.5cm}
\subsubsection{Adaptive Force Softening for Collisionless Particles: The Need for Additional Timestep Criteria}
\label{sec:resolution:time:ags}

As described in \S~\ref{sec:methods:gravity}, for a subset of our tests presented in \S~\ref{sec:resolution:spatial:collisionless.tests}, we use adaptive softening for collisionless (DM and stellar) particles. Variations of these methods have been explored in a number of studies on collisionless, self-gravitating systems \citep{athanassoula:2000.optimal.force.softening.collisionless.sims,price:2007.lagrangian.adaptive.softening,bagla:2009.adaptive.treepm,iannuzzi:2011.collisionless.adaptive.softening.gadget,barnes:2012.softening.is.smoothing,iannuzzi:2013.no.need.adaptive.softening.for.dm}. 
We have validated our numerical implementation with all tests presented in \citet{price:2007.lagrangian.adaptive.softening} and \citet{iannuzzi:2011.collisionless.adaptive.softening.gadget}. 

However, introducing these adaptive softenings requires stronger timestep criteria for collisionless particles, especially in dense halo centers where DM particles on highly-radial orbits may ``plunge'' and interact with particles of widely-differing $\epsilon$ over the course of their orbits. 

Fig.~\ref{fig:ags.timesteps} shows the results of using adaptive softening in DM-only simulations ({\bf m10v}) of varying resolution, {\em without} imposing any additional timestep limiter beyond what is described above for collisionless particles with {\em fixed} softening $\epsilon$. Clearly, the central sub-kpc regions of the DM profile exhibit a spurious density ``cusp'' well in excess of the {converged} (much higher-resolution) results from constant-$\epsilon_{\rm DM}$ runs. We confirm that this feature is artificial by simply re-running the simulations enforcing a factor of $30$ smaller timestep; this agrees well with the fixed-$\epsilon$ runs. We have also confirmed that both sets of runs maintain global conservation (as expected) -- the error appears to be associated with local integration errors between particle neighbors when the timesteps are too large. 

With fixed softening, there is no Courant-like condition required for collisionless particles: since the equations of motion depend only on the collective long-range forces (the gravitational potential), if the potential is sufficiently smooth ($|{\bf a}|$ is small), particles can safely ``move through'' one another (Monte Carlo-sampling the phase-space distribution function). But this is not true with adaptive softening (again, for collisionless particles), because the softening length (hence self-gravity) of a particle depends on the local neighbor configuration, introducing local correction terms that must be integrated smoothly as particles move through one another -- otherwise these terms are under-sampled and effectively scatter the orbits. 

For gas, this error does not occur because of the Courant condition. The solution for collisionless particles with adaptive softening is therefore straightforward: we should implement a similar timestep criterion. We require: 
\begin{align}
\label{eqn:dt.ags} \Delta t_{\rm AGS}^{a} &< 0.25\,{\rm MIN}\left\{ \frac{1}{|\langle \tilde{\nabla}\cdot {\bf v} \rangle_{a}|}\ ,\  \frac{\epsilon_{a}}{v_{{\rm sig,\,AGS}}^{\,a}} \right\} \\ 
\nonumber v_{{\rm sig,\,AGS}}^{a} &\equiv {\rm MAX}_{(b: |{\bf x}|_{ba} < H_{a},\,H_{b})}\,\left\{ \left| ( {\bf v}_{b}-{\bf v}_{a} )\cdot \hat{\bf x}_{ba} \right| \right\} \\ 
\nonumber  \langle \tilde{\nabla}\cdot {\bf v} \rangle_{a} &\equiv \frac{\sum_{b: |{\bf x}|_{ba} < H_{a}}\,\left( {\bf v}_{b}-{\bf v}_{a} \right)  \cdot \nabla W({\bf x}_{ba},\,H_{a}) }{\Omega_{a}\,\sum_{b}W({\bf x}_{ba},\,H_{a})}
\end{align}
Here $H_{a}\approx 2\,\epsilon_{a}$ (with $\epsilon_{a} = h_{a}$) is the domain of the nearest-neighbor search around particle $a$, and the sums over $b$ represent sums over all interacting neighbors {\em of the relevant particle type} such that they contribute to defining the inter-particle spacing and softening length $\epsilon$ of particle $a$. By analogy to the Courant condition, we define $v_{\rm sig,\,AGS}^{a}$ as the maximum approach/recession velocity of any neighbor within this interaction kernel -- this requirement is simply that two particles cannot ``cross'' more than $\sim 1/2$ their relative softening lengths in a single timestep. The particle-divergence $\langle \tilde{\nabla} \cdot {\bf v} \rangle_{a}$ is defined exactly the same as the traditional SPH velocity divergence,\footnote{With adaptive softening for DM, in 3D, we {\em define} $\epsilon_{a} = h_{a} = (3\,N_{\rm eff}/4\pi)^{1/3}\,H_{a}$ such that, in 3D, $h_{a}^{3}\,\bar{n}_{a} = 1$ and $4\pi/3\,H_{a}^{3}\,\bar{n}_{a} = N_{\rm eff}$ (with $N_{\rm eff}=32$ for our standard cubic spline kernel), where $\bar{n}_{a} \equiv \sum W({\bf x}_{ba},\,H_{a})$ is a kernel-averaged particle neighbor number density. With some straightforward algebra, this gives an {\em exact} discrete equation for the Lagrangian derivative of $h_{a}$, $Dh_{a}/Dt = -(h_{a}/3\Omega_{a}\bar{n}_{a})\, \nabla \bar{n}_{a} \cdot \partial ({\bf x}_{b}-{\bf x}_{a})/\partial t = -(h_{a}/3)\,\langle \tilde{\nabla}\cdot {\bf v} \rangle_{a}$, where $\Omega_{a} \equiv 1 + (h_{a}/3\bar{n}_{a})\,\partial \bar{n}_{a}/\partial h_{a}$ \citep[for derivations, see][]{price:2012.sph.review,hopkins:lagrangian.pressure.sph}. So if we do not want the softening $h_{a}$ to change by more than a factor $|(\Delta h_{a})/h_{a}| < \alpha$ in one timestep, we require a timestep $\Delta t < 3\alpha/|\langle \tilde{\nabla}\cdot {\bf v} \rangle_{a}|$.} in such a manner that this requirement prevents $\epsilon_{a}$ from changing by more than $\sim 10\%$ within a single timestep. Because we use adaptive timesteps, we also enforce a ``wakeup'' condition identical to that used for the hydrodynamics \citep[see][]{saitoh.makino:2009.timestep.limiter,durier:2012.timestep.limiter}; specifically, if an ``active'' particle in a sub-step interacts with an ``inactive'' particle in a timestep $>4$ times larger ($v_{{\rm sig,\,AGS}}$ from its previous active step $>4$ times larger), the inactive particle is stopped from taking the larger timestep and moved to the smallest active timebin in the active hierarchy. This prevents particles moving very rapidly from artificially moving ``through'' a particle with a long timestep.

Fig.~\ref{fig:ags.timesteps} demonstrates that this timestep limiter cures the errors seen before. The added timestep criterion does add a significant cost to the DM-only run, though far less costly than the uniform factor $\sim30$ smaller-timestep case, since only a small number of particles are affected at each time. In any case the agreement between adaptive softening with appropriate timesteps and fixed-softening runs is excellent down to $r_{0.06}$. 
We have also recently learned that other authors who have implemented adaptive DM softening following \citet{price:2007.lagrangian.adaptive.softening} have reached the same conclusions and found it necessary to include similar timestep criteria to maintain numerical stability (V.\ Springel, private communication).

\begin{figure}
\hspace{-0.3cm}
\plotonesize{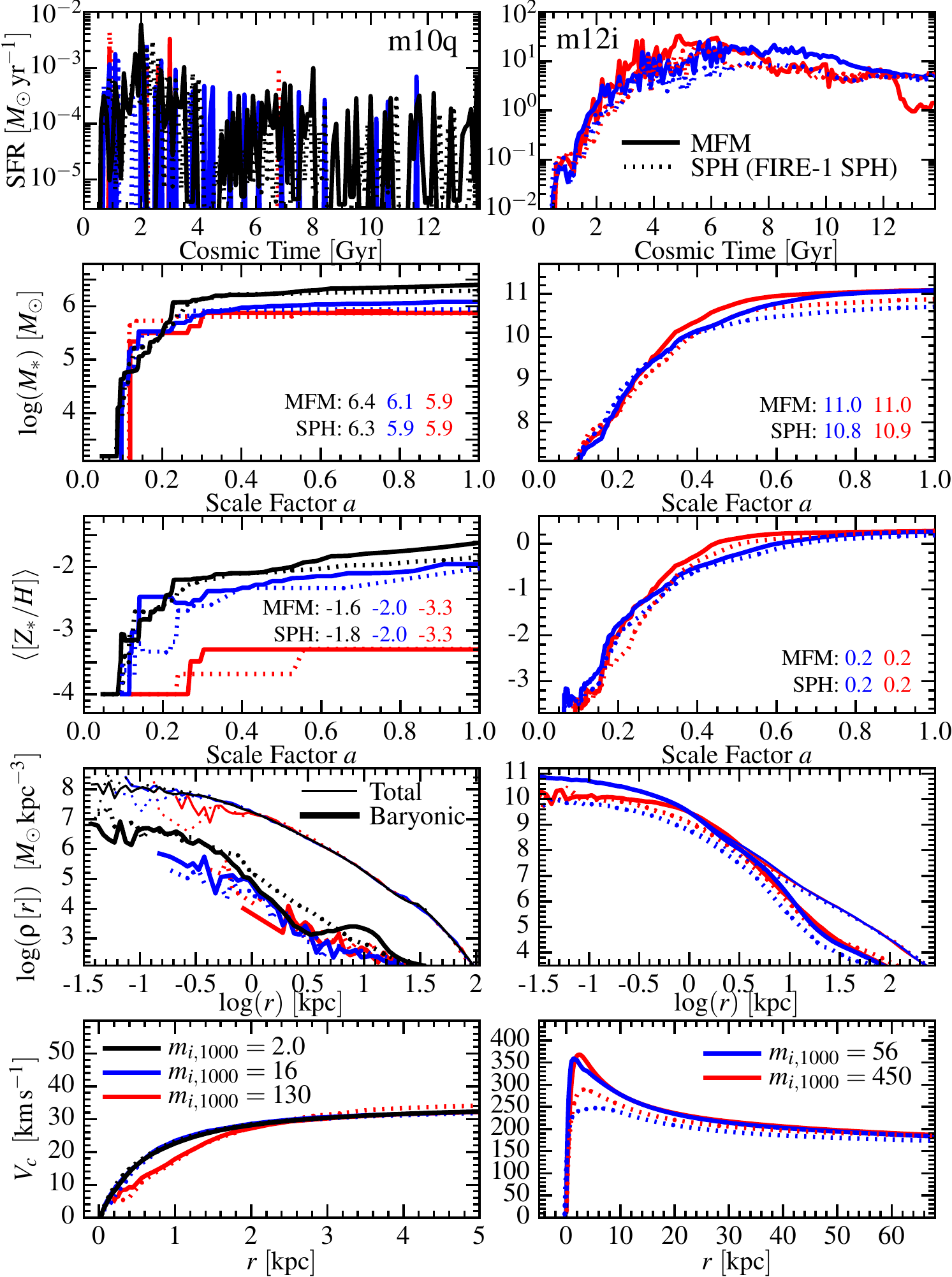}{1.02}
    \vspace{-0.25cm}
    \caption{Effects of the method for solving the hydrodynamic equations, in cosmological simulations of both a dwarf ({\bf m10q}; {\em left}) and MW-mass galaxy ({\bf m12i}; {\em right}), as \demofigcosmo. We compare our default method in {\small GIZMO} --  MFM (solid lines), a higher-order accurate, mesh-free finite-volume Godunov method -- to the {\small GIZMO} implementation of smoothed-particle hydrodynamics (SPH) which was used for FIRE-1 (dotted lines). {\small GIZMO} is a multi-method code so we can change the hydro solver while keeping {\em all} other physics and numerics identical here. We repeat each comparison at multiple resolution levels. For dwarfs, we see excellent agreement between MFM and SPH, at every resolution level. For MW-mass systems, on the other hand, FIRE-1 SPH predicts somewhat lower SFRs, stellar masses, and central $V_{c}$. The critical difference between dwarfs and massive galaxies is likely to be the presence of the ``hot gaseous halo'' around massive systems (absent in dwarfs), which determines the cooling rate onto the galaxy and into which galactic winds propagate. Known issues in SPH can suppress fluid mixing in the halo, even with ``state of the art'' SPH formulations, which in turn leads to easier escape of winds, less efficient cooling, and (in turn) lower masses.
    \label{fig:sph.vs.time}}
\end{figure}

\begin{figure}
\hspace{-0.3cm}
\includegraphics[width=0.55\columnwidth]{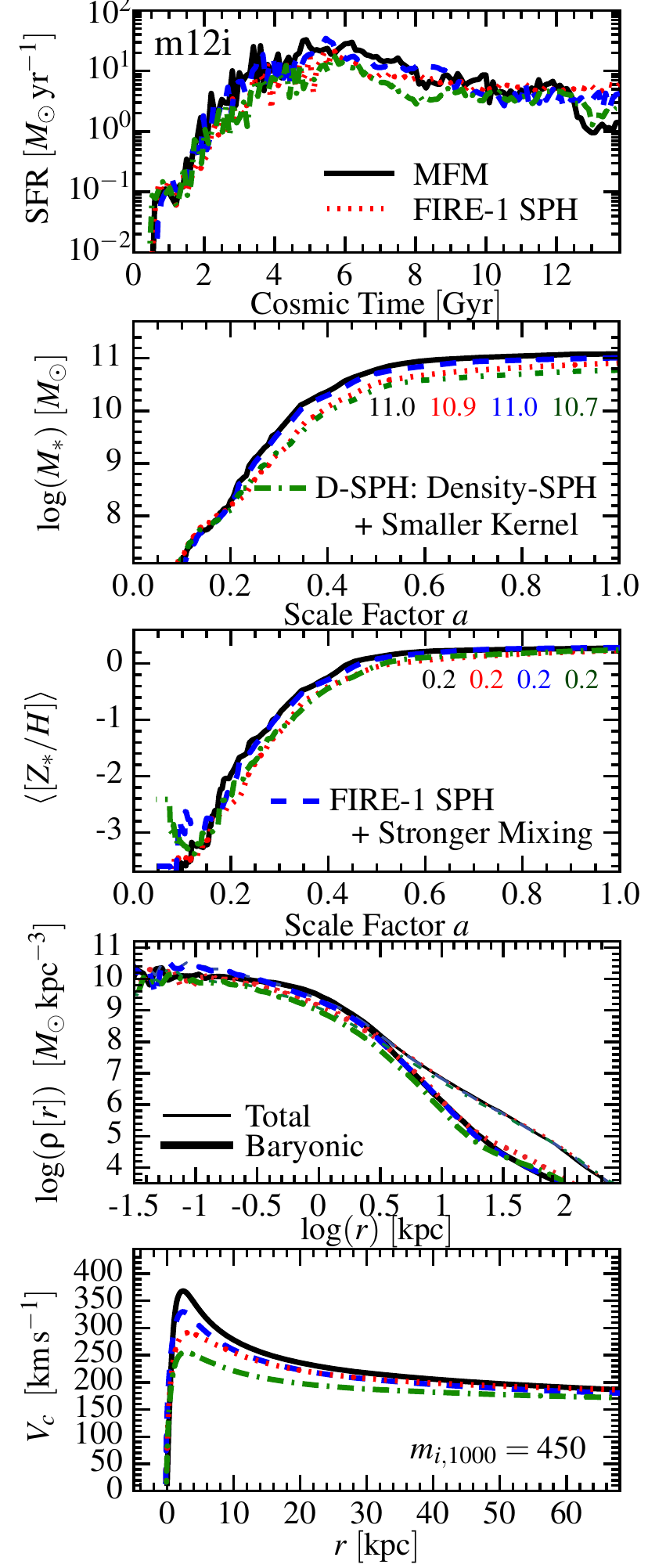} 
\hspace{-0.75cm}
\includegraphics[width=0.55\columnwidth]{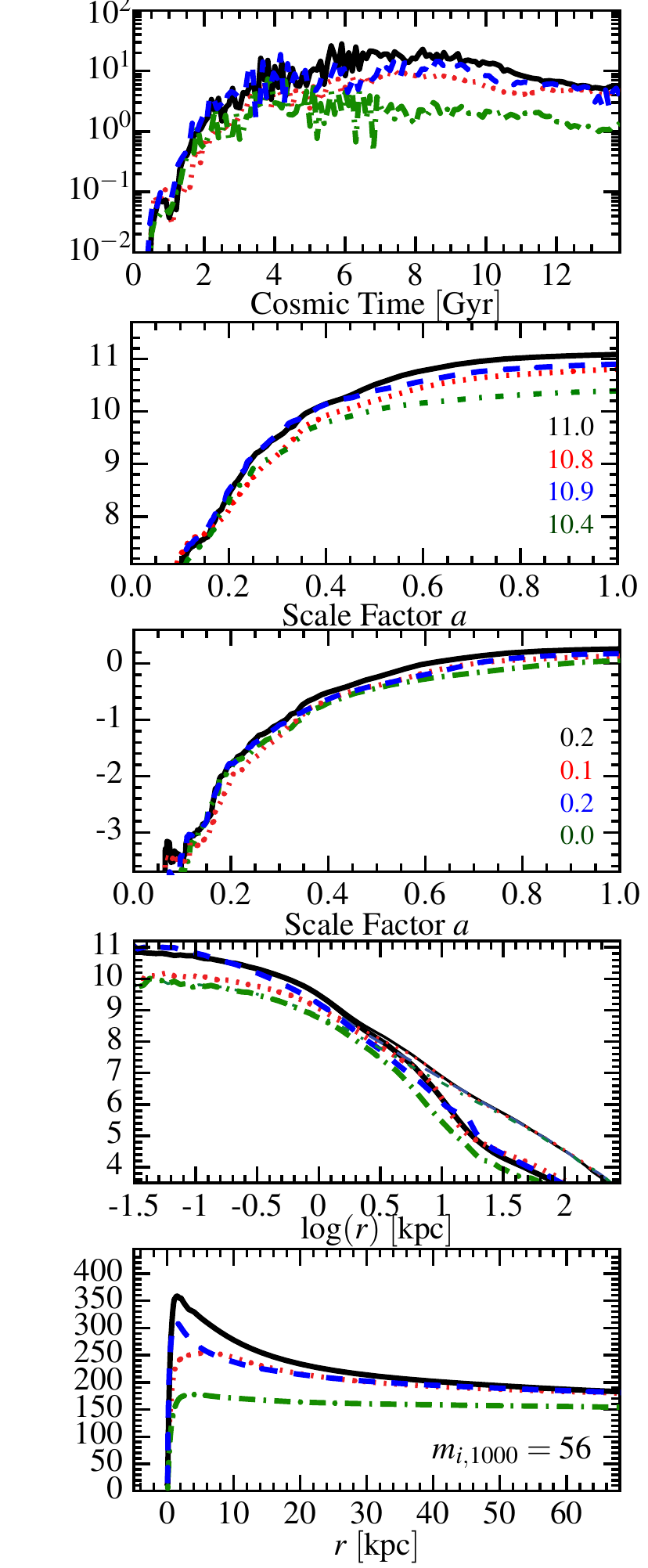} 
    \vspace{-0.52cm}
    \caption{Effects of the hydrodynamic method in our cosmological {\bf m12i} simulation at two resolution levels ({\em left} and {\em right}), as Fig.~\ref{fig:sph.vs.time}. Here we compare different ``flavors'' of SPH (see \S~\ref{sec:sph.flavors}): (a) the FIRE-1 implementation from Fig.~\ref{fig:sph.vs.time}, which uses the pressure-energy formulation and a larger smoothing kernel, designed to reduce fluid-mixing errors; (b) the FIRE-1 SPH model with ``stronger mixing'' -- an explicitly increased thermal energy/entropy mixing term (larger ``artificial conductivity''); (c) D-SPH: a simpler SPH implementation which uses the ``density-energy'' formulation of the equations and a smaller kernel (but still uses the higher-order artificial viscosity and conductivity switches of FIRE-1 SPH), which strongly suppresses the ability of the method to capture fluid mixing instabilities. The ``FIRE-1 SPH + Stronger Mixing'' run agrees well with MFM; the ``D-SPH'' run much more strongly suppresses the SFR, stellar mass, and central $V_{c}$, in a manner which appears to {\em diverge} with resolution (consistent with the fact that the SPH errors are zeroth-order). This demonstrates that the effects of SPH on fluid mixing physics dominate the differences between runs in Fig.~\ref{fig:sph.vs.time}. 
    \label{fig:sph.vs.time.added.diffusion}}
\end{figure}

\begin{figure}
\plotonesize{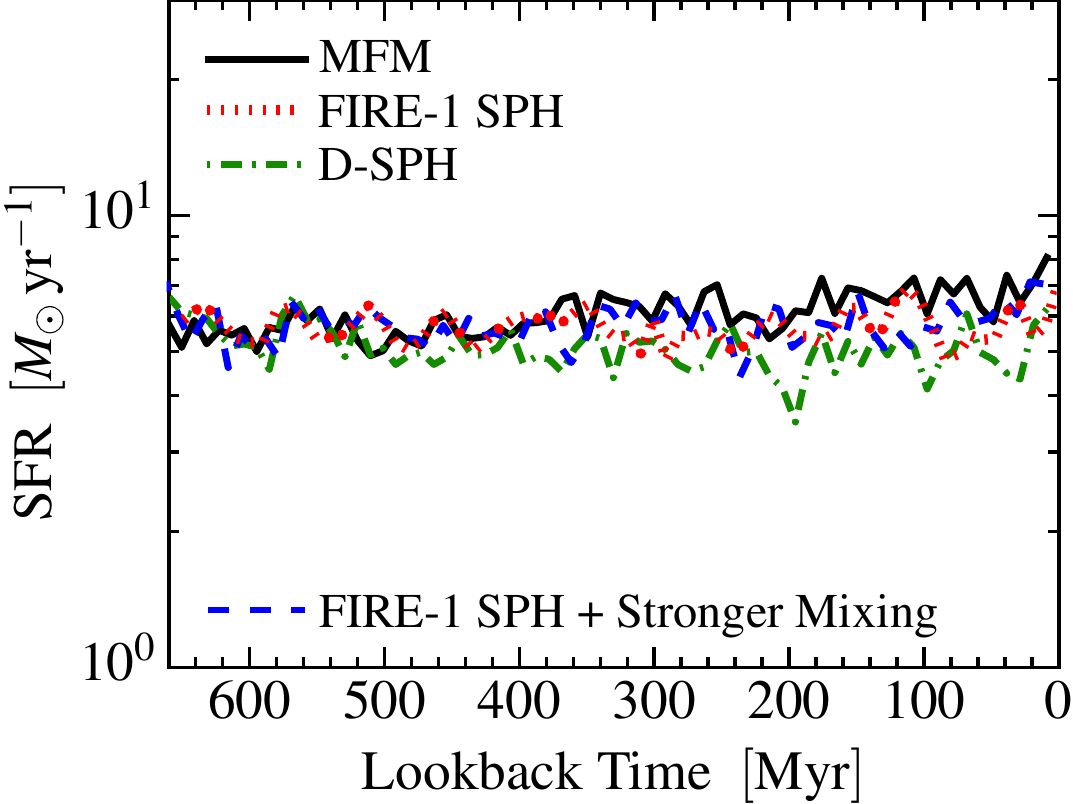}{0.95}
    \vspace{-0.25cm}
    \caption{Effects of the hydrodynamic method on the SFR at low-redshift of a re-started MW-mass ({\bf m12i}; $m_{i,\,1000}=56$) simulation as \demofigrestart. We compare MFM and various SPH ``flavors'' as Fig.~\ref{fig:sph.vs.time}. For fixed initial conditions (early times) the SFR is feedback-regulated and identical. The two slowly diverge by $\sim 30\%$ over $\sim$\,Gyr as the SPH flavors with less accurate fluid-mixing treatments are able to more easily eject gas from the galaxy. However the effect is small compared to the cosmologically time-integrated effects seen in Fig.~\ref{fig:sph.vs.time}. This is consistent with the idea that the mixing and re-cycling of these outflows, and subsequent hot gas cooling (which occurs on timescales of order the Hubble time), not the {\em generation} of outflows or self-gravitating fragmentation and star formation in the dense galactic disk, is the dominant reason for the difference between certain SPH flavors and MFM in massive galaxies in Fig.~\ref{fig:sph.vs.time}. 
    \label{fig:sph.vs.sfr}}
\end{figure}

\vspace{-0.5cm}
\section{Effects of the Hydrodynamic Method}
\label{sec:hydro}

\begin{figure}
\begin{tabular}{cc}
\hspace{-0.25cm}
\includegraphics[width=0.49\columnwidth]{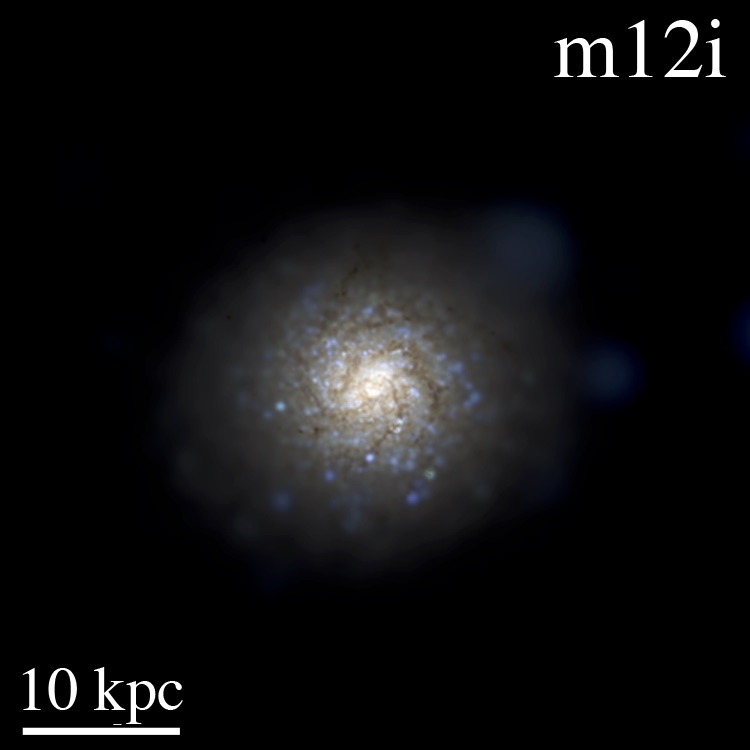} &
\hspace{-0.50cm}
\includegraphics[width=0.49\columnwidth]{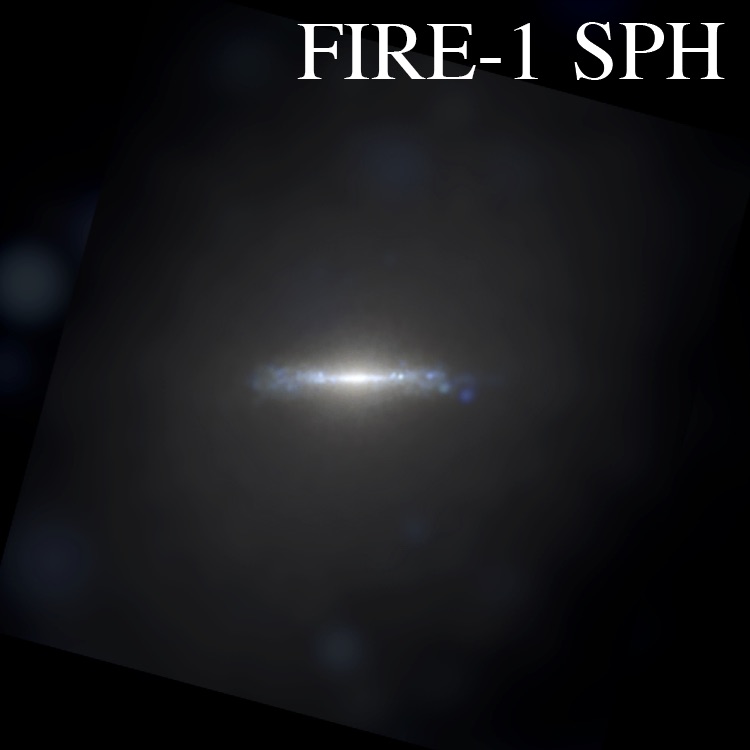} \\
\hspace{-0.25cm}
\includegraphics[width=0.49\columnwidth]{figs_images/m12i_ref12_s0600_t000_star_N3c_res.jpg} &
\hspace{-0.50cm}
\includegraphics[width=0.49\columnwidth]{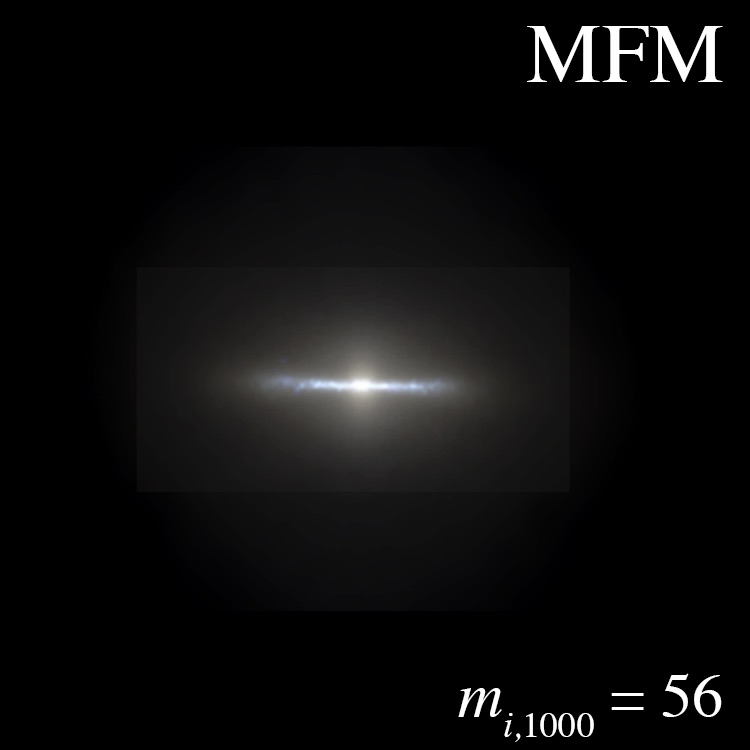} \\
\end{tabular}
    \vspace{-0.25cm}
    \caption{Mock images, as in Fig.~\ref{fig:images.dwarfs}, showing the effects of the hydrodynamic solver on the $z=0$ morphology of the simulated galaxies, for the highest-resolution ($m_{i,\,1000}=56$) MW-mass ({\bf m12i}) systems in Fig.~\ref{fig:sph.vs.time}. Dwarfs are not shown, as they have irregular morphologies independent of the hydrodynamics or resolution (as expected). The large-scale visual morphology is similar, although the MFM simulation exhibits a slightly more compact, thinner disk and more well-ordered spiral structure. The major differences in stellar mass and central circular velocity are not obvious in the stellar morphology. The same is true for all SPH flavors in Fig.~\ref{fig:sph.vs.time.added.diffusion}.
    \label{fig:image.morph.sph}}
\end{figure}

\begin{figure*}
\includegraphics[width=0.33\textwidth]{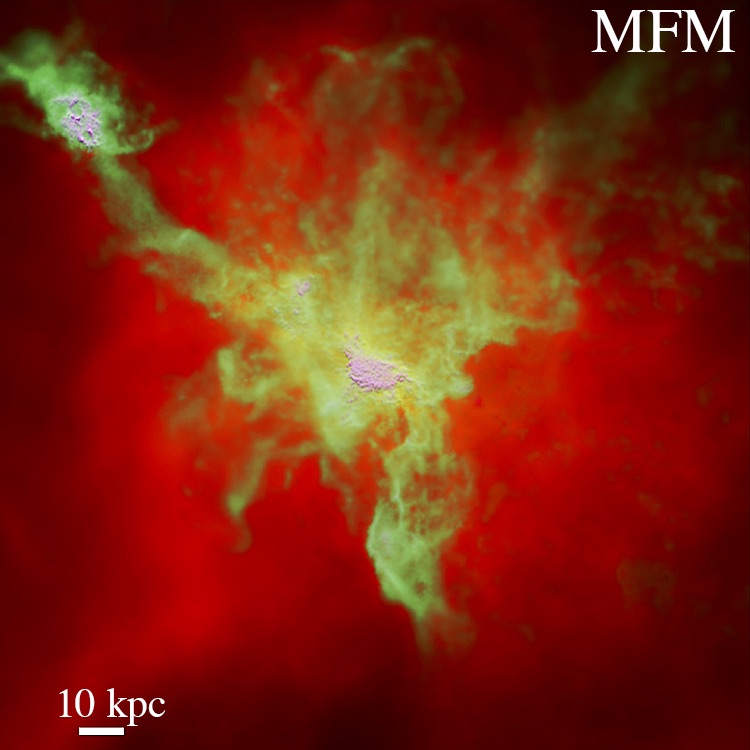}
\includegraphics[width=0.33\textwidth]{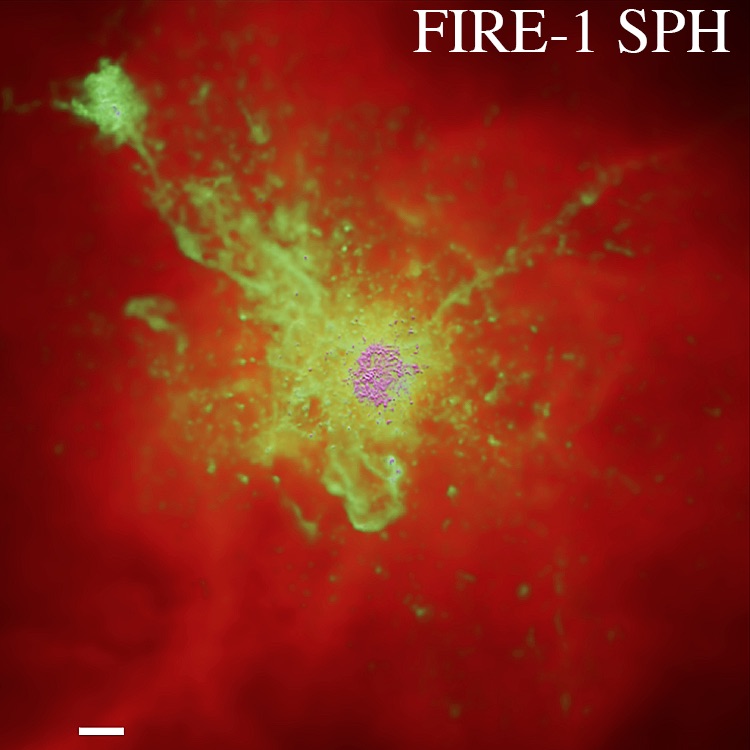}
\includegraphics[width=0.33\textwidth]{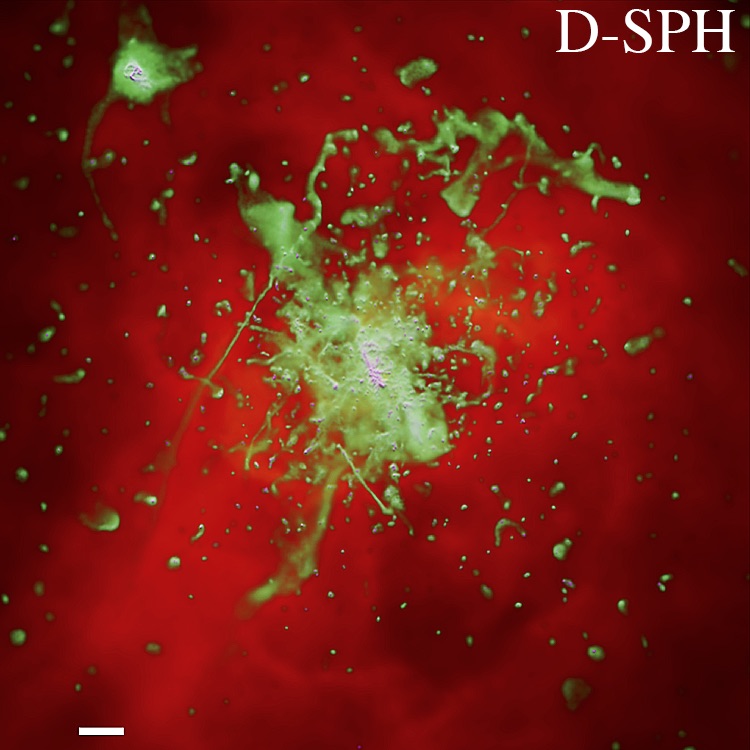} 
    \vspace{-0.25cm}
    \caption{Gas morphology around the galaxy in the {\bf m12i} simulations (with $m_{i,1000}=56$) from Fig.~\ref{fig:sph.vs.time}, at $z=0.9$ in $200$\,kpc boxes, with different hydrodynamic methods (as Fig.~\ref{fig:sph.vs.time}; labeled). Images are logarithmically-weighted surface-density projections, with red/green/magenta showing hot ($>10^{6}\,$K), warm ionized ($\sim 10^{4}-10^{5}\,$K), and cool neutral ($<8000\,$K) gas.
    The most dramatic differences appear in the CGM; we choose a time where a merger has triggered violent outflows of cool gas to maximize these differences. 
    In some less-accurate SPH formulations, such as ``D-SPH'' here ({\em right}; density-SPH with smaller smoothing kernel), the outflow has broken into the well-known artificial ``SPH blobs'' that result from errors treating fluid mixing interfaces. In our non-SPH, finite-volume MFM method ({\em left}), these are absent. The FIRE-1 implementation of SPH (``P-SPH'' with larger kernel) was specifically formulated to reduce the fluid-mixing errors in SPH; it dramatically reduces (but does not completely eliminate) the ``blobs'' (FIRE-1 SPH+Stronger Mixing closely resembles this, but with slightly fewer ``blobs'').
    Also, in MFM the hot halo gas is more compact/dense, with a sharper shock front, while in SPH it is lower-density with an extended boundary, owing to difficulties in shock-capturing and numerical dissipation/mixing \citep{bauer:2011.sph.vs.arepo.shocks}. These effects in SPH make it ``easier'' for cold winds to be ejected and avoid mixing in the halo, lowering the CGM gas cooling rate, explaining the suppression in SF in Figs.~\ref{fig:sph.vs.time}-\ref{fig:sph.vs.sfr}. \vspace{-0.4cm}
    \label{fig:image.sph.cgm}}
\end{figure*}


\begin{figure*}
\plotsidesize{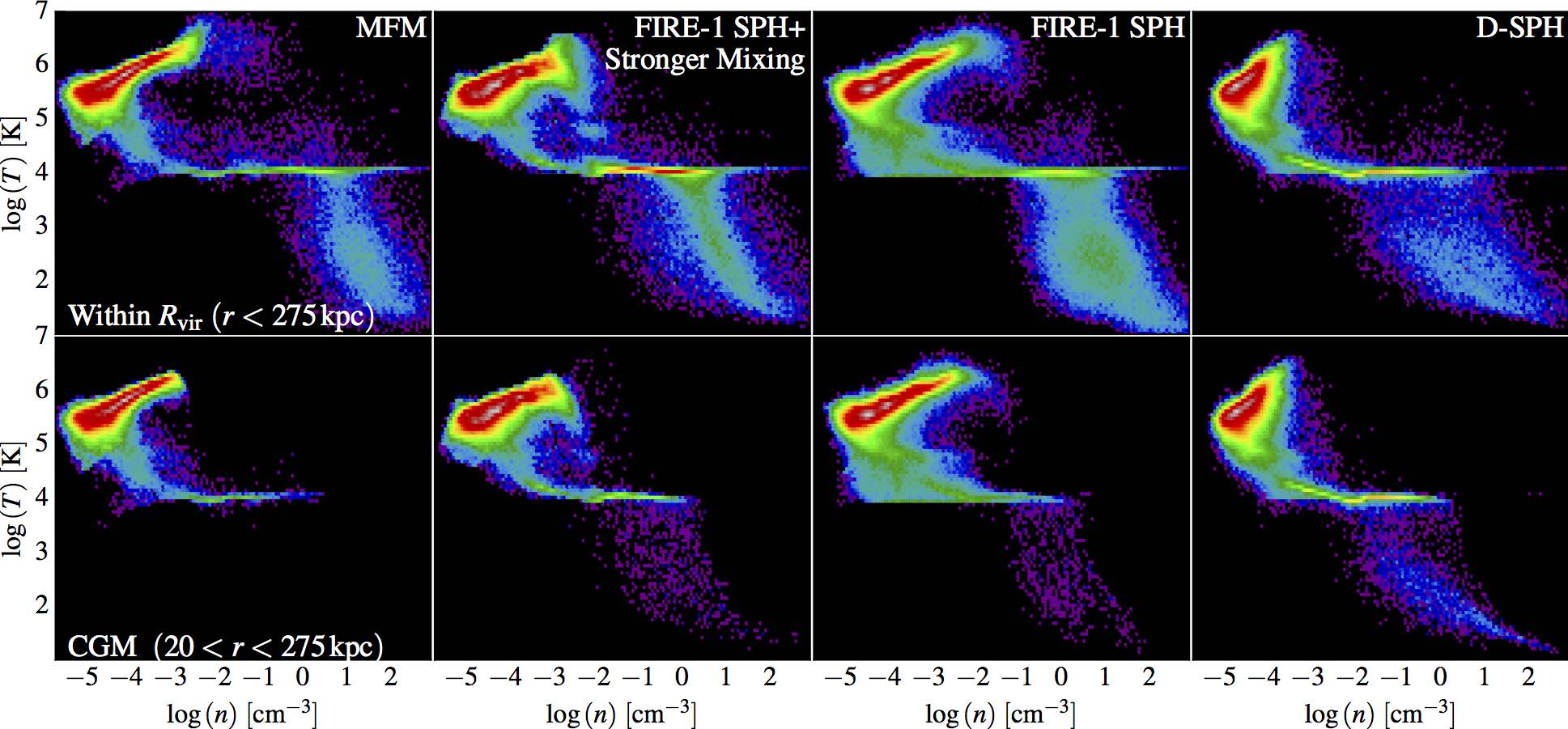}{0.99}
    \vspace{-0.25cm}
    \caption{Comparison of hydrodynamic methods (MFM and SPH ``flavors'' from Fig.~\ref{fig:sph.vs.time.added.diffusion}) in the temperature-density phase diagram (at $z=0$) in the MW-mass ({\bf m12i}, with $m_{i,\,1000}=56$) systems in Figs.~\ref{fig:sph.vs.time}-\ref{fig:image.sph.cgm}. 
    We compare all gas inside the virial radius ({\em top}) and gas within the CGM (excluding the galaxy; {\em bottom}). Colors are a mass-weighted heat map with the density of gas mass in the space shown increasing logarithmically from black-blue-green-yellow-red-white. The ``D-SPH'' (simpler, less accurate fluid-mixing) flavor produces no hot halo gas with densities $n \gg 10^{-4}\,{\rm cm^{-3}}$, owing to a combination of poor shock-capturing, numerical dissipation, and artificial ``ease'' with which cold ``blobs'' in shredded galactic winds escape the halo (instead of shocking); there is also a substantial population of gas with temperatures $T\sim 10-1000\,$K and densities $n\sim 10^{-2} - 10^{3}\,{\rm cm^{-3}}$ in the CGM, from the same ``blobs.'' 
    The more accurate fluid mixing treatment in the FIRE-1 SPH mostly eliminates the cold CGM blobs and restores hot dense gas -- but there is still a significant amount of gas at $\sim 10^{4}$\,K even at very low densities which disappears when we add ``stronger mixing,'' in agreement with our MFM simulations.
    \label{fig:temperature.density.sph.phase.diagram}}
\end{figure*}

%
%

\subsection{Finite-Volume Godunov Methods (MFM) vs.\ SPH}
\label{sec:hydro:overview}

As noted in \S~\ref{sec:methods:hydro}, a range of idealized test problems demonstrate that our default, finite-volume MFM method for the hydrodynamics is significantly more accurate than SPH, at fixed mass resolution. However, given that gravity and feedback overwhelmingly dominate over pure hydrodynamic forces in the simulations here, does the hydrodynamic accuracy matter? There have been many studies arguing that the treatment of feedback is more important than details of the hydrodynamic solver in this context \citep{scannapieco:2012.aquila.cosmo.sim.compare,power:2013.sphs.entropy.cores,hu:2014.psph.galaxy.tests,few:2016.disk.spiral.arm.sims.gizmo.vs.ramses.vs.sph,dave.2016:mufasa.fire.inspired.cosmo.boxes,zhu:2016.sph.vs.gizmo.cosmo.sims.mw.mass.galaxy,stewart:scylla.halo.accretion.vs.codes,kim:agora.isolated.disk.test}, although in some of these cases it is difficult to isolate the effects of the hydrodynamics within the context of a single feedback model. Fortunately, {\small GIZMO} is an inherently multi-method code, so we can compare simulations with otherwise identical physics and numerics, replacing {\em only} the hydrodynamic solver. 

Figs.~\ref{fig:sph.vs.time}, \ref{fig:sph.vs.time.added.diffusion}, \ref{fig:sph.vs.sfr}, \&\ \ref{fig:image.morph.sph} compare {\small GIZMO} simulations using both our (1) default (non-SPH) MFM method,\footnote{We have also considered a limited comparison of the ``meshless finite volume'' (MFV) method in {\small GIZMO}. This differs from MFM only in a second-order mass-flux term, and gives nearly-identical results in test problems \citep{hopkins:gizmo,hopkins:mhd.gizmo}, so (unsurprisingly) the results are very similar to MFM. We prefer MFM as our ``default'' method because it maintains element masses, minimizing particle-splitting and reducing $N$-body noise.} (2) various ``flavors'' of SPH.

\vspace{-0.5cm}
\subsubsection{``Flavors'' of SPH}
\label{sec:sph.flavors}

There is an extensive literature of various SPH ``flavors'' which attempt to reconcile the differences between SPH and finite-volume methods like our MFM here \citep[see e.g.][and references therein]{wadsley:2008.sph.mixing.cosmology,price:2008.sph.contact.discontinuities,saitoh:2012.dens.indep.sph,hopkins:lagrangian.pressure.sph,read:2012.sph.w.dissipation.switches,hu:2014.psph.galaxy.tests,rosswog:2014.sph.accuracy}. In this section we will consider three such flavors: 

\begin{itemize}

\item{\bf FIRE-1 SPH:} This is the SPH formulation used in FIRE-1, and described in \citet{hopkins:2013.fire}. Briefly, we use the ``pressure-energy'' formulation of SPH (P-SPH) from \citet{hopkins:lagrangian.pressure.sph} to eliminate the spurious ``surface tension'' error present in ``density-energy'' (D-SPH) formulations at contact discontinuities; a higher-order kernel (the quintic spline; with $\sim 64$ effective neighbors instead of $\sim 32$ for the cubic spline in MFM) to reduce zeroth-order SPH errors \citep[see][]{dehnen.aly:2012.sph.kernels,zhu:2014.sph.convergence}; more accurate gradient estimators using the moving-least-squares approach \citep[as in][]{garciasenz:2012.integral.sph,rosswog:2014.sph.accuracy}; higher-order switches to minimize the artificial viscosity following \citet{cullen:2010.inviscid.sph} with further improvements described in \citet{hu:2014.psph.galaxy.tests} and \citet{hopkins:gizmo}; and added ``artificial conductivity'' terms to allow entropy/thermal energy diffusion similar to artificial viscosity \citep{wadsley:2008.sph.mixing.cosmology,price:2008.sph.contact.discontinuities}. These improvements to SPH are designed to improve behavior in fluid-mixing and shock capturing, while reducing noise and artificial numerical diffusivity away from shocks (see references above). 

Readers interested in more details of the SPH method here should consult the public version of {\small GIZMO}: this is the default SPH implementation.

\item{\bf FIRE-1 SPH + Stronger Mixing:} A common feature in most of the SPH flavors above is the use of an ``artificial conductivity'' term (as in FIRE-1 SPH) which allows for diffusion of entropy between particles: without this, entropy becomes ``particle-locked'' leading to artificial resolution-scale discontinuities that suppress fluid mixing instabilities. 

With these ``artificial diffusion'' terms in SPH, there is considerable freedom to adjust the form and normalization of the diffusivity. While the prescriptions are usually tuned to give some desired balance of accuracy on a mix of different idealized test problems, there is usually no single ``correct'' prescription. We therefore consider the effects of adjusting this diffusivity term to give ``stronger mixing'' in SPH, in order to eliminate the spurious ``blobs'' discussed below. Specifically, we take the artificial conductivity in the {\small GIZMO} SPH implementation to be:
\begin{align}
\frac{d\,E_{a}^{\rm SPH}}{dt} =& \sum_{b} \alpha_{ab}\,\tilde{v}_{ab}^{\rm sig}\,\frac{m_{a}\,m_{b}}{\bar{\rho}_{a}+\bar{\rho}_{b}}\,
\left( u_{a} - u_{b} \right)\,\tilde{W}_{ba}\\
\nonumber \tilde{W}_{ba} \equiv &
\frac{\partial W(|{\bf x}_{ba}|,\,h_{a})}{\partial |{\bf x}_{ba}|} + \frac{\partial W(|{\bf x}_{ba}|,\,h_{b})}{\partial |{\bf x}_{ba}|} \\ 
\nonumber \tilde{v}_{ab}^{\rm sig} \equiv& c_{s,\,a}  + c_{s,\,b} - 3\,({\bf v}_{b}-{\bf v}_{a})\cdot \hat{\bf x}_{ba} \\ 
\nonumber \alpha_{ab} \equiv& \Theta(\tilde{v}_{ab}^{\rm sig})\,\frac{| P_{a} - P_{b} |}{P_{a}+P_{b}} \,
\begin{cases}
	{\displaystyle \frac{\alpha_{a}^{V} + \alpha_{b}^{V}}{8} \ \  \hfill { ({\rm ``Default\ SPH''})}} \\
	{\displaystyle 1\ \ \ \ \ \ \ \  \hfill { ({\rm ``Stronger\ Mixing''})}}
\end{cases}\\
\nonumber \Theta(\tilde{v}_{ab}^{\rm sig}) \equiv& 
\begin{cases}
	{\displaystyle 0\ \ \ \ \ \ \ \ \ \  \hfill { (\tilde{v}_{ab}^{\rm sig} \le 0)}} \\
	{\displaystyle 1\ \ \ \ \ \ \ \ \ \  \hfill { (\tilde{v}_{ab}^{\rm sig} > 0)}}
\end{cases}
\end{align}
where ${\bf x}_{ba} \equiv {\bf x}_{b} - {\bf x}_{a}$ is the particle separation, $m$ mass, $\bar{\rho}$ the SPH-estimated density, $W$ the SPH smoothing kernel as a function of ${\bf x}_{ba}$ and smoothing length $h_{a}$, $c_{s}$ the sound speed, ${\bf v}$ the velocity, $P$ the pressure, and $\alpha_{a}^{V}$ is the artificial viscosity coefficient defined in \citealt{hopkins:gizmo}, Appendix~F2, Eq.~F16-F17. This $\alpha^{V}$ varies between $0.05-2$ depending on the velocity divergence $\nabla\cdot {\bf v}$ and its time-derivative, reaching large values when particles approach increasingly rapidly ($\nabla\cdot {\bf v} < 0$ and $d[\nabla\cdot {\bf v}]/dt < 0$) and decaying rapidly otherwise. 

The difference between our ``default'' FIRE-1 SPH and ``Stronger Mixing'' simulations, therefore, is that ``FIRE-1 SPH'' (by design) only allows for mixing/diffusion of entropy and thermal energy when particles are effectively inter-penetrating or ``move through one another.'' In the ``Stronger Mixing'' case, we make the effective diffusivity/mixing stronger in those cases (by a factor $\approx 4$), but more importantly, we allow mixing in shear flows. The latter case has been shown to improve the accuracy of SPH in modeling fluid mixing instabilities such as the Kelvin-Helmholtz instability \citep{price:2008.sph.contact.discontinuities} and more accurately represents the ``unresolved turbulent mixing'' terms advocated in SPH by \citet{wadsley:2008.sph.mixing.cosmology}; it is also numerically closer to the implementation advocated by \citet{read:2012.sph.w.dissipation.switches,hobbs:2013.cooling.instab.sphs}, who have argued this gives better convergence modeling cores of cooling-flow halos \citep{power:2013.sphs.entropy.cores}. On the other hand, the ``Stronger Mixing'' implementation tends to ``smear out'' shock fronts and Keplerian shear flows much more significantly, further reducing the ``effective resolution'' of SPH \citep{hu:2014.psph.galaxy.tests,few:2016.disk.spiral.arm.sims.gizmo.vs.ramses.vs.sph} -- this is why it was not our ``default'' approach in FIRE-1. Note, though, that even in the ``Stronger Mixing'' case, there is still a ``switch'' $\alpha_{ab}$ which suppresses diffusion in supersonically receding flows and sharp phase discontinuities.

\item{\bf Density-SPH + Smaller Kernel (D-SPH):} Alternatively, we can compare an implementation of SPH where we {\em remove} some of the improvements used in our FIRE-1 SPH method. Specifically, in these simulations, we use the ``density-energy'' SPH (D-SPH)  formulation instead of the ``pressure-energy'' SPH formulation, which introduces a non-convergent (sub-zeroth-order) error that has the functional appearance of a surface tension force at phase discontinuities \citep[see][]{saitoh:2012.dens.indep.sph,hopkins:lagrangian.pressure.sph}. We also reduce the size of the SPH smoothing kernel, using a cubic spline with $\sim 32$ effective neighbors; this increases the zeroth-order SPH errors and reduces the ability of the code to capture fluid-mixing instabilities. This is the ``T-SPH'' formulation studied in \citet{hopkins:gizmo}. We retain all other aspects of FIRE-1 SPH (specifically the higher-order gradient, artificial viscosity, and artificial conductivity estimators). We consider this implementation because it allows us to specifically highlight the effects of SPH errors that suppress fluid mixing at contact discontinuities.

\end{itemize}

\vspace{-0.5cm}
\subsection{(Weak) Effects in Dwarfs \&\ the ISM of Massive Galaxies}
\label{sec:hydro:dwarfs}

Fig.~\ref{fig:sph.vs.time} shows that at all resolution levels, our dwarf galaxy simulations are very similar in SPH and MFM methods (at all redshifts). It also clearly shows that SPH and MFM agree well on the properties of the main progenitor of massive (MW-mass) galaxies {\em while the progenitor is still a dwarf} with stellar mass $M_{\ast}\lesssim 10^{10}\,\msun$ at redshifts $z\gtrsim 1.5-2$. This holds for all properties plotted and all others we have examined: e.g.\ the distribution of mass in different phases, CGM gas morphology, covering fraction of absorbers, metallicity gradients, and galaxy rotation.  

This is not surprising: such galaxies are in the ``cold accretion'' regime where there is little or no ``hot halo'' of virial shock-heated gas surrounding the galaxy \citep[e.g.][]{keres:hot.halos}. Rather, gas falls onto galaxies on a free-fall time before being expelled by feedback. As such the properties are simply regulated by a combination of feedback and self-gravity; subtleties of fluid mixing (where SPH differs most dramatically from other hydrodynamic methods) are unimportant. This is consistent with previous studies in FIRE-1 using different ``flavors'' of SPH, which had no effect on dwarf galaxy properties \citep{hopkins:2013.fire}.

Figs.~\ref{fig:sph.vs.time}-\ref{fig:sph.vs.sfr} show that for MW-mass galaxies, which do have hot halos, the qualitative behavior is similar in MFM and SPH, but some significant quantitative differences appear (see below). However, Fig.~\ref{fig:sph.vs.sfr} shows that if we re-start an identical initial condition in MFM and SPH at low redshift, the SFRs are initially identical, and only slowly drift apart (here by just $\sim30\%$ over $\sim1\,$Gyr). Moreover Fig.~\ref{fig:image.morph.sph} shows that the visual galaxy morphologies in SPH and MFM are very similar. 

These results are consistent with a series of studies of {\em non-cosmological} simulations of MW-mass galaxies, comparing different SPH flavors \citep{hopkins:stellar.fb.winds,hopkins:2013.merger.sb.fb.winds,hu:2014.psph.galaxy.tests,kim:agora.isolated.disk.test} and SPH versus moving-mesh codes \citep{hayward:arepo.gadget.mergers}. These studies showed that {\em within galaxies} known SPH errors have little effect on predictions, since the turbulence is primarily super-sonic, cooling is fast compared to dynamical times, and phase structure is primarily driven by gravitational collapse and feedback -- all limits where SPH performs well \citep[see][]{kitsionas:2009.grid.sph.compare.turbulence,price:2010.grid.sph.compare.turbulence,hopkins:gizmo}.

\vspace{-0.5cm}
\subsection{(Significant) Effects in the CGM of Massive Galaxies' ``Hot Halos''}
\label{sec:hydro:massive.cgm}

However, despite the similarity in the SFRs in re-started massive galaxies and non-cosmological simulations, Figs.~\ref{fig:sph.vs.time}-\ref{fig:sph.vs.time.added.diffusion} clearly show that the {\em cosmological} SFHs of MW-mass systems diverge between some SPH flavors and MFM, as they reach stellar masses $M_{\ast}\gtrsim 10^{10}\,\msun$, corresponding to halo masses $\gtrsim 10^{11.5}\,\msun$. In the D-SPH runs (which maximize the difference), the final stellar mass is suppressed relative to the MFM run by a factor of $\sim 1.5$ (low resolution) or $\sim 3$ (intermediate resolution), giving a correspondingly smaller peak circular velocity.

Similarly, \citet{dave.2016:mufasa.fire.inspired.cosmo.boxes} compare low-resolution cosmological large-volume simulations using a simplified, non-FIRE sub-grid ISM and feedback model and MFM and ``P-SPH'' (FIRE-1 SPH) methods in {\small GIZMO}. Over the mass range simulated ($M_{\ast}\sim 10^{9}-10^{12}\,M_{\sun}$), the stellar mass functions agree very well at $z> 1$, but then begin to differ, with SPH tending towards smaller masses by $\sim 0.2\,$dex by $z=0$. Yet another study using still different ISM and feedback models in {\small GIZMO} presented in \citet{zhu:2016.sph.vs.gizmo.cosmo.sims.mw.mass.galaxy}, who see a similar ``divergence'' between MFM and some SPH flavors. 

Not coincidentally, this epoch where SPH and MFM diverge corresponds precisely to the formation of the ``hot halo'' (where virial-shocked gas has a cooling time longer than the dynamical time and establishes a steady-state atmosphere within the halo; see \citealt{keres:hot.halos}). In fact, many studies have shown that much, if not most, of the fuel supply for massive galaxies at later times owes to recycled wind material trapped in the CGM in these hot halos \citep[see][]{oppenheimer:recycled.wind.accretion,dave:2011.mf.vs.z.winds,faucher-giguere:2011.halo.inflow.properties,christensen:2016.baryon.cycle,muratov:2016.fire.metal.outflow.loading,angles.alcazar:particle.tracking.fire.baryon.cycle.intergalactic.transfer}. But this requires following the interaction of multi-phase winds moving sub-sonically or trans-sonically through a pressure-supported medium -- precisely the regime where the known SPH errors are most problematic \citep{oshea:sph.tests,agertz:2007.sph.grid.mixing,read:2010.sph.mixing.optimization,bauer:2011.sph.vs.arepo.shocks,sijacki:2011.gadget.arepo.hydro.tests,torrey:2011.arepo.disks}. 

Fig.~\ref{fig:image.sph.cgm} shows a visual comparison of the gas morphology and phase structure in the CGM at redshifts $z\sim1$, where the differences between methods are most apparent. In D-SPH -- which, we emphasize, is a formulation of SPH known to introduce larger errors compared to our ``FIRE-1 SPH'' -- cool/cold gas in the CGM is primarily locked into numerically spurious ``blobs.'' These are a well-known result of numerical errors in SPH, specifically its inability to capture multi-phase fluid mixing interfaces near the resolution scale \citep[see][]{agertz:2007.sph.grid.mixing,keres:2011.arepo.gadget.disk.angmom,saitoh:2012.dens.indep.sph,power:2013.sphs.entropy.cores,hu:2014.psph.galaxy.tests,few:2016.disk.spiral.arm.sims.gizmo.vs.ramses.vs.sph}. This causes inflows and outflows to ``shred'' into blobs, rather than properly mixing. The FIRE-1 ``P-SPH'' formulation is specifically designed to remove the specific ``surface tension'' error in SPH that allows the blobs to be self-insulating and long-lived. P-SPH therefore reduces, but does not completely eliminate, the spurious ``blobs.'' We note that all SPH variants shown include the ``artificial conductivity'' (entropy diffusion term) above, which has been argued to eliminate such spurious structures \citep[see][]{price:2008.sph.contact.discontinuities,wadsley:2008.sph.mixing.cosmology,read:2012.sph.w.dissipation.switches}; we will discuss this further below.

Fig.~\ref{fig:image.sph.cgm} also shows that in SPH, the halo is more extended with more diffuse boundaries. This is also related to well-known SPH issues, specifically grid-scale heating/noise and difficulty shock-capturing, as well as excessive numerical dissipation of sub-sonic and trans-sonic turbulence \citep[see e.g.][]{bauer:2011.sph.vs.arepo.shocks,sijacki:2011.gadget.arepo.hydro.tests,keres:2011.arepo.gadget.disk.angmom}. Both \citet{springel:arepo} and \citet{hopkins:gizmo} show that SPH (even with state-of-the-art artificial viscosity switches from \citealt{cullen:2010.inviscid.sph}) produces larger velocity noise around cosmological shocks (compared to MFM or moving-mesh methods), damps turbulence strongly below Mach numbers $\mathcal{M}\sim 1$, and requires $\sim 8-10$ inter-particle spacings ($\sim 30-50\,$kpc at the virial radius for the low resolution in Fig.~\ref{fig:image.sph.cgm}) to fully capture shock jumps.

The differences are more striking when we examine the $z=0$ temperature-density phase diagram in Fig.~\ref{fig:temperature.density.sph.phase.diagram}. While both SPH and MFM produce a significant amount of warm ($\sim 10^{5}\,$K) and cool ($\sim 10^{4}\,$K) CGM gas, we see clearly that in D-SPH there is a large amount of cold ($T\ll 10^{4}\,$K) gas which has  survived in the CGM being ``protected'' in cold, dense lumps, despite having very short physical mixing times (without magnetic fields -- not included here -- to ``protect'' the clouds). Also, in D-SPH, there is almost no hot, intermediate-density gas ($T\gtrsim 10^{6}\,$K with $n_{H}\gtrsim 10^{-4}\,{\rm cm^{-3}}$), owing to the difficulties of shock-capturing. 

Together, these issues suppress cooling in some SPH flavors, especially from hot gas in more massive halos \citep[again consistent with previous studies; see][]{keres:2011.arepo.gadget.disk.angmom,torrey:2011.arepo.disks,hobbs:2013.cooling.instab.sphs,zhu:2016.sph.vs.gizmo.cosmo.sims.mw.mass.galaxy}, and they make it relatively ``easier'' for cold outflows to escape from the galaxy. 
Perhaps most troubling, however, in the simplest SPH formulations (e.g.\ D-SPH here), the ``blobs'' and associated under-mixing errors do not converge away at higher resolution, but persist at the resolution scale and actually reach larger density contrasts and contain {\em more} mass at higher resolution \citep[this is because they owe to sub-zeroth-order errors; see][]{agertz:2007.sph.grid.mixing,read:2010.sph.mixing.optimization}. This is why our D-SPH runs at MW-mass deviate {\em more} dramatically from MFM at high-resolution, and do not appear to be converging.
Of course in MFM (as with all Godunov methods) there is almost certainly some numerical {\em over}-mixing at low resolution; a key difference is that this converges away with increasing resolution in a numerically well-defined manner \citep{Harten:1983:hll.riemann.solver}.

We stress, however, that the FIRE-1 SPH (used in all published FIRE-1 simulations) dramatically reduces these discrepancies.

\vspace{-0.5cm}
\subsection{Comparing SPH ``Flavors'': Adding Diffusion Explains Differences in the CGM}
\label{sec:hydro:massive.diffusion}

%
%

Fig.~\ref{fig:sph.vs.time.added.diffusion} also shows the results of using the ``FIRE-1 + Stronger Mixing'' SPH flavor instead of our default ``FIRE-1 SPH'' in the cosmological {\bf m12i} simulation. With this enhanced mixing, the results agree well with our MFM simulations. In contrast, as noted above, the D-SPH simulations using the ``density-energy'' formulation and a smaller kernel (both of which increase the SPH errors on fluid-mixing problems) show the largest disagreement with MFM.

This demonstrates that the MFM-SPH differences, where present, owe to the treatment of fluid mixing, particularly how winds in ``hot halos'' do or do not mix and recycle onto the galaxy. This is a challenging numerical problem, and is almost certainly incompletely resolved in {\em any} numerical galaxy formation simulation. Therefore it is difficult to say which method is ``more correct'' at low resolution -- rather we simply urge caution in interpreting these results at present. The major caveat which must be borne in mind for SPH is that, if the answer depends on an arbitrarily adjustable parameter (e.g.\ the artificial conductivity), and is non-convergent (as we find here), then SPH has unfortunately limited predictive power -- we are forced to calibrate the SPH method at each resolution level to calculations with other codes, rather than simply increase the SPH resolution directly and trust that the errors should converge away \citep[for more discussion, see][]{hobbs:2013.cooling.instab.sphs,zhu:2014.sph.convergence,zhu:2016.sph.vs.gizmo.cosmo.sims.mw.mass.galaxy}.

This also suggests that the large central $V_{c}$ ``spikes'' we see in some of our low-resolution MFM runs, which are also sensitive to resolution, may be related to the same wind-mixing physics in hot halos.  

For additional systematic comparison of ``improved'' SPH and MFM methods in recent cosmological hydrodynamic simulations of galaxy formation, we refer readers to \citet{zhu:2016.sph.vs.gizmo.cosmo.sims.mw.mass.galaxy}. Although the simulations there use a completely different treatment of feedback, cooling, and star formation, many of the conclusions -- most importantly regarding the effects of fluid mixing in different methods -- are identical.

\vspace{-0.5cm}
\section{Effects Of ``Artificial Pressure'' Terms}
\label{sec:hydro:artificial.pressure}

\begin{figure}
\plotonesize{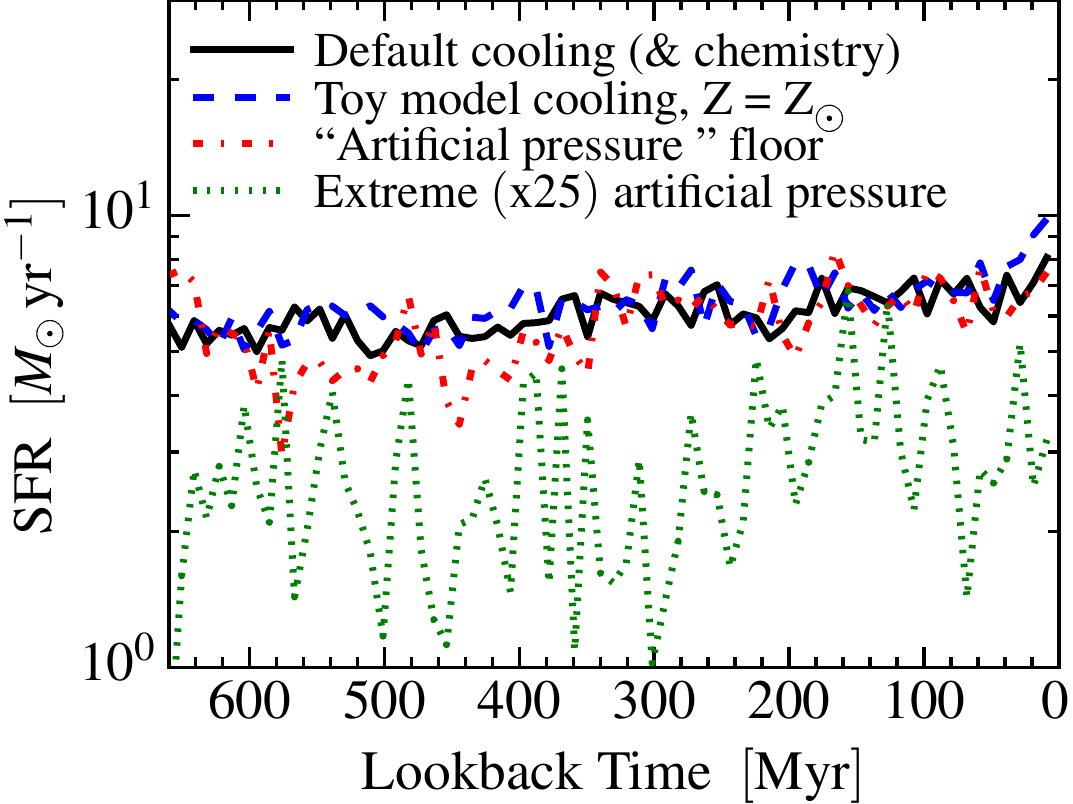}{1.0}
    \vspace{-0.25cm}
    \caption{Effects of cooling physics details (\S~\ref{sec:cooling}) and artificial ``pressure floors'' (\S~\ref{sec:hydro:artificial.pressure}) on a restart of our MW-mass {\bf m12i} simulation near $z=0$, as \demofigrestart. Our default model is full-physics, with no ``artificial pressure.'' 
    {\bf (1)} {\em Toy Model Chemistry/Cooling:} We replace our default, detailed cooling physics with the toy model in \S~\ref{sec:cooling} which puts all gas on a single, solar-metallicity cooling curve, artificially turns off self-shielding (preventing cooling to $T\ll 10^{4}\,$K), and removes the requirement that SF be in molecular gas. Because the cooling time is much faster than the dynamical time in all cases, details of the cooling functions have almost no effect on SF. 
    {\bf (2)} {\em Artificial Pressure Floor:} We add an artificial numerical ``pressure floor'' as described in \S~\ref{sec:hydro:artificial.pressure}, with ``modest'' values designed to artificially inflate the thermal Jeans length to always be equal to four softening lengths. For these values this produces some spurious artifacts in e.g.\ the GMC and star cluster mass function, but does not change any galaxy-scale results, since super-sonic turbulence and resolved collapse from much larger scales dominate SF.
    {\bf (3)} {\em Extreme Artificial Pressure:} We now inflate the artificial pressure so the thermal Jeans length is always $>20$ softening lengths. This is far larger than reasonable and makes the entire disk thermally-supported with $Q\gg1$, shutting down  substructure. Star formation only occurs when the disk becomes globally unstable leading to ``mini-bursts'' -- the dynamics are clearly unphysical.
    \label{fig:sf.z0.cooling}}
\end{figure}

In order to follow ISM structure and star formation self-consistently, we specifically avoid the use of artificial or effective equations of state in our simulations -- i.e.\ in FIRE-2 we do {\em not} adopt, as some previous studies (including FIRE-1) have, an  ``artificial pressure floor,'' numerically forcing the \citet{truelove:1997.jeans.condition} criterion to have some value. In those approaches, an additional pressure of the form $P_{\rm artificial} \sim G\,(N\,h\,\rho)^{2}/\gamma$ is added to the equations of motion, where $h$ is the resolution (local {\em gravitational} softening), $\rho$ the density, and $N$ is approximately the ratio of the Jeans length to $h$ (after the artificial pressure is added) -- this artificially moves the thermal Jeans mass to a desired scale (and in Lagrangian codes, forms an adiabat with $P_{\rm artificial} \propto \rho^{4/3}$). 

A common misconception is that this ``prevents artificial fragmentation,'' but this is not accurate \citep[see][]{teyssier:grid.methods.review}; in cases where $P_{\rm artificial}$ dominates, the medium is {\em by definition} Jeans unstable, so fragmentation (usually down to unresolved scales) is physically correct. Rather, what the artificial pressure does is move the fragmentation scale up to a larger scale, set by $\sim N\,h$, and suppress fragmentation on smaller scales. This is numerically convenient in some circumstances. However, given our star formation prescription, it is not appropriate to suppress fragmentation. Indeed, a cold (weakly pressure-supported), Jeans-unstable, self-gravitating region, with a Jeans scale that is small (stellar-mass scale), is {\em precisely} that which should be identified as star forming by our algorithm. The un-resolved fragmentation to small scales is already handled, in our method, by assigning the mass to a ``star particle'' which represents an aggregate of stars (formed via the un-resolved fragmentation to solar-mass scales).\footnote{We note that \citet{truelove:1997.jeans.condition} (or \citealt{bate:1997.sph.res.reqs}) did not propose an ``artificial pressure.'' Rather, their conclusion was that in order to {\em explicitly} resolve fragmentation {\em down to the thermal Jeans mass scale} (the brown dwarf scale, $\sim 0.1\,\msun$, in the cold ISM), one needed to maintain $\gtrsim 4$ resolution elements per Jeans mass. They suggested this as a local refinement criterion, for AMR-type simulations of proto-star formation. Later, \citet{federrath:2010.sink.particles} and many subsequent authors \citep[e.g.][]{padoan:2011.new.turb.collapse.sims,myers:2013.trulove.with.mhd,gong:2013.sink.particles.athena,federrath:2014.low.sfe,skinner:2015.cloud.sf.frag} pointed out that one can instead use this criterion to determine when local fragmentation is unresolved, and assign the mass to sink particles instead of following it explicitly (standard practice in GMC-scale simulations) -- this is, of course, our star formation prescription.}

Furthermore, ``artificial pressure'' prescriptions require careful treatment, as they: (a) can violate energy conservation (providing an infinite source of ``$PdV$ work''); (b) are designed for thermally-supported disks, and it is not clear how to generalize them when turbulent or magnetic pressure dominates \citep[see][]{myers:2013.trulove.with.mhd,lee:2014.mhd.bondihoyle.stellar.accretion}; and (c) are not necessary given our numerical methods. Specifically \citet{kratter:grav.instab.review,nelson:2006.artificial.fragmentation.in.lagrangian.codes}, and \citet{chiaki:2015.particle.splitting.truelove.criterion.not.correct.for.lagrangian.codes} show (with very different approaches) that in Lagrangian methods, as long as (1) the Toomre mass is resolved (our own criterion in \S~\ref{sec:resolution:mass:firephysics}), (2) the gas disk scale-height is resolved (identical to the Toomre criterion if $Q\sim1$), and (3) fully-adaptive gravitational softenings are used for gas ($\epsilon_{\rm gas}=h_{i}$), the fragmentation cascade converges accurately and is numerically stable (resolution will truncate the lower-limit of the cascade, but all the power is on large scales, just as in turbulent motions). This appears to be confirmed by our tests in Figs.~\ref{fig:sf.z0.mass.resolution} \&\ \ref{fig:clumps.res}. 



That said, because our FIRE-1 simulations used such a prescription, we have checked whether adding an artificial pressure term in the simulations makes a qualitative difference to our results; in Fig.~\ref{fig:sf.z0.cooling}, we implement the standard prescription above with $N=4$. For $N\sim 1-5$, the difference in bulk galaxy properties (SFR, sizes, stellar masses) are small. This is because, for $N=4$, the artificial pressure will stabilize an otherwise unstable spherical overdensity with radius $R \lesssim 30\,{\rm pc}\,m_{i,\,1000}^{1/3}\,(n/10\,{\rm cm^{-3}})^{-1/3}$; so for the resolution and typical gas densities inside the central $\sim 4$\,kpc (the half-light radius) in Fig.~\ref{fig:sf.z0.cooling}, the largest GMCs can still collapse. \citet{manuel:2016.no.effects.from.truelove.criterion} reach similar conclusions (the added terms have little effect) in GMC-scale simulations. However, we caution that there are some potentially un-physical small-scale artifacts introduced by an artificial pressure term, including: noise in the low-temperature gas thermal properties, a ``bump'' in the mass function of GMCs/self-gravitating clouds (owing to ``pileup'' of clouds at the scale where the fragmentation cascade is truncated), and spurious stellar clusters \citep[see][]{nelson:2006.artificial.fragmentation.in.lagrangian.codes,lukat.banerjee:sink.particle.truelove.double.count}. If we adopt $N\gtrsim 10$ (at the resolution shown here), then we are, by construction, forcing the gas to be smooth on scales larger than the gas disk scale height -- i.e.\ we suppress {\em any} substructure in the galactic disks, which in turn artificially suppresses star formation everywhere except at galaxy centers. 

\vspace{-0.5cm}
\section{Do the Details of Cooling \&\ Chemistry Matter?}
\label{sec:cooling}

\begin{figure}
\plotonesize{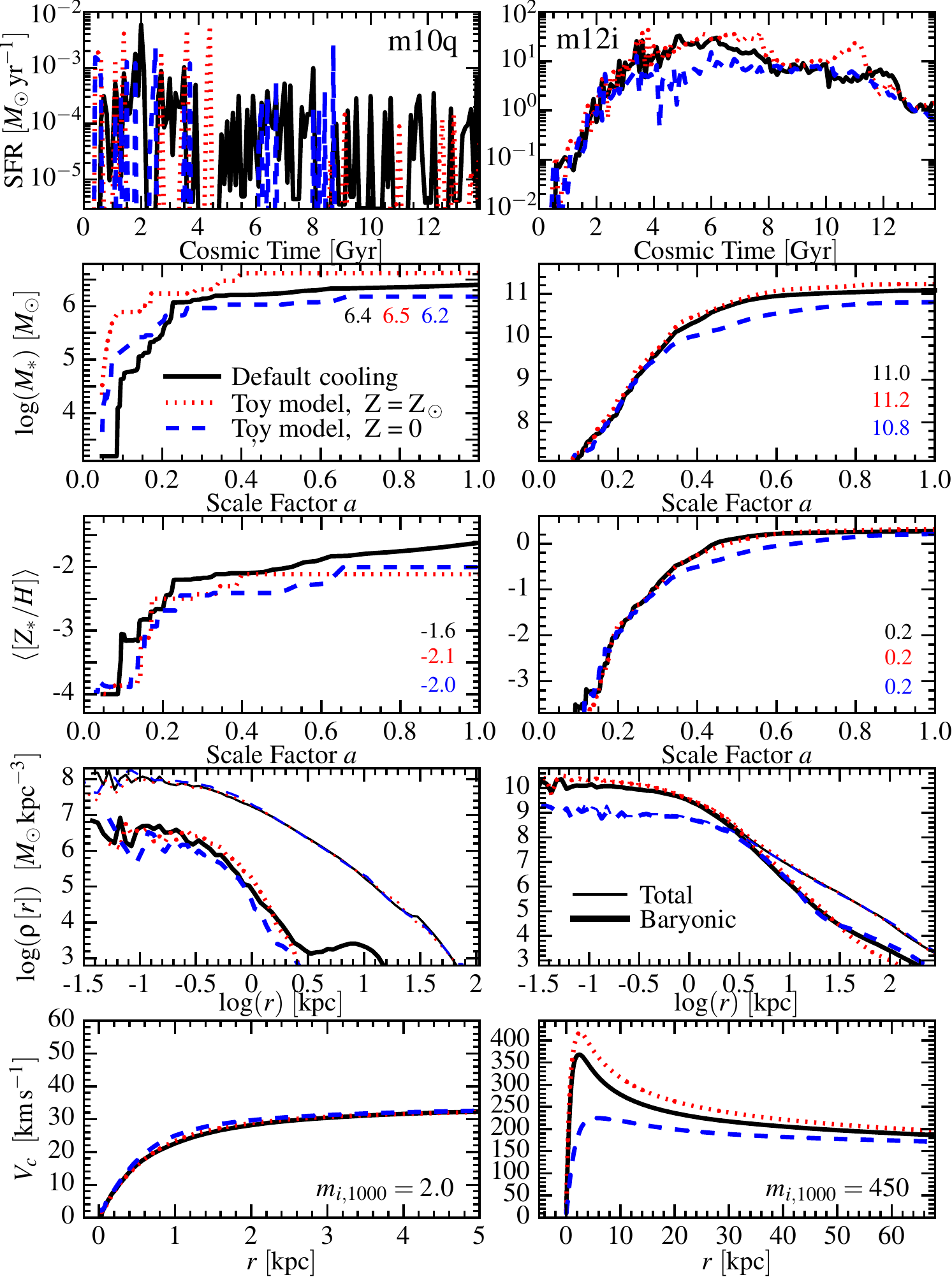}{0.99}
    \vspace{-0.25cm}
    \caption{Effects of the details of cooling \&\ chemistry (as Fig.~\ref{fig:sf.z0.cooling}), on cosmological simulations of a dwarf ({\bf m10q}) and MW-mass ({\bf m12i}) galaxy, as \demofigcosmo. We compare our default, full-physics \&\ chemistry treatment, to the extremely simplified ``toy'' cooling model from Fig.~\ref{fig:sf.z0.cooling} \&\ \S~\ref{sec:cooling} (single, solar metallicity $Z=Z_{\sun}$ cooling curve, with no self-shielding so no low-temperature cooling or chemistry). We also compare a version of the toy model assuming $Z=0$ (primordial abundances only, no metal-line cooling at low or high temperatures). The simplified cooling model produces remarkably similar galaxies to the full-physics model, provided the metallicity is similar ($Z=0$ in {\bf m10q} and $Z=Z_{\sun}$ in {\bf m12i}). Because cooling times within galaxies are often much shorter than dynamical times so as long as some cooling channel exists, the details of cooling have weak effects. The stellar masses are higher with the $Z=Z_{\sun}$ toy model, owing to solar metallicity being assumed in all gas -- this is higher than the metallicity of large-scale super-bubbles in {\bf m10q} or outer parts of the extended ``hot halo'' of {\bf m12i}, artificially enhancing cooling at $T\sim 10^{5}-10^{7}\,$K. Conversely, with $Z=0$ cooling only, SNe bubbles and the hot halo cool less efficiently; notably in {\bf m12i} this reduces the SFR around $z\sim1$, where the dense central bulge forms, flattening the rotation curve substantially. Note that taking our ``full cooling'' model and {\em only} disabling high-temperature ($T>10^{4}\,$K) metal-line cooling produces nearly identical results here to the ``toy'' $Z=0$ model. 
    \label{fig:sf.history.cooling}}
\end{figure}

\begin{figure*}
\plotsidesize{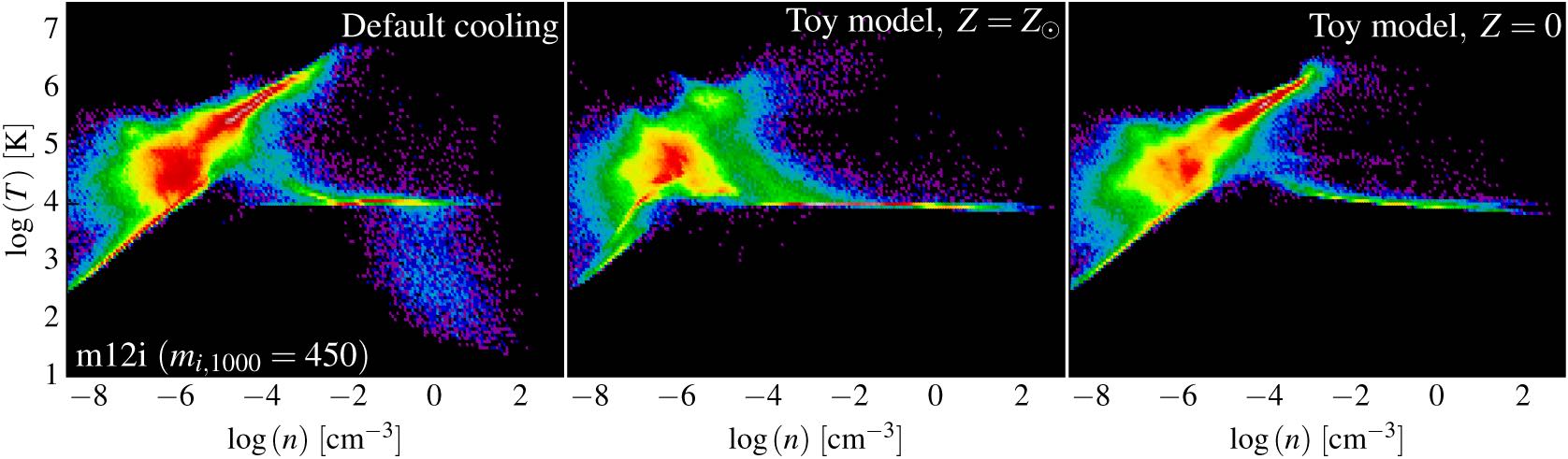}{0.99}
    \vspace{-0.25cm}
    \caption{Temperature-density phase diagram of all gas in the simulation box at $z=0$, for the {\bf m12i} simulations in Fig.~\ref{fig:sf.history.cooling} with our default, full cooling physics implementation and the ``toy model'' which puts all gas on a single cooling curve, with no self-shielding, at a uniform metallicity $Z=Z_{\sun}$ or $Z=0$. As expected, without self-shielding, the toy models produce no cold gas ($T\ll 10^{4}\,$K). However, this is still ``cold'' as far as providing dynamical support to the disk (thermal $Q<1$) is concerned, and $t_{\rm cool}\ll t_{\rm dyn}$ remains true, and so even at $\sim 10^{4}\,$K gas can and should physically fragment efficiently. Therefore, this produces little or no dynamical effect in the simulations. The $Z=Z_{\sun}$ run noticeable suppresses the hot gas at $T\sim 10^{6}\,$K and densities $n\sim 10^{-4}-10^{-2}\,{\rm cm^{-3}}$ -- much of this is gas in the extended CGM and nearby IGM (around $R_{\rm vir}$) which has been shocked (by SNe \&\ O-star/AGB winds but also the halo virial shock), and is highly {\em sub-solar} in metallicity, so the toy model over-estimates the cooling rates by a factor $\sim 10-30$. However, this does not contribute much to the galaxy accretion rate. The ``primordial''  ($Z=0$) model, on the other hand, is closer to correct in the hot halo at $\sim R_{\rm vir}$, but significantly under-estimates the cooling rate for SNe bubbles in the gas disk and inner halo (while the shape of the warm/hot gas distribution is similar, there is much less gas in the warm phase compared to the Default cooling run).
    \label{fig:phase.diagram.coolingphysics}}
\end{figure*}

\subsection{Cooling, Heating, and Self-Shielding Physics}
\label{sec:cooling:cooling.curve}

We now consider how the {\em details} of cooling alter our predictions. Of course, it goes without saying that cooling is critical for galaxy and star formation -- ``turning off cooling'' would produce no galaxies! Moreover, detailed chemistry and cooling physics are obviously critical for some {\em observables} such as line diagnostics \citep[e.g.\ fine-structure or CO excitation; see][]{richings:2016.chemistry.uvb.photoelec.fx}, but it is not clear whether this in turn produces any {\em dynamical} effects on bulk galaxy properties studied here. Our interest here is therefore in exploring whether the more subtle details of cooling, which are often uncertain at the factor $\sim 2$ level and treated differently or incompletely in different codes (for example, some include chemical networks for low-temperature and/or non-equilibrium cooling, some include only pre-tabulated cooling tables, some include additional sub-dominant cooling processes), ultimately have significant effects on our galaxy-scale predictions (masses, SFHs, morphologies, sizes, etc.). 

Figs.~\ref{fig:sf.z0.cooling}-\ref{fig:sf.history.cooling} show the results of turning on or off different portions of the cooling physics in our simulations.\footnote{For the {\bf m12i} example in Fig.~\ref{fig:sf.history.cooling}, we show results at very low resolution, which maximizes the contrast between different cooling models. At higher resolution ($m_{i,\,1000}=56$) the results are similar but with slightly weaker differences.} Specifically, we compare our default FIRE-2 cooling physics described in \S~\ref{sec:methods:cooling} to simulations adopting an extremely simplified toy model. In the toy model, we place all gas on a single, invariant cooling curve $\Lambda(T,\,Z,\,I_{\nu},\,n) \rightarrow \Lambda(T)$ for fully ionized, non-self-shielded gas (so there is essentially no cooling to $T\ll 10^{4}\,$K), with a single metallicity $Z=Z_{\sun}$ or $Z=0$ (primordial cooling only). We also remove the requirement that stars form in molecular/self-shielded gas, since this is not self-consistently computed in the toy model. This toy model is not intended to be physically correct. Rather, it is intended only to illustrate how even radical changes to cooling physics (much more extreme than typical model differences in cooling) have relatively modest impacts on galaxy formation.

For dwarf galaxies, with either $Z=0$ or $Z=Z_{\sun}$ (Fig.~\ref{fig:sf.history.cooling}), we see there is no large effect on the galaxy formation history. 
There is some enhanced early cooling and star formation at $Z=Z_{\sun}$ as expected (this is much higher than the mean metallicity in {\bf m10q}). For MW-mass galaxies, we again find only modest changes to the dynamics in both full cosmological simulations (Fig.~\ref{fig:sf.history.cooling}) or re-starts near $z=0$ with fixed initial conditions (Fig.~\ref{fig:sf.z0.cooling}), {\em if} we adopt approximately the ``correct'' metallicity in and around the galaxy, $Z=Z_{\sun}$. However for MW-mass systems, the primordial cooling-only ($Z=0$) run does differ more dramatically. We have also examined the visual galaxy morphology, and wind outflow rates and covering factors, in each run and find similar results. 

Fig.~\ref{fig:phase.diagram.coolingphysics} shows the $z=0$ temperature-density phase diagram of the {\bf m12i} runs: clearly, details of the phase structure differ, as expected. With the toy model, there is no dense, cold gas ($T\ll 10^{4}\,$K), by construction. For the toy model with $Z=Z_{\sun}$, the hot, denser gas has also cooled more efficiently -- much of this was gas in the halo with lower metallicities which now radiates more rapidly than it should. But this does not appear to have a strong {\em dynamical} effect (there is some increase in the late-time SFR, stellar mass, and circular velocity evident in Fig.~\ref{fig:sf.history.cooling} in this run, owing to additional cooling, but the effect is weak). 

This weak dependence of galaxy properties on the more subtle details of cooling physics at $T\lesssim 10^{4}$\,K (for approximately the correct metallicity) is consistent with previous studies. For example, both \citet{hopkins:rad.pressure.sf.fb} and \citet{hu:photoelectric.heating} added/removed photo-electric heating, and  \citet{hopkins:fb.ism.prop} added/removed molecular cooling at low temperatures, and all found this had essentially no effect on SFRs, wind outflow rates, and other galaxy properties. A more detailed study on sub-galactic scales by \citet{glover:2011.molecules.not.needed.for.sf} considering each microphysical cooling mechanism in turn showed that as long as {\em some} cooling channel exists, the details of that cooling are largely irrelevant in the dense ISM. 

The reason for this is simple: for almost all gas (by mass) in galaxies, the cooling time is much shorter than the dynamical time ($t_{\rm cool} \ll t_{\rm dyn}$), and the gas is super-sonically turbulent. Under these conditions, the details of the cooling physics have little effect on dynamics -- cooling is not a ``rate-limiting'' step. We find our predicted SFRs are similar so long as the cooling model captures two key effects: (1) $t_{\rm cool} \lesssim t_{\rm dyn}$ for gas with $T \ll 10^{7}\,{\rm K}\,(n/0.01\,{\rm cm^{-3}})$ (although for SNe blastwaves, the more relevant comparison is to the remnant expansion time, giving $T\lesssim 10^{6}\,$K), and (2) gas in this rapidly-cooling regime has equilibrium temperatures $\lesssim 10^{4}$\,K or colder (the colder gas has no significant thermal pressure support, so assigning it $T\sim 10^{4}$ does not appreciably change its large-scale dynamics). 

Basically, in gas with temperatures $\lesssim 10^{6}-10^{7}$\,K, details of the cooling physics are not important. Above this threshold, $t_{\rm cool} \gtrsim t_{\rm dyn}$ and cooling can be important; this can matter for (1) super-bubbles and hot galactic winds, and (2) ``hot halos'' of virial-shocked gas around massive galaxies. For winds (1), thermal pressure-driven winds generally cool adiabatically as they expand \citep{chevalier:1974.sne.breakout.conditions}, so the exact radiative cooling is not important so long as the transition from $t_{\rm cool} \ll t_{\rm dyn}$ to $t_{\rm cool} \gg t_{\rm dyn}$ happens at more or less the correct temperature (see \paperone). In hot halos (2), the cooling rate in the halo center determines accretion rates onto galaxies. This is irrelevant in dwarfs ($T_{\rm vir} \ll 10^{7}\,$K), but important in MW-mass halos, hence we see significant effects in {\bf m12i} if we adopt the $Z=0$ toy model. We have re-run this turning on/off each piece of cooling physics in turn, and verified that almost the entire change in Fig.~\ref{fig:sf.history.cooling} owes to the $Z=0$ toy model having no high-temperature ($T>10^{4}\,$K) metal-line cooling. This suppresses the SFR around $z\sim1-2$, as the hot halo forms. In the ``default cooling'' run, accretion onto the galaxy at this time is dominated by recycled wind material from earlier episodes, mixed with other less metal-rich CGM material (leading to additional cooling in that gas) -- the additional cooling from metal-rich wind material is absent in the toy model. Similar conclusions have also been reached in previous studies \citep[e.g.][]{choi:2009.mf.vs.metalcooling,piontek:feedback.vs.disk.form.sims,schaye:2010.cosmo.sfh.sims}.

\begin{figure*}
\plotsidesize{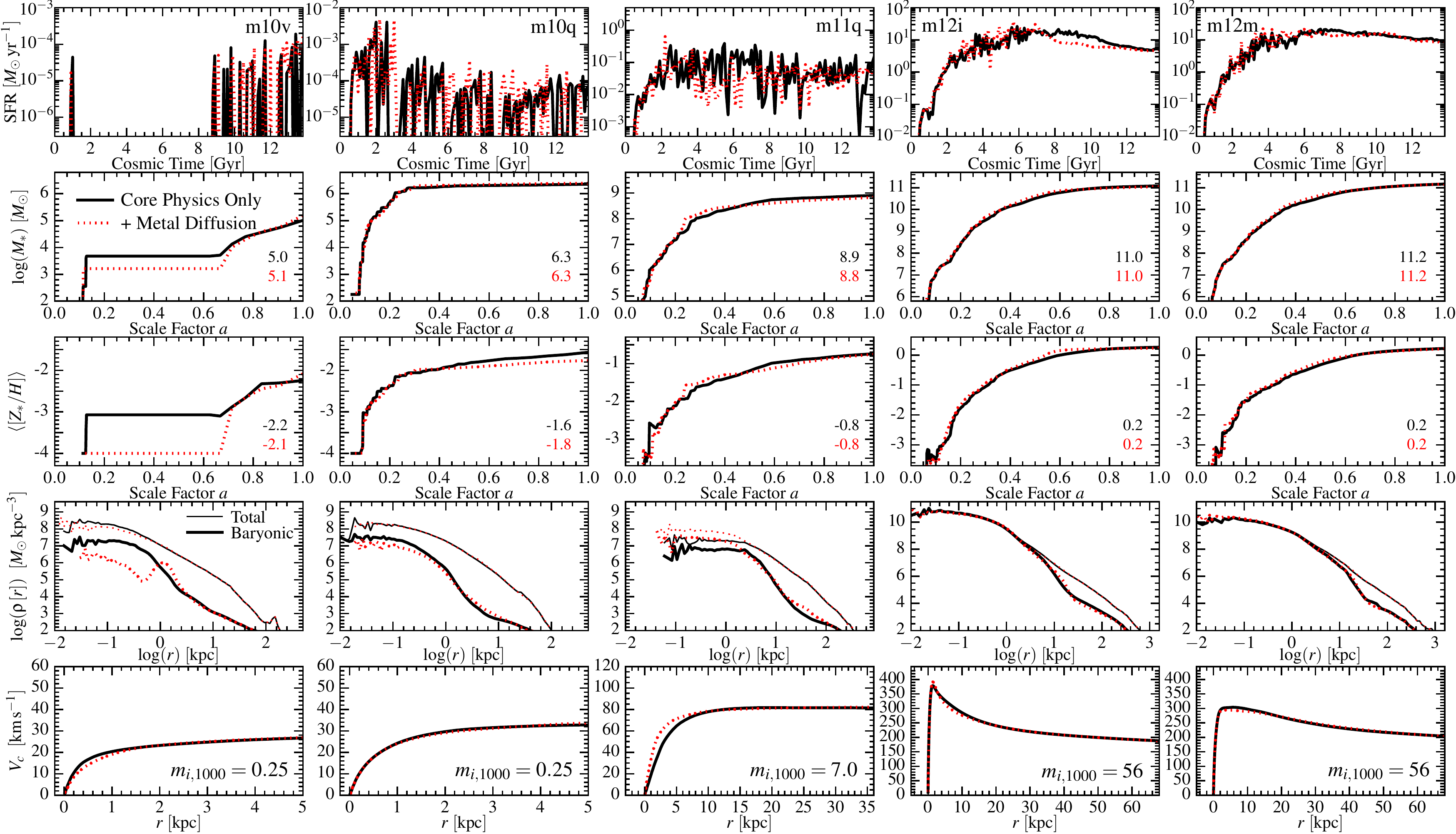}{0.99}
    \vspace{-0.25cm}
    \caption{Comparison of predicted galaxy formation histories as \demofigcosmo, with and without an explicit numerical ``metal diffusion'' term. Our ``core physics only'' runs using the MFM hydrodynamic solver follow fixed-mass elements (so there is no advection between elements and metals are strictly ``locked'' to the element they were injected into via stellar mass-loss). The ``metal diffusion'' runs add an explicit passive-scalar diffusion term to account for sub-resolution diffusion and turbulent mixing of metals between neighboring elements (see Appendix~\ref{sec:turbulent.diffusion}). This is flux-limited so that the diffusivity is never larger than would be obtained from simply numerical diffusivity in our MFV (moving-mesh or AMR-like) hydrodynamic solver. There is no systematic effect at any mass scale. In addition to the runs shown we have also compared {\bf m09} and {\bf m12b,c,f,q} and find the same. \citet{su:2016.weak.mhd.cond.visc.turbdiff.fx} examine non-cosmological disks and our {\bf m10q} and {\bf m12i} runs and show the same, also for the gas and stellar morphology, ISM density distribution in different phases, galactic wind mass-loadings and velocity distributions, and ISM turbulent velocities.
    \label{fig:sf.metaldiff}}
\end{figure*}

\begin{figure}
\hspace{0.5cm}
\vspace{0.1cm}
\includegraphics[width=0.457\columnwidth]{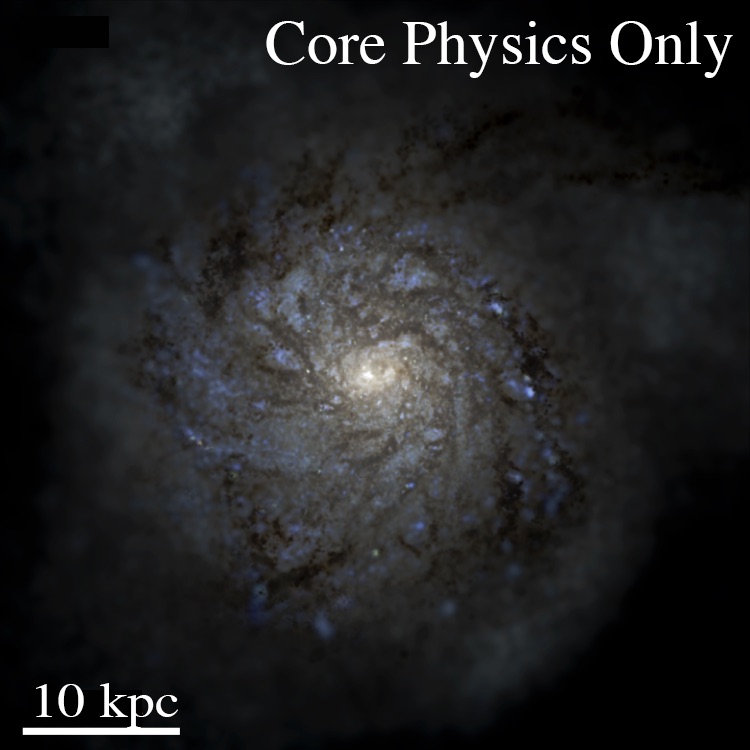} 
\includegraphics[width=0.457\columnwidth]{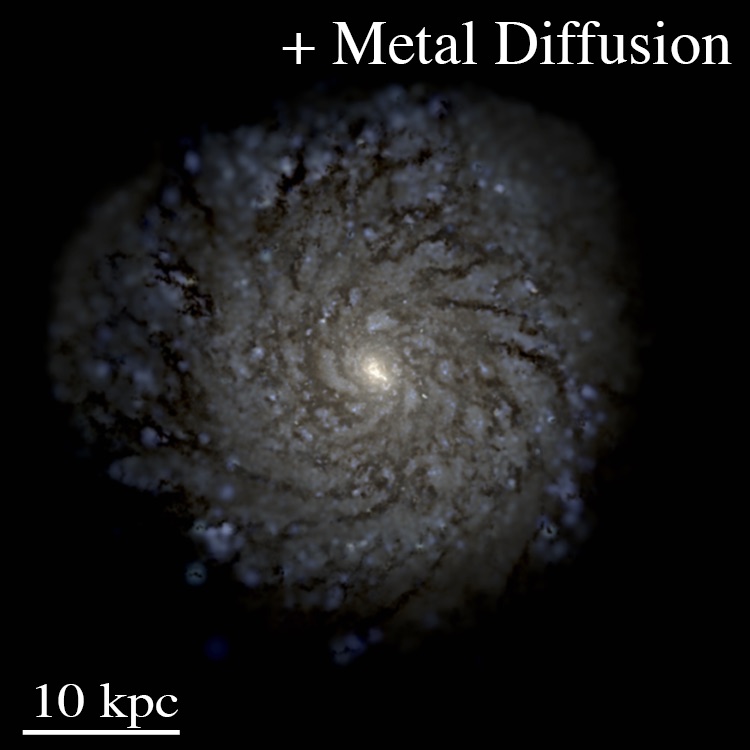} \\
\hspace{-0.1cm}
\plotonesize{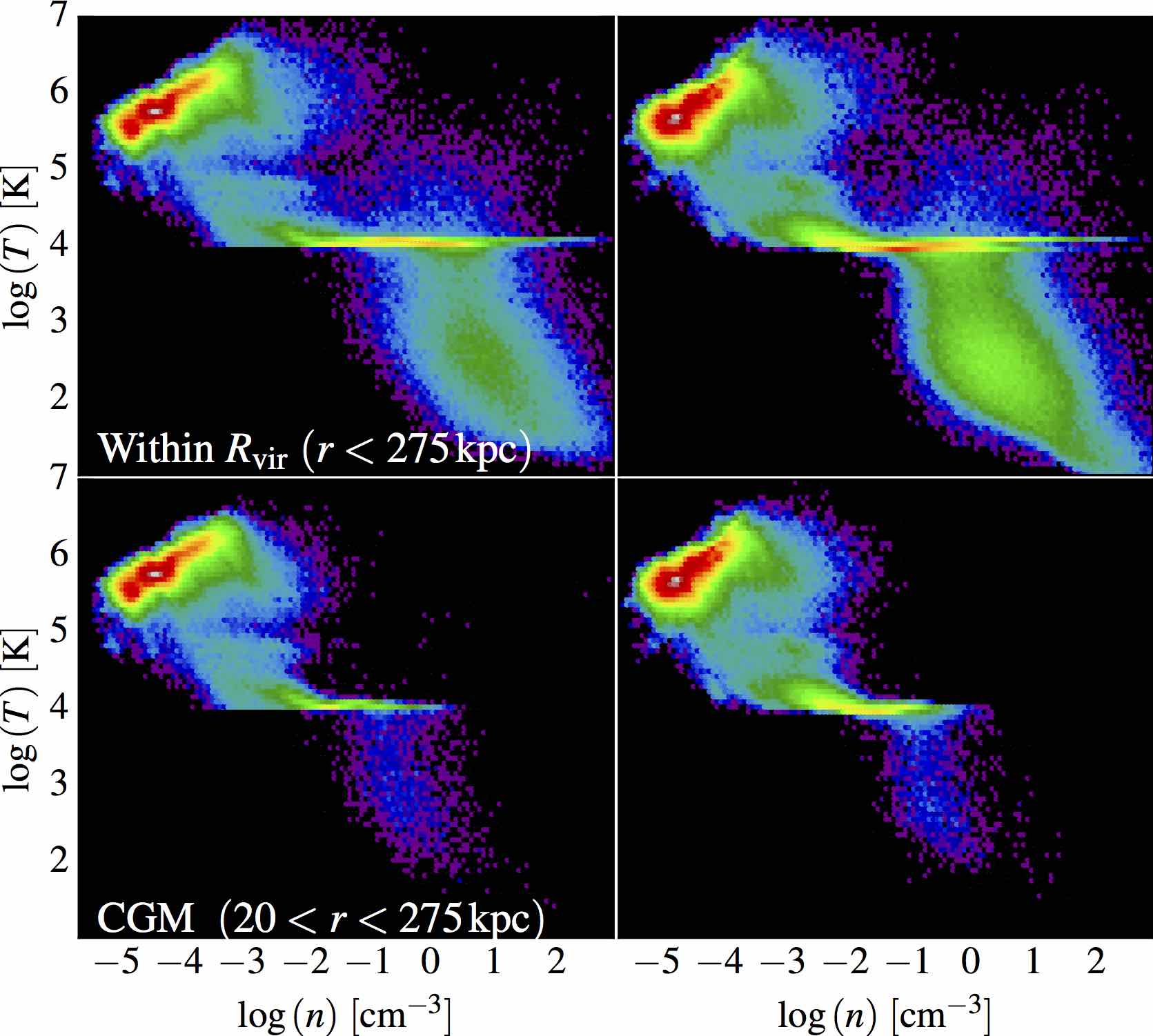}{0.99} \\
    \vspace{-0.25cm}
    \caption{{\em Top:} Visual morphology (as Fig.~\ref{fig:images.resolution}), at $z=0$, for the {\bf m12f} ($m_{i,1000}=56$) simulations run with our ``core physics only'' ({\em left}) or core physics with the additional metal diffusion term as Fig.~\ref{fig:sf.metaldiff} ({\em right}). 
    {\em Bottom:} Phase diagram of the ISM \&\ CGM (as Fig.~\ref{fig:temperature.density.sph.phase.diagram}) for the same.
    The diagrams are qualitatively very similar but there is slightly more cool gas in the ISM at $z=0$ in the runs with metal diffusion, owing to its allowing marginally enhanced cooling from the hot halo that forms at late times in massive galaxies. There is no detectable difference at dwarf masses.
    \label{fig:phase.metaldiff}}
\end{figure}

\begin{figure}
\plotonesize{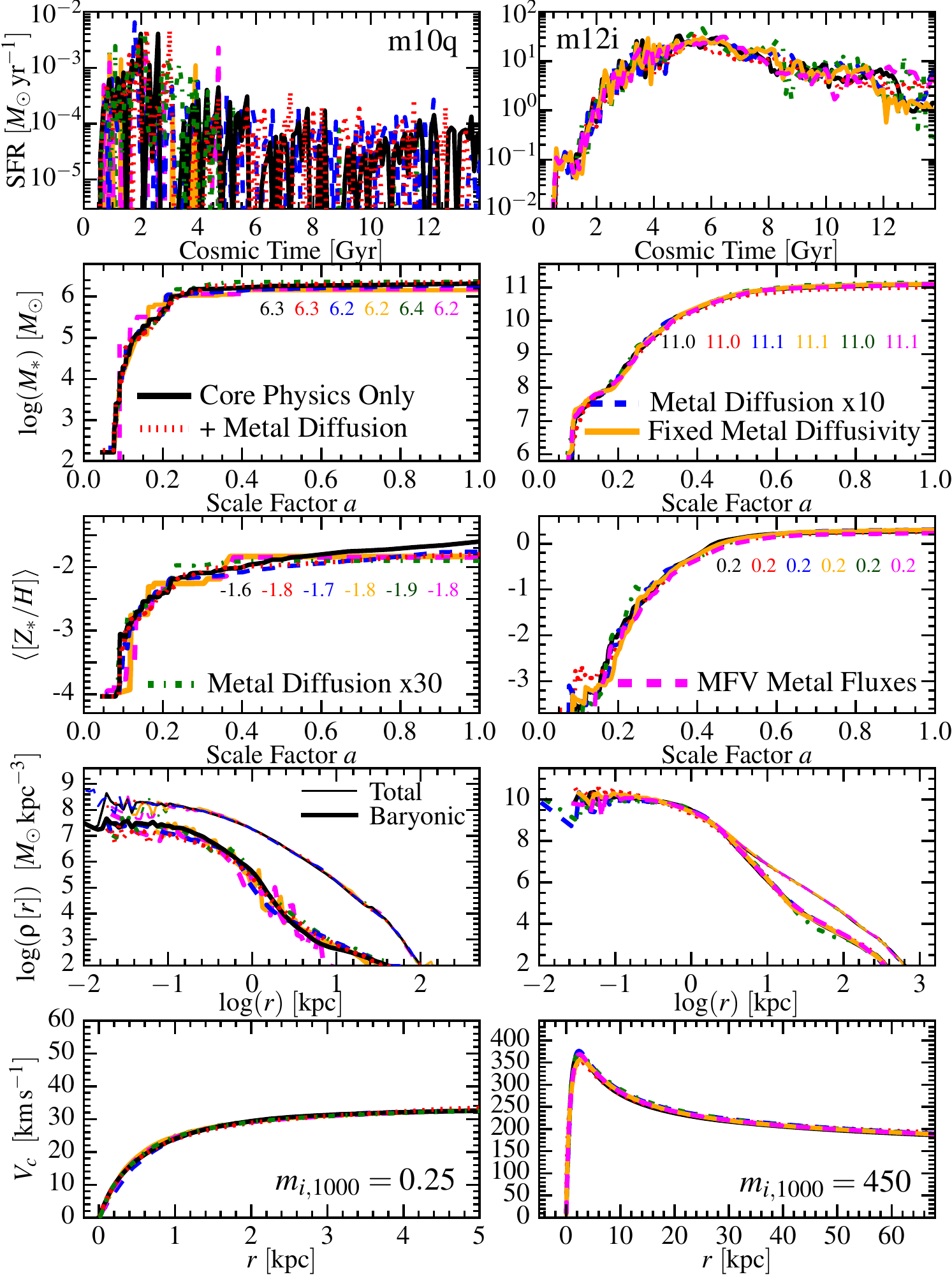}{0.99}
    \vspace{-0.25cm}
    \caption{Comparison of the effects of explicit numerical metal diffusion, as Fig.~\ref{fig:sf.metaldiff}, at dwarf ({\bf m10q}) and MW ({\bf m12i}) mass scales, as a function of the diffusivity. We compare: (1) {\em Core Physics Only}: metals strictly locked to original injection element. (2) {\em Metal Diffusion}: the default numerical implementation in Appendix~\ref{sec:turbulent.diffusion}. (3) {\em Metal Diffusion x10}: as ``Metal Diffusion'' but with the numerical diffusion coefficient increased by a factor $10$. (4) {\em Metal Diffusion x30}: as ``Metal Diffusion'' but with the numerical diffusion coefficient increased by a factor $30$. (5) {\em Fixed Metal Diffusivity}: replacing the adaptive numerical diffusion prescription with a pure isotropic diffusion equation of fixed diffusivity $\kappa_{\rm turb}=10^{24}\,{\rm cm^{2}\,s^{-1}}$, comparable to the large-scale eddy diffusivity for ISM turbulence with $\sigma\sim10\,{\rm km\,s^{-1}}$ and driving scale $\sim 100\,{\rm pc}$. (6) {\em MFV Metal Fluxes}: explicitly adding the cell-to-cell metal advection fluxes that {\em would} appear if we solved the hydrodynamic equations with {\small GIZMO}'s ``meshless finite volume'' (MFV) method, instead of our default MFM method (which has vanishing advective fluxes). This is approximately the ``numerical metal diffusion'' that would un-avoidably appear in a moving-mesh type code. In all cases, the effects are weak.
    \label{fig:sf.metaldiff.coeff}}
\end{figure}

\vspace{-0.5cm}
\subsection{Effects of Numerical Metal Mixing}
\label{sec:turbulent.diffusion.tests}

In \S~\ref{sec:methods:additional.physics} we note that, in our ``core physics only'' simulations, using our default MFM hydrodynamic solver, passive scalars such as metals are ``locked'' to the gas resolution elements into which they are initially deposited (e.g.\ the nearest few elements to each SN). This occurs because the MFM method is a finite-mass method, i.e.\ advective fluxes vanish {\em identically}. But since we have finite resolution, this fails to capture e.g.\ small-scale turbulent eddies and microphysical diffusion of metals between the boundaries of two neighboring resolution elements. In finite-volume methods such as grid-based Eulerian codes, or moving-mesh codes, or the meshless finite-volume ``MFV'' method in {\small GIZMO} (which is similar to MFM, differing only in how the face is assumed to deform when solving the Riemann problem between cells) there is an explicit advective (mass) flux, hence there is always some {\em un-avoidable} mixing at the boundary (in fact, it is well known that such methods will tend to over-mix, relative to a converged solution, at finite resolution). 

Under some circumstances, it may be desirable to explicitly model this microphysical metal transport between resolution elements. In Appendix~\ref{sec:turbulent.diffusion} we describe how this is implemented as an ``additional physics'' option in FIRE-2. Essentially this follows the \citet{smagorinsky.1963:eddy.approximation.for.diffusion.terms} approximation, where we explicitly solve a diffusion equation between cells with the diffusivity $\kappa_{\rm turb}$ set to the resolution-scale ``eddy diffusivity'' ($\sim \ell \,|\Delta v_{\rm turb}(\ell)|$; i.e.\ assuming the mixing time scales with the eddy turnover time, based on the local spatial resolution and velocity field). This is tested directly in {\small GIZMO} with our same MFM solver in idealized, converged ``turbulent box'' simulations in \citet{colbrook:passive.scalar.scalings} and Rennehan et al. (in prep.). An upper limit to the mixing between cells is enforced, equal to what would be obtained by the {\small GIZMO} MFV solver for the same cells and same timestep -- in other words, the {\em maximum possible} diffusivity of the added term in MFM is simply the un-avoidable diffusivity inherent to the MFV method (itself very similar to that in moving-mesh codes). This allows us to study how this particular numerical diffusion term, which is always present in hydrodynamic methods with explicit mass fluxes, alters our predictions.

Fig.~\ref{fig:sf.metaldiff} shows the effects of adding this term, in several of our simulations at both dwarf and MW mass scales, on the galaxy formation history and $z=0$ mass profiles. Fig.~\ref{fig:phase.metaldiff} shows the effects on the $z=0$ ISM and CGM gas phase distribution and visual morphologies in a MW-mass galaxy (there is no detectable difference in dwarfs, consistent with other properties we have surveyed). We see essentially no systematic effect in any gross galaxy property (star formation history, stellar mass, mass profile, rotation curves, visual morphology of gas or stars), including the mean metallicity itself (both in gas and stars). This is true at all mass scales we have explored and at all resolution levels. There is {\em slightly} more cool gas at $z=0$ in the disk of the {\bf m12f} simulation with metal diffusion, which could owe to metal mixing promoting additional cooling from the hot halo, but we caution that this effect is similar in magnitude to stochastic variations between runs. It is worth noting that the cold gas within the disk ($T\lesssim 10^{4}\,$K, $n_{H}\gtrsim 0.1$ cm$^{-3}$) forms a slightly tighter (more visually obvious) cooling sequence; this is because the metal diffusion smooths local particle-to-particle variations in gas-phase metallicity in e.g.\ the same GMCs, and the metals are the dominant coolant in this regime. 

\citet{su:2016.weak.mhd.cond.visc.turbdiff.fx} explore in greater detail the effects of this metal-mixing term in the ISM of both isolated (non-cosmological) disks and cosmological simulations with the FIRE-2 physics. They specifically compare the bivariate distribution of gas temperatures and densities in the ISM (at several redshifts); galactic wind outflow rates, phases, density distributions, and velocity distributions; turbulent velocity distribution functions in the ISM; and star formation histories. They conclude in all cases that the metal-mixing terms have weak or negligible effects. In \citet{escala:turbulent.metal.diffusion.fire} we present the actual abundance ratio {\em distribution functions} within the galaxies here; there, it is clear that the mixing terms do have important effects (as expected). However because the mixing terms conserve (by construction) the mean metallicity and total metal mass, and we have already shown that the metallicity enters only relatively weakly into dynamical effects via its effect on the cooling rates (see \S~\ref{sec:cooling}), the dynamical effect of these mixing terms on other galaxy properties is weak. 

Fig.~\ref{fig:sf.metaldiff.coeff} further demonstrates this by considering variations to the standard metal-mixing term, at both dwarf and MW mass scales. Specifically we consider (in addition to the ``core physics only'' and ``metal diffusion'' runs) cases where we add the metal diffusion term but arbitrarily multiply the diffusivity $\kappa_{\rm turb}$ by an additional factor of $10$ or $30$. We also consider a case where we replace the standard adaptive eddy diffusivity (defined in Appendix~\ref{sec:turbulent.diffusion}) with a constant $\kappa_{\rm turb}=10^{24}\,{\rm cm^{2}\,s^{-1}}$ (i.e.\ simply treat the metals as if they obey a strict, constant-diffusivity diffusion equation, and we also remove the MFV-based limiter above). This value of $\kappa_{\rm turb}$ corresponds crudely to typical values of the turbulent diffusivity estimated by Eq.~\ref{eqn:turb.diffusivity} in Appendix~\ref{sec:turbulent.diffusion} on the maximal turbulent scales (for $\ell \sim 100\,$pc and $|\Delta v_{\rm turb}(\ell)|\sim 10\,{\rm km\,s^{-1}}$). Finally, we consider a case where we include no diffusion equation, but instead we solve the metal flux {\em exactly} as we would if we used the MFV hydrodynamic method (with explicit mass fluxes), and include the resulting metal fluxes (even though all {\em other} fluxes use the MFM solution, identical to our ``standard'' runs). In other words we always transfer the metals that would be ``numerically diffused'' in a finite-volume code. In each of these cases, we see no effect on the properties in Fig.~\ref{fig:sf.metaldiff.coeff}.

\vspace{-0.5cm}
\subsection{Yields and Explicit Abundance-Ratio Tracking}
\label{sec:cooling:yields}

As noted above, we track 11 species on-the-fly with yield tables for SNe (Ia \&\ II) and stellar mass-loss rates given in Appendix~\ref{sec:stellar.evolution.approximations}. But nucleosynthetic yields (even IMF-averaged) have very large uncertainties, especially at progenitor metallicities far from solar. For example, the \citet{woosley.weaver.1995:yields} yields differ from those in \citet{nomoto2006:sne.yields} even at $Z=Z_{\sun}$ by $\sim0.4\,$dex for {\small Mg} and {\small Ne}. However, given the weak dependence of our results on the details of the cooling curve shape (especially at low temperatures; \S~\ref{sec:cooling:cooling.curve}) and detailed metal mixing (\S~\ref{sec:turbulent.diffusion.tests}), we do not expect this to have large dynamical effects on other galaxy properties. Moreover, even high-temperature metal-line cooling (which is important to include) is dominated by {\small O}, which constitutes most of the metal mass and therefore is better constrained (differing by $<10\%$ between the \citealt{woosley.weaver.1995:yields} and \citealt{nomoto2006:sne.yields} models, for example). We have verified this directly by re-running our {\bf m10q} ($m_{i,\,1000}=0.25$) and {\bf m12i} ($m_{i,\,1000}=56$) simulations, replacing the default \citet{nomoto2006:sne.yields} yields for core-collapse SNe with those from \citet{woosley.weaver.1995:yields}. In all properties surveyed here, we see no detectable difference. 

Given this and our comparisons in \S~\ref{sec:cooling:cooling.curve}, it appears that as long as the total metallicity is approximately correct, differences in the detailed abundance ratios introduce no major dynamical effects. We have therefore re-run the same simulations, ignoring all detailed abundance information and simply tracking the total metal abundance, then assuming solar abundance ratios in the cooling and other relevant routines. We find this produces nearly identical results to our default simulations (in which the cooling is explicitly solved species-by-species). 

To leading order, then, detailed abundance patterns are important as {\em tracers} for predictions of various observables, but do not introduce important dynamical effects.

\vspace{-0.5cm}
\section{Effects of the Star Formation Algorithm}
\label{sec:star.formation}

\begin{figure*}
\plotsidesize{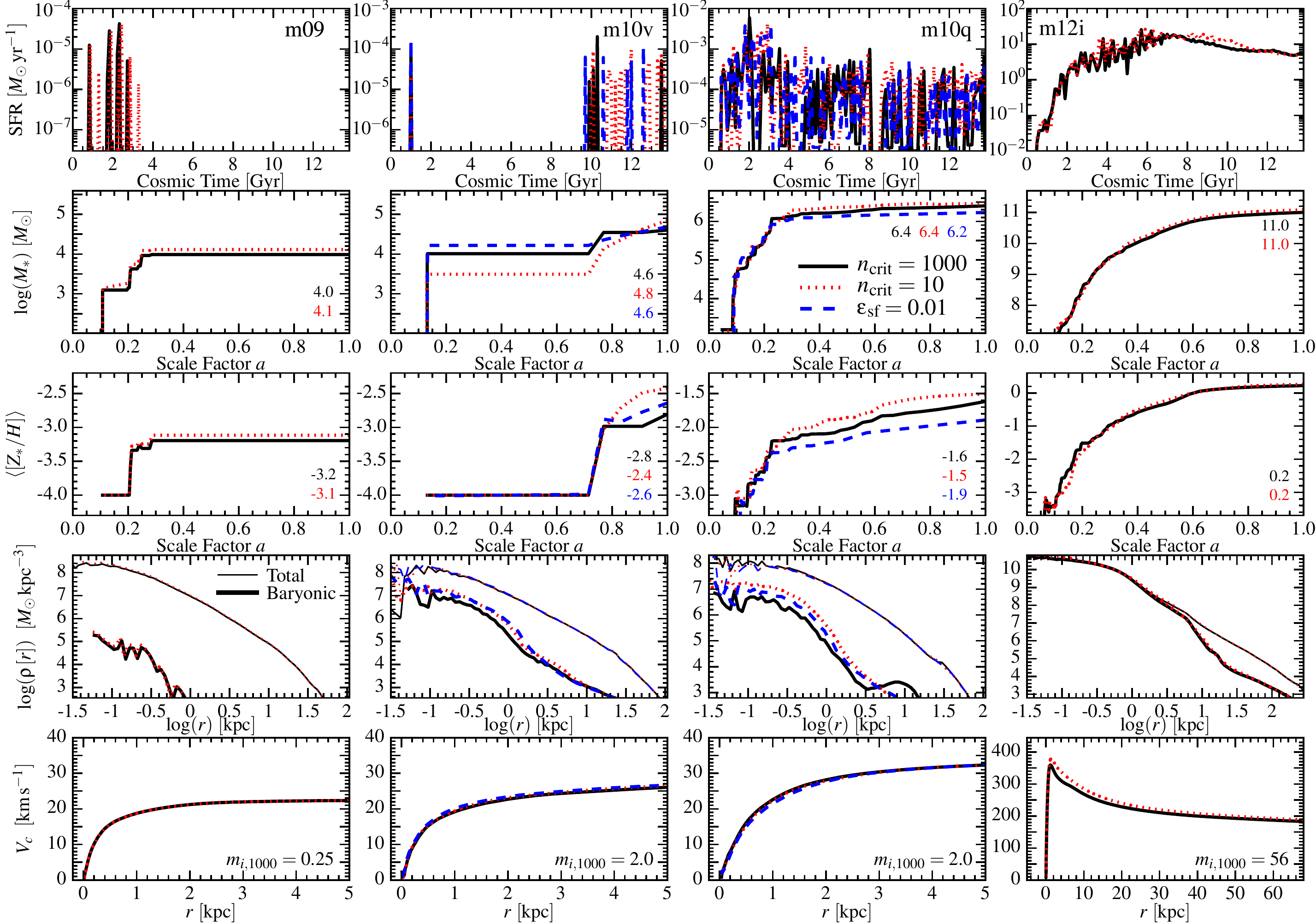}{0.99}
    \vspace{-0.25cm}
    \caption{Comparison of galaxy properties in cosmological FIRE-2 simulations (as \demofigcosmo) as a function of the resolution-scale assumptions for individual star particle formation. We compare {\bf m09}, {\bf m10v}, {\bf m10q}, and {\bf m12i} (Table~\ref{tbl:sims}). 
    For each, we compare a run with a minimum density for star formation of $n_{\rm crit}=10$ or $n_{\rm crit}=1000\,{\rm cm^{-3}}$; recall, in all cases the SF is still only allowed in self-gravitating, self-shielding gas. We have repeated this experiment for {\bf m10q} and {\bf m12i} at lower resolution, and for {\bf m11q}, {\bf m11v}, {\bf m10y} with $n_{\rm crit}=100$ and $=1000$, and reach the same conclusion in every case. For {\bf m10v} and {\bf m10q}, we also compare a simulation in which we arbitrarily multiply the SFR in the ``eligible'' gas (sufficiently dense, self-shielding/molecular, locally self-gravitating) by a factor $\epsilon_{\rm SF}=0.01$ (i.e.\ artificially ``slow down'' the SF in the collapsing gas by a factor of $\sim100$, relative to its dynamical time). Consistent with extensive studies in the FIRE-1 and previous simulations, these choices have no effect on our predictions.
    \label{fig:sf.threshold.zooms}}
\end{figure*}

\begin{figure}
\plotonesize{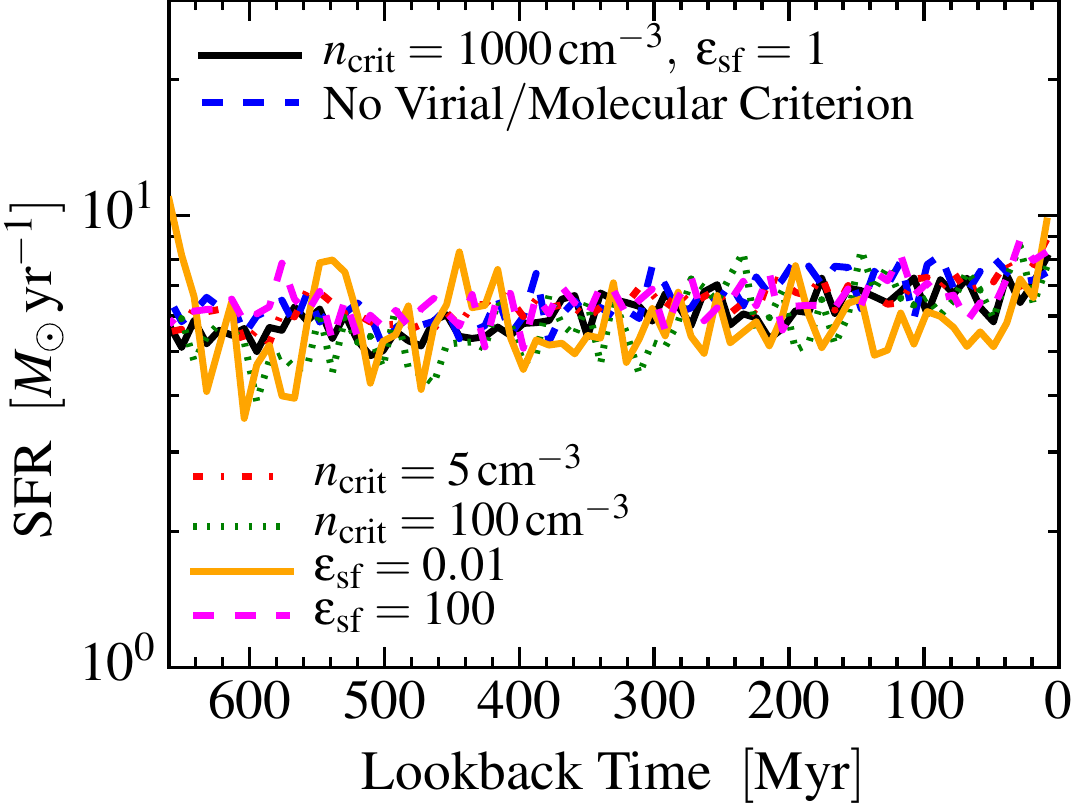}{1.0}
    \vspace{-0.25cm}
    \caption{Effects of the resolution-scale assumptions for individual star formation, in re-starts of a MW-mass galaxy ({\bf m12i}) at late times, as Fig.~\ref{fig:sf.z0.mass.resolution}, keeping fixed mass resolution $m_{i,\,1000}=56$ but changing the SF algorithm. In our ``default'' model gas which is self-gravitating (sub-virial), self-shielding/molecular, dense ($n>n_{\rm crit}=1000\,{\rm cm^{-3}}$), and Jeans unstable forms stars at a rate $\dot{\rho}_{\ast} = \rho_{\rm mol}/t_{\rm ff}$. We compare: (1) removing the self-gravity \&\ molecular restrictions (all dense gas can form stars), (2-3) lowering/raising $n_{\rm crit}=5-1000\,{\rm cm^{-3}}$, and (4-5) multiplying $\dot{\rho}_{\ast}$ by an arbitrary factor $\epsilon_{\rm sf}=0.01-100$. As in Fig.~\ref{fig:sf.threshold.zooms} and our previous studies, this has no effect on the galaxy-scale SFR, because it is feedback-regulated.
    \label{fig:restart.sfmodel}}
\end{figure}

It has been extensively demonstrated in the literature that, provided the Toomre scale is resolved and stellar feedback is treated explicitly (as in our simulations), the exact resolution-scale SF model has essentially no effect on predicted galaxy-scale SFRs. For the sake of completeness, we demonstrate this here, but refer readers to \citet{hopkins:rad.pressure.sf.fb,hopkins:virial.sf,hopkins:dense.gas.tracers} and \citet{kim:disk.self.reg} for more extensive discussions (see also \S~\ref{sec:resolution:mass:firephysics}). 

Figs.~\ref{fig:sf.threshold.zooms}-\ref{fig:restart.sfmodel} show the effects of varying the resolution-scale star formation model in the simulations, in both full cosmological runs at different mass scales and resolution, and re-starts of our MW-mass simulation at late times (guaranteeing an identical initial condition for the comparison). We compare our default model from \S~\ref{sec:methods:star.formation} (requiring gas be self-gravitating, self-shielding/molecular, Jeans unstable, and have density $n>n_{\rm crit}$ with $n_{\rm crit}=1000\,{\rm cm^{-3}}$), to variations with $n_{\rm crit}=5-1000\,{\rm cm^{-3}}$; turning on/off the self-gravity (virial), molecular, and Jeans criteria; and arbitrarily multiplying the SFR per free-fall time $\dot{\rho}_{\ast}$ in the gas which meets these criteria by a constant factor $\epsilon_{\rm sf} = 0.01 - 100$. 

In every case, the predicted SFR and all other galaxy properties we examine here are essentially identical. This has now been shown in many different contexts, with simulations using different detailed implementations of star formation and stellar feedback; mass resolution ranging from sub-solar to $\sim 10^{6}\,\msun$; isolated (non-cosmological), galaxy-merger, and fully-cosmological simulations; circum-nuclear simulations of star formation around AGN; and simulations using very different feedback mechanisms (removing SNe, stellar mass-loss, or HII photo-heating); and in different codes \citep[see e.g.][]{saitoh:2008.highres.disks.high.sf.thold,shetty:2008.sf.feedback.model,hopkins:stellar.fb.winds,hopkins:2013.merger.sb.fb.winds,hopkins:rad.pressure.sf.fb,hopkins:stellar.fb.mergers,hopkins:virial.sf,hopkins:dense.gas.tracers,hopkins:qso.stellar.fb.together,federrath:2012.sfr.vs.model.turb.boxes,agertz:2013.new.stellar.fb.model,kim:2011.sf.selfreg.disks,kim:disk.self.reg,kim.ostriker:sne.momentum.injection.sims}. We specifically showed the same was true in our FIRE-1 simulations in \citet{hopkins:2013.fire,orr:ks.law}. This is also consistent with simulations of individual GMCs (which track star formation on much smaller scales), and analytic ``multi-free-fall'' models of star formation, in which independent clumps collapse on their own local free-fall times \citep[see e.g.][]{federrath:2012.sfr.vs.model.turb.boxes,federrath:2012.sfe.pwrspec.vs.time.sims,cafg:sf.fb.reg.kslaw,guszejnov:gmc.to.protostar.semi.analytic,grudic:sfe.cluster.form.surface.density}. 

This is because the rate-limiting step in star formation does not occur at the resolution limit of the simulations; rather it is the formation and efficient destruction of the largest self-gravitating objects (large GMC complexes, which have dynamical times $\sim 100\,$Myr). Because these are {\em resolved},  and our force softening is fully-adaptive (and we do not force some artificial lower-limit to the softening), a self-gravitating cloud or sub-clump will continue to fragment, as it should, on its local free-fall time, until it either forms stars or is destroyed by feedback (see below). This means that absent star formation or sufficient feedback, the densities in such a clump will (correctly) become arbitrarily high and the internal dynamical times will become arbitrarily short, within a finite physical time (of order the original parent cloud free-fall time). So whatever value of $n_{\rm crit}$ we set will eventually be exceeded, and even if $\epsilon_{\rm sf}\ll1$ (so the cloud collapses faster than it forms stars), the actual physical SF timescale will become arbitrarily short within the clump as it collapses. Provided star formation {\em can} occur, then the simulations above have shown that feedback self-regulates the level of star formation. Stars form until sufficient numbers of young stars are present that their feedback destroys the parent clouds, which may occur after just a small fraction of the cloud turns into stars \citep[e.g.][]{grudic:sfe.cluster.form.surface.density}, giving a low time-averaged efficiency per free-fall time. This self-regulating behavior depends on the strength of feedback -- changing the strength of feedback immediately changes the equilibrium SFR in a galactic disk -- but is independent of how individual stars form in dense gas.

Of course, the ratio of SFR to dense gas {\em at the resolution limit} depends, by construction, on the SF model; comparison to dense gas tracers within galaxies (e.g.\ HCN as opposed to CO) in previous work appears to favor the ``default'' normalization and SF criteria we adopt here \citep[see the comparison in][]{hopkins:dense.gas.tracers}. But this is only relevant within the densest resolved gas clumps.


\vspace{-0.5cm}
\section{Stellar Feedback}
\label{sec:feedback}

\begin{figure}
\plotonesize{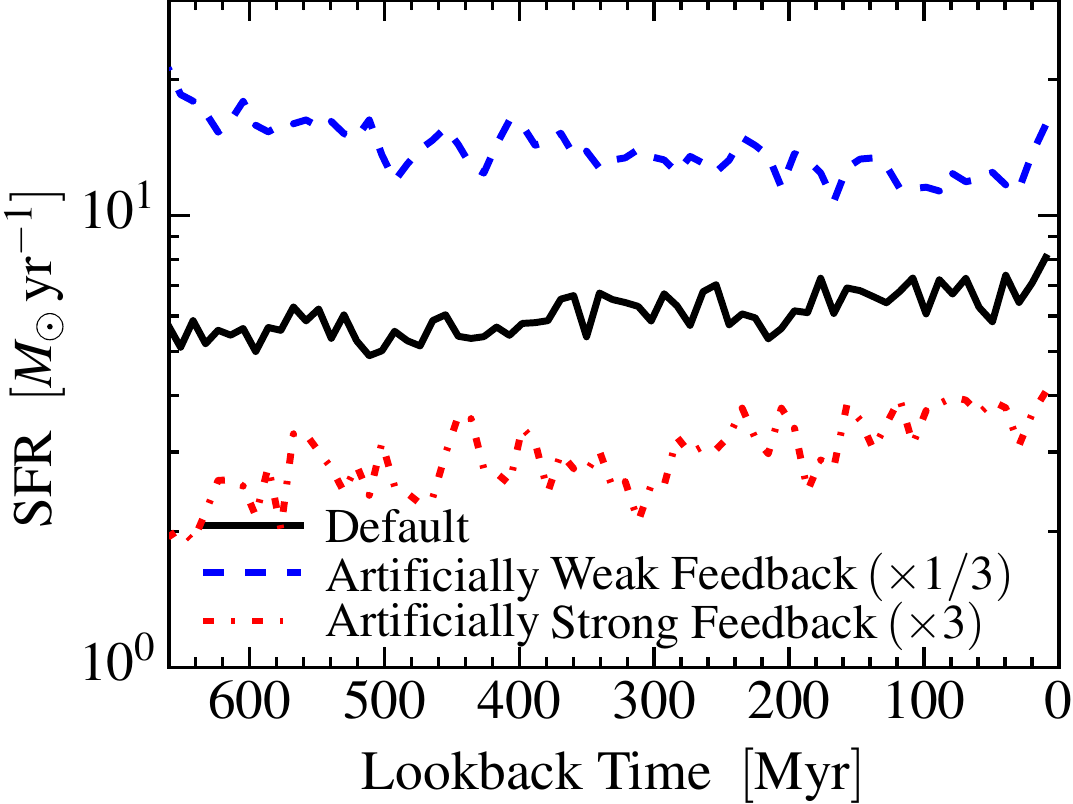}{0.99}
    \vspace{-0.25cm}
    \caption{Effects of arbitrarily changing the strength of stellar feedback on a re-start of an identical {\bf m12i} simulation (with $m_{i,\,1000}=56$) at low redshift (i.e.\ guaranteeing identical initial galaxy properties), as \demofigrestart. Here, we compare our ``default'' model (in which all feedback physics, rates and energetics are taken from stellar evolution models without re-adjustment), and compare it to two models where we arbitrarily multiply {\em all} feedback rates (e.g.\ SNe rates, wind kinetic luminosities and mass loss rates, stellar luminosities at all frequencies) by a factor of $\sim 3$ (``strong'') or $\sim 1/3$ (``weak''). These are much larger than actual physical uncertainties in these quantities, we simply show it for illustrative purposes. Clearly SF is instantaneously feedback-regulated: the steady-state SFR is inversely proportional to the feedback strength, as expected.
    \label{fig:sf.fb.strength.restart}}
\end{figure}

\begin{figure*}
\begin{tabular}{cc}
\includegraphics[width=0.63\textwidth]{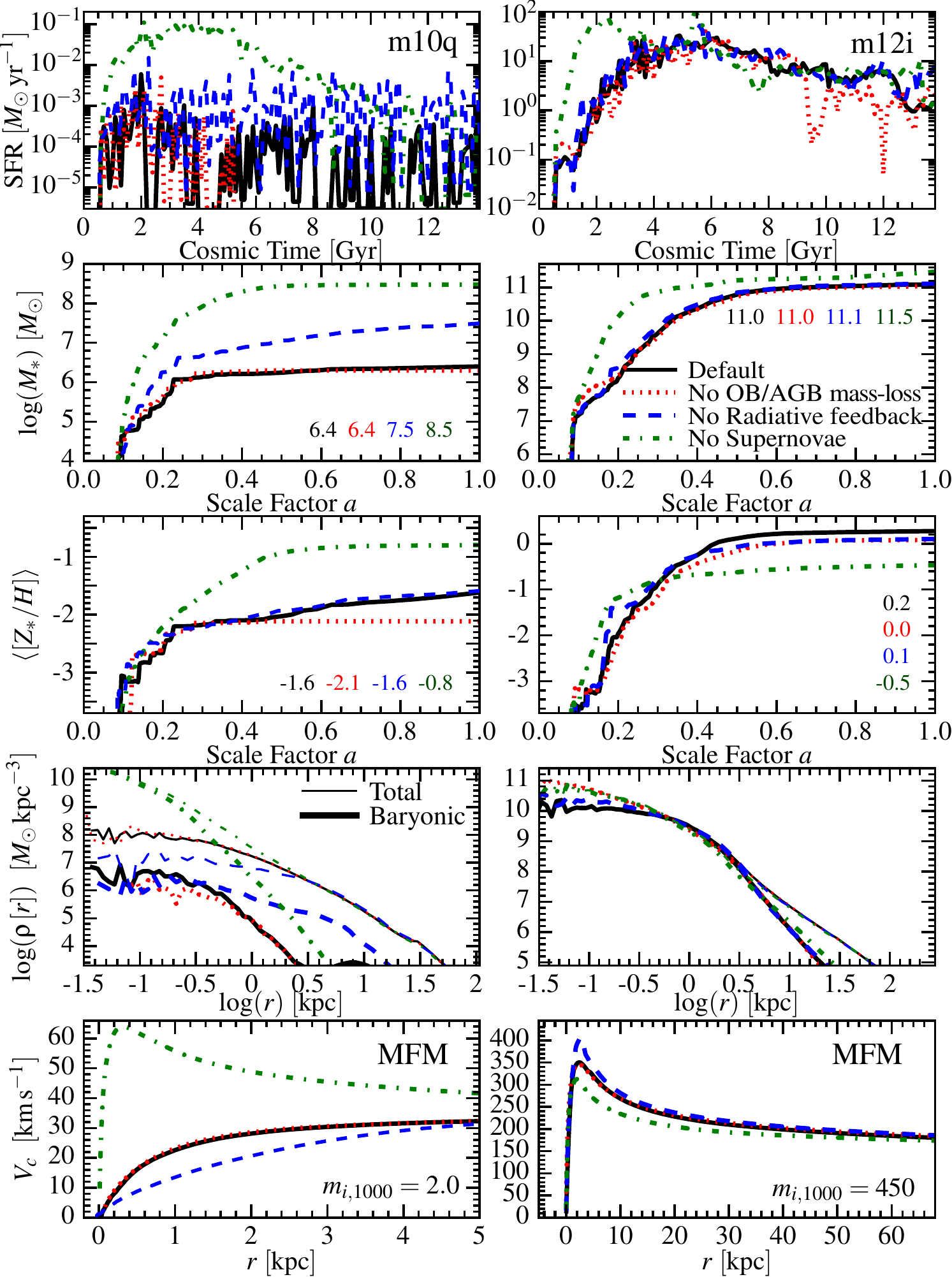} &
\includegraphics[width=0.35\textwidth]{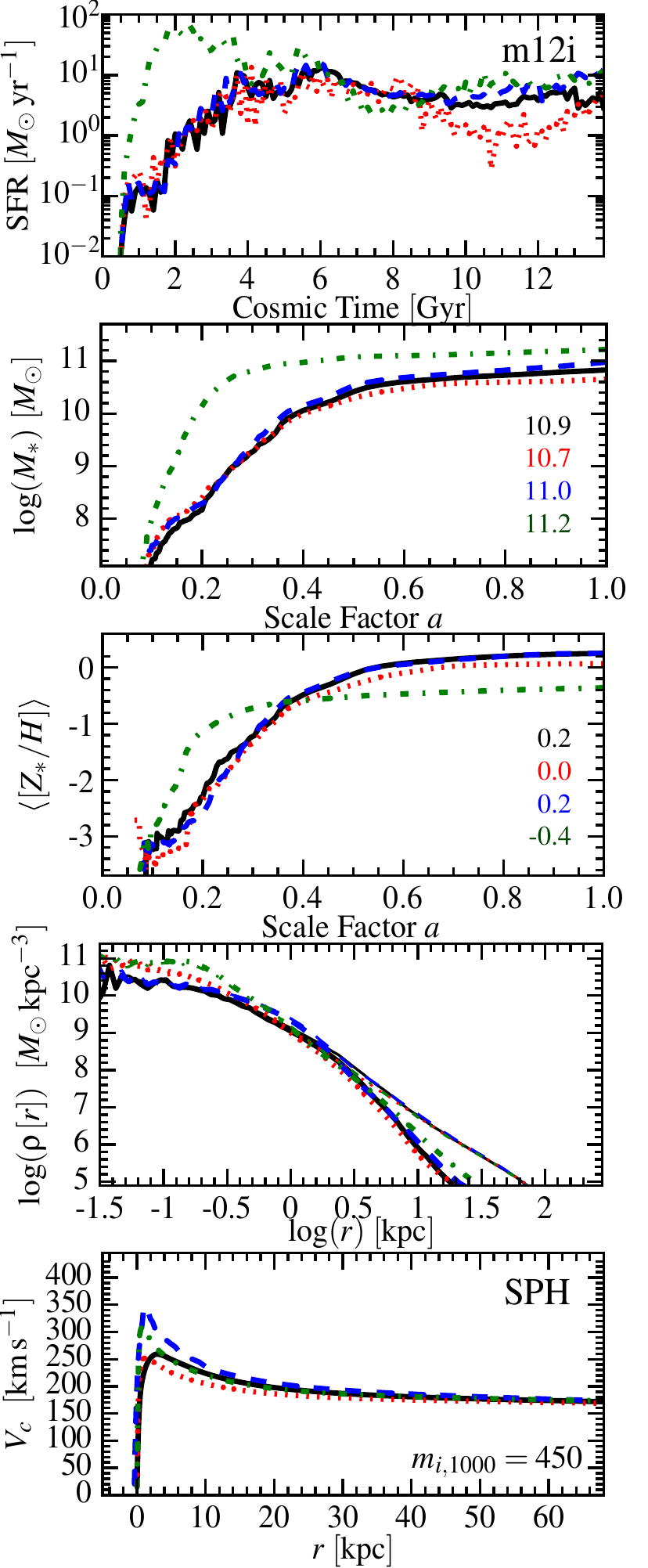}
\end{tabular}
    \vspace{-0.25cm}
    \caption{Effects of different stellar feedback physics on cosmological galaxy formation histories, as \demofigcosmo. 
    We compare a dwarf ({\bf m10q}) and MW-mass galaxy ({\bf m12i}), both run with our default, higher-order accurate hydrodynamic solver (MFM; {\em left}); but we also show a set of simulations of {\bf m12i} with the same physics variations using the smoothed-particle hydrodynamics (SPH; {\em right}) method, to demonstrate that even though the hydrodynamic solvers produce differences in massive galaxies, the qualitative effects of different feedback mechanisms are identical (independent of the hydrodynamic method). 
    {\bf (1)} {\em Default:} Our default models include all standard stellar evolution processes.
    {\bf (2)} {\em No OB/AGB mass-loss:} Removing continuous stellar mass loss (both OB and AGB-star winds) produces slightly lower metallicities (owing to the lack of recycling) and significantly lower late-time SFRs -- it appears the primary role of stellar mass-loss is to provide an additional source of gas fueling late-time SF in both dwarfs and MW-mass systems. 
    {\bf (3)} {\em No Radiative feedback:} This removes all radiative feedback (radiation pressure as well as photo-ionization and photo-electric heating by local particles and the meta-galactic background). In dwarfs (even with $V_{\rm max}\sim 40\,{\rm km\,s^{-1}}$, shown here), removing the photo-ionization heating (dominated by the UVB) produces $\sim 10\times$ larger SFRs and stellar masses (producing large bursts that make a core and lower $V_{c}$). In massive galaxies, the effects are weaker but removing radiation pressure produces significantly {\em higher} central densities (more strongly-peaked rotation curves in the central $\sim 5\,$kpc).
     {\bf (4)} {\em No Supernovae:} SNe clearly dominate on cosmological scales, as removing them produces orders-of-magnitude higher SFRs at early times, giving rise to runaway collapse to extremely high densities until the gas is depleted.
    \label{fig:sf.history.fb.mechanisms}}
\end{figure*}

\begin{figure}
\plotonesize{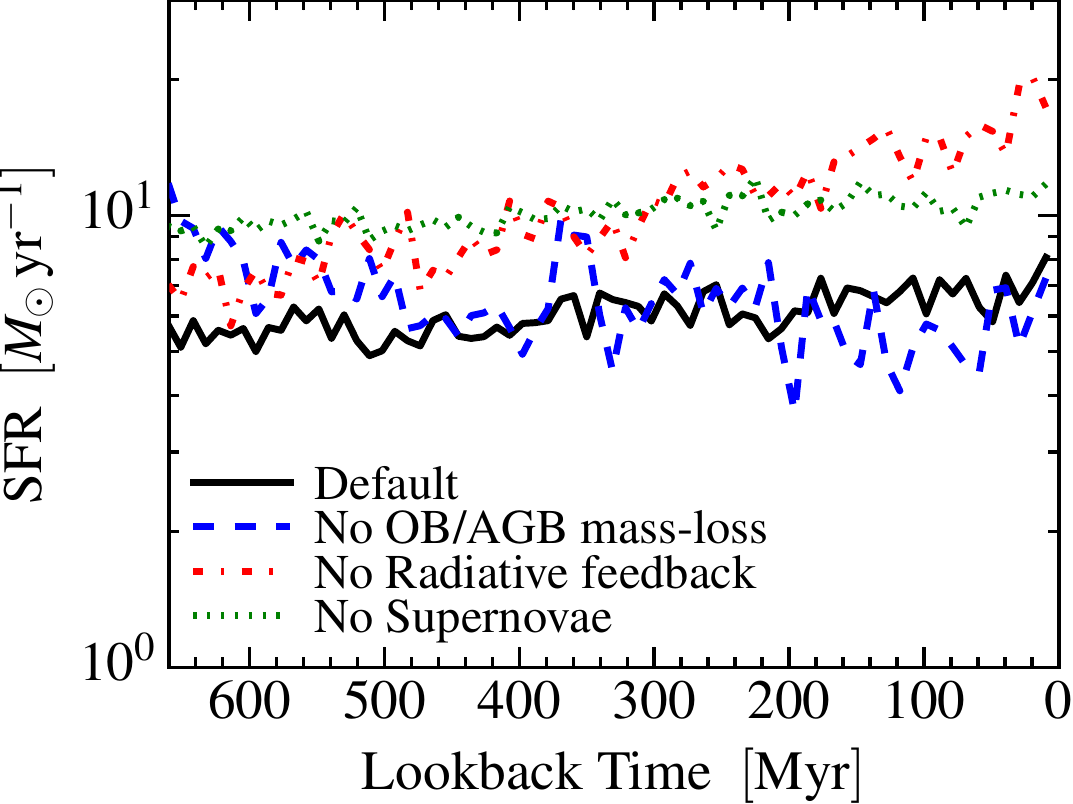}{0.99}
    \vspace{-0.25cm}
    \caption{Effects of different stellar feedback physics on a re-start of an identical MW-mass simulation ({\bf m12i}; as Fig.~\ref{fig:sf.fb.strength.restart}); we disable different mechanisms in turn as in Fig.~\ref{fig:sf.history.fb.mechanisms}. {\bf (1)} {\em Default:} All physics included. {\bf (2)} {\em No OB/AGB mass-loss:} Since the simulations start from identical initial conditions (identical gas supply), removing stellar mass-loss initially leads to a higher SFR (the OB winds cannot act as a source of feedback) -- but by $\sim 500\,$Myr later, the net effect of removing the winds is to slightly decrease the SFR, as the winds cannot act continuously over this time to supply new gas mass to the disk. {\bf (3)} {\em No (Local) Radiative feedback:} Here we still include a UV background, but remove only local (star particle) radiative feedback. We see a steady rise in SFR until this is forming more stars than the no-SNe case! Two effects occur: (a) GMCs are no longer efficiently disrupted by radiative feedback before SNe explode, therefore the SFR needed to self-regulate within a disk of a given surface density rises (see e.g.\ \citealt{hopkins:rad.pressure.sf.fb}), and (b) the outer atomic disk, which was stabilized by photo-heating from young stars in the galaxy, can now cool efficiently and form stars. {\bf (4)} {\em No Supernovae:} The SFR is systematically larger, as expected, but by a relatively modest factor $\sim 2$. This is because other feedback mechanisms instantaneously can regulate the SFR of gas in the galactic disk. However, removing SNe eliminates almost all galactic winds -- on cosmological timescales, this means the disk will increase in gas mass and the SFR will run away, as we saw in Fig.~\ref{fig:sf.history.fb.mechanisms}. Critically, this demonstrates that how feedback regulates SF {\em instantaneously within the disk} (e.g.\ the Kennicutt-Schmidt relation) is {\em not} the same as how feedback regulates SF {\em integrated over cosmological growth timescales} (e.g.\ the stellar mass-halo mass relation). 
    \label{fig:sf.fb.mechanisms.restart}}
\end{figure}

Many previous studies have argued that feedback is the most important determinant of star formation in galaxies (see references in \S~\ref{sec:intro}). In this section we therefore explore how basic feedback physics changes our predictions; in \paperone-\papertwo, we examine more subtle physical and numerical aspects of the {\em implementation} of said feedback.

\vspace{-0.5cm}
\subsection{Feedback ``Strength''}
\label{sec:feedback:strength}

In Fig.~\ref{fig:sf.fb.strength.restart}, we simply re-start our {\bf m12i} simulation near $z=0$ as \demofigrestart, but arbitrarily multiply {\em all} feedback rates (e.g.\ SNe rates, wind kinetic luminosities and mass loss rates, stellar luminosities) by a factor of $\sim 3$ (``strong'') or $\sim 1/3$ (``weak''), relative to the ``default'' values (which, recall, are taken directly from stellar evolution models without adjustment). We stress that there is no physical motivation for this -- this is a larger shift than most uncertainties in stellar evolution -- but we consider it purely for illustrative purposes. Exactly as expected in ``feedback-regulated'' scenarios, this produces a corresponding direct shift in the SFR -- the SFR is self-regulating at the level where feedback offsets gravitational collapse, so if feedback is $3\times$ stronger ``per star,'' then $3\times$ fewer stars form when the equilibrium is realized. This is almost exactly what we see (the shift in SFR is very slightly sub-linear, owing to non-linear effects). This is consistent with many previous simulation and analytic studies \citep[see][and references therein]{shetty:2008.sf.feedback.model,kim:2011.sf.selfreg.disks,hopkins:rad.pressure.sf.fb,hopkins:fb.ism.prop,cafg:sf.fb.reg.kslaw,orr:ks.law}.

\vspace{-0.5cm}
\subsection{The Role of Different Feedback Mechanisms: Galactic Winds \&\ Cosmological Timescales}
\label{sec:feedback:cosmo}

In Fig.~\ref{fig:sf.history.fb.mechanisms} we consider how turning off each feedback mechanism in turn alters galaxy evolution in fully cosmological simulations. In both dwarf and MW-mass galaxies, we clearly see SNe have the most dramatic effect. This is because what ultimately regulates galactic star formation efficiencies over a Hubble time is the competition between inflow and galactic outflows, and the high-speed outflows that are not simply recycled quickly are predominantly SNe-driven \citep[e.g.][and references therein]{dave:2011.outflow.metal.fgas}. Without SNe, galaxies form far too many stars (factor $>100$ in our {\bf m10q} dwarf; factor $\sim 2$ in the MW-mass system, going from $\sim 30\%$ to $\sim 60\%$ baryon-to-star ``conversion'' efficiency); the stars form too early (everything collapses rapidly into early dwarf halos, and has mostly turned into stars by redshift $z\sim 3-4$); the resulting galaxy is too dense (rotation curves peak strongly, even in small dwarfs, and are never ``gently rising''); and the metallicities are essentially constant across stellar mass because galaxies have similar formation histories and baryon conversion efficiencies (late-time metallicities are lower for massive galaxies because there is no SNe enrichment, only RSG/AGB enrichment).\footnote{Note the surprisingly lower $V_{c}$ for the no-SNe, MFM run owes to very early gas exhaustion leading to the galaxy growing only via ``dry'' gas-poor mergers at late times, which puff up the central dispersion.} Likewise properties of the ISM (gas phases) and CGM (covering factors of outflows and metals, especially) are grossly discrepant with observations. We also note that because all of our galaxies are star-forming, the SNe Type II rate dominates over the Ia rate at all times. All of these results are consistent with many previous studies (see references in \S~\ref{sec:intro}). We discuss the effects of SNe in much greater detail in \paperone\ -- our purpose here is only to confirm they are indeed critical.

Turning off continuous stellar mass-loss (OB/AGB winds), the most obvious effect is that the late-time SFRs of both dwarfs and MW-mass galaxies are {\em suppressed}, relative to their ``default'' values. Despite the fact that OB winds carry (roughly) comparable energy and momentum flux to SNe, and AGB winds can do the same if the relative star-gas velocity is sufficiently large (see Appendix~\ref{sec:mechanical.fb.implementation}), their {\em net} effect seems to be increasing the supply of gas mass within the galaxy which can eventually cool and form stars at late times. This is also consistent with previous studies, many of which have noted that OB winds tend to not be sufficiently well-confined to build up comparable momentum (after coupling) to SNe \citep{jungwiert:stellar.mass.loss.model,ciottiostriker:cooling.flow.selfreg.2,ciotti:recycling.with.feedback.rad.vs.mech,schaye:2010.cosmo.sfh.sims,leitner:2011.stellar.mass.loss.fueling.late.sf,novak:2011.bh.feedback.cycles,gan:2014.mixed.feedback.models.isolated.elliptical,choi:agn.fb.massive.elliptical.bal.winds}.

Removing all radiative feedback -- including the effects of the meta-galactic UV background -- has a dramatic effect on dwarf stellar masses, even at halo masses $\gtrsim 10^{10}\,\msun$ with $V_{\rm max}\gtrsim 40\,{\rm km\,s^{-1}}$ that should be well above the threshold for UVB ``quenching'' star formation. This will be explored in more detail in future work (Wheeler et al., in prep.). In short, the stellar masses increase by a factor $\sim 10$ \citep[consistent with our FIRE-1 results;][]{hopkins:2013.fire}. Unlike removing SNe, the qualitative shape of the SFH is similar to the ``full feedback'' case -- the stars do not form uniformly early, and the SFH is still ``bursty,'' but the bursts reach much larger amplitude, owing to the much larger gas supply. Removing only {\em local} radiative feedback (keeping the UVB, but removing radiative feedback from explicit star particles) has more subtle effects: the SF becomes much more ``bursty,'' the stellar mass changes but by a smaller factor, and the metallicity is suppressed; we discuss these in detail in \papertwo. In MW-mass halos, the UVB has weak effects, but removing local radiative feedback produces a clear, significant increase in the central $V_{c}$ (perhaps expected, as the galaxy center is exactly the region where we expect massive, dense GMCs which must be pre-processed by radiative feedback in order to reach low densities where subsequent SNe can efficiently expel the material, see \citealt{thompson:rad.pressure,murray:momentum.winds,hopkins:rad.pressure.sf.fb}). The dependence is similar in FIRE-1, although there we saw a stronger dependence of stellar mass on radiative feedback; the difference owes to the improvements in the treatment of SNe which allow, in massive GMCs, for the first SNe to play a similar role to radiative feedback and ``pre-process'' the cloud for subsequent SNe. 

As Fig.~\ref{fig:sf.history.fb.mechanisms} clearly demonstrates, our conclusions about the systematic effects of different feedback mechanisms are independent of the hydrodynamic method (MFM vs.\ SPH), even where the methods themselves produce significant systematic differences. Likewise, we have checked the conclusions above are independent of resolution, star formation prescription, and details of the cooling functions.

\vspace{-0.5cm}
\subsection{The Role of Different Feedback Mechanisms: Self-Regulation Within Galaxies and the Kennicutt-Schmidt Relation}
\label{sec:feedback:restarts}

In Fig.~\ref{fig:sf.fb.mechanisms.restart} we repeat the experiment from Fig.~\ref{fig:sf.history.fb.mechanisms} turning on and off different feedback mechanisms, but in restarts of the $z\sim0$ MW-mass galaxy as \demofigrestart. This allows us to separate non-linear, long-timescale cosmological effects on the SFR (e.g.\ galactic wind generation and recycling), from the {\em instantaneous} self-regulation of star formation {\em within} a galaxy.

As expected, removing SNe produces a systematically higher SFR. Removing OB/AGB winds leads, initially, to higher SFRs, as the additional feedback (gas heating via shocked winds) is no longer present (but the gas mass of the disk is still essentially fixed at its initial value); however the SFR then declines as the missing mass-loss is unable to ``re-supply'' gas lost to star formation. Most interesting, without radiative feedback, the SFR systematically rises with time, comparable to or even exceeding the SFR of the no-SNe test! This is consistent with a number of studies of isolated (non-cosmological) galaxy simulations \citep[see][]{hopkins:rad.pressure.sf.fb,hopkins:fb.ism.prop,kim:disk.self.reg,agertz:2013.new.stellar.fb.model,benincasa:2016.selfreg.sf.disk.modes}.

Why does the SFR appear more sensitive to radiative feedback in Fig.~\ref{fig:sf.fb.mechanisms.restart} and these isolated galaxy simulations, as opposed to Fig.~\ref{fig:sf.history.fb.mechanisms}? On cosmological spatial and temporal scales, a simple steady-state argument (e.g., \citealt{2010ApJ...718.1001B,2012MNRAS.421...98D,2013MNRAS.433.1910F,2013ApJ...772..119L,2015MNRAS.449.3274F,2017MNRAS.464.2766M}) implies the sum of the time-average galactic SFR $\langle \dot{M}_{\ast} \rangle $ and superwind mass-loss rates $\langle \dot{M}_{\rm wind} \rangle \equiv \eta\,\langle \dot{M}_{\ast} \rangle$ (where $\eta$ is the wind mass-loading, which can depend on arbitrary galaxy properties) should equal the inflow rate $\dot{M}_{\rm in}$ (we assume gas mass in the disk is steady-state; see \citealt{dave:2011.outflow.metal.fgas}). Thus $\langle \dot{M}_{\ast} \rangle \sim (1+\eta)^{-1}\,\dot{M}_{\rm in}$. However, on scales {\em within} the galactic disk (and timescales of order the galaxy dynamical time, much shorter than the Hubble time), the SFR is set by self-regulation via feedback, at the point where the momentum flux per unit area from feedback, $\sim (p_{\ast}/m_{\ast})\,\dot{\Sigma}_{\ast}$ (where $p_{\ast}/m_{\ast}\sim 3000\,{\rm km\,s^{-1}}$ is the time and IMF-averaged flux of momentum {\em into the dense, star forming gas} per unit stellar mass formed) offsets the gravitational force per unit area $\sim G\,\Sigma_{\rm disk}\,\Sigma_{\rm gas}$ \citep[an identical scaling is obtained assuming feedback must destroy star-forming clouds, and/or offset turbulent dissipation, in a $Q\sim 1$ disk; see][]{shetty:2008.sf.feedback.model,kim:2011.sf.selfreg.disks,cafg:sf.fb.reg.kslaw,hayward.2015:stellar.feedback.analytic.model.winds}. Thus $\langle \dot{M}_{\ast} \rangle \sim (p_{\ast}/m_{\ast})^{-1}\,G\,\Sigma_{\rm disk}\,\langle M_{\rm gas} \rangle$. Of course, these two scalings must be consistent: if the ``instantaneous'' $\dot{M}_{\ast}$ is ``too low'' compared to the cosmological scaling above, then $\dot{M}_{\rm in}$ is not offset by star formation+outflow and gas ``piles up,'' raising $M_{\rm gas}$ until the SFR ``catches up.'' If the radiative feedback contributes significantly to $p_{\ast}/m_{\ast}$ (either directly or via pre-processing GMCs before SNe), then removing it lowers $p_{\ast}/m_{\ast}$ and in turn increases the SFR given {\em fixed} $\Sigma_{\rm disk}$ and $M_{\rm gas}$, as in Fig.~\ref{fig:sf.fb.mechanisms.restart}. However, if $\eta$ is less sensitive to radiative feedback, the winds from higher $\dot{M}_{\ast}$ would be stronger, and so the galaxy will self-regulate at the same $\langle \dot{M}_{\ast} \rangle$. Thus, the instantaneous self-regulation (the Kennicutt-Schmidt relation) is not the same as self-regulation of cosmological growth (e.g.\ the stellar mass-halo mass relation).


\vspace{-0.5cm}
\section{Summary \&\ Conclusions}
\label{sec:discussion}

\subsection{Overview}

We present the FIRE-2 simulations (Table~\ref{tbl:sims}), a suite of cosmological simulations of galaxy formation using the FIRE physics modules in the {\small GIZMO} code. The FIRE-2 suite represents an update over FIRE-1, primarily in the use of a newer, more accurate hydrodynamic method, together with other numerical improvements to the physics algorithms and resolution -- but no significant change in the actual {\em physics} simulated (\S~\ref{sec:methods}). This includes high-resolution simulations run to $z=0$ with full baryonic physics of the cold, multi-phase ISM, at both the dwarf and MW mass scales (reaching sub-pc spatial and $\sim 100\,\msun$ mass scales; see Table~\ref{tbl:res} for details). 

In all properties investigated here, our primary conclusions from FIRE-1 appear qualitatively robust to these improvements in numerical accuracy (see \S~\ref{sec:results:overview}, Figs.~\ref{fig:fire1vs2}-\ref{fig:mgal.mhalo}), with typically less than factor of a few changes in global galaxy properties such as stellar mass, metallicity, and star formation rates. With {\em explicit} treatment of stellar feedback from SNe Types Ia \&\ II, OB \&\ AGB stellar winds, radiation pressure (UV, optical, and IR), photo-ionization and photo-electric heating, and explicit resolution of the multi-phase ISM, galactic winds emerge naturally, producing galaxies with morphologies (Figs.~\ref{fig:images.m12} \&\ \ref{fig:images.dwarfs}), internal ISM structure (Fig.~\ref{fig:images.fisheye.m12f}), flat or rising star formation histories and flat rotation curves (Figs.~\ref{fig:demo}-\ref{fig:fire1vs2}), metallicities (Fig.~\ref{fig:demo}), and stellar masses (Fig.~\ref{fig:mgal.mhalo}) apparently consistent with observations. Of course, it is impossible to consider an exhaustive list of galaxy properties here, and many more remain to be investigated in future work. Still, this is remarkable, considering that there is no fine-tuning or direct calibration of any parameters in the simulations to match these observations. Each of these properties has been investigated in more detail in previous work, described in \S~\ref{sec:intro}.

It is worth noting that, as previous papers have pointed out \citep{hopkins:2013.fire,ma:2016.disk.structure,wetzel.2016:latte}, there is no tension between even very thin stellar disks at $z=0$ (e.g.\ Fig.~\ref{fig:images.m12}), and strong stellar feedback which drives bursty star formation and strong galactic outflows at high redshift, strongly suppressing the stellar masses at $z=0$ relative to no-feedback models. This owes to a combination of resolved venting of hot winds through a multi-phase ISM, which remove mass without disturbing galaxy morphologies, and a rapid transition from the bursty star-forming mode and more ``gentle'' steady-state thin-disk mode as gas fractions and specific star formation rates decline in massive, low-redshift halos \citep[see][]{muratov:2015.fire.winds,hayward.2015:stellar.feedback.analytic.model.winds,cafg:bursty.sf.toymodel, ma:2016.disk.structure}.

We have provided a complete algorithmic description of all aspects of the FIRE simulations (see \S~\ref{sec:methods} \&\ Appendices~\ref{sec:stellar.evolution.approximations}-\ref{sec:additional.physics.methods}). We also make our hydrodynamics+gravity code {\small GIZMO}, and all initial conditions for the runs here, public.

We have considered an extensive study varying numerical and physical aspects of our simulations, to identify the most important ``ingredients'' and physical effects in the simulations. Table~\ref{tbl:summary} summarizes our conclusions; below we discuss the most important.

\vspace{-0.5cm}
\subsection{Resolution}

\begin{enumerate}

\item Mass resolution is most important for the physics that can be captured in Lagrangian methods; in \S~\ref{sec:resolution:mass} we present a number of new resolution criteria relevant for simulations of resolved galaxy formation, phase structure, and galactic winds. For example, for dwarf galaxies, stellar masses are robust to resolution once the galaxies contain just a few star particles (Fig.~\ref{fig:res.summary}). Metallicity and the shape of the SFH converge more slowly, requiring $\sim 100$ star particles to sample the self-enrichment history and a baryonic particle mass $\lesssim 10^{-6}\,M_{\rm halo}$ (Eq.~\ref{eqn:mbursty}) to avoid numerically enhanced burstiness (from single star particles representing ``too many'' SNe at once). Accurately capturing phase structure in the ISM and gravitational fragmentation requires resolving the Toomre mass in the ISM, i.e.\ the largest GMCs, which dominate the galactic SFR (Eq.~\ref{eqn:mtoomre}; Fig.~\ref{fig:clumps.res}). Morphology is very robust to resolution for dwarfs; for massive galaxies, it appears to depend on galaxy formation history -- some massive systems are disky (with similar sizes and rotation curves) at all resolution levels, others (where the disk forms somewhat later) are more sensitive (Fig.~\ref{fig:images.resolution}). Details of galactic winds, in particular hot gas ``venting'' and recycling in the CGM, are most sensitive to resolution of the properties we consider, with true ``convergence'' likely requiring at least the ability to resolve the cooling radii of individual SNe (Eq.~\ref{eqn:mcool.sne}). 

\item Force softenings for {\em collisionless} (DM+stellar) particles have little effect on our predictions (including galaxy baryonic properties, DM halo mass profiles, sub-structure mass and velocity distribution functions), provided they are chosen within a broad range (\S~\ref{sec:resolution:spatial}, Figs.~\ref{fig:spatial.res.history}-\ref{fig:dm.force.resolution.substructure}). For DM this is between $\sim (0.005 - 0.5)\,r_{0.06}$, where $r_{0.06}$ is approximately the radius containing $\sim 200$ DM particles at the halo center at $z=0$ (\S~\ref{sec:resolution:spatial:optimal}-\ref{sec:resolution:spatial:gas.tests}). Our ``default'' DM and stellar softening choices maximize the integration accuracy and convergence of our runs; larger softenings can artificially suppress central densities of DM halos.

\item We show that the radius $r_{0.06}$ is the radius to which DM profiles are converged to $\sim10\%$ or better -- this is smaller than the commonly quoted ``convergence radius'' of \citet{power:2003.nfw.models.convergence}, owing to more accurate integration and timestepping, and smoother kernels (Figs.~\ref{fig:dm.mass.resolution}-\ref{fig:dm.plus.baryons.mass.resolution}). At these spatial and mass resolution scales, $N$-body heating is negligible (Eq.~\ref{eqn:nbody.heat.gas}-\ref{eqn:nbody.heat.dm}). Owing to this insensitivity, it makes little difference whether we adopt constant softenings or fully-adaptive softenings (set to a fixed multiple of the local inter-particle separation) for collisionless particles.

\item Force softenings for {\em gas} should be set adaptively, so that the {\em same} mass distribution is being treated in the hydrodynamic equations and gravity equations (\S~\ref{sec:resolution:spatial}). With such a choice, turbulent fragmentation can be resolved down to small spatial scales, and no additional resolution criteria need be applied; of course, the meaning of ``spatial resolution'' is adaptive, so Table~\ref{tbl:res} summarizes the effective spatial resolution of our simulations in different regimes. In dense, star-forming gas, the effective resolution reaches $\sim1-10$\,pc in MW-mass systems and $\sim0.1-1$\,pc in dwarfs. Because the medium is super-sonically turbulent, accurately capturing turbulent fragmentation and galactic star formation does {\em not} require that the {\em thermal} Jeans mass/length be resolved.

\item If, for some reason, constant gravitational force softening for gas is desired in simulations with a resolved multi-phase ISM, we show (Figs.~\ref{fig:spatial.res.history}-\ref{fig:sf.z0.spatial.res}) that it should be chosen sufficiently small that (1) the {\em Toomre} length scale (molecular gas disk scale height) is resolved ($\epsilon \ll 100\,$pc), and (2) the minimum ``threshold'' density for star formation is resolved {\em gravitationally}, so that actual self-gravitating regions are what form stars ($\epsilon \lesssim 7\,{\rm pc}\,m_{i,\,1000}^{1/3}\,(n_{\rm SF,\,min}/100\,{\rm cm^{-3}})^{-1/3}$). If the latter criterion is violated then star formation is driven (incorrectly) entirely by {\em global} (galaxy-wide) contraction or shock-compression of gas, and has nothing to do with local fragmentation and collapse of GMCs. 

\item In addition to standard timestep criteria, guaranteeing proper time-resolution of feedback requires a limiter (Eq.~\ref{eqn:stellar.dt.limiter}) such that star particles cannot ``skip'' stages of stellar evolution or experience $\gg1$ SNe per particle per timestep (\S~\ref{sec:resolution:time}). 

\item Moreover, if one adopts adaptive gravitational softening for collisionless particles, we show that standard timesteps based on the acceleration are numerically unstable. The reason is that the accelerations and potential now depend explicitly on the local particle configuration which changes as particles move through one another, not just on the total potential gradient; this means a Courant-like condition {\em must} be used to ensure stability (Fig.~\ref{fig:ags.timesteps}; Eq.~\ref{eqn:dt.ags}). 

\end{enumerate}

\vspace{-0.5cm}
\subsection{Hydrodynamic Methods}

\begin{enumerate} 

\item Comparing our new mesh-free finite-volume Godunov hydrodynamic methods (``MFM'') to SPH (\S~\ref{sec:hydro}), we show that at all resolution levels, the properties of dwarf and high-redshift galaxies are insensitive to the details of the hydrodynamic solver (Fig.~\ref{fig:sph.vs.time}). However, for MW-mass and larger galaxies at late times, when a ``hot halo'' develops, the details of the hydrodynamics become important for the late-time steady-state SFR of the galaxies, and their gas disk sizes (and subsequent circular velocity profiles). Similar conclusions have been reached in previous comparisons of moving-mesh and fixed-grid codes to SPH, but with simpler treatments of ISM cooling/phases and feedback \citep{torrey:2011.arepo.disks,sijacki:2011.gadget.arepo.hydro.tests,keres:2011.arepo.gadget.disk.angmom,vogelsberger:2011.arepo.vs.gadget.cosmo}. We show that the gross morphology and instantaneous SFR given some ISM gas supply are not particularly sensitive (Figs.~\ref{fig:sph.vs.sfr}-\ref{fig:image.morph.sph}). Instead, the difference owes to the mixing and recycling of galactic winds in the CGM and initial accretion and shocking of hot gas, and its subsequent non-linear effect on cooling and re-accretion (Figs.~\ref{fig:image.sph.cgm}-\ref{fig:temperature.density.sph.phase.diagram}). SPH -- even in our state-of-the-art implementation -- can produce some spurious non-mixing ``blobs''  in the CGM, smear out accretion shocks, and suppress sub-sonic turbulence, leading to more cold and less hot, dense gas in the CGM. 
We explicitly show that by adding stronger ``artificial conductivity'' to our SPH runs, we can essentially reproduce our MFM results (Fig.~\ref{fig:sph.vs.time.added.diffusion}), indicating that details of numerical diffusion are important in the CGM of ``hot halos.'' 

\item We note that it is not appropriate to use ``artificial pressure floors'' when a sink-particle model (like ours) for star formation (which identifies self-gravitating regions to turn into stars) is already present (\S~\ref{sec:hydro:artificial.pressure}). However, for modest  values often used in the literature, we show such a floor has weak effects -- but making the pressure floors too large produces clear numerical artifacts and prevents proper physical convergence in treating turbulent fragmentation (Fig.~\ref{fig:sf.z0.cooling}). 

\item In finite-mass numerical methods without explicit advective fluxes (MFM), passive scalars (e.g.\ metals) remain locked (by default) to gas elements where they are first injected; in methods with such fluxes (e.g.\ AMR) there is a significant numerical mixing between elements. At finite resolution, these tend to under and over-estimate (respectively) true physical mixing by un-resolved turbulent eddies and microphysical diffusion between neighbor cells. For AMR codes reducing the error requires higher resolution; in MFM one can add an explicit metal diffusivity. However, while this can be important for predictions of detailed abundance pattern distributions within galaxies, we show that this has essentially no effect on any properties studied here (\S~\ref{sec:turbulent.diffusion.tests}; Fig~\ref{fig:sf.metaldiff}-\ref{fig:phase.metaldiff}), regardless of the numerical implementation (Fig.~\ref{fig:sf.metaldiff.coeff}).

\item Given the sensitivity of the CGM in hot halos to fluid mixing details, it is likely that additional physics -- magnetic fields, anisotropic Spitzer-Braginskii conduction and viscosity, cosmic ray transport -- may produce larger changes in the late-time cooling rates from hot halos than the difference between hydro solvers.

\end{enumerate}

\vspace{-0.5cm}
\subsection{Cooling, Star Formation, \&\ Feedback}

\begin{enumerate} 

\item Consistent with previous work \citep{piontek:feedback.vs.disk.form.sims,hopkins:rad.pressure.sf.fb,hopkins:fb.ism.prop,glover:2011.molecules.not.needed.for.sf,hu:photoelectric.heating}, we find that most {\em details} of radiative cooling and gas chemistry generally have little effect on galaxy dynamics, star formation, and galactic winds (\S~\ref{sec:cooling}; Fig.~\ref{fig:sf.z0.cooling}-\ref{fig:sf.history.cooling}). This is because almost all gas within the ISM has cooling times much shorter than dynamical times, so the exact cooling time and/or temperature is dynamically irrelevant. This is especially true in the ``cold'' and ``cool'' phases of the ISM/CGM/IGM ($T \lesssim 10^{5}\, $K). Detailed variations in yields, which species are tracked, and numerical metal-mixing produce correspondingly weak effects. 

\item However, in hot gas ($T\gg 10^{6}\,$K) in the CGM or SNe-heated bubbles, the cooling time can be longer than the dynamical time, and so high-temperature metal-line cooling has a significant effect on the phase structure of the CGM (Fig.~\ref{fig:phase.diagram.coolingphysics}) and cooling onto (hence SF in) the galaxy from the halo (Fig.~\ref{fig:sf.history.cooling}; see also \citealt{choi:2009.mf.vs.metalcooling,schaye:2010.cosmo.sfh.sims}). To leading order the total metallicity is the important quantity here, while detailed abundance-ratio variations have a much smaller effect.

\item As we have shown in previous work, provided the largest (Toomre) scales of fragmentation are resolved ({\em gravitationally} and {\em hydrodynamically}), and star formation occurs in that fragmenting gas, the resolution-scale star formation prescription has essentially no detectable effects on our predictions for galaxy-integrated SFRs, stellar masses, sizes, positions on the Schmidt-Kennicutt law (both galaxy-wide and spatially-resolved), galactic winds, and more \citep[see \S~\ref{sec:star.formation}, Figs.~\ref{fig:sf.threshold.zooms}-\ref{fig:restart.sfmodel}, and][]{saitoh:2008.highres.disks.high.sf.thold,shetty:2008.sf.feedback.model,hopkins:rad.pressure.sf.fb,hopkins:virial.sf,kim:2011.sf.selfreg.disks,agertz:2013.new.stellar.fb.model}. This includes orders-of-magnitude variations in the density threshold and resolution-scale ``rate per free-fall time.'' The key criteria are that these densities can (of course) be resolved, and that the SF occur only in self-gravitating gas above the mean galaxy density, so that the SF is automatically clustered (since the natural clustering of star formation non-linearly influences the likelihood of e.g.\ SNe bubbles overlapping and driving superwinds). Of course, predictions {\em within the dense gas} (corresponding to the densities where our sub-grid SF models take over) will be sensitive to the exact choices made; our default model is chosen based on previous work arguing it best reproduced observations of dense gas tracers (HCN and CO(3-2)) in idealized (non-cosmological) simulations \citep{hopkins:dense.gas.tracers}.

\item Provided the above criteria are met, SF is instead {\em feedback-regulated}. Uniformly increasing/decreasing the strength of feedback directly shifts both the integrated stellar mass, and position of galaxies on the KS law (Fig.~\ref{fig:sf.fb.strength.restart}). 

\item Stellar mass-loss (OB/AGB winds) is primarily important as a late-time fuel source for star formation (\S~\ref{sec:feedback}; Figs.~\ref{fig:sf.history.fb.mechanisms}-\ref{fig:sf.fb.mechanisms.restart}); its net effect is to {\em increase}, not decrease, late-time SFRs, especially in massive galaxies. 

\item Radiative feedback has a strong effect regulating the {\em instantaneous} SFRs of galaxies (i.e.\ the Kennicutt-Schmidt law), though less so on cosmologically-averaged SFRs (which are also regulated by the availability of fresh gas from IGM accretion). 

\item The most important form of radiative feedback for dwarfs is {\em external} radiative feedback, i.e.\ the UV background. In \papertwo, we show that removing the UVB has a much larger effect on the stellar masses of dwarfs than removing internal radiative feedback, even at mass scales as large as $M_{\ast}\gtrsim 10^{7}\,\msun$. We will explore these effects further in future work.

\item On cosmological scales, SNe (specifically core-collapse and prompt Ia) are the most important feedback mechanism regulating galaxy properties. Without SNe (even given other feedback mechanisms), galaxies (especially dwarfs) form too many stars, and form these stars far too early. In \paperone, we therefore investigate numerical SNe treatments in more detail.

\end{enumerate}

\vspace{-0.5cm}
\subsection{Summary of Ingredients}

To summarize, we find that the following criteria are essential for physically realistic high-resolution (multi-phase ISM) zoom-in simulations of galaxy formation:

\begin{enumerate}

\item The Toomre mass/length is mass and force-resolved; disk scale heights and at least the largest scales of fragmentation are resolved, with gravitational softening able to follow the mass. 

\item In addition to standard (optically thin, primordial) cooling, some accounting for self-shielding and low-temperature ($T\ll 10^{4}\,$K) cooling is included (to enable fragmentation), and high-temperature metal-line cooling is included to account for faster cooling in metal-enriched halo gas.

\item Star formation is restricted to gas at (force-resolved) densities significantly larger than the mean (ideally, to self-gravitating gas with a sink-particle type approach), so that it will occur in said fragments and therefore naturally be clustered, as observed.

\item Stellar mass loss is included, to provide a continuous additional fuel supply to the galaxy. 

\item Radiative feedback is included, particularly heating from the meta-galactic UV background, as well as photo-ionization heating and single-scattering photon momentum (radiation pressure) from stars in the simulation.

\item SNe (Ia \&\ II) are included, with a coupling algorithm that carefully ensures manifest conservation of energy, mass, and momentum. More importantly, a careful accounting of both the energy and momentum budget of the coupled terms that properly treats whether the coupled radii are inside or outside the cooling radius, is necessary to obtain converged solutions (see \paperone). 

\item All feedback quantities follow standard stellar evolution models for a standard IMF.

\end{enumerate}

\vspace{-0.5cm}
\subsection{Future Work}

In companion papers, we will study in more detail how mechanical feedback (SNe \&\ stellar mass-loss; \paperone) and radiative feedback (photo-heating \&\ photon momentum; \papertwo) influence our predictions -- both the details of their physics {\em and} numerical implementations. Because we argue that feedback is more important than many numerical details, it is extremely important to treat it as accurately as possible. 

As discussed above, more work is clearly warranted to investigate how additional fluid microphysics (e.g.\ magnetic fields) alters fluid mixing and subsequent cooling from the CGM. A major motivation of our switch to the new hydrodynamics solver in FIRE-2 is that it allows us to incorporate such physics in future work. A preliminary investigation of some of these physics is presented in \citet{su:2016.weak.mhd.cond.visc.turbdiff.fx}. However this was primarily focused on the ISM inside galaxies; detailed, higher-resolution CGM studies are clearly necessary.

In this paper, we study only systems with halo masses $\lesssim 2\times10^{12}\,\msun$. This is because it is widely believed that feedback from supermassive black holes (not included in our simulations here) is critical to explain the properties (especially the quenching of star formation, and further quiescence) of more massive ``red and dead'' systems \citep[see][]{croton:sam,hopkins:qso.all}. Our preliminary studies from FIRE-1 support the idea that stellar feedback alone cannot explain all the observed properties of the most massive galaxies \citep{hopkins:2013.fire,feldmann:colors.highz.quiescent.massivefire.gals}. In future work, we study the effects of black hole feedback on galaxy properties (for preliminary results see \citealt{daa:BHs.on.FIRE}). 

Finally, we focus here almost exclusively on numerical studies. A series of papers will present the scientific predictions of these simulations for current and future observations.

\vspace{-0.7cm}
\acknowledgments 
We thank our editor and referee, Matthieu Schaller, for a number of helpful suggestions. 
Support for PFH and co-authors was provided by an Alfred P. Sloan Research Fellowship, NASA ATP Grant NNX14AH35G, and NSF Collaborative Research Grant \#1715847 and CAREER grant \#1455342. 
ARW was supported by a Caltech-Carnegie Fellowship, in part through the Moore Center for Theoretical Cosmology and Physics at Caltech.
CAFG was supported by NSF through grants AST-1412836 and AST-1517491, and by NASA through grant NNX15AB22G. The Flatiron Institute is supported by the Simons Foundation. MBK and AF were partially supported by the NSF through grant AST-1517226. MBK also acknowledges support from NASA through HST theory grants (programs AR-12836, AR-13888, AR-13896, and AR-14282) awarded by STScI. DK was supported by NSF Grant AST1412153 and a Cottrell Scholar Award from the Research Corporation for Science Advancement. RF was supported by the Swiss National Science Foundation (grant No. 157591). 
Numerical calculations were run on the Caltech compute cluster ``Wheeler,'' allocations TG-AST130039 \&\ TG-AST150080 granted by the Extreme Science and Engineering Discovery Environment (XSEDE) and PRAC NSF.1713353 supported by the NSF, and the NASA HEC Program through the NAS Division at Ames Research Center and the NCCS at Goddard Space Flight Center. \\

\vspace{-0.2cm}
\bibliography{/Users/phopkins/Dropbox/Public/ms}

\begin{appendix}

\section{Approximate Stellar Evolution Tabulations}
\label{sec:stellar.evolution.approximations}

Here we present simple fits to the stellar evolution models and yields used in the FIRE simulations. We note that in several cases the simulations in this paper utilize a more detailed look-up table; however for all practical purposes the fits here are sufficiently accurate that the differences are negligible. Stellar evolution results are obtained from {\small STARBURST99} \citep{starburst99} assuming a \citet{kroupa:2001.imf.var} IMF. SNe Ia rates follow \citet{mannucci:2006.snIa.rates} including both prompt and delayed populations. Yields for core-collapse SNe are IMF-averaged for the same IMF, from the tables (including hypernovae) in \citet{nomoto2006:sne.yields}. Yields for SNe Ia follow \citet{iwamoto:1999.sneIa.yields}. Yields for OB/AGB winds are taken from the synthesis of the models from \citet{vandenhoek:1997.agb.yields}, \citet{marigo:2001.agb.yields}, and \citet{izzard:2004.agb.yields} as synthesized in \citet{wiersma:2009.enrichment}, appropriately re-averaged over the IMF. 

\begin{enumerate}

\item{\bf SNe Ia:} These occur with rate-per-unit-stellar mass $R_{\rm Ia} = d N_{\rm Ia}/dt = 0$ for $t_{\rm Myr}<37.53$ (where $t_{\rm Myr}$ is the age of the star particle in Myr), then $R_{\rm Ia}/({\rm SNe\,Myr^{-1}\,M_{\sun}^{-1}}) = 5.3\times10^{-8} + 1.6\times10^{-5}\,\exp{\{-[(t_{\rm Myr}-50)/10]^{2}/2\}}$ for $t_{\rm Myr}\ge37.53$. Since the rate is per-stellar-mass, the expectation value of the number of SNe for a star particle of mass $m_{b}$ and timestep $\Delta t_{b}$ is $=R_{\rm Ia}(t_{\rm Myr}^{b})\,m_{b}\,\Delta t_{b}$, and the trial for ``success'' (an explosion) is drawn from a binomial distribution. Each SN Ia has ejecta mass $M_{\rm ej} = 1.4\,\msun$ and energy $E_{\rm ej}=(1/2)\,M_{\rm ej}\,v_{\rm ej}^{2} = 10^{51}\,{\rm erg}$. The ejecta yield mass for the tracked species is: ({\small Z, He, C, N, O, Ne, Mg, Si, S, Ca, Fe}) $=(1.4,$ $0,$ $0.049,$ $1.2\times10^{-6},$ $0.143,$ $0.0045,$ $0.0086,$ $0.156,$ $0.087,$ $0.012,$ $0.743)\,\msun$.

\item{\bf SNe II:} The core-collapse SNe rate can be surprisingly well-fit by a simple piecewise-constant function, $R_{\rm II}=0$ for $t_{\rm Myr}<3.401$ or $t_{\rm Myr}>37.53$; $R_{\rm II}/({\rm SNe\,Myr^{-1}\,M_{\sun}^{-1}})=5.408\times10^{-4}$ for $3.4<t_{\rm Myr}<10.37$; and $R_{\rm II}/({\rm SNe\,Myr^{-1}\,M_{\sun}^{-1}})=2.516\times10^{-4}$ for $10.37<t_{\rm Myr}<37.53$. The IMF-averaged ejecta mass per explosion is $M_{\rm ej}=10.5\,\msun$, with ejecta energy $E_{\rm ej}=(1/2)\,M_{\rm ej}\,v_{\rm ej}^{2} = 10^{51}\,{\rm erg}$. These are normalized so that the ejecta mass and energy exactly match the integrated totals from {\small STARBURST99}; the time-averaged energy injection rate ($\langle R_{\rm II}\,E_{\rm ej} \rangle$) is within $\sim 10\%$ of the tabulated {\small STARBURST99} rate at all times. The IMF-averaged yields are ({\small He, C, N, O, Ne, Mg, Si, S, Ca, Fe}) $=(3.87,$ $0.133,$ $0.0479\,\tilde{N},$ $1.17,$ $0.30,$ $0.0987,$ $0.0933,$ $0.0397,$ $0.00458,$ $0.0741)\,\msun$, where $\tilde{N} = {\rm MAX}(Z/Z_{\sun},\,1.65)$ accounts for the strongly progenitor-metallicity dependent {\small N} yields. The total metal yield $Z$ is given by summing the explicitly-followed species, with an additional $\sim 2\%$ added to account for un-tracked species ($Z=1.02\,\sum\,Z_{i,\,{\rm followed}}$); the remaining ejecta is H.\footnote{For reference, the yields from \citet{woosley.weaver.1995:yields}, used in FIRE-1, are ({\small He, C, N, O, Ne, Mg, Si, S, Ca, Fe}) $=(4.03,$ $0.117,$ $0.0399\,\tilde{N},$ $1.06,$ $0.169,$ $0.0596,$ $0.0924,$ $0.0408,$ $0.00492,$ $0.0842)\,\msun$. In both FIRE-1 and FIRE-2, we choose to omit the progenitor stellar metallicity dependence of the predicted yields (using the yields for solar-metallicity progenitors instead) if the standard deviation of the metallicity dependence $|d M_{Z}({\rm species})/d Z_{\rm progenitor}|$ between the models of \citet{chieffi:2004.sne.yields}, \citet{woosley.weaver.1995:yields}, and \citet{nomoto2006:sne.yields} is larger than the magnitude of the actual predicted trend $| dM_{Z}({\rm species})/d Z_{\rm progenitor}|$ from \citealt{nomoto2006:sne.yields}. The only tracked species which passes this test is N. We do set the progenitor metallicity as a ``floor'' to the yields, following exactly the algorithm in \citet{wiersma:2009.enrichment}.}

\item{\bf OB/AGB Mass-Loss:} We include all non-SNe mass-loss channels here, but this is dominated by OB/AGB-star winds. The IMF-integrated mass-loss rate for a stellar population/particle of mass $M_{\ast}$ is $\dot{M}_{\rm w} = M_{\ast}\,f_{\rm w}\,{\rm Gyr^{-1}}$ with $f_{\rm w}=4.763\,(0.01 + Z/Z_{\sun})$ for $t_{\rm Myr}<1$; $f_{\rm w}=4.763\,(0.01 + Z/Z_{\sun})\,t_{\rm Myr}^{1.45+0.8\,\ln(Z/Z_{\sun})}$ for $1<t_{\rm Myr}<3.5$; $f_{\rm w}=29.4\,(t_{\rm Myr}/3.5)^{-3.25} + 0.0042$ for $3.5<t_{\rm Myr}<100$; and $f_{\rm w} = 0.42\,(t_{\rm Myr}/1000)^{-1.1} / (19.81 - \ln{(t_{\rm Myr})})$ for $t_{\rm Myr}>100$. The total (IMF-averaged) kinetic luminosity of the mass-loss is given by $L_{\rm kinetic} = (1/2)\,\dot{M}_{\rm w}\,\langle v_{\rm w}^{2} \rangle = \psi\times10^{12}\,{\rm erg\,g^{-1}}\,\dot{M}_{\rm w}$, with $\psi = (5.94\times 10^{4})/(1 + (t_{\rm Myr}/2.5)^{1.4} + (t_{\rm Myr}/10)^{5.0}) + 4.83$ for $t_{\rm Myr}<100$ and $\psi = 4.83$ for $t_{\rm Myr}>100$. The yields are given by the maximum of either the progenitor stellar surface abundances or, for the light species ({\small He, C, N, O}), mass fractions $=({0.36,\,0.016,\,0.0041\,\tilde{N},\,0.0118})$.

\item{\bf Radiation:} We define the light-to-mass ratio in a given band $\Psi_{\rm band}$, with units $L_{\sun}/\msun$. Then the bolometric $\Psi_{\rm bol} =1136.59$ for $t_{\rm Myr}<3.5$, and $\Psi_{\rm bol}=1500\,\exp{[-4.145\,x + 0.691\,x^{2} - 0.0576\,x^{3}]}$ with $x\equiv \log_{10}(t_{\rm Myr}/3.5)$ for $t_{\rm Myr} > 3.5$. For the bands used in our radiation hydrodynamics, we have the following {intrinsic} (before attenuation) bolometric corrections. In the mid/far IR, $\Psi_{\rm IR}=0$. In optical/NIR, $\Psi_{\rm opt}=f_{\rm opt}\,\Psi_{\rm bol}$ with $f_{\rm opt}=0.09$ for $t_{\rm Myr}<2.5$; $f_{\rm Opt}=0.09\,(1 + [(t_{\rm Myr}-2.5)/4]^{2})$ for $2.5 < t_{\rm Myr} < 6$; $f_{\rm Opt}=1-0.841/(1+[(t_{\rm Myr}-6)/300])$ for $t_{\rm Myr}>6$. For the photo-electric FUV band $\Psi_{\rm FUV} = 271\,[1+(t_{\rm Myr}/3.4)^{2}]$ for $t_{\rm Myr}<3.4$; $\Psi_{\rm FUV}=572\,(t_{\rm Myr}/3.4)^{-1.5}$ for $t_{\rm Myr}>3.4$. For the ionizing band $\Psi_{\rm ion} = 500$ for $t_{\rm Myr}<3.5$;  $\Psi_{\rm ion}=60\,(t_{\rm Myr}/3.5)^{-3.6} + 470\,(t_{\rm Myr}/3.5)^{0.045 - 1.82\,\ln{t_{\rm Myr}}}$ for $3.5<t_{\rm Myr}<25$; $\Psi_{\rm ion}=0$ for $t_{\rm Myr}>25$. The remaining UV luminosity, $\Psi_{\rm bol}-(\Psi_{\rm IR}+\Psi_{\rm opt}+\Psi_{\rm FUV}+\Psi_{\rm ion})$ is assigned to the NUV band $\Psi_{\rm NUV}$. The flux-mean dust opacities adopted are $(\kappa_{\rm FUV},$ $\kappa_{\rm NUV},$ $\kappa_{\rm opt},$ $\kappa_{\rm IR})$ $=(2000,\,1800,\,180,\,10)\,(Z/Z_{\sun})\,{\rm cm^{2}\,g^{-1}}$. The photo-ionization rate (and corresponding $\kappa_{\rm ion}$) is calculated from the neutral hydrogen density as described in Appendix~\ref{sec:radiative.fb.implementation} and Appendix~\ref{sec:cooling.approximations} below.

\end{enumerate}

\vspace{-0.5cm}
\section{Approximate Cooling Functions}
\label{sec:cooling.approximations}

Here we provide simple fitting-function approximations to the complete set of cooling functions used in our FIRE simulations. Note that for several of these, we use somewhat more accurate look-up tables in the simulations (as a function of temperature, density, and metallicity), but we provide functions accurate to $\sim10\%$ over the relevant dynamic range in the simulations ($\sim 10-10^{9}\,$K), so that interested readers can reproduce our full cooling physics treatment.

The instantaneous cooling+heating rate per unit volume is given by the sum over all processes, 
\begin{align}
\frac{d e_{\rm thermal}^{a}}{dt} = -n_{{\rm H},\,a}^{2}\,\Lambda_{\rm net}^{a} = -n_{{\rm H},\,a}^{2}\,\sum_{i}\,\Lambda_{i}
\end{align}
where $e_{\rm thermal}^{a}$ is the thermal energy density of a gas element $a$. Here we will use $\Lambda$ to denote both cooling and heating rates, but with opposite signs (a positive sign here denotes cooling). Also, define $\tilde{n}_{x} \equiv n_{x} / n_{\rm H}$ as the number of species $x$ per hydrogen nucleus (e.g.\ $\tilde{n}_{e}$, $\tilde{n}_{\Hn}$, and $\tilde{n}_{\Hep}$ denote the free electron, neutral hydrogen, and neutral helium numbers). Below, all units are cgs ($[T]={\rm K}$, $[n_{\rm H}]={\rm cm^{-3}}$, $[\Lambda]={\rm erg\,s^{-1}\,cm^{3}}$). The processes we track include:

\begin{enumerate}

\item{\bf Collisional Excitation}: from \citet{katz:treesph} (incorporating earlier fits from \citealt{cen:1992.cosmo.cooling.methods}):
\begin{align}
\Lambda_{\rm CE} =& \left( \beta_{\Hn}\,\tilde{n}_{\Hn} + \beta_{\Hep}\,\tilde{n}_{\Hep}  \right)\,\tilde{n}_{e} \\ 
\beta_{\Hn} &= 7.50\times10^{-19}\,\tau_{5}\,\exp{\left( - \frac{118348}{T}\right)} \\
\beta_{\Hep} &= 5.54\times10^{-17}\,\tau_{5}\,T^{-0.397}\,\exp{\left( - \frac{473638}{T}\right)} \\
\tau_{5} &\equiv \left[ 1 + \left( \frac{T}{10^5} \right)^{1/2} \right]^{-1} 
\end{align}

\item{\bf Collisional Ionization}: also from \citet{katz:treesph}: 
\begin{align}
\Lambda_{\rm CI} =& 10^{-11}\,\tilde{n}_{e} \times  \\ 
\nonumber & \left( 2.18\,\gamma_{\Hn}\,\tilde{n}_{\Hn} +  3.94\,\gamma_{\Hen}\,\tilde{n}_{\Hen} +  8.72\,\gamma_{\Hep}\,\tilde{n}_{\Hep}  \right) \\ 
\gamma_{\Hn} &= 5.85\times10^{-11}\,T^{1/2}\,\tau_{5}\,\exp{\left(-\frac{157809.1}{T}\right)} \\ 
\gamma_{\Hen} &= 2.38\times10^{-11}\,T^{1/2}\,\tau_{5}\,\exp{\left(-\frac{285335.4}{T}\right)} \\ 
\gamma_{\Hep} &= 5.68\times10^{-12}\,T^{1/2}\,\tau_{5}\,\exp{\left(-\frac{631515.0}{T}\right)} 
\end{align}

\item{\bf Recombination}: from \citet{verner.ferland:recombination.rates}:
\begin{align}
\Lambda_{\rm Rec} =& 1.036\times10^{-16}\,T\,\tilde{n}_{e}\, \times \\ 
\nonumber &\left(\alpha_{\Hp}\,\tilde{n}_{\Hp}+\left[ \alpha_{\Hep}+\frac{629922.78}{T}\,\alpha_{\rm di} \right]\,\tilde{n}_{\Hep}+\alpha_{\Hepp}\,\tilde{n}_{\Hepp}\right) \\
\alpha_{\Hp} &= 7.982\times10^{-11}\,\left(\frac{1.774}{T^{0.5}}\right) \times \\
& \left( 1 + \frac{T^{0.5}}{1.774} \right)^{-0.252}\left(1 + \frac{T^{0.5}}{838.81} \right)^{-1.748} \nonumber \\ 
\alpha_{\Hep} &= 9.356\times10^{-10}\,\left(\frac{0.2065}{T^{0.5}}\right) \times \\
& \left( 1 + \frac{T^{0.5}}{0.2065} \right)^{-0.2108}\left(1 + \frac{T^{0.5}}{6063.0} \right)^{-1.7892} \nonumber \\ 
\alpha_{\Hepp} &= 1.5964\times10^{-10}\,\left(\frac{2.5092}{T^{0.5}}\right) \times \\
& \left( 1 + \frac{T^{0.5}}{2.5092} \right)^{-0.252}\left(1 + \frac{T^{0.5}}{1677.6} \right)^{-1.748} \nonumber \\ 
\alpha_{\rm di} &= 1.9\times10^{-3}\,T^{1.5}\,\exp{\left(-\frac{470000}{T} \right)} \times \\ 
& \left(1 + 0.3\,\exp{\left[-\frac{94000}{T}\right]}\right) \nonumber
\end{align}
Note the $\alpha_{\rm di}$ term here comes from dielectric recombination.

\item{\bf Free-free emission}: from \citet{rybicki.lightman:1986.radiative.processes.book}:
\begin{align}
\Lambda_{\rm FF} =& \beta_{\rm ff}(T) \, \left(\tilde{n}_{\Hp} + \tilde{n}_{\Hep} + 4\,\tilde{n}_{\Hepp} \right)\,\tilde{n}_{e} \\ 
\beta_{\rm ff}(T) &= 1.43\times10^{-27}\,T^{1/2}\,\times \\ 
& \left[1.1+0.34\,\exp{\left\{ -(5.5 - \log_{10}[T])^{2}/3\right\}}\right] \nonumber 
\end{align}

\item{\bf High-Temperature Metal-Line Cooling}: this refers to metal-line cooling processes in mostly ionized gas, with temperatures $\gtrsim 10^{4}\,$K. We use the public look-up tables from \citet{wiersma:2009.coolingtables}, for which:
\begin{align}
\Lambda_{\rm Metal} =& \sum_{{\rm species}\ i}\,\Lambda_{\rm Metal}^{i} = \sum_{i}\,\tilde{n}_{e}\,\xi_{i}(n_{\rm H},\,T,\,z)\,\frac{Z^{i}}{Z_{\sun}^{i}}  
\end{align}
where this refers to the sum over all tracked metal species $i$ (here {\small C, N, O, Ne, Mg, Si, S, Ca, Fe}), and $Z^{i}/Z_{\sun}^{i}$ is the abundance of species $i$ relative to solar. We adopt solar abundances 
({\small Z, He, C, N, O, Ne, Mg, Si, S, Ca, Fe}) $=(0.02,$ $0.28$, $3.26\times10^{-3}$, $1.32\times10^{-3}$, $8.65\times10^{-3}$, $2.22\times10^{-3}$, $9.31\times10^{-4}$, $1.08\times10^{-3}$, $6.44\times10^{-4}$, $1.01\times10^{-4}$, $1.73\times10^{-3}$), which give abundance ratios matching \citet{asplund:2009.solar.composition} scaled to the total metallicity $=0.02$ (used because it matches the assumed values in the cooling computations and stellar evolution models). The functions $\xi_{i}$ depend on density, temperature, and redshift $z$ (because they assumes photo-ionization by a redshift-dependent UV background); they are taken from the look-up tables provided by \citet{wiersma:2009.coolingtables}, at abundances $Z=Z_{\sun}$ (defined above).\footnote{Available at \coolingurl}

\item{\bf Low-Temperature Metal Line, Fine-Structure, \&\ Molecular Cooling}: this combines the gas-phase low-temperature cooling (including molecular and atomic processes) in mostly neutral gas below $\lesssim 10^{4}\,$K. From our compilation of {\small CLOUDY} runs \citep{ferland:1998.cloudy}, fitting the resulting look-up tables, we obtain approximately:
\begin{align}
\Lambda_{\rm Cold} =& 2.896\times10^{-26}\,{\Bigl \{} 
\left( \frac{T}{125.215} \right)^{-4.9202} + \\
\nonumber & \left( \frac{T}{1349.86} \right)^{-1.7288} + 
\left( \frac{T}{6450.06} \right)^{-0.3075} 
{\Bigr \}}^{-1}\times \\ 
\nonumber & \left(\frac{1 + (Z/Z_{\sun})}{1+0.00143\,n_{\rm H}}\right)\,
\left(1 - f_{\rm selfshield} \right)\,
\times \\
\nonumber & \left(0.001 + \frac{0.10\,n_{\rm H}}{1+n_{\rm H}} + 
\frac{0.09\,n_{\rm H}}{1+0.1\,n_{\rm H}}
+ \frac{(Z/Z_{\sun})^{2}}{1+n_{\rm H}} \right) \times \\ 
\nonumber & \exp{\left(-\left[ \frac{T}{158000} \right]^{2} \right)}
\end{align}
where $f_{\rm selfshield}$ accounts for the local radiation environment by applying a simple (fitted) local shielding correction for UV/ionizing photons, $f_{\rm selfshield}\equiv\exp{(-\tilde{\tau}_{a}^{\rm ion})}$ with $\tilde{\tau}_{a}^{\rm ion} \equiv \sigma^{H}_{\nu_{0}}\,n_{{\rm H},\,a}\,\ell_{a}^{\rm fit}$ where $\sigma^{H}_{\nu_{0}}\equiv 6\times10^{-18}\,{\rm cm^{-2}}$ and $\ell_{a}^{\rm fit}\equiv4.4\,{\rm pc}\,(T/10^{4}\,K)^{-0.173}\,\Gamma_{-12}^{-2/3}$ ($\Gamma_{-12}$ is the ionization rate in units of $10^{-12}\,{\rm s^{-1}}$, including both the UV background and local sources assuming they have the same spectral shape, as defined below for photo-ionization heating).  

\item{\bf Dust Collisional Heating/Cooling}: from \citet{meijerink.spaans:xray.cooling.models}:
\begin{align}
\Lambda_{\rm Dust} =& 1.12\times10^{-32}\,\left(T-T_{\rm dust}\right)\,T^{1/2}\,\times\\
\nonumber & \left(1 - 0.8\,\exp{\left[-\frac{75}{T}\right]} \right)\,\left(\frac{Z}{Z_{\sun}}\right)
\end{align}
where we take $T_{\rm dust}=30\,$K here, and the $Z/Z_{\sun}$ term comes from assuming a constant dust-to-metals ratio.

\item{\bf Compton Heating/Cooling}: from the CMB, gives \citep{rybicki.lightman:1986.radiative.processes.book}:
\begin{align}
\Lambda_{\rm Compton} =& 5.65\times10^{-36}\,\tilde{n}_{e}\,(T - T_{\rm CMB}[z])\,(1+z)^{4}\,n_{\rm H}^{-1}
\end{align}

\item{\bf Photo-Ionization Heating}: from the UVB and local (in-simulation stellar sources) gives a heating rate, hence negative $\Lambda$, of 
\begin{align}
\Lambda_{\rm ion} =& -\tilde{f}\left( \epsilon_{\Hn}\,\tilde{n}_{\Hn} + \epsilon_{\Hen}\,\tilde{n}_{\Hen} + \epsilon_{\Hep}\,\tilde{n}_{\Hep}  \right)\,n_{\rm H}^{-1} \\ 
\tilde{f} &\equiv \left( 1 + \frac{e_{\nu,\,{\rm ion}}^{\rm local}}{e_{\nu,\,{\rm ion}}^{\rm UVB}} \right)\,f_{\rm selfshield} \\ 
\log_{10}(\epsilon_{\Hn}) &\approx -24.6 + 1.62\,x + 14.9\,x^{2} - 45.5\,x^{3} \\ 
\nonumber & + 46.2\,x^{4}\,-16.7\,x^{5} - \exp{[50\,(x-1.05)]} \\ 
\nonumber \log_{10}(\epsilon_{\Hen}) &\approx \log_{10}(\epsilon_{\Hn}) - 0.0366 + 0.376\,x \\ 
\nonumber \log_{10}(\epsilon_{\Hep}) &\approx -26.3 + 0.816\,x + 78.2\,x^{2} - 837\,x^{3} + 4770\,x^{4} \\
\nonumber &- 15600\,x^{5} + 29600\,x^{6} - 32400\,x^{7} + 18900\,x^{8}-4550\,x^{9}\\
 x &\equiv \log_{10}(1 + z)
\end{align}
where $\epsilon_{\Hn}$, $\epsilon_{\Hen}$, and $\epsilon_{\Hep}$ are pre-tabulated for the assumed UV background magnitude and shape, as a function of redshift in \citet{faucher-giguere:2009.ion.background}\footnote{See \uvburl} -- values above are simple polynomial fits good to $\sim 10\%$ up to $z\sim 10$. The factor $\tilde{f}$ accounts for both self-shielding (reducing the effective incident flux by $f_{\rm selfshield}$) and the contribution from local sources, where $e_{\nu,\,{\rm ion}}^{\rm local}$ is the ionizing band radiation energy density calculated explicitly from the radiation-hydrodynamic treatment in the code (Appendix~\ref{sec:radiative.fb.implementation}), and $e_{\nu,\,{\rm ion}}^{\rm UVB}$ is the meta-galactic UV background (UVB) energy density integrated in the same band for the same wavelength range (H-ionizing frequencies). Note this means we assume the spectral slope of escaping, ionizing radiation from resolved stars in the simulation is the same as the UVB.

\item{\bf Cosmic Ray Heating}: from \citet{guo.oh:cosmic.rays}:
\begin{align}
\Lambda_{\rm CR} =& -1.0\times10^{-16}\,(0.98 + 1.65\,\tilde{n}_{e}\,X_{H})\,e_{\rm CR}\,n_{\rm H}^{-1}
\end{align}
where we assume an approximately uniform MW-like cosmic ray background, $e_{\rm CR}\approx 9.0\times10^{-12}\,f_{\rm CR}$. Here $f_{\rm CR} = n_{\rm H}/(0.01+n_{\rm H})$ when $n_{\rm H}$ exceeds $1000\times$ the mean baryonic density of the Universe, and $f_{\rm CR}=0$ otherwise, to avoid an artificially high CR heating rate in extremely low-density regions (e.g.\ outside galaxies) or at very high redshifts (before star formation).


\item{\bf Photo-Electric Heating}: from \citet{wolfire.2003:neutral.atomic.cooling}:
\begin{align}
\Lambda_{\rm PE} =& -1.3\times10^{-24}\,\tilde{e}_{\nu}^{\rm pe}\,n_{\rm H}^{-1}\,\left(\frac{Z}{Z_{\sun}} \right) \times \\ 
\nonumber & \left( \frac{0.049}{1 + (x_{\rm pe}/1925)^{0.73}} + \frac{0.037\,(T/10^{4})^{0.7}}{1 + (x_{\rm pe}/5000)} \right) \\ 
x_{\rm pe} &\equiv \frac{\tilde{e}_{\nu}^{\rm pe}\,T^{0.5}}{0.5\,\tilde{n}_{e}\,n_{\rm H}} 
\end{align}
where $\tilde{e}_{\nu}^{\rm pe}$ is the photon energy density in the photo-electric band, normalized to the \citet{habing:1968.uv.intensity.unit} MW units, $\tilde{e}_{\nu}^{\rm pe} \equiv e_{\nu}^{\rm pe} / (3.9\times10^{-14}\,{\rm erg\,cm^{-3}})$. The $Z/Z_{\sun}$ term comes from assuming a constant dust-to-metals ratio. Here the field $e_{\nu}^{\rm pe}$ is the FUV band radiation energy density calculated explicitly from the radiation-hydrodynamic treatment in the code, described in Appendix~\ref{sec:radiative.fb.implementation}. 

\item{\bf Magneto-Hydrodynamic Work \&\ Shocks}: from the MHD equations, we obtain some fluid-dynamic change to the temperature (owing to compression/expansion, shocks, etc). We include this self-consistently in the fully-implicit temperature update:
\begin{align}
\Lambda_{\rm MHD} =& -\mu\,\frac{\partial u_{\rm thermal}}{\partial t}{\Bigr|}_{\rm MHD}\,n_{\rm H}^{-1}
\end{align}
where $u_{\rm thermal}$ is the specific internal energy (internal energy per unit mass). 

\item{\bf Optically-Thick Cooling}: lacking a full radiative transfer solution for cooling radiation, we approximate the effects of optically-thick cooling using the method from \citet{rafikov:2007.convect.cooling.grav.instab.planets}, which captures the most important effects by approximating each element as a ``slab'' with column density estimated via the Sobolev approximation and integrating a vertical atmosphere through to its photosphere to determine the net photon escape. This amounts to first summing the contributions above to determine the net heating/cooling rate $\Lambda_{\rm Net}$, and then restricting this to the cooling rate of said slab: 
\begin{align}
|\Lambda_{\rm Net}| <& \Lambda_{\rm BB} \\ 
\Lambda_{\rm BB} & \equiv 5.67\times10^{-5}\,T^{4}\,\left(\frac{\mu}{\Sigma_{\rm eff}} \right)\frac{1}{1 + \kappa_{\rm eff}\,\Sigma_{\rm eff}}\,n_{\rm H}^{-1} 
\end{align}
where $\Sigma_{\rm eff} = \langle \Sigma_{\rm column}^{a,\,{\rm Sobolev}} \rangle_{\phi,\,\theta} = \rho_{a}\,(h_{a} + \rho_{a}/|\nabla \rho_{a} |)$ uses the local Sobolev approximation to estimate the column density to infinity and is defined in  Appendix~\ref{sec:radiative.fb.implementation}, and $\kappa_{\rm eff}$ is the effective opacity.\footnote{We take $\kappa_{\rm eff}$ for dust ($T<1500$) from the detailed tables in \citet{semenov:2003.dust.opacities}, assuming the dust, gas, and radiative temperatures are in equilibrium (true in the optically thick limit at these temperatures), which is approximately well-fit by $\kappa_{\rm eff}=5$ for $150\le T \le 1500$ and $\kappa_{\rm eff}=0.0027\,T^{3/2}$ for $T<150$. At higher temperatures the system is rarely optically thick, but for completeness we compute $\kappa_{\rm eff}$ from the gas-phase using standard approximations for stellar atmospheres: $\kappa_{\rm eff}^{-1} = \kappa_{\rm rad}^{-1} + \kappa_{\rm cond}^{-1}$, with $\kappa_{\rm cond}=2.6\times10^{-7}\,\tilde{n}_{e}\,T^{2}\,\rho^{-2}$, $\kappa_{\rm rad}=\kappa_{\rm mol} + 1/(\kappa_{H^{-}}^{-1} + [\kappa_{e} + \kappa_{\rm Kr}]^{-1})$, $\kappa_{\rm mol}=0.1\,Z$, $\kappa_{e}=0.2\,(1+X_{H})$, $\kappa_{H^{-}}=1.1\times10^{-25}\,(Z\,\rho)^{1/2}\,T^{7.7}$, $\kappa_{\rm Kr}=4.0\times10^{25}\,(1+X_{H})\,Z\,\rho\,^{3.5}$.}

\end{enumerate}

As noted in the text, the actual heating/cooling step is solved fully implicitly for each gas element on its own timestep.

\vspace{-0.5cm}
\section{Algorithmic Implementation of Star Formation}
\label{sec:sf.algorithm}

\begin{enumerate}
\item{\bf Self-Gravitating:} First, following standard sink-particle approaches, we calculate the virial parameter $\alpha$ (ratio of thermal plus kinetic energy to potential energy) in a resolution element and allow only star formation in bound particles with $\alpha < 1$. From \citet{hopkins:virial.sf}, 
\begin{align}
\alpha &\equiv \frac{\| \nabla \otimes {\bf v}\|_{a}^{2} + (c_{s,\,a}/h_{a})^{2}}{8\pi\,G\,\rho_{a}} 
\end{align}
where $\| \nabla \otimes {\bf v} \|_{a}$ the Frobenius norm ($\| {\bf A} \|^{2} \equiv \sum_{\alpha\beta\gamma...}\,A^{2}_{\alpha\beta\gamma...}$) of the velocity gradient tensor ($\otimes$ is the outer product), $c_{s,\,a}$ is the sound speed, $\rho_{a}$ the density, and $h_{a}$ the usual resolution scale (inter-particle spacing). This has the advantage that it converges to an explicitly resolution-independent expression in the super-sonic turbulence limit \citep[see][]{hopkins:virial.sf}.
Note that in {\small GIZMO}, we always use the higher-order accurate, matrix-based gradient estimators described in \citet{hopkins:gizmo}, which remain second-order accurate, consistent, and robust despite arbitrary particle configurations within the stencil \citep[see also][]{maron:2003.gradient.particle.mhd,luo:2008.compressible.flow.galerkin,lanson.vila:2008.meshfree.consistency,mocz:2014.galerkin.arepo,pakmor.2016:improving.arepo.convergence}. Using the zeroth-order inaccurate SPH gradient estimator, in contrast, gives similar results statistically, but makes identification of individual physically collapsing clouds much more noisy.

\item{\bf Self-Shielding:} If $\alpha<1$, we next calculate the shielded/molecular fraction $f_{\rm shielded}^{({\rm sf})}$, which is the fraction of the mass that should be able to self-shield and so cool efficiently (hence fragment to stellar mass scales). The expression for the shielded fraction from \citet{krumholz:2011.molecular.prescription} is: 
\begin{align}
f_{\rm shielded}^{({\rm sf})} &\equiv 1 - \frac{3}{1+4\,\tilde{\psi}_{a}} \\ 
\tilde{\psi}_{a} &\equiv \frac{0.6\,\tilde{\tau}_{a}\,(0.01 + Z_{a}/Z_{\odot})}{\ln{(1+0.6\,\tilde{\phi}_{a}+0.01\,\tilde{\phi}_{a}^{2})}} \\ 
\tilde{\phi}_{a} &\equiv 0.756\,(1+3.1\,Z_{a}/Z_{\sun})^{0.365} \\ 
\tilde{\tau}_{a} &\equiv 434.8\,{\rm cm^{2}\,g^{-1}}\,\rho_{a}\,\left( h_{a} + \frac{\rho_{a}}{|\nabla\rho|_{a}} \right)
\end{align}
We require $f_{\rm shielded}^{({\rm sf})}>0$ for star formation.

\item{\bf Jeans-Unstable:} If $\alpha<1$ and $f_{\rm shielded}^{({\rm sf})}>0$, we calculate the Jeans mass $m_{J}$, and only allow star formation in Jeans-unstable particles, specifically those where $m_{J} < m_{J,\,{\rm crit}} \equiv {\rm MAX}(m_{a},\,10^{3}\,\msun)$, where $m_{a}$ is the particle mass. We calculate the Jeans mass as 
\begin{align}
m_{J} &= 2\,\msun\,\left(\frac{c_{s,\,a}}{0.2\,{\rm km\,s^{-1}}}\right)^{3}\,\left( \frac{n_{a}}{10^{3}\,{\rm cm^{-3}}} \right)^{-1/2}
\end{align}
where $n_{a} \equiv \rho_{a}/\mu_{a}$ is the gas number density.

\item{\bf Sufficiently-Dense:} If $\alpha<1,\,f_{\rm shielded}^{({\rm sf})}>0$, and $m_{J}<m_{J,\,{\rm crit}}$, we check if $n_{a} > n_{\rm crit}$, where $n_{\rm crit}=1000\,{\rm cm^{-3}}$ is a minimum density (and $n_{a} \equiv n_{a,\,H} = X_{H}\,\rho_{a}/m_{p}$), to prevent spurious triggering of the above criteria in low-density gas.
\end{enumerate}

We then assign the gas particle a volume-integrated SFR: 
\begin{align}
\dot{m}_{\ast}^{a} &= \Theta(\alpha_{a},\,f_{{\rm shielded},a},\,n_{a},\,m_{J,\,a})\,\frac{f_{\rm shielded}^{({\rm sf})}\,m_{{\rm gas},\,a}}{t_{{\rm freefall},\,a}} \\ 
t_{{\rm freefall},a} &\equiv \sqrt{\frac{3\pi}{32\,G\,\rho_{a}}} \\ 
\label{eqn:sfr.limiter} \Theta &= 
\begin{cases}
      {\displaystyle  1}\ \ \ \ \ \ \ \hfill{\tiny ( \alpha<1,\ f_{\rm shielded}^{({\rm sf})}>0,\ m_{J}<m_{J,\,{\rm crit}},\ n_{a}>n_{\rm crit} )} \\ 
      {\displaystyle 0}\ \ \ \ \ \hfill {\tiny {\rm otherwise}} \\ 
\end{cases}
\end{align}

Because we wish to maintain equal stellar and gas element masses, at each timestep $\Delta t_{a}$ we assign the gas particle a probability $p_{a} = 1-\exp{(-\dot{m}_{\ast}^{a}\,\Delta t_{a}/m^{a}_{\rm gas})}$ of turning into a star particle that timestep; we draw a uniform random variable $0<x<1$ and if $x<p_{a}$, we convert the gas particle to a star particle. It inherits all relevant properties of its parent particle.

\vspace{-0.5cm}
\section{Algorithmic Implementation of Mechanical Feedback}
\label{sec:mechanical.fb.implementation}

Here we describe our implementation of mechanical feedback, used for SNe (Types Ia \&\ II) and stellar mass-loss. This algorithm was first developed and presented in a series of papers, beginning with \citet{hopkins:fb.ism.prop}, and the version used in FIRE-1 (which contains most of the important features here up to some specific numerical improvements for FIRE-2) was presented in detail in \citet{hopkins:2013.fire}. Similar aspects of that algorithm -- in particular the treatment of SNe momentum accounting for the terminal momentum -- have been recently developed for other codes by \citet{kimm.cen:escape.fraction}, \citet{martizzi:sne.momentum.sims}, and \citet{rosdahl:2016.sne.method.isolated.gal.sims}. 

In \paperone, we discuss each piece of this algorithm in detail, and consider a large suite of idealized test problems and cosmological simulations, to test and validate each and show how it influences our predictions. However for the sake of completeness, we include the full algorithm here.

\begin{enumerate}

\item Every timestep $\Delta t_{a}$, for each star particle $a$ (at position ${\bf x}_{a}$), we first determine whether an ``event'' occurs: a SN Ia, SN II, and/or non-zero stellar mass-loss. This follows the rates and algorithms in Appendix~\ref{sec:stellar.evolution.approximations}. If an event occurs, it has some initial ejecta (or wind) mass $m_{\rm ej}$, metal mass $m_{Z,\,{\rm ej}}$ (defined for each species we track), momentum $p_{\rm ej}=m_{\rm ej}\,v_{\rm ej}$, and energy $E_{\rm ej}$. These also are given in Appendix~\ref{sec:stellar.evolution.approximations} (for winds, $m_{\rm ej} = \Delta t\,\dot{M}_{w}$ from the star, for SNe it is the ejecta mass).

\item Identify gas elements surrounding the star particle: in a grid code this is straightforward, but in our mesh-free method, we define an effective neighbor number $N_{\rm ngb} = (4\pi/3)\,H_{a}^{3}\,\bar{n}_{a}(H_{a})$ in the same manner as for hydrodynamics, where $W$ is the kernel function, $\bar{n}_{a}=\sum W({\bf x}_{ba}\equiv {\bf x}_{b}-{\bf x}_{a},\,H_{a})$, and $H_{a}$ is the search radius. The equation for $N_{\rm ngb}(H_{a})$ is non-linear so is solved iteratively in the neighbor search; see \citet{springel:gadget}.  Thus we obtain all gas elements $b$ within a radius $|{\bf x}_{ba}| < H_{a}$ (where ``$a$ sees $b$''); we {\em also} identify all neighbors with $|{\bf x}_{ba}| < H_{b}$ (i.e.\ ``$b$ sees $a$''). We show in \paperone\ that this is important to ensure the ``effective faces'' close and the resulting distribution of ejecta is isotropic, in regions with highly disordered gas element positions.

\item Assign ``vector weights'' to each neighbor, by first boosting to the rest-frame of the star (${\bf x}_{a}={\bf 0}$, ${\bf v}_{a} \equiv d{\bf x}_{a}/dt = {\bf 0}$; in which the ejecta should be isotropic), then calculating the ``effective face'' that would be seen by the star particle (using the {\em same} definitions of inter-cell faces used in the hydrodynamics) and integrating the ejecta over solid angle through to each face. This amounts to defining the vector weight function $\bar{\bf w}_{ba}$
\begin{align}
\label{eqn:vector.weight.normalized} \bar{\bf w}_{ba} &\equiv \frac{{\bf w}_{ba}}{\sum_{c}\,|{\bf w}_{ca}|} \\ 
\label{eqn:vector.weight.normalized.sub1} {\bf w}_{ba} &\equiv \omega_{ba}\, \sum_{+,\,-}\,\sum_{\alpha}\,(\hat{\bf x}_{ba}^{\pm})^{\alpha}\,\left( f_{\pm}^{\alpha} \right)_{a} \\ 
\label{eqn:vectornorm} \left( f_{\pm}^{\alpha} \right)_{a} &\equiv \left\{ \frac{1}{2}\,\left[1 +  \left( \frac{\sum_{c}\,\omega_{ca}\,|\hat{\bf x}_{ca}^{\mp}|^{\alpha}}{\sum_{c}\,\omega_{ca}\,|\hat{{\bf x}}_{ca}^{\pm}|^{\alpha}} \right)^{2}\right]\right\}^{1/2} \\
\label{eqn:solidangle}\omega_{ba} &= \frac{\Delta\Omega_{ba}}{4\pi} \equiv \frac{1}{2}\,\left(1-\frac{1}{\sqrt{1+({\bf A}_{ba}\cdot \hat{\bf x}_{ba})/(\pi\,|{\bf x}_{ba}|^{2})}}\right)
\end{align}
where ${\bf A}_{ba}$ is the effective vector face between elements $b$ and $a$ used in the finite-volume hydrodynamic calculations,\footnote{For our MFM hydrodynamic method, the face ${\bf A}_{ba}$ is defined as (see \citealt{hopkins:gizmo}):
\begin{align}
\label{eqn:mfm.face.area.def} {\bf A}_{ba} &\equiv  \bar{n}_{a}^{-1}\,\bar{\bf q}_{b}({\bf x}_{a}) + \bar{n}_{b}^{-1}\,\bar{\bf q}_{a}({\bf x}_{b})\\ 
\label{eqn:mfm.face.area.def.sub1} \bar{\bf q}_{b}({\bf x}_{a}) &\equiv {\bf E}_{a}^{-1} \cdot {\bf x}_{ba}\, W({\bf x}_{ba},\,H_{a}) \\
\label{eqn:mfm.face.area.def.sub2} {\bf E}_{a} &\equiv \sum_{c}\,({\bf x}_{ca} \otimes {\bf x}_{ca}) \,W({\bf x}_{ca},\,H_{a}) 
\end{align}
For SPH, the face is defined by the simpler relation ${\bf A}_{ba} = [\bar{n}_{a}^{-2}\,\partial W(|{\bf x}|_{ba},\,H_{a})/\partial |{\bf x}|_{ba} + \bar{n}_{b}^{-2}\partial W(|{\bf x}|_{ba},\,H_{b})/\partial |{\bf x}|_{ba} ]\ \hat{\bf x}_{ba}$. In moving-mesh or fixed-grid finite-volume codes, the face ${\bf A}_{ba}$ is the explicit geometric mesh face between cells.} 
and the $\hat{\bf x}_{ca}^{\pm}$ are the positive or negative (singly-signed) projection vectors: 
\begin{align}
\label{eqn:vector.weight.def} \hat{\bf x}_{ba} &\equiv \frac{{\bf x}_{ba}}{|{\bf x}_{ba}|} = \sum_{+,\,-}\,\hat{\bf x}_{ba}^{\pm} \\ 
\label{eqn:vector.weight.def.sub1} (\hat{\bf x}^{+}_{ba})^{\alpha} &\equiv {|{\bf x}_{ba}|^{-1}}\,{\rm MAX}({\bf x}_{ba}^{\alpha},\,0)\,{\Bigr|}_{\alpha=x,\,y,\,z}\\
\label{eqn:vector.weight.def.sub2} (\hat{\bf x}^{-}_{ba})^{\alpha} &\equiv {|{\bf x}_{ba}|^{-1}}\,{\rm MIN}({\bf x}_{ba}^{\alpha},\,0)\,{\Bigr|}_{\alpha=x,\,y,\,z}
\end{align} 

These expressions are complicated but are derived in detail in \paperone. Their important properties are (1) they maintain {\em manifest} conservation of mass, momentum, and energy (see below). (2) They give fluxes which are statistically isotropic in the rest frame of the star, i.e.\ the ejecta are not systematically biased in one direction or another, even if there is a global density gradient such that there are, on average, more gas elements in one direction. We demonstrate this in numerical tests explicitly in \paperone, and show that many simpler prescriptions lead to systematic, unphysical biases in the ejecta deposition, e.g.\ if there is a thin, dense disk such that more gas neighbors are ``in the disk,'' simple weighting $\bar{\bf w}_{ba}$ proportional to, say, the SPH kernel, leads to almost all the ejecta being coupled in the disk, driving an expanding ring, with almost no ejecta going into the vertical direction -- when in fact the converged solution to this problem is exactly the opposite (hot gas ``vents'' in the vertical direction). (3) They approximate, as closely as possible without an expensive numerical quadrature, the exact integral of the ejecta through and into the ``domains'' of each gas neighbor determined by the hydrodynamic volume partition. 

\item Assign initial fluxes in the rest-frame of the star: 
\begin{align}
\label{eqn:flux.m} \Delta m_{b} &= |\bar{\bf w}_{b}|\,m_{\rm ej} \\ 
\label{eqn:flux.z} \Delta m_{Z,\,b} &= |\bar{\bf w}_{b}|\,m_{Z,\,{\rm ej}} \\ 
\label{eqn:flux.e} \Delta E_{b} &= |\bar{\bf w}_{b}|\,E_{\rm ej} \\ 
\label{eqn:flux.p} \Delta {\bf p}_{b} &= \bar{\bf w}_{b}\,p_{\rm ej}
\end{align}
It is easy to see that our definition of $\bar{\bf w}_{ba}$ guarantees {\em exact} conservation of mass, energy, and linear momentum, and that the correct total radial (outward) momentum is assigned, e.g.:
\begin{align} 
\label{eqn:flux.m.conservation} \sum\,\Delta m_{b} &= m_{\rm ej} \\ 
\label{eqn:flux.z.conservation} \sum\,\Delta m_{Z,\,b} &= m_{Z,\,{\rm ej}} \\ 
\label{eqn:flux.e.conservation} \sum\,\Delta E_{b} &= E_{\rm ej} \\ 
\label{eqn:flux.p.conservation1} \sum\,|\Delta {\bf p}_{b}| &= p_{\rm ej} \\ 
\label{eqn:flux.p.conservation2} \sum\,\Delta {\bf p}_{b} &= {\bf 0} 
\end{align}

\item Boost back to the simulation (``lab'') frame: if the star is moving with velocity ${\bf v}_{a}$, then this boost transforms the momentum and energy fluxes: 
\begin{align}
\label{eqn:flux.p.framecorr} \Delta {\bf p}_{b}^{\prime} &\equiv \Delta {\bf p}_{b} + \Delta m_{b}\,{\bf v}_{a} \\ 
\label{eqn:flux.e.framecorr} \Delta E_{b}^{\prime} &\equiv \Delta {E}_{b} + \frac{1}{2\,\Delta m_{b}}\,\left( |\Delta {\bf p}_{b}^{\prime}|^{2} - |\Delta {\bf p}_{b}|^{2} \right)
\end{align}
(the mass fluxes are unchanged, $\Delta m_{b}^{\prime} = \Delta m_{b}$, $\Delta m_{Z,\,b}^{\prime}=\Delta m_{Z,\,b}$). Of course this maintains manifest conservation: the total momentum added to the neighbors via the $\Delta m_{b}\,{\bf v}_{a}$ term exactly cancels that lost by the star, since its mass decreases by $\sum \Delta m_{b} = m_{\rm ej}$. 

\item Account for $PdV$ (mechanical) work: consider that we have a particle $b$ representing a volume domain with mass $m_{b}$ around our source, with some mean distance in the volume element $|{\bf x}_{ba}|$ (which we call the ``coupling radius''). The ejecta, in order to reach that point, must sweep up the mass $m_{b}$ (in e.g.\ a shock or shell) -- it cannot simply ``spread uniformly'' throughout the volume. This means some $PdV$ work must have been done, converting thermal energy into kinetic energy. Thus the correct momentum ($\Delta {\bf p}_{b}^{\prime\prime}$) to couple into the domain $b$ is {\em not} the initial ejecta momentum $\Delta {\bf p}_{b}^{\prime}$. Rather, if the shock is energy-conserving (neglecting second-order terms in the ratio of particle velocity to ejecta velocity, discussed in \paperone), it is trivial to show that the correct momentum is $\Delta {\bf p}_{b}^{\prime}\,(1 + m_{b}/\Delta m_{b})^{1/2}$. In the early stages of SNe expansion, the shocks are indeed energy-conserving to high accuracy. Of course, at sufficiently long times (or equivalently large radii and/or large entrained masses), the shock becomes radiative, the residual thermal energy is lost, and the shock asymptotically reaches a final ``terminal momentum'' $p_{\rm t}$ (and becomes momentum, rather than energy conserving). Therefore we must impose an upper limit $\Delta {\bf p}_{b}^{\prime}\,p_{\rm t}/p_{\rm ej}$. We therefore have:
\begin{align}
\label{eqn:dp.subgrid.sub1} \Delta {\bf p}_{b}^{\prime\prime} &\equiv \Delta {\bf p}_{b}^{\prime}\ {\rm MIN}\left[ \sqrt{1 + \frac{m_{b}}{\Delta m_{b}} }\  , 
\ \frac{p_{\rm t}}{p_{\rm ej}} \right] \\ 
\label{eqn:terminal.p}\frac{p_{\rm t}}{\msun\,{\rm km\,s^{-1}}} &\approx 4.8\times10^{5}
\left(\frac{E_{\rm tot,\,ej}}{10^{51}\,{\rm erg}}\right)^{\frac{13}{14}}
\left(\frac{n_{b}}{{\rm cm^{-3}}}\right)^{-\frac{1}{7}}
f(Z_{b})^{\frac{3}{2}}\\
\label{eqn:terminal.p.zdep} f(Z) &= 
\begin{cases}
	{\displaystyle 2 \, \ \ \ \ \ \ \ \ \ \ \ \ \ \ \ \ \ \ \ \ \ \ \ \hfill { (Z/Z_{\sun}<0.01)}} \\
	{\displaystyle (Z/Z_{\sun})^{-0.14}\ \ \ \ \ \hfill { (0.01 \le Z/Z_{\sun})}} 
\end{cases}
\end{align}

The expression for $p_{\rm t}$ comes from high-resolution simulations of individual SNe explosions \citep[see e.g.][]{cioffi:1988.sne.remnant.evolution,draine:1991.snr.with.xrays,slavin:snr.expansion,thornton98,martizzi:sne.momentum.sims,walch.naab:sne.momentum,kim.ostriker:sne.momentum.injection.sims,haid:snr.in.clumpy.ism,iffrig:sne.momentum.magnetic.no.effects,hu:photoelectric.heating,li:multi.sne.sims}, in media with different densities and metallicities. We discuss this at length in \paperone, and show that (1) it is the correct expression for a single SN explosion in a homogeneous background, given the {\em same} cooling functions and all other physics implemented in FIRE, (2) it appears to be remarkably robust, across many numerical studies, and (3) our conclusions are robust to variations in the exact value of $p_{\rm t}$ much larger than its actual physical uncertainty. It is easy to verify, given the form of Eq.~\ref{eqn:dp.subgrid.sub1}, that at sufficiently high resolution ($m_{b} \ll 1000\,\msun$), the $p_{\rm t}$ term simply never enters our equations -- in other words, the SNe cooling radii are always resolved. This motivates our SNe explosion resolution criteria in the text. However the design of the expressions here is such that our coupling scheme automatically correctly treats each of e.g.\ the ejecta free-streaming, Sedov-Taylor, and snowplow phases. Note that since $\Delta E$ represents the {\em total} energy, this is not directly modified by changing $\Delta {\bf p}$ (the correct thermal-kinetic breakdown is automatic). 

Also note that Eq.~\ref{eqn:dp.subgrid.sub1} is an approximation if the gas surrounding the star particle is moving at a non-uniform velocity (with non-negligible velocities relative to the ejecta); in this limit the exact expression is given in \paperone\ (Appendix~E).

\item Add final fluxes to the neighboring gas elements, in a fully-conservative manner:
\begin{align}
\label{eqn:flux.m.coupling} m_{b}^{\rm new} &= m_{b} + \Delta m_{b}^{\prime} \\ 
\label{eqn:flux.z.coupling} (Z\,m_{b})^{\rm new} &= Z^{\rm new}\,m_{b}^{\rm new} = (Z\,m_{b}) + \Delta m_{Z,\,b}^{\prime} \\ 
\label{eqn:flux.p.coupling} {\bf p}_{b}^{\rm new} &= m_{b}^{\rm new}\,{\bf v}_{b}^{\rm new} = {\bf p}_{b} + \Delta {\bf p}_{b}^{\prime\prime} \\ 
\label{eqn:flux.e.coupling} E_{b}^{\rm new} &= E_{\rm kinetic}^{\rm new} + U_{\rm internal}^{\rm new} = E_{b} + \Delta E_{b}^{\prime}
\end{align}
So (like our hydrodynamic update), we add conserved quantities ($m$, ${\bf p}$, $E$) and from those update primitive quantities ($Z$, ${\bf v}$, internal energy $U$, etc.). We check that any residual momentum or mass (from e.g.\ round-off error) is re-assigned to the star so conservation is always machine-accurate. 

To be fully consistent with the radiative losses described above (when the cooling radius is un-resolved), we must also modify $U_{\rm internal}^{\rm new}$. Following \citet{thornton98}, the thermal post-shock energy outside $R_{\rm cool}$ decays rapidly as $\propto (r/R_{\rm cool})^{-6.5}$; so we estimate the effective $R_{\rm cool}$ via the requirement that, at the end of the energy-conserving phase, $(1/2)\,(m_{\rm ej} + m_{\rm swept}[R_{\rm cool}])\,v_{f}^{2} = (1/2)\,m_{\rm ej}\,v_{\rm ej}^{2}$ and $p_{\rm t} = m_{\rm swept}[R_{\rm cool}]\,v_{f}$ (where $m_{\rm swept}$ is the enclose mass ``swept up'' by the shell), giving $R_{\rm cool} \approx 28.4\,{\rm pc}\,(n_{b}/{\rm cm^{-3}})^{-3/7}\,(E_{\rm tot,\,ej}/10^{51}\,{\rm erg})^{2/7}\,f(Z_{b})$ for $p_{\rm t}$ in Eq.~\ref{eqn:terminal.p}. If $|{\bf x}_{ba}| < R_{\rm cool}$, we leave $U_{\rm internal}^{\rm new}$ un-modified. If $|{\bf x}_{ba}| > R_{\rm cool}$, we calculate the increase in internal energy from shock-heating, ignoring cooling: $\Delta U_{b} \equiv U_{\rm internal}^{\rm new} - U_{\rm internal}^{\rm old}$, and then modify it to determine the correct internal energy: $U_{\rm internal} = U_{\rm internal}^{\rm old} + \Delta U_{b}\,(|{\bf x}_{ba}|/R_{\rm cool})^{-6.5}$. We show in \paperone\ that this extra step has a negligible effect, since (by definition), when we couple the ejecta to a size/mass scale larger than $R_{\rm cool}$, it will radiate its energy rapidly, so in practice we find that if we simply leave $\Delta U_{b}$ un-modified, the residual energy is (correctly) radiated away in the next timestep. But for the sake of physical consistency and accuracy, we adopt the full expression here.

\end{enumerate}

\vspace{-0.5cm}
\section{Algorithmic Implementation of Radiative Feedback}
\label{sec:radiative.fb.implementation}

Now we describe the implementation of radiative feedback, used for radiation pressure (in all wavebands, UV-through-IR), photo-ionization, and photo-electric heating. The algorithm here was first developed in \citet{hopkins:fb.ism.prop}, and the version used in FIRE-1, which is almost exactly identical to that here,\footnote{The only difference between the radiation algorithms in FIRE-1 and FIRE-2, as noted in the main text, is that in FIRE-2 we allow ionizing photons to propagate outside of the numerical domain boundaries. In FIRE-1, for numerical convenience, their propagation was ``truncated'' at these (large-scale) boundaries. The effects of this are negligible in our simulations here, because almost all the ``work'' done by ionizing photons is on nearby gas.} was presented in \citet{hopkins:2013.fire}. As noted in the text, we for convenience denote the radiative transport algorithm as the ``LEBRON'' (Locally Extincted Background Radiation in Optically-thin Networks) approximation.

We emphasize that this is {\em not} the same as the algorithm used in some earlier work \citep[e.g.][]{hopkins:rad.pressure.sf.fb}. That algorithm was developed for very specific simulations which followed {\em only} infrared multiple-scattering radiation pressure (ignoring single-scattering radiation pressure, photo-heating, SNe, and OB/AGB winds), with much higher resolution than the FIRE simulations here (following star formation down to protostellar cores with densities $\sim 10^{6}\,{\rm cm^{-3}}$ and size scales $<0.1\,$pc). 

In \papertwo, we discuss each piece of the FIRE radiative feedback algorithm in detail, and consider a suite of both idealized test problems and cosmological simulations, to test and validate each and show how they influence our predictions. We also compare to a  set of radiation-hydrodynamics simulations using alternative approximations to the radiation-hydrodynamics equations, specifically the flux-limited diffusion (FLD), optically-thin variable Eddington tensor (OTVET), first-moment (M1), and full Monte Carlo methods. We show that these give similar conclusions, provided care is taken with the alternative methods to ensure the radiation pressure terms are not artificially suppressed. 

The complete algorithm is: 

\begin{enumerate}

\item Determine Background Radiation (Source Luminosities): every timestep $\Delta t_{a}$ for each star particle $a$, we take the luminosity $L_{\nu}^{a}= \Psi_{\nu}^{a}\,m_{\ast,\,a}$ as a function of the star particle age, metallicity, and mass, directly from the stellar evolution models. This follows the tabulation given in Appendix~\ref{sec:stellar.evolution.approximations}, for each of five broad bands we follow: ionizing ($L_{\rm ion}$, $\lambda<912\,$\AA), far-UV ($L_{\rm FUV}$, $912\,$\AA$<\lambda<1550\,$\AA), near-UV ($L_{\rm UV}$, $1550<\lambda<3600\,$\AA), optical/near-IR ($L_{\rm Opt}$, $3600\,$\AA$<\lambda < 3\,\mu$), and mid/far-IR ($L_{\rm IR}$, $\lambda > 3\,\mu$). The FUV band is used for photo-electric heating, while NUV/optical bands dominate the single-scattering radiation pressure (bolometric luminosity), and the IR band is reserved for light re-radiated by dust. 
 
\item Locally Extinct: We now process the absorption/extinction in the vicinity of each source. Along a sightline, the optical depth seen by the source is $\tau_{\nu}^{a} = \kappa_{\nu}\,\Sigma_{\rm column}^{a}$, where $\kappa_{\nu}$ is the flux-weighted opacity in each band (given in Appendix~\ref{sec:stellar.evolution.approximations}; these are calculated for the mean un-obscured spectrum in the stellar populations models, integrated over each band). Since we are interested in {\em local} extinction, we approximate $\Sigma_{\rm column}^{a}$ using the Sobolev approximation for the isotropic (angle-averaged) column density integrated outward from the source: 
\begin{align}
\label{eqn:tau.sobolev} \tau_{\nu}^{a} &= \langle \kappa \rangle_{\nu} \times \langle \Sigma_{\rm column}^{a,\,{\rm Sobolev}} \rangle_{\phi,\,\theta} \\ 
\langle \kappa \rangle_{\nu} &\equiv \frac{\int_{\rm band}\,\kappa_{\nu}\,\langle L_{\nu} \rangle^{\rm unabsorbed}\,d\nu}{\int_{\rm band}\,\langle L_{\nu} \rangle^{\rm unabsorbed}\,d\nu} \\ 
\label{eqn:sigma.sobolev} \langle \Sigma_{\rm column}^{a,\,{\rm Sobolev}} \rangle_{\phi,\,\theta} &\equiv \rho_{a} \left[ h_{a} + \frac{\rho_{a}}{|\nabla \rho_{a} |}  \right]
\end{align}
Here $\rho_{a}$, $\nabla \rho_{a}$, and $h_{a}$ are the gas density, density gradient and inter-element spacing, evaluated at location ${\bf x}_{a}$ (with the same algorithm as our usual hydrodynamics). The $\rho_{a}/|\nabla \rho_{a}|$ term accounts for the gas column integrated to infinity -- it is exact for e.g.\ a density distribution which declines exponentially with distance from the source -- while the $\rho_{a}\,h_{a}$ term is just the column through the local cell. Since this is isotropic, the absorbed luminosity (in a narrow band) is just $L_{{\rm abs},\,\nu}^{a} = (1 - \exp{(-\tau_{\nu}^{a})})\,L_{\nu}^{a}$ and the surviving un-absorbed luminosity is:
\begin{align}
\label{eqn:lum.emergent} L_{{\rm emergent},\,\nu}^{a} &= \exp{\left(-\tau_{\nu}^{a}\right)}\,L_{\nu}^{a}
\end{align}
Recall our spectral templates include negligible ``initial'' luminosity in the mid/far IR band: but we assume the luminosity absorbed by {\em dust} -- that from the FUV, NUV, and optical/near-IR bands -- is immediately re-radiated in the mid/far-IR band, giving an emergent IR luminosity:
\begin{align}
L_{\rm IR}^{a} &= \sum_{\nu = {\rm FUV,\,UV,\,Opt}}\,L_{{\rm abs},\,\nu}^{a} = \sum_{\nu = {\rm FUV,\,UV,\,Opt}}\,(1 - \exp{(-\tau_{\nu}^{a})})\,L_{\nu}^{a}
\end{align}

For the ionizing band, the opacity comes from neutral hydrogen, and we must jointly solve for the ionizing state and photon absorption; we therefore treat this separately using a simple Stromgren approximation. From Step (i) above (the stellar evolution models) we have $\dot{N}_{a}^{\rm ion}$, the rate of production of ionizing photons ($\propto L_{\rm ion}$). Now, we take all gas elements $b$ in the vicinity of $a$ and sort them by increasing distance $|{\bf x}_{ba}|$. Beginning with the closest, we test whether it is already ionized (either because $T_{b} > 10^{4}\,$K, or because it is already tagged as a member of another HII region), and if so we move on to the next-closest particle. If it is not ionized, we calculate the ionization rate needed to fully ionize it as: $\Delta \dot{N}_{b} =  N(H)_{b}\,\beta\,n_{e,\,b}$ (where $N(H)_{b}=X_{H}\,m_{b}/\mu\,m_{p}$ is the number of H atoms in $b$, $\beta\approx3\times10^{-13}\,{\rm cm^{3}\,s^{-1}}$ is the recombination coefficient, and $n_{e,\,b}$ is the electron density assuming full ionization). If $\Delta \dot{N}_{b} \le  \dot{N}_{a}^{\rm ion}$, then particle $b$ is tagged as being within an HII region, and the photons are ``consumed,'' so $\dot{N}_{a}^{\rm ion}\rightarrow \dot{N}_{a}^{\rm ion} - \Delta \dot{N}_{b}$. We then proceed to the next particle and repeat. If we reach a particle which is not ionized but for which $\Delta \dot{N}_{b} >  \dot{N}_{a}^{\rm ion}$, we determine whether or not to ionize it randomly, with probability $p=  \dot{N}_{a}^{\rm ion} / \Delta \dot{N}_{b}$, and consume the remaining photons (guaranteeing the correct total mass is ionized, on average). Any particle tagged as ``within an HII region'' is fully ionized and not allowed to cool to temperatures lower than $<10^{4}$\,K within that same timestep. If we reach the end of the local computational domain, or distance from the source where the optically-thin flux density falls below the meta-galactic ionizing background, we stop the iteration and the remaining photons are ``emergent'' as in Eq.~\ref{eqn:lum.emergent}. Tests of this algorithm (both static but also dynamic tests of D-Type ionization front expansion) are also shown in \citet{hu:2017.rad.fb.model.photoelectric}. 

\item Account for Momentum of Locally-Absorbed Photons: Over a timestep $\Delta t$, the absorbed photons in these bands impart their single-scattering momentum to the surrounding gas, radially directed away from the star particle, with total momentum 
\begin{align}
\Delta p = \frac{L_{\rm abs}^{a}}{c}\,\Delta t = \frac{\Delta t}{c}\,\sum_{\nu}\,L_{{\rm abs},\,\nu}^{a}
\end{align}
This momentum flux is distributed among the neighbors, directed radially away from the star particle, as described in Appendix~\ref{sec:mechanical.fb.implementation}. 

\item Transport the Locally-Extincted Radiation via an Optically-thin Network: 
We now have an ``emergent'' spectrum {\em after local attenuation} around each star, $L_{{\rm emergent},\,\nu}$. Since we assume the absorption is dominated by the gas/dust local to the star, and the emission (from the star) is isotropic, the incident flux ${\bf F}_{\nu}^{b}$ and photon energy density $e_{\nu}^{b}$ at a distant gas element $b$ are just
\begin{align}
\label{eqn:fluxes.fire} {\bf F}_{\nu}^{b} &=  \sum_{a}\,{\bf F}_{\nu,\,{\rm emergent}}^{a} = \sum_{a}\,\frac{L_{{\rm emergent},\,\nu}^{a}}{4\pi\,|{\bf x}_{b} - {\bf x}_{a}|^{2}}\,\frac{{\bf x}_{b} - {\bf x}_{a}}{|{\bf x}_{b} - {\bf x}_{a}|} \\ 
\label{eqn:longrange.optically.thin.photon.energy.density} 
e_{\nu}^{b} &= \sum_{a}\,\frac{L_{{\rm emergent},\,\nu}^{a}}{4\pi\,c\,|{\bf x}_{b} - {\bf x}_{a}|^{2}}
\end{align}
This is identical in form to the equation for gravity, so is computed in the same pass in the gravity tree (we ``soften'' the sources identical to how we soften gravity, in fact, to prevent a $1/r^{2}$ divergence and reflect the physical fact that each star particle really represents many stars distributed within the softening length).\footnote{The softening kernel, following \citet{hopkins:gizmo}, is given by replacing Eqs.~\ref{eqn:fluxes.fire}-\ref{eqn:longrange.optically.thin.photon.energy.density} with ${\bf F}_{\nu,\,{\rm emergent}} = \sum_{a}\,(1/4\pi)\,L_{{\rm emergent},\,\nu}^{a}\,({\bf x}_{b}-{\bf x}_{a})\,H_{a}^{-3}\,\mathcal{F}_{s}(u_{ba})$ and $\epsilon_{\nu}^{b} = \sum_{a}\,(1/4\pi\,c)\,L_{{\rm emergent},\,\nu}^{a}\,|{\bf x}_{b}-{\bf x}_{a}|\,H_{a}^{-3}\,\mathcal{F}_{s}(u_{ba})$ where $u_{ba} \equiv |{\bf x}_{b} - {\bf x}_{a}| / H_{a}$, $H_{a}= (24/\pi)^{1/3}\,h_{a} = (24/\pi)^{1/3}\,\epsilon_{a}$ is the maximum kernel search radius, and 
\begin{align}
\mathcal{F}_{s}(u) &\equiv 
\begin{cases}
	{\displaystyle \frac{32}{15}\,\left[5 + 3\,u^{2}\,(5\,u-6) \right]\ \ \ \ \ \ \ \  \hfill { \left( u \le \frac{1}{2} \right)}} \\
	\\
	{\displaystyle \frac{32}{15}\,\left[10 - \frac{45}{2}\,u + 18\,u^{2} - 5\,u^{3} - \frac{1}{32\,u^{3}} \right]\ \ \ \hfill { \left( \frac{1}{2} < u < 1 \right)}} \\
	\\
	{\displaystyle \frac{1}{u^{3}}\ \ \ \ \ \ \ \  \hfill { \left( u \ge 1 \right)}}
\end{cases}
\end{align}
This becomes exactly inverse-square at $r>H$, and but prevents a $1/r^{2}$ divergence as $r\rightarrow 0$ (with flux $\sim 1/h_{a}^{2}$ instead of $1/r^{2}$). 
}

\item Calculate incident radiative acceleration from long-range fluxes: 
For a gas element $b$ with effective face area $A_{b}$ and mass $m_{b}$ (hence surface density $\Sigma_{b}\equiv m_{b}/A_{b}$) seeing an incident flux ${\bf F}_{\nu}^{b}$, the exact radiative acceleration is given by 
\begin{align}
\label{eqn:rad.accel} \frac{d{\bf v}_{b}}{dt}{\Bigr|}_{\nu} &= \frac{1}{m_{b}}\,\frac{d{\bf p}_{b}}{dt}{\Bigr|}_{\nu}= \frac{{\bf F}_{\nu}^{b}}{c}\,\frac{A_{b}}{m_{b}}\,\left[ 1 - \exp{\left(-\kappa_{\nu}\,\frac{m_{b}}{A_{b}}\right)} \right]
\end{align}
In the optically thin limit ($\kappa_{\nu}\,\Sigma_{b} \ll 1$), this reduces to the common expression ${\bf a} = \kappa_{\nu}\,{\bf F}_{\nu}/c$, but in the optically thick limit ($\kappa_{\nu}\,\Sigma_{b} \gg 1$), the force saturates at $m_{b}\,{\bf a} = ({\bf F}_{\nu}\,A_{b})/ c$, i.e.\ the element absorbs all the flux across its subtended area (but no more).
For simplicity here we take the effective area to be that of a sphere with the same volume as the element ($=m_{b}/\rho_{b} = (4\pi/3)\,h_{b}^{3}$), i.e.\ $\pi\,h_{b}^{2}$; using the more complicated hydrodynamic face areas introduces negligible ($\ll 10\%$) differences in the accelerations here. We adopt $\kappa_{\nu} = \langle \kappa \rangle_{\nu}$ for each band.

\item Self-shield and pass incident fluxes to cooling routines: Because we have accounted for shielding around the emitter, but not the absorber, we include an additional shielding pass at absorption for the photo-heating terms: at a gas element $b$, we take the photon energy density $\langle e^{b}_{\nu} \rangle = e^{b}_{\nu}\,\exp{(-\tau_{\nu}^{b})}$, where $e^{b}_{\nu}$ is the photon energy density given by Eq.~\ref{eqn:longrange.optically.thin.photon.energy.density}, and  $\tau_{\nu}^{b}$ is the optical depth estimated using the Sobolev approximation in Eq.~\ref{eqn:tau.sobolev}, but now at the location of the absorbing {\em gas} element (instead of around the emitting star). The resulting, shielded $e_{\nu}^{b}$ are passed to the cooling/heating routines, to compute photo-ionization and photo-electric heating rates as in Appendix~\ref{sec:cooling.approximations}. 


\end{enumerate}

\vspace{-0.5cm}
\section{Additional Fluid Physics: Magnetic Fields, Conduction, Viscosity, Turbulent Diffusion}
\label{sec:additional.physics.methods}

Here, we describe the numerical implementations of additional physics {\em not} included in the ``core physics only'' FIRE simulations, but studied either here or in companion papers \citep[e.g.][]{su:2016.weak.mhd.cond.visc.turbdiff.fx} which take standard FIRE-2 simulations and add, e.g.\ magnetic fields. We emphasize again that these physics are not used in the ``default'' or ``core physics'' runs in this paper. However, because we wish to present a complete, thorough, and fully-consistent set of numerical methods, we summarize them here, referring to the appropriate methods papers for more details.

\vspace{-0.5cm}
\subsection{Magnetic Fields}
\label{sec:mhd}

In simulations with magnetic fields, we solve the equations of ideal magnetohydrodynamics (MHD) as implemented in {\small GIZMO} in \citet{hopkins:mhd.gizmo}. The exact numerical formulation of the equations is presented there along with an extensive series of several dozen test problems, as well as tests of full galaxy simulations using our FIRE physics. The tests demonstrate that the implementation in our MFM solver is accurate and converges at second order. In particular, \citet{hopkins:mhd.gizmo},  \citet{hopkins:cg.mhd.gizmo}, and \citet{hopkins.2016:dust.gas.molecular.cloud.dynamics.sims} show that our implementation correctly captures traditionally difficult phenomena such as the magneto-rotational instability (MRI), magnetic jet launching in disks, magnetic fluid-mixing instabilities, and sub-sonic and super-sonic MHD turbulent dynamos. The accuracy and convergence order appears comparable to state-of-the-art grid codes (e.g.\ {\small ATHENA}) on the problems of interest and greatly superior to the P-SPH implementation in {\small GIZMO}, especially in problems where angular momentum, super-sonic advection, strong shocks, and fluid mixing instabilities appear (typical of cosmological simulations). Non-ideal MHD effects (Ohmic resistivity, ambipolar diffusion, and the Hall effect) are also implemented and well-tested in {\small GIZMO} \citep{hopkins:gizmo.diffusion}, but these are not expected to be important on galactic scales.

Readers interested in more details of our numerical implementation of MHD should consult \citet{hopkins:mhd.gizmo} and the public {\small GIZMO} source code.

\vspace{-0.5cm}
\subsection{Anisotropic Spitzer-Braginskii Viscosity and Conduction}
\label{sec:conduction.viscosity}

The implementation of anisotropic Spitzer-Braginskii viscosity and conduction in {\small GIZMO} is described and tested in \citet{hopkins:gizmo.diffusion}. In addition to the usual MHD fluxes, this adds an anisotropic viscous stress-energy tensor $\boldsymbol{\Pi}$ to the momentum flux (${\bf F}_{\bf p}=\boldsymbol{\Pi}$) and energy flux (${\bf F}_{e}=\boldsymbol{\Pi}\cdot {\bf v}$), and a conductive energy flux ${\bf F}_{e} = {\bf K}\cdot \nabla T$. The appropriate anisotropic tensor expressions for MHD are given by \citep{spitzer:conductivity,braginskii:viscosity}: 
\begin{align}
{\bf K} &\equiv \kappa_{\rm cond}\,\hat{B}\otimes \hat{B} \\
\kappa_{\rm cond} &= \frac{0.96\,f_{i}\,(k_{B}T)^{5/2}\,k_{B}}{m_{e}^{1/2}e^{4}\ln{\Lambda}}\,\left({1+4.2\,\ell_{e}/\ell_{T}}\right)^{-1} \\
\boldsymbol{\Pi} &\equiv 3\,\nu_{\rm visc}\,\left( \hat{B}\otimes\hat{B} - \frac{1}{3}{\bf I}\right)\,\left[ \left( \hat{B}\otimes\hat{B} - \frac{1}{3}{\bf I} \right) : \left( \nabla\otimes {\bf v} \right) \right] \\ 
\nu_{\rm visc} &= \frac{0.406\,f_{i}\,m_{i}^{1/2}\,(k_{B}T)^{5/2}}{(Z_{i}\,e)^{4}\ln{\Lambda}}\,\left({1+4.2\,\ell_{i}/\ell_{v}}\right)^{-1} 
\end{align}
where $\otimes$ denotes the outer product; $\hat{B}$ is the direction of the magnetic field vector; ${\bf I}$ is the identity matrix; ${\bf v}$ the velocity; ``$:$'' denotes the double-dot-product (${\bf A}:{\bf B} \equiv {\rm Trace}({\bf A}\cdot{\bf B})$); $\ln{\Lambda}\approx 37.8$ is the Coulomb logarithm \citep{sarazin:coulomb.log}; $m_{e}$, $e$, $m_{i}$, $Z_{i}\,e$ are the electron mass and charge and ion mass and charge; $f_{i}$ the ionized fraction (calculated self-consistently in our cooling routines); $k_{B}$ the Boltzmann constant; $\ell_{e}$ ($\ell_{i}$) is the electron (ion) mean-free path, and $\ell_{T} = T/|\nabla T|$ ($\ell_{v} = |{\bf v}|/||\nabla \otimes {\bf v}||$) is the temperature (velocity) gradient scale length (this term correctly accounts for saturation of $\kappa_{\rm cond}$ or $\nu_{\rm visc}$ when electrons/ions have long mean free paths, by not allowing the gradient scale length to be shorter than $\ell_{e,\,i}$). In these equations, $\kappa_{\rm cond}$ is the conductivity, and $\nu_{\rm visc}$ the viscosity. Additional details of the coefficients, and a study of their effects, are in \citet{su:2016.weak.mhd.cond.visc.turbdiff.fx}. In \citet{hopkins:gizmo.diffusion}, we show that the numerical implementation of these fluxes is accurate, able to handle arbitrarily large anisotropies, converges comparably to higher-order fixed-grid codes, and is able to correctly capture complicated non-linear instabilities sourced by anisotropic diffusion such as the magneto-thermal and heat-flux bouyancy instabilities.

Readers interested in more details of our numerical implementation of anisotropic diffusion should consult \citet{hopkins:gizmo.diffusion} and the public {\small GIZMO} source code.

\vspace{-0.5cm}
\subsection{Sub-Grid Turbulent Eddy Diffusivity}
\label{sec:turbulent.diffusion}

In some models for turbulence (e.g.\ mixing-length theory), the effects of unresolved (small-scale) eddies and microphysical processes transporting passive scalars (such as metals) are treated as diffusive processes. The implementation and tests of generic diffusion operators in {\small GIZMO} are presented in \citet{hopkins:gizmo.diffusion}; the solver is an explicit finite-volume scheme which converges at second-order accuracy (comparable to higher-order grid codes) and manifestly conserves metal mass. For a passive scalar, the transport equation is: $\partial (\rho\,Z)/\partial t = \nabla\cdot(\kappa_{\rm turb}\,\rho\, \nabla Z)$, where $Z$ is the abundance per unit mass of the scalar (i.e.\ the metallicity) and the ``eddy diffusivity'' $\kappa_{\rm turb}\sim \lambda_{\rm eddy}\,v_{\rm eddy}$ is the product of the scale length and rms velocity of the largest un-resolved eddies (those at the resolution scale), which dominate the transport on unresolved scales (larger eddies are, of course, resolved). In other words, one simply assumes that the diffusion or mixing time at scale $\lambda_{\rm eddy}$ scales with the eddy turnover time. Following the common \citet{smagorinsky.1963:eddy.approximation.for.diffusion.terms} approximation, we can approximate the ``eddy diffusivity'' as 
\begin{align}
\label{eqn:turb.diffusivity} \kappa_{\rm turb}^{a} &\equiv \sqrt{2}\,C^{2}\,\|{\bf S}_{a}\|\,h_{a}^{2}
\end{align}
where $C\sim 0.05-0.15$ is a constant calibrated to numerical simulations, motivated by a Kolmogorov cascade in  \citet{smagorinsky.1963:eddy.approximation.for.diffusion.terms}, $h_{a}$ is the grid scale (for our MFM method, this is equal to the rms inter-element separation), and ${\bf S}_{a}\equiv [(\nabla \otimes {\bf v})_{a} + (\nabla \otimes {\bf v})_{a}^{T}] - {\rm Trace}(\nabla \otimes {\bf v})_{a}/3$ is the symmetric shear tensor (and $\| {\bf S}\|$ denotes the Frobenius norm). Note that we use our higher-order matrix-based gradient formalism from \citet{hopkins:gizmo} to calculate ${\bf S}_{a}$; this is much more accurate and less noisy compared to common SPH or pure ``face-based'' mesh gradient estimators, which is important to reduce artificial numerical diffusivity \citep[see e.g.][]{maron:2003.gradient.particle.mhd,luo:2008.compressible.flow.galerkin,lanson.vila:2008.meshfree.consistency,mocz:2014.galerkin.arepo,pakmor.2016:improving.arepo.convergence}. 

The key assumption here -- namely, the assumption that the diffusion timescale scales with eddy turnover time -- has been verified in many experiments on ISM scales \citep[see][and references therein]{pan:2010.turbulent.mixing.times,petit:metal.mixing.turbulence.galactic.disks}, and the scaling in Eq.~\ref{eqn:turb.diffusivity} has been used in many applications in galaxy simulations \citep[e.g.][]{wadsley:2008.sph.mixing.cosmology}. In \citet{colbrook:passive.scalar.scalings}, we have performed our own study of the turbulent mixing, using 3D, high-resolution supersonic turbulent box simulations (with and without magnetic fields and/or shear), and verify that this prescription, with $C\approx0.05$, is reasonable specifically in our identical MFM code with the definitions of $h$ and ${\bf S}$ here (although such simple prescriptions do fail to capture some potentially important non-Gaussian features which emerge from real, resolved turbulent mixing). An independent, more extensive study (including a range of more complex problem setups) will be presented in Rennehan et al.\ (in prep.), but also finds $C\approx0.03-0.05$. We therefore adopt $C=0.05$. However, in the main text (\S~\ref{sec:turbulent.diffusion.tests}) and in \citet{escala:turbulent.metal.diffusion.fire}, we show that order-of-magnitude variation in $C$ produces no significant effects on our predictions.

We stress that a term like Eq.~\ref{eqn:turb.diffusivity} is ``built into'' many numerical hydrodynamic methods. Specifically, it is well known that in finite-volume methods with advective mass fluxes (e.g.\ traditional grid-based methods or moving-mesh codes), an intrinsic numerical diffusivity in advection with magnitude $\sim h_{a}\,\Delta v(h_{a})$ appears; this automatically produces scalar/metal diffusion via ``numerical mixing.'' It is straightforward to show that if the necessary assumptions of the Smagorinsky model (Eq.~\ref{eqn:turb.diffusivity}) are true, then the artificial numerical mixing in these methods is always {\em larger} than the ``true'' effective turbulent diffusivity. If we used such a method, it would therefore not be necessary to explicitly solve Eq.~\ref{eqn:turb.diffusivity}. However, in our default MFM hydrodynamic method, we follow fixed-mass elements (i.e.\ there are no advective mass fluxes, hence no artificial ``numerical diffusivity'' of passive scalars). While the methods will converge to an identical solution at sufficiently high resolution \citep[][]{hopkins:gizmo}, the concern is that at fixed resolution, MFM will under-estimate the metal-mixing owing to un-resolved small-scale eddies that should mix between the boundaries of neighboring resolution elements. If we resolved individual stars, the stars would draw mass from many resolution elements (each with their own abundances) and this would still not be a problem, but at our resolution single star particles inherit the abundances of their (single) parent gas particle, so this effect can artificially introduce ``shot noise'' in the abundances of stars forming from neighboring gas elements if we do not include an explicit numerical mixing term. One therefore can view Eq.~\ref{eqn:turb.diffusivity} as a purely numerical term which ``restores'' the desirable aspect of the numerical diffusivity present in certain numerical methods. 

We caution, however, that simple diffusion prescriptions such as Eq.~\ref{eqn:turb.diffusivity}, naively applied, can substantially over-estimate the diffusivity. The critical assumption is that the resolution-scale motion $\Delta v \sim \|{\bf S}\|\,h \sim v_{\rm eddy}$ is entirely due to turbulence; if there is any bulk motion included in ${\bf S}$, this will over-estimate $\kappa_{\rm turb}$. This can be particularly problematic if e.g.\ differential rotation in a disk or shear in CGM outflows is poorly-resolved, in which case the naively inferred $\kappa_{\rm turb}$ can over-estimate by an order of magnitude the true turbulent motion. For example, if the disk scale height $H_{\rm disk}\approx \sigma_{\rm turb}/\Omega$ is unresolved, $h_{a} \gg H_{\rm disk}$, then Eq.~\ref{eqn:turb.diffusivity} will return $\kappa_{\rm turb} \sim h_{a}\,(h_{a}\,\Omega)$ instead of the correct maximum diffusivity for disk-scale eddies, $\kappa_{\rm turb} \sim H_{\rm disk}\,\sigma_{\rm turb} \sim H_{\rm disk}^{2}\,\Omega$. There is no obvious universal ``switch'' to cure these pathologies; however we can limit the magnitude of the errors. In \citet{hopkins:gizmo} we develop two mesh-free finite-element hydrodynamic methods, our default MFM method here, and a ``meshless finite-volume'' (MFV) method, the latter of which includes advective mass fluxes (more similar to a moving-mesh code), but otherwise is identical to MFM. The MFV method therefore includes the inherent numerical diffusivity described above. In the diffusion step, therefore, we can first calculate the absolute value of the metal flux that would have been calculated in MFV (owing simply to the advection term; see \citealt{hopkins:gizmo} for the exact values of these terms), and then impose this as an upper limit to the diffusive flux. Since MFV is a second-order, quasi-Lagrangian method, this eliminates the most egregious errors in Eq.~\ref{eqn:turb.diffusivity}. We find that in idealized test problems, this correction is negligible, but in realistic cosmological simulations it prevents the most severe pathological situations. Essentially, then, our implementation of ``unresolved turbulent diffusion'' (Eq.~\ref{eqn:turb.diffusivity}) is guaranteed to -- {\em at most} -- produce the same metal mixing we would have obtained had we simply run our simulations using a finite-volume (MFV or moving-mesh) hydrodynamic method. A more detailed study of various (more sophisticated) turbulent and numerical mixing models will be presented in Rennehan et al.\ (in prep.); preliminary results indicate that the alternative methods give similar results in galaxy-scale simulations.

Readers interested in more details of our numerical implementation of eddy diffusivity should consult \citet{hopkins:gizmo.diffusion} and the public {\small GIZMO} source code.

\vspace{-0.5cm}
\section{Computational Scaling \&\ Runtime Requirements}
\label{sec:scaling.details}
\begin{figure*}
\centering
\includegraphics[width = 0.33 \textwidth]{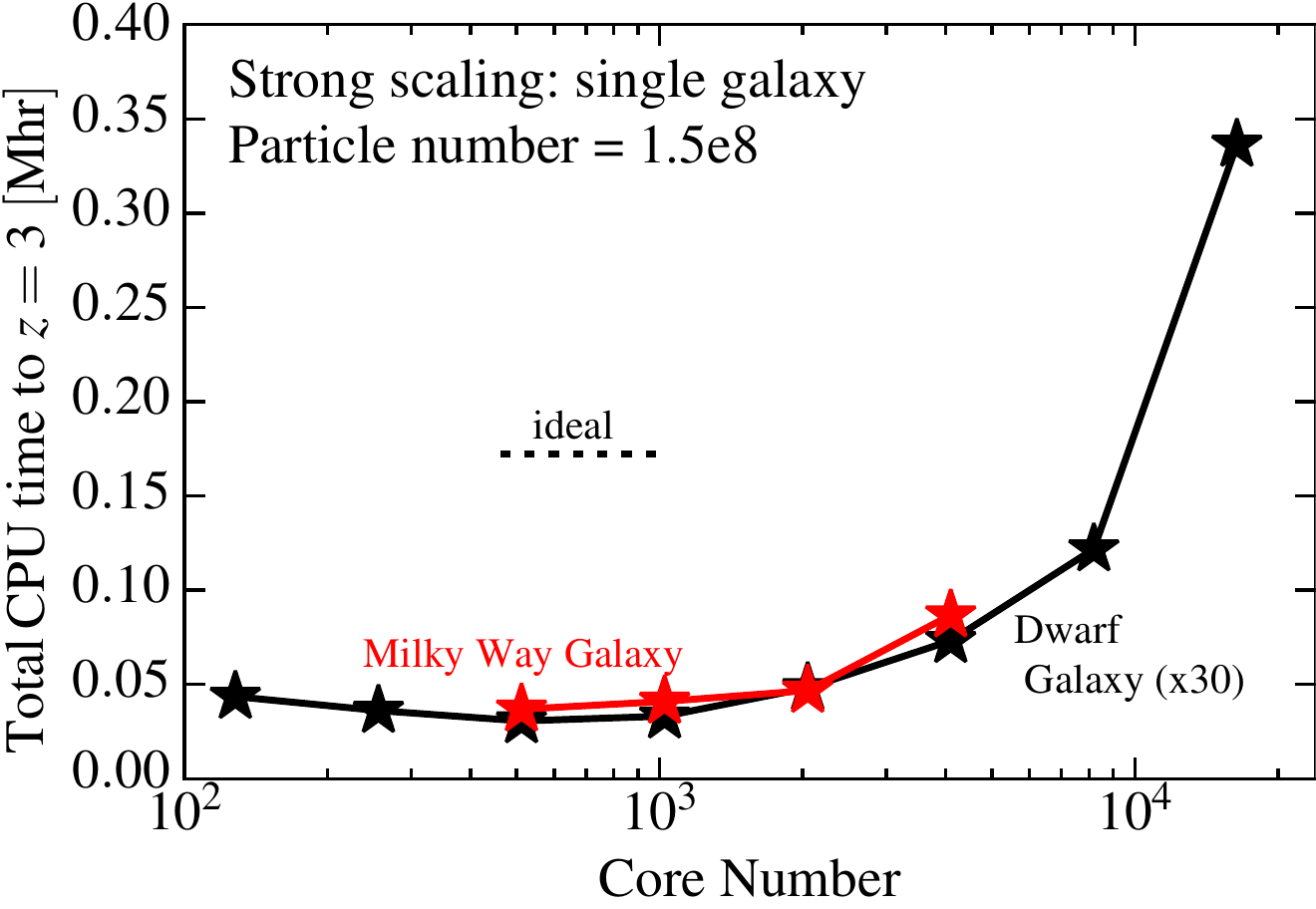} 
\includegraphics[width = 0.33 \textwidth]{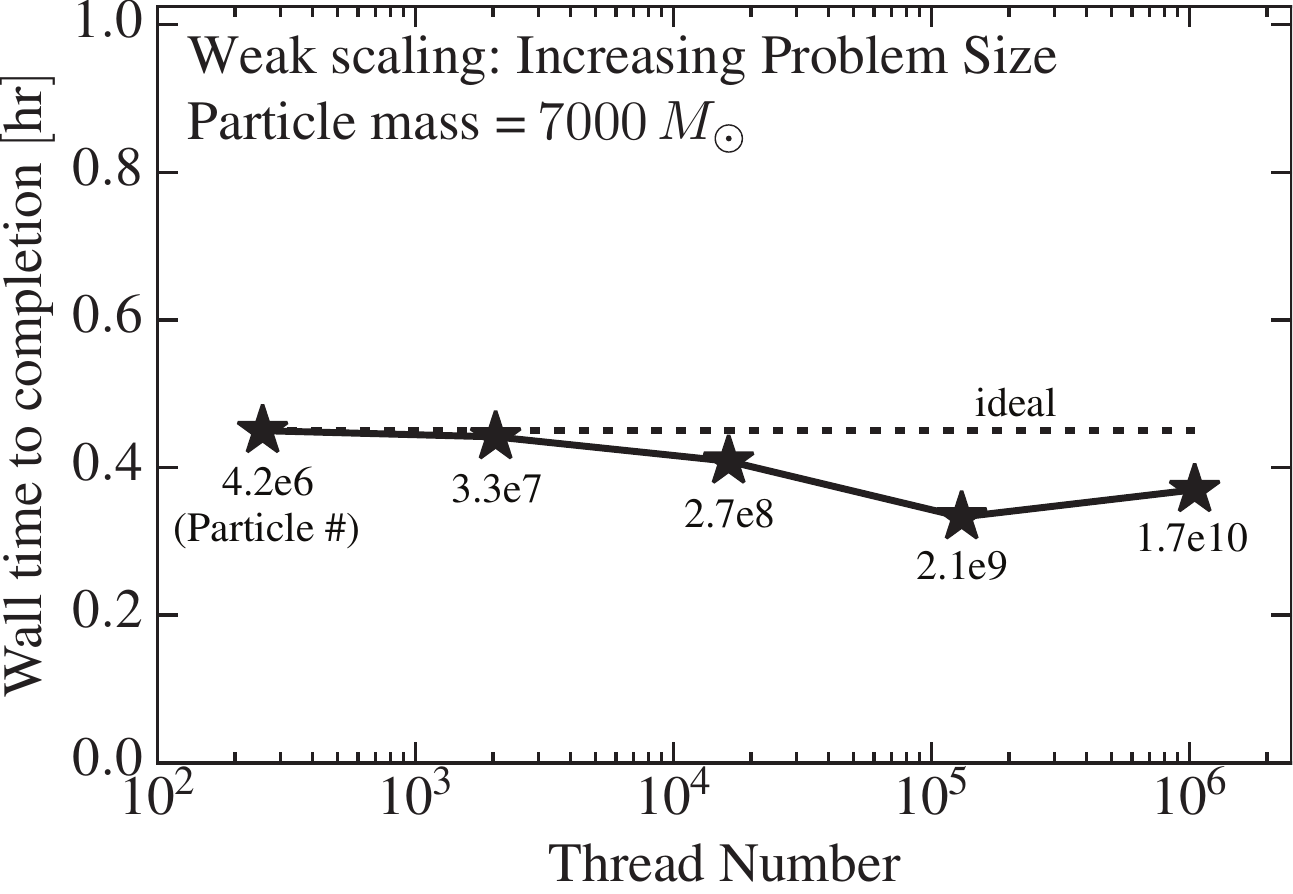} 
\includegraphics[width = 0.33 \textwidth]{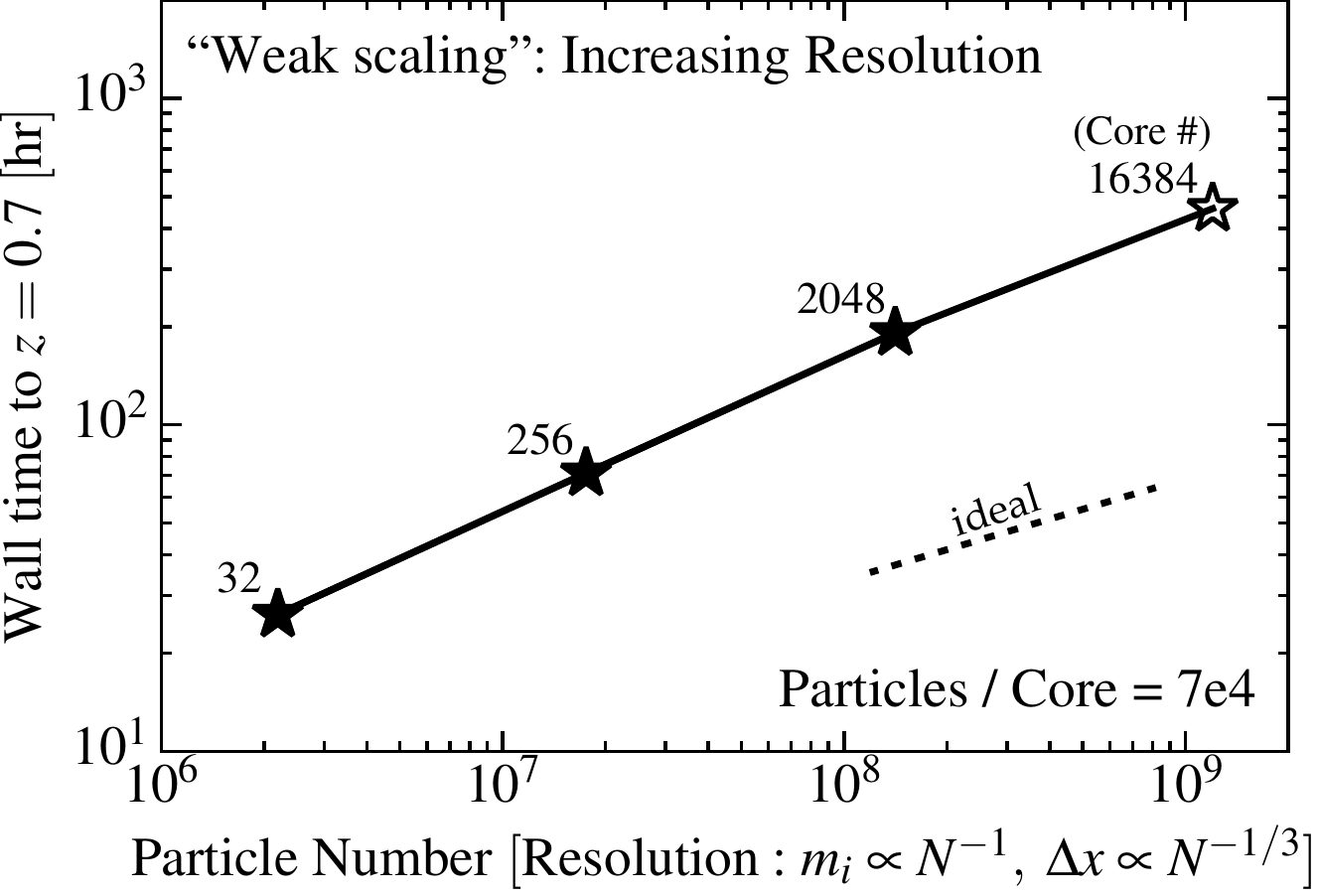}
\caption{
Code scalings of {\small GIZMO} in full production-quality FIRE-2 simulations, at our production resolution, with the full, identical physics of gravity, hydrodynamics, cooling, star formation, and feedback to our published simulations. 
{\em Left:} Strong scaling for a zoom-in of a MW-mass ({\bf m12i}) or dwarf galaxy ({\bf m10q}) halo, each using $1.5\times10^{8}$ particles, run to $25\%$ of the age of the Universe, using the optimal MPI+OpenMP hybrid configuration at each core number. Our  optimizations allow us to maintain near-ideal strong scaling to $\sim 14,000$ cores per billion particles ($2048$ for the problem shown).
{\em Center:} Weak scaling, for a full cosmological box, populated with the same high-resolution particles, run for a short fraction of the age of the Universe ($z\sim10$).
Here, we keep resolution fixed at baryonic particle mass $7000 \msun$, identical to our high-resolution MW simulation at {\em left}, but we increase the cosmological volume from 2 to $10^{4}$ comoving ${\rm Mpc}^{3}$.
The weak scaling of GIZMO's gravity+MHD algorithm is near-ideal (actually slightly {\em better} at intermediate volume, owing to fixed overheads and increasing statistical homogeneity of the box at larger sizes), to greater than a million threads (here $2^{20}$ threads, $2^{18}$ cores, $2^{17}$ MPI tasks, $2^{14}$ nodes). 
{\em Right:} ``Weak scaling'' test, increasing the resolution instead of the problem size (specifically, increasing the particle number for the same MW-mass galaxy). Because the resolution increases (hence timestep decreases) with particle number here, the ideal weak scaling for a converged solution is wall-clock time $\propto t_{\rm Hubble} / \Delta t \propto (\Delta x / c_{s})^{-1} \propto N^{1/3}$, shown. 
Our achieved scaling is only slightly worse than this, because new, dense structures such as star clusters appear at higher resolution.
\label{fig:scaling}
}
\vspace{-0.5cm}
\end{figure*}

The work here was made possible by extensive optimizations and improvements to the code scaling for ``zoom-in'' simulations. The challenge in these high-dynamic range problems is that small, dense regions (e.g.\ dense GMCs or star clusters), which occupy an extremely small fraction of the total mass and volume of the simulation (and therefore cannot be ``broken up'' over too many processors) require extremely small timesteps. But, especially with strong feedback present, the rest of the simulation (which is free to take much larger timesteps) cannot advance until they ``catch up.'' We have addressed this with several improvements.

\begin{enumerate}

\item{We have made optimizations to the feedback sub-routines (for example, long-range radiation forces) which require significant neighbor communication but can be efficiently included in other operations such as tree construction, reducing their portion of the runtime from $\approx 40\%$ in previously-published work to $<10\%$ in the current code. More generally, the entire {\small GIZMO} code has been line-by-line optimized, manually unrolling or (where possible) vectorizing certain expensive operations, replacing expensive functions in neighbor loops with look-up tables, eliminating redundant operations, and pre-computing additional quantities outside of loops; this has produced an additional factor $\sim 2$ speed improvement.}

\item{We have optimized the structure of the domain decomposition to make it more spatially flexible and separately parallelize each level of the timestep hierarchy (increasing the memory imbalances by factors of $\sim2-3$, but extending the strong scaling to $\sim2$ times as many cores at fixed resolution). We have also more aggressively implemented a problem-specific particle weighting scheme, where e.g.\ dense, star-forming gas, and young stars (as opposed to old stars) are given larger weights in the domain decomposition so that their future cost (via gravitational collapse and/or feedback) is more accurately predicted. This allows for a further factor of $\sim 2-3$ reduction in load imbalances at the smallest timesteps.}

\item{We employ a hybrid tree-particle mesh gravity solver, following {\small GADGET-3}, to efficiently reduce the cost of the gravity solution for the low-resolution regions outside of the zoom-in area.}

\item{We use the adaptive individual-timestep integration scheme from {\small GADGET-3} with a hierarchical power-of-two subdivision, updated such that in each timestep we calculate pairwise updates of all fluxes of conserved quantities at interfaces (maintaining exact conservation, and eliminating all redundant pair-wise interations; see \citealt{springel:arepo}).}

\item{We adopt adaptive gravitational softenings as described above. This imposes essentially no cost for gas (since $h_{i}$ must be computed already for hydrodynamics), but allows us to take much larger timesteps for low-density particles, and (more important for this problem) avoid over-softening particles in dense regions (where a too-large softening radius might encompass thousands of neighbor particles, which imposes a substantial cost in the tree-gravity calculation).}

\item{We have developed and use a hybrid OpenMP-MPI parallelization of the code, which allows us to extend the weak scaling of the code considerably further as we go to large processor number (where we previously found communication costs between neighbors, which are alleviated by the shared-memory structure of OpenMP, were beginning to dominate the run time).}

\item{Our MFM hydrodynamic solver is actually slightly faster than the P-SPH algorithm used for FIRE-1. \citet{hopkins:gizmo} compare run-times on three problems: a 3D Kelvin-Helmholtz instability, an isolated (non-cosmological) galaxy with star formation and feedback, and the cosmological but strictly ideal-gas (no star formation or cooling) Santa Barbara cluster test. They show that compared to the P-SPH formulation of SPH from \citet{hopkins:lagrangian.pressure.sph} (which incorporated improvements in artificial diffusion terms as well as a larger neighbor number needed in SPH to capture certain instabilities), the speedup on these tests ranged from a factor $\sim 1.3 - 2.5$, mostly owing to the larger neighbor number needed in P-SPH to achieve comparable accuracy.}

\item{We also stress that the inclusion of realistic feedback itself greatly speeds up the code (for galaxy formation simulations) -- perhaps more than any purely numerical optimization. Dense regions which would otherwise slow down the computation (as described above) tend to be quickly destroyed by stellar feedback. Without feedback, it would be impossible to run the simulations here with the same resolution and cooling physics below redshift $z\sim 2$, because the extremely dense relic star clusters would require constant, extremely short timesteps.}

\end{enumerate}

Readers interested in more details should consult the public source code.

Figure~\ref{fig:scaling} demonstrates the scaling of the code {\small GIZMO} on a production quality set of FIRE-2 simulations, including all the physics of our production runs (full cosmological integration with self-gravity, baryonic physics including cooling, star formation, and stellar feedback, etc.). This is a ``real world'' test, as opposed to the scaling on idealized test problems (which can, of course, be much better). All runs were run with an otherwise identical version of the code; at each CPU number the optimal OpenMP-MPI configuration was used. Strong and weak-resolution scaling tests were run on the XSEDE Stampede machine, weak-problem size runs on the DOE ALCF Mira machine.

Our optimizations allow us to extend good strong scaling, at our modest ``typical FIRE-1'' resolution, to $\sim 1024$ cores. Even more strikingly, the optimizations we have made allow us to maintain good weak scaling up to $\sim 4096$ cores for a simulation with $3\times10^{8}$ particles -- our ``Latte'' resolution -- and $\sim 16,384$ cores for a zoom-in simulation with $>10^{9}$ particles (and as many as $\sim10^{6}$ cores for large-volume simulations with $\sim 10^{10}$ particles, which naturally exhibit superior weak scaling). This is especially non-trivial for these sorts of problems, since the gravitational softening is adaptive, so higher particle number implies smaller force softening and hence smaller timesteps in the dense regions. 

For comparison, if we compare similar runs to the strong-scaling tests in Fig.~\ref{fig:scaling} with the public {\small GADGET-2} code (using the simpler \citet{springel:models} sub-grid model for ISM physics and feedback, with wind mass-loading set by-hand to produce a similar mass as our {\small GIZMO} runs), we find the scaling saturates at $\sim 64-128$ cores. 

\vspace{-0.5cm}\section{Congratulations!}\label{sec:congratulations} \begin{figure}\plotonesize{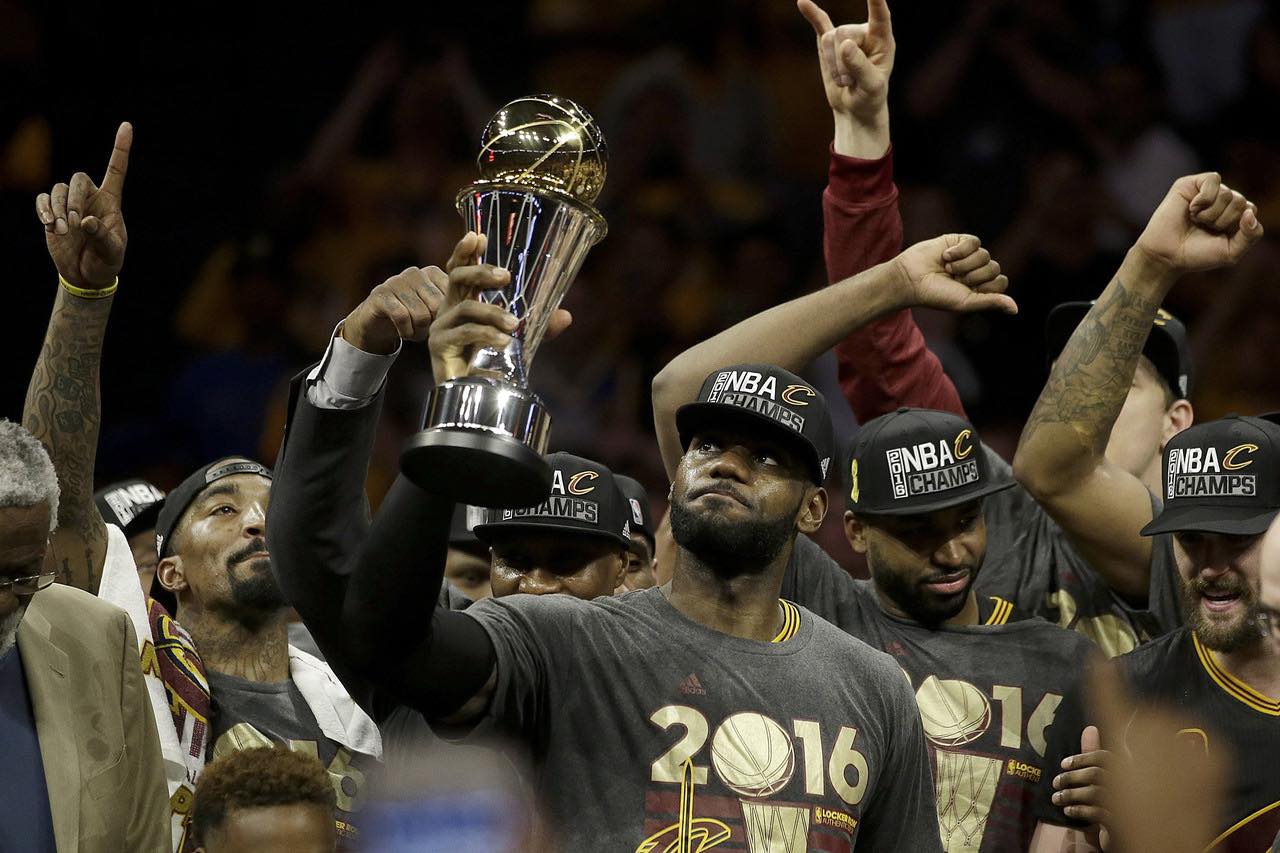}{1.0}\vspace{-0.1cm}\caption{Believeland.\label{fig:cleveland}}\vspace{-0.5cm}\end{figure} You made it to the end of this paper. Fig.~\ref{fig:cleveland} is your reward. Go Cleveland!

\end{appendix}

\end{document}